\documentclass[journal,twocolumn]{IEEEtran}
\usepackage{graphicx, amsmath, amsfonts, epsfig}
\usepackage{psfrag}
\usepackage{dsfont}
\usepackage{color}
\usepackage{cancel}
\usepackage{subcaption}
\usepackage{multirow}
\usepackage{nicefrac}
\usepackage{manfnt}
\usepackage{amssymb,eucal}
\usepackage{exscale}
\usepackage{verbatim}
\usepackage{amstext}
\usepackage{latexsym}
\usepackage{ifthen}
\usepackage{fancybox}
\usepackage{cancel}
\usepackage{cite}
\usepackage{algorithm}
\usepackage{algorithmicx}
\usepackage{algpseudocode}
\usepackage{dblfloatfix}

\usepackage[outline]{contour}
\contourlength{3pt}

\usepackage{tikz}
\usepackage{pgfplotstable}
\usepackage{rotate}

\definecolor{rubblue}{cmyk}{1,0.5,0,0.6}
\definecolor{rubgreen}{cmyk}{0.5,0,1,0}
\definecolor{rubgray}{cmyk}{0.03,0.03,0.03,0.1}

\usepgfplotslibrary{units}

\usetikzlibrary{%
patterns,%
calc,%
fit,%
arrows,%
plotmarks,%
shadows,%
chains,%
shapes%
}

\tikzstyle{smalldot} = [draw, circle, minimum size=0.1pt,scale=0.2,fill=black,black]

\newtheorem{theorem}{Theorem}
\newtheorem{lemma}{Lemma}

\newtheorem{remark}{Remark}
\newtheorem{example}{Example}
\newtheorem{corollary}{Corollary}

\newtheorem{definition}{Definition}
\newcommand{\z}[0]{\ensuremath{\boldsymbol{z}}}
\newcommand{\x}[0]{\ensuremath{\boldsymbol{x}}}
\newcommand{\y}[0]{\ensuremath{\boldsymbol{y}}}
\newcommand{\X}[0]{\ensuremath{\boldsymbol{X}}}
\newcommand{\Y}[0]{\ensuremath{\boldsymbol{Y}}}
\newcommand{\Z}[0]{\ensuremath{\boldsymbol{Z}}}
\newcommand{\U}[0]{\ensuremath{\boldsymbol{U}}}

\newcommand{\nt}[0]{\ensuremath{n_{\rm t}}}
\newcommand{\nr}[0]{\ensuremath{n_{\rm r}}}

\newcommand{\G}[0]{\ensuremath{\mathbb{G}}}
\renewcommand{\U}[0]{\ensuremath{\mathbb{U}}}

\algdef{SE}[DOWHILE]{Do}{doWhile}{\algorithmicdo}[1]{\algorithmicwhile\ #1}%

\newcommand{\ac}[1]{\textcolor{black}{#1}}
\newcommand{\acc}[1]{\textcolor{black}{#1}}

\begin{document}
\IEEEoverridecommandlockouts
\title{On the Capacity of Intensity-Modulation Direct-Detection Gaussian Optical Wireless Communication Channels: A Tutorial}
\author{
\IEEEauthorblockN{Anas Chaaban, {\it Senior Member, IEEE}, Zouheir Rezki, {\it Senior Member, IEEE}, \\and Mohamed-Slim Alouini, {\it Fellow, IEEE}}\\
\IEEEauthorblockA{}
\thanks{% 
A. Chaaban is with the School of Engineering, University of British Columbia, Kelowna, BC  Canada V1V 1V7. Email: anas.chaaban@ubc.ca. 

Z. Rezki is with the Electrical and Computer Engineering Department, University of California Santa Cruz, Santa Cruz, CA, USA. Email: zrezki@ucsc.edu.

M.-S. Alouini is with the Division of Computer, Electrical, and Mathematical Sciences and Engineering (CEMSE) at King Abdullah University of Science and Technology (KAUST), Thuwal 23955-6900, Saudi Arabia. Email: slim.alouini@kaust.edu.sa.

This material is based upon work supported by the National Science Foundation (CAREER) under Grant No. 2114779.
}
}

\maketitle

\begin{abstract}
Optical wireless communication (OWC) using intensity-modulation and direct-detection (IM/DD) has a channel model which possesses unique features, due to the constraints imposed on the channel input. The aim of this tutorial is to overview results on the capacity of IM/DD channels with input-independent Gaussian noise as a model of OWC channels. It provides the reader with an entry point to the topic, and highlights some major contributions in this area. It begins with a discussion on channel models and how this IM/DD Gaussian channel model comes about, in addition to an explanation of input constraints. Then, it discusses the capacity of the single-input single-output channel, its computation, and capacity bounds and asymptotic capacity results. Then, it extends the discussion to the multiple-input multiple-output setup, and reviews capacity bounds for this channel model. Finally, it discusses multi-user channels modelled as a broadcast channel (downlink) or a multiple-access channel (uplink), with their associated capacity bounds. 
\end{abstract}

\begin{IEEEkeywords}
Optical wireless; intensity modulation; channel models; Gaussian channel; input-independent; capacity; capacity bounds; multi-user channels.
\end{IEEEkeywords}

\section{Introduction}
As our daily lives become ever more dependent on data-connectivity, the load on wireless networks continues to grow. Future networks are expected to have a great increase in machine-to-machine communications and smartphone traffic, and it is expected that wireless and mobile traffic will constitute 71\% of the total IP traffic by 2022 \cite{CISCO_2017_2022}. Consequently, wireless networks' capabilities have to continuously improve in order to cope with the mounting pressure. To realize this goal, there is continuous need for more bandwidth, which has triggered the research on millimetre waves \cite{RappaportEtAlMMWave} and terrahertz communications \cite{AkyildizJornetHan} recently, and has also revived the interest in optical wireless communications (OWC). 

OWC has a vast unlicensed bandwidth spanning around $3$PHz. It can be used to create point-to-point links using lasers in what is known as free-space optics (FSO), and to create point-to-multipoint links using LEDs. FSO links can be used for front-haul and back-haul links between base-stations or for front-haul links in cloud radio-access networks for instance. The advantage compared to optical-fibre links is that FSO is less demanding in terms of deployment and infrastructure. LEDs can be used to realize visible-light communications (VLC) \cite{VLC:BeyondP2P} for LiFi (Light-Fidelity) access-points \cite{HaasYinWangChen,ElgalaMeslehHaas}, which is useful for indoor (house and office lights) and outdoor (street lights), combining illumination and communication for increased energy efficiency. This is in addition to other applications such as non-line-of-sight links using ultraviolet light \ac{for civilian and military applications requiring enhanced security} \cite{Drost_2014,Wang:17,Sun:18,XuSadler}, underwater communications \ac{for subsea monitoring applications} \cite{Oubei_2018,KedarArnon}, car-to-car \cite{LuoGhassemlooyLeMinh}, and on-chip communications \cite{SikderKodiKennedy}. All these application \ac{benefit from the wide bandwidth and \acc{license-free} nature of the optical spectrum,} and have been the topic of investigation over the last decade. See \cite{KaushalKaddoum,KhalighiUysal,KarunatilakaZafarKalavallyParthiban,
PathakFenHuMohapatra,BorahBoucouvalasDavis,DemersYanikomerogluHilaire} for excellent surveys on the topic.

While OWC can be realized using coherent \ac{communication techniques} where one can modulate \ac{and detect} the amplitude and phase of the optical carrier \ac{(heterodyne detection)} \cite{DeLange}, a more favoured operation mode is incoherent OWC using intensity-modulation and direct-detection (IM/DD) due to its simplicity and low-cost \cite{KahnBarry}. In IM/DD, the light intensity is modulated as an information bearing signal, and information is recovered at the receiver side by detecting the intensity of received light. As a consequence of this operation, the modulating signal (current) is real-valued and positive. This is a fundamental distinguishing factor from radio-frequency (RF) coherent communications, where the modulated signal is complex-valued. Moreover, in IM/DD, the modulated signal may be peak-constrained and/or average constrained due to operational, safety, and illumination considerations \cite{Kartalopoulos_FSO}. Several models exist for IM/DD OWC including the Poisson channel \cite{Wyner_Poisson}, the square-root Gaussian channel \cite{TsiatmasWillemsBaggen,TsiatmasWillemsBaggen_SCVT}, the Gaussian channel with input-dependent noise \cite{Moser,Safari_DependentNoise,LeitingerGeigerWitrisal,YuanShaLiangJiangWangZhao}, and the Gaussian channel with input-independent noise \cite[Ch. 7]{AdvancedOpticalWireless}.

Due to this fundamental difference, the performance of \ac{IM/DD OWC} is evaluated using different techniques than \ac{coherent} RF communications. \ac{We focus on performance in terms of \acc{capacity} in this tutorial, i.e., the highest rate of information transmission under which the error rate can be made vanishingly small by increasing the code length \cite{CoverThomas}.} Capacity analysis of \ac{the aforementioned channel models for IM/DD OWC} can be found in \cite{Wyner_Poisson,Wyner_Poisson_II,Moser,LapidothMoserWigger,
FaridHranilovic_SelectedAreas,FaridHranilovic,WangHuWang_Dimmable,
WangHuWang_ICC,ChanHranilovicKschischang,AdvancedOpticalWireless}. In this tutorial, we focus on the Gaussian channel with input-independent noise \ac{and with a real-valued, nonnegative, peak- and average-constrained input. This channel} is suitable for modelling \ac{IM/DD} OWC with strong ambient light and/or thermal noise \cite{KahnBarry}. \ac{For brevity, we call this channel an {\it IM/DD Gaussian channel}, where we use `IM/DD' to emphasize the input constraints (real-valued, nonnegative, peak- and average- constrained) and to discern this channel from the popular Gaussian channel used for modelling coherent RF communications, and we use `Gaussian' to emphasize the noise characteristics of the channel.} 

\ac{The IM/DD Gaussian channel has been the focus of many studies lately due to its applicability in the areas of VLC and FSO. In particular, an IM/DD Gaussian channel can be used to model static OWC channels, and is a building block in modelling time-varying OWC channels (due to turbulence and pointing errors e.g.) \cite{FaridHranilovic_OutageMIMO,IT_Limits_FSO_OWC_An_Emerging_Tech}. For instance, an FSO channel can be modelled as a Gaussian channel during each coherence interval. This Gaussian channel model has been used to study many aspects of OWC including} the performance of single-hop and multi-hop FSO systems was studied in 
\cite{AkellaYukselKalyanaraman,KazemlouHranilovicKumar,ZediniAnsariAlouini,
GarciaZambranaCastilloVazquez,KashaniUysal,HuangMehrpoorSafari} and the performance of various modulation schemes for VLC systems was studied in~
\cite{GaoWangXuHua,KazemiHaas,ChenHaas,KizilirmakRowellUysal,
BawazirSofotasiosMuhaidat,DangZhang,LeeJungKwon,GaoHuGongXu,ZuoZhangZhangChen}.

While the capacity of the \ac{IM/DD Gaussian} channel is still unknown in closed-form, existing results show properties of the capacity achieving distribution (discreteness) \cite{ChanHranilovicKschischang} in addition to capacity bounds and asymptotics\cite{HranilovicKschischang,LapidothMoserWigger,FaridHranilovic,ChaabanMorvanAlouini}. In \cite{LapidothMoserWigger}, capacity lower bounds were derived using Exponential and truncated Exponential input distibutions, and capacity upper bounds were derived using the dual-capacity expression studied in \cite{Moser_Dissertation}. In \cite{FaridHranilovic,FaridHranilovic_SelectedAreas}, capacity lower bounds were derived using Geometric and truncated Geometric input distributions, and capacity upper bounds were derived using sphere-packing and the Steiner-Minkowski formula for polytopes \cite{BergerGeometryII,Morvan,SteinerMorvan}. In \cite{ChaabanMorvanAlouini}, capacity lower bounds were derived using truncated Gaussian input distributions, and capacity upper bounds were derived using a new sphere-packing approach. Further bounds were given in \cite{WangHuWang_Dimmable,WangHuWang_ICC}. 

The advantage of these bounds is that they enable a better understanding of the performance limits of IM/DD systems beyond schemes which are commonly used in the literature. Such schemes include on-off keying (OOK) and binary pulse-position modulation (PPM) \cite{GarciaZambranaCastilloVazquez,FaridHranilovic_Outage,AkellaYukselKalyanaraman,
KazemlouHranilovicKumar,KazemiMostaaniUysalGhassemlooy,KashaniUysal,HsiehShiu,
NuwanpriyaZhangGrant}, pulse-amplitude modulation (PAM) and higher-order constellations \cite{NuwanpriyaHoZhang,KaroutAgrellSzczerbaKarlsson,DrostSadler}, various types of unipolar orthogonal frequency-division multiplexing (OFDM) schemes \cite{DissanayakeArmstrong,LiVucicJungnickelArmstrong,ArmstrongLowery,YouKahn,
FernandoHongViterbo,ElgalaLittle,LamWilsonElgalaLittle,MazahirChaaban,
ZhouZhang,YuBaxleyZhou,DimitrovHaas,WuBar-Ness,YangSunGao}, and PAM discrete multi-tone (PAM-DMT) 
\cite{LeeRandelBreyerKoonen}. The performance of these schemes has been extensively studied in the literature from difference perspective. For instance, performance in terms of error and outage probability has been studied in \cite{FaridHranilovic_Outage,FaridHranilovic_OutageMIMO,LetzepisFabregas,
KhalighiSchwartzAitamerBourennane,KashaniUysal,AkellaYukselKalyanaraman,
KazemlouHranilovicKumar,MonteiroHranilovic,TsiftsisSandalidisKaragiannidisUysal,
SapenovChaabanRezki_WCL}. Space-time block code (STBC) \cite{Alamouti,TarokhJafarkhaniCalderbank} designs for IM/DD Gassian channels was studied in \cite{BayakiSchober,ZhangYuZhangZhuWang,ZhangYuZhangZhu,SafariUysal}, and in \cite{ChaabanSapenovRezkiAlouini} which shows that space-only coding is quasi-optimal within the class of DC-offset STBC. While all are practical schemes, they generally fall short of achieving the channel capacity due to limitations in their construction (cf. \cite{ChaabanHranilovic} for an example on OFDM schemes). 

This tutorial serves to shed light on existing bounds on the capacity of \ac{IM/DD Gaussian channels modelling OWC through the following steps:}
\begin{itemize}
\item \ac{Describing various IM/DD channel models in detail;}
\item \ac{Defining capacity and describing how it is evaluated numerically;}
\item \ac{Reviewing capacity bounds and asymptotics for the single-input single-output (SISO) IM/DD Gaussian channel;}
\item \ac{Reviewing capacity bounds and asymptotics for the multiple-input multiple-output (MIMO) IM/DD Gaussian channel; and}
\item \ac{Reviewing capacity bounds for the IM/DD Gaussian broadcast channel and the multiple-access channel.}
\end{itemize}
The tutorial starts with the single user channel for which the channel model is first discussed, and then channel capacity bounds are reviewed for a single-input single-output (SISO) system. Then, it covers multiple-input multiple-output (MIMO) system. A MIMO \ac{IM/DD OWC system} can be realized by using an array of LEDs and detectors \cite{YangKhalighiVirieuxBourennaneGhassemlooy,ZengObrienFaulkner,FathHaas}, or using multiple LED colors (e.g. color-shift keying or wave-division multiplexing) \cite{GaoWangXuHua,MonteiroHranilovic,CiaramellaArimotoContestabile,
WangNirmalathasLimSkafidas}. \ac{Transmission schemes for MIMO IM/DD OWC} systems have been studied in \cite{ButalaElgalaLittle,KazemiMostaaniUysalGhassemlooy,
FaridHranilovic_OutageMIMO,TsiftsisSandalidisKaragiannidisUysal,
RiedigerSchoberLampe,GaoGongXu}. In general, MIMO schemes have benefits in terms of error and outage probability compared to their SISO counterpart \cite{KhalighiSchwartzAitamerBourennane,LetzepisFabregas,Safari,BushuevArnon,
YangKhalighiVirieuxBourennaneGhassemlooy,BasarPanayirciUysalHaas,XuYuZhuCai,
ButalaElgalaLittle,ZengObrienFaulkner,ParkKoAlouini,ArarYongacoglu}. 
\ac{We call the channel that models a MIMO IM/DD OWC system a MIMO IM/DD Gaussian channel.} \ac{The capacity of this channel does not coincide with that of the standard MIMO Gaussian channel used to model multi-antenna coherent RF communications whose capacity is well-known \cite{TseViswanath}. Thus, the capacity of the MIMO IM/DD Gaussian channel deserves special attention.} The capacity of the related MIMO Poisson channel was studied in \cite{HaasShapiro_MIMOPoissonCap,HaasShapiro}. The capacity of the MIMO IM/DD Gaussian channel with no crosstalk, i.e., parallel IM/DD Gaussian channels, was studied in \cite{ChaabanRezkiAlouini_ParallelOWC_TCOM,ChaabanRezkiAlouini_ParallelOWC_ICC,ChaabanRezkiAlouini_LowSNR_CommL}, relying on capacity bounds for the SISO channel in addition to intensity allocation algorithms. The capacity of the MIMO IM/DD Gaussian channel with crosstalk was studied in \cite{ChaabanRezkiAlouini_MIMOOWC_ICC,MoserMylonakisWangWigger,ChaabanRezkiAlouini_MIMOOWC_TWC,
ChaabanRezkiAlouini_MIMOOWC_ICC,ChaabanRezkiAlouini_MIMOOWC_LowSNR_ISIT,ChaabanRezkiAlouini_MIMO_IMDD_Low_SNR_TCOM,
LiMoserWangWigger,MoserWangWigger_IT,MoserWangWigger}. Some capacity bounds and asymptotic capacity results for MIMO IM/DD Gaussian channels in these papers are reviewed in this tutorial. 

Then, the tutorial discusses multi-user IM/DD Gaussian channels, in particular, broadcast channels (BC) and multiple-access channels (MAC). The BC and MAC model scenarios when an OWC access point communicates with multiple users and vice versa, respectively.  Such scenarios have been studied from error rate and achievable data rate perspectives under various transmission schemes in \cite{KahnBarryAudeh,TsonevVidevHaas,MarshoudDawoudKapinas,ElmirghaniCryan,GhaffariMatinfarSalehi,GrinerArnon,
MarshoudKapinasKaragiannidisMuhaidat,PhamPham,LianBrandtPearce,
AbdelhadyAminChaabanAlouini,AbdelhadyAminChaabanShihadaAlouini_Pareto,
PathakFenHuMohapatra,KashefAbdallahQaraqe,MaLampeHranilovic}. While several works study the performance of orthogonal codes \ac{in multi-user IM/DD OWC} \cite{ElmirghaniCryan,GhaffariMatinfarSalehi,DangZhang}, such codes are generally suboptimal in terms of capacity. To assess the performance of such codes and other schemes compared to capacity, one needs to derive or bound the capacity of multi-user IM/DD Gaussian channels. The study of BC and MAC dates back to the 70's with the seminal works in \cite{Cover,Ahlswede, Liao}. Works on the IM/DD BC and MAC in the literature aim to derive capacity bounds and asymptotic capacity expressions which are specific to the IM/DD channel. The capacity of the Poisson BC and MAC have been studied in \cite{LapidothMoser_Poisson,LapidothShapiro_Poisson,KimNachmanElGamal_Journal,
LapidothShamaiPoissonMAC}. The capacity of the SISO IM/DD Gaussian BC and MAC has been studied in \cite{ChaabanRezkiAlouini_TWC,ChaabanIbraheemyNaffouriAlouini_TWC,
SoltaniRezkiChaaban}. This tutorial overviews results on the capacity of the IM/DD Gaussian BC and MAC.

Note that in addition to the BC and MAC, the capacity of other multi-terminal IM/DD \ac{channels} has been studied in the literature. This includes the IM/DD wiretap channel which has been studied in \cite{MostafaLampeTSP,MostafaLampeJSAC,SoltaniRezki,ArfaouiZaidRezki,
ZhangChaabanLampe,WangLiuWangaWuLinCheng}, and the IM/DD interference channel which has been studied in \cite{ZhangChaaban}, for instance. 

The rest of the tutorial is organized as follows. Sec. \ref{Model} discusses the IM/DD channel models. Then Sec. \ref{Sec:Constraints} discusses constraints of IM/DD channel inputs. Sec. \ref{Sec:SISO_Cap}, \ref{Sec:NumCapacity}, and \ref{Sec:CapacityBounds} discuss the capacity of the IM/DD SISO Gaussian channel, its evaluation, and capacity bounds, respectively. Sec. \ref{Sec:MIMO} discusses parallel and MIMO IM/DD Gaussian channels, and Sec. \ref{Sec:MultiUser} discusses the IM/DD Gaussian BC and MAC, \ac{in addition to a brief overview of some works on other multi-terminal IM/DD channels such as the interference channel and the wiretap channel}. Finally, Sec. \ref{Sec:Summary} summarizes the paper. To assist the reader, a summary of the paper notation is given in Table \ref{Tab:Notation}.

\begin{table}
\centering
\caption{A summary of paper notation.}
\label{Tab:Notation}
\begin{tabular}{c||c}
\hline
%$\lambda$					& 		Wavelength\\\hline
%${\sf e}_{\lambda}$			& 		Photon energy at wavelength $\lambda$, equal to $\frac{{\sf hc}}{\lambda}$  (${\sf h}$: Planck constant; ${\sf c}$: Speed of light)\\\hline
%${\sf c}$ 						&		Speed of light\\\hline
%${\sf h}$							& 		The Plack constant\\\hline
%$n_{\rm t}$, $r_{\rm t}$	& 		Transmitted photon number, rate (resp.)\\\hline
%$n_{\rm r}$, $r_{\rm r}$	& 		Received photon number, rate (resp.)\\\hline
%$p_{\rm t}$, $p_{\rm r}$	&		Transmit power, received power (resp.)\\\hline
%$d_{\rm p}$					& 		Propagation distance from transmitter to receiver\\\hline
%$a_{\rm d}$					& 		Detector area\\\hline
%$\theta_{\rm e}$			&		Emission solid angle\\\hline
%$t_{\rm s}$					& 		Symbol duration\\\hline
$\mathbb{R}, \mathbb{R}_+, \mathbb{C}$		&		Real, nonnegative real, and complex sets\\\hline
\ac{$\mathbb{N}, \mathbb{N}_+$}		&		\ac{Integer and nonnegative integer sets}\\\hline
$\mathbb{P}\{\cdot\}$		&		Probability of an event\\\hline
$X\sim \mathbb{P}_X$		&		Random variable $X$ follows the distribution $\mathbb{P}_X$\\\hline
i.i.d.						& 		Independent and identically distributed\\\hline
Bern$(\alpha)$				&		Bernoulli distribution with parameter $\alpha$\\\hline
$\mathcal{N}(\alpha,\beta)$	&	Gaussian distribution with mean $\alpha$ and variance $\beta$\\\hline
$\mathcal{CN}(\alpha,\beta)$	&	Circularly symmetric complex Gaussian\\
& distribution with mean $\alpha$ and variance $\beta$\\\hline
$Q(x)$ 						&	Standard Gaussian tail function\\\hline 
$\log(x)$					&		Natural logarithm of $x$\\\hline
$\mathcal{F}\{\cdot\}$		&		Fourier transform\\\hline
$H(X)$,	$H(X|Y)$			& 		Discrete entropy and  conditional entropy (resp.) \\\hline
$h(X)$,	$h(X|Y)$			& 		Differential entropy and conditional entropy (resp.) \\\hline
$I(X;Y)$					& 		Mutual information between $X$ and $Y$\\\hline
$D(\mathbb{P}_X\|\mathbb{P}_Y)$	& 		Relative entropy between $\mathbb{P}_X$ and $\mathbb{P}_Y$\\\hline
$\|\cdot\|_p$				&		$\ell_p$-norm of a vector\\\hline
%$\mathbb{I}_{x\in\mathcal{X}}$			&		Indicator function\\\hline
$\mathbb{E}_X[\cdot]$		&		Expectation with respect to $X$ \\\hline
\end{tabular}
\end{table}

\section{Channel Model and Main Assumptions}
\label{Model}
The most common channel used to model IM/DD OWC in the literature is the input-independent Gaussian noise channel. This channel is described by an input $X\geq 0$ which is subject to peak and average constraints $X\leq \mathcal{A}$ and $\mathbb{E}[X]\leq\mathcal{E}$, and an output 
\begin{align}
Y=gX+Z,
\end{align}
where $g\geq 0$ is a channel gain, and $Z\sim\mathcal{N}(0,\sigma^2)$. \ac{In this model, several system parameters (such as transmitter and receiver responsiveness, geometric loss, background noise power, etc.) are `lumped into' the channel gain $g$ and the noise variance $\sigma^2$ as we shall see.}

How do we arrive at this channel model in an optical channel described by 'discrete' photon transmission? To answer this question, in the following subsections, we describe the basic physical aspects of a simple transmitter-receiver system with some idealized assumptions. Then, step-by-step, we develop the Gaussian channel model given above which will be the main focus of this tutorial. Along the way, we will arrive at various channel models that have been studied throughout the history of IM/DD OWC. \ac{A summary of various IM/DD channel models that have been studied in the literature along with some works that study their capacities is given in Table \ref{Tab:Models} for reference.}

\begin{table}\centering
\caption{\ac{Some IM/DD channel models and works that study their capacities.}}\label{Tab:Models}
\begin{tabular}{l|l}
Channel model & Capacity-related works\\\hline\hline
Discrete-time Poisson channel & \cite{PiercePosnerRodemich,Shamai_DT_Poisson,
LapidothMoser_Poisson,
LapidothShapiro_Poisson,Martinez,CaoHranilovicChen1} \\\hline
Continuous-time Poisson channel &  \cite{Wyner_Poisson,
HaasShapiro,HaasShapiro_MIMOPoissonCap,Wyner_Poisson_II}\\
&\cite{Davis,PiercePosnerRodemich,Kabanov}\\\hline
Input-dependent Gaussian noise channel & \cite{Moser,WangYangWangChen,WangWangChenWang,
AminianGhourchianGohariMirmohseniNasiri-Kenari,GhourchianAminianGohariMirmohseniKenari,
DadamahallehHodtani}\\\hline
Input-independent Gaussian noise channel & \cite{LapidothMoserWigger,FaridHranilovic,FaridHranilovic_SelectedAreas,
ChaabanMorvanAlouini,WangHuWang_Dimmable,XuYuZhuSun}\\\hline
\end{tabular}
\end{table}

\subsection{An Ideal Optical Channel}
\subsubsection{Geometric Loss}
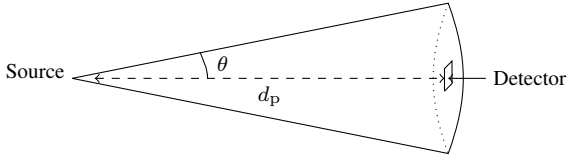
\begin{figure}
\centering
\begin{tikzpicture}
\footnotesize  
\node (s) at (0,0) [above] {};
\node at ($(s)+(0,.1)$) [left] {Source};
\node (d) at (5,0) [above] {};
\node (d2) at ($(d)+(.5,.0)$) [right] {Detector};
\draw[<-] (d.center) to (d2);
\draw ($(d)+(-.05,-.17)+(0,0)$) -- ($(d)+(-.05,-.17)+(0.1,0.1)$) -- ($(d)+(-.05,-.17)+(0.1,0.4)$)
 -- ($(d)+(-.05,-.17)+(0,0.3)$) -- ($(d)+(-.05,-.17)+(0,0)$);
\draw ($(s)+(0,0)$) -- ($(d)+(0,1)$) to [out=-70,in=70] ($(d)+(0,-1)$) -- ($(s)+(0,0)$);
\draw[dotted] ($(d)+(0,1)$) to [out=-110,in=110] ($(d)+(0,-1)$);
\node at ($(s)+(2.0,.2)$) {$\theta$};
\draw[dashed,<->] ($(s)+(0.3,0)$) to node[below] {$d_{\rm p}$} ($(d)+(-.05,0)$);
\draw[solid] ($(s)+(1.7,.35)$) to [out=-70,in=90] ($(s)+(1.8,0)$);
%\node at ($0.5*(s)+0.5*(d)-0.5*(0,1.1)$) [below] {$d_{\rm p}$};
\end{tikzpicture}
\caption{An illustration of Example \ref{Eg:Photons}. In a static and lossless system, the number of photons received at the detector depends on the propagation geometry. For a uniform beam profile with a conic propagation, this number will depend on the cone apex angle and distance.}
\label{Fig:PhysicalModel}
\end{figure}

Consider a transmitter-receiver system consisting of a light source and a detector as shown in Fig. \ref{Fig:PhysicalModel}. The light source (laser, LED) transmits a number $n_{\rm t}$ of photons with wavelength $\lambda$.\footnote{We will focus on a specific wavelength. The analysis can be readily generalized to multiple wavelengths.} At the receiver side, a detector captures some of the transmitted photons. In a static and lossless propagation medium, the number of photons received at the detector will depend on the geometrical parameters of the system, such as propagation distance, beam profile (Gaussian, uniform, etc.), beam divergence, transmitter-receiver angle, detector area, and optical filters and concentrators.\footnote{Other propagation effects that are not essential for the current analysis will be introduced later in Sec. \ref{Sec:Fading}.} As far as the OWC channel model is concerned, all these parameters can be abstracted into a {\it geometric loss} coefficient which we will denote by $g_{\rm g}<1$. The received number of photons will be approximately equal to 
\begin{align}
\label{InOut:Ideal}
n_{\rm r}=g_{\rm g}n_{\rm t}. 
\end{align}
This is an approximation because $n_{\rm r}$ my not be an integer, and actually represents an expectation of the number of received photons which is a random variable in general.

The geometric loss depends on the system. In an indoors system using VLC or IR, it is common to use the Lambertian model to calculate $g_{\rm g}$ \cite{KahnBarry}. In an FSO system, $g_{\rm g}$ is calculated by taking into account the propagation distance, beam divergence, beam profile, in addition to transmitter and receiver optics \cite{Kartalopoulos_FSO}. The following example calculates $g_{\rm g}$ for a simple system.

\begin{example}[Geometric Loss]
\label{Eg:Photons}
Let the transmitter send $n_{\rm t}$ photons uniformly in directions that form a cone with apex angle $2\theta$. At a distance $d_{\rm p}$ meters, the photon density will be $\frac{n_{\rm t}}{2\pi(1-\cos(\theta)) d_{\rm p}^2}$ photons/m$^2$ since the cone's solid angle is $\phi=2\pi (1-\cos(\theta))$ Steradians and the surface area of the spherical cap suspended by this cone is $\phi d_{\rm p}^2$ m$^2$. Thus, a detector with area $a_{\rm d}$ m$^2$ will receive approximately $n_{\rm r}=\frac{a_{\rm d}n_{\rm t}}{2\pi(1-\cos(\theta))d_{\rm p}^2}$ photons if $n_{\rm t}$ is large enough, and $g_{\rm g}=\frac{a_{\rm d}}{2\pi(1-\cos(\theta))d_{\rm p}^2}$.
\end{example}

\subsubsection{Information Transmission} 
To convey information to the receiver, the transmitter discretizes the time axis to intervals of duration $\Delta t$, and varies the number of photons it sends in each time interval $[(i-1)\Delta t,i\Delta t)$, $i=1,2,\ldots$. We denote the number of photons sent in interval $i$ by $n_{\rm t}(i)$. The received signal in the same interval becomes\footnote{Recall that we assume a static system only with geometric loss. We shall deviate from this assumption later.}
\begin{align}
\label{InOut:Ideal_t}
n_{\rm r}(i)=g_{\rm g}n_{\rm t}(i).
\end{align}
%Note that by multiplying this expression by the photon energy ${\sf e}=h\frac{c}{\lambda}$ where $h$ is the Planck constant and $c$ is the speed of light, and dividing by $t_{\rm s}$ we can express this system in terms of transmit and received optical power as
%\begin{align}
%\label{InOut:Ideal_t_P}
%p_{\rm r}(t)=g_{\rm g}p_{\rm t}(t) \text{\ \  Watts}. 
%\end{align}

This ideal model is our entry point to a statistical channel model, the Poisson channel, which will then lead us to the Gaussian model of OWC channels. But before transitioning to these models, we discuss few aspects about the ideal system in \eqref{InOut:Ideal_t}.

In the ideal system described by \eqref{InOut:Ideal_t}, information can be conveyed from the transmitter to the receiver at an extremely high rate in bits per second (bps). The following OOK example illustrates this point.

\begin{example}[Ideal OOK]
\label{Eg:Noiseless} 
Consider a binary source that generates i.i.d. bits $B\sim$Bern$(\nicefrac{1}{2})$ denoted $b_i$, $i=1,2,\ldots$. The transmitter sends $n_{\rm t}(i)=n\in\mathbb{N}_+$ photons if $b_i=1$ and $n_{\rm t}(i)=0$ otherwise (OOK), subject to an optical power constraint $\lim_{\tau\to\infty}\frac{1}{\tau}\sum_{i=1}^\tau p_{\rm t}(i)\leq \bar{p}_{\rm t,max}$ where $p_{\rm t}(i)=\frac{{\sf e}_{\lambda}n_{\rm t}(i)}{\Delta t}$ and ${\sf e}_{\lambda}$ is the photon energy. Since $B\sim$Bern$(\nicefrac{1}{2})$, we must have $\frac{n{\sf e}_{\lambda}}{2\Delta t}\leq \bar{p}_{\rm t,max}$. Let the geometric loss be $g_{\rm g}\approx 0.01$,\footnote{This corresponds to setting $a_{\rm d}=1$ cm$^2$, $\theta=1^{\circ}$, and $d_{\rm p}=10$ m in Example \ref{Eg:Photons}.} and let $\bar{p}_{\rm t,max}=1$~mW and $\lambda=850$~nm (infrared). The receiver declares $b_i=0$ if $n_{\rm r}(i)=0$, and $b_i=1$ otherwise. In the ideal model \eqref{InOut:Ideal_t}, if $b_i=1$, $n_{\rm r}(i)$ will be nonzero with high probability if $n\gg 100$. Letting $n=2000$ and $\frac{1}{\Delta t}=\frac{\bar{p}_{\rm t,max}}{1000{\sf e}_{\lambda}}$, the bit rate of the system will be close to $\frac{1}{\Delta t}=\frac{\bar{p}_{\rm t,max}}{1000{\sf e}_{\lambda}}\approx 4.3$~Tbps {since ${\sf e}_{\lambda}=2.34\times10^{-19}$~Joules}.\footnote{{Interpreting $g_{\rm g}$ as the probability of a photon's landing on the detector,} the probability that no photon lands on the detector when $b_i=1$ is $(1-g_{\rm g})^n\approx1.8\times10^{-9}$. This leads to a very small detection error probability of mistaking a $1$ for a $0$, which can be corrected using channel coding at (almost) no cost.}
\end{example}

%\begin{remark}
%The bit rate in Example \ref{Eg:Noiseless} is `close to' $\frac{1}{\Delta t}$ because photons might miss the detector sometimes leading to misdetecting a `1' as a `0'. However, since $n=2000$ and $g_{\rm g}\approx 0.01$, the probability of this event is low. The error probability will be low, and, with a little bit of redundancy, errors can be corrected.
%\end{remark}
%

The example above involves several idealistic assumptions which can be combined under two main categories:
\begin{enumerate}
\item Infinite amplitude resolution: In time interval $i$, the source can send precisely $n_{\rm t}(i)$ photons and $n_{\rm r}(i)$ is equal to $g_{\rm g}n_{\rm t}(i)$. % recall that this is not number of received photons, but an average.
\item Infinite temporal resolution: The source can switch from sending $n_{\rm t}(i)$ to $n_{\rm t}(i+1)$, and the detector can discern $n_{\rm r}(i)$ and $n_{\rm r}(i+1)$ no matter how small $\Delta t$ is (note that $\Delta t=\frac{1000{\sf e}_{\lambda}}{\bar{p}_{\rm t,max}}\approx233$fs in Example \ref{Eg:Noiseless}).
\end{enumerate}
Both points do not hold in practice. We will start by discussing the amplitude resolution followed by the temporal resolution.

\subsection{The Poisson Channel: A Noisy Optical Channel}
In practice, we can not adjust the number of photons $n_{\rm t}(i)$ transmitted by a laser or an LED with infinite resolution. Instead, what we can adjust is the photon transmission rate, i.e., the expected number of photons per second by modulating the optical power of the source. At a transmit optical power of $p_{\rm t}(i)$ Watts, the source sends $\frac{p_{\rm t}(i)}{{\sf e}_{\lambda}}$ photons/second. Thus, the expected number of photons that we will send in $\Delta t$ seconds will be 
$$\bar{n}_{\rm t}(i)=\frac{p_{\rm t}(i)}{{\sf e}_{\lambda}}\Delta t \text{ photons}.$$ 
We connect this fact with Example \ref{Eg:Noiseless} in the following example.

\begin{example}
\label{Eg:PowerTransmission}
In Example \ref{Eg:Noiseless}, instead of sending $2000$ photons in a time interval when $b_i=1$, we send at a power $p_{\rm t}(i)=\frac{2000{\sf e}_{\lambda}}{\Delta t}=2$~mW. Clearly, the average power constraint is satisfied since $B\sim$Bern$(\nicefrac{1}{2})$.
\end{example}

Similarly, the average number of photons that reach the detector in $\Delta t$ seconds is not $n_{\rm r}(i)=g_{\rm g}n_{\rm t}(i)$. Instead, photons will reach the detector at a rate of $\bar{n}_{\rm r}(i)=g_{\rm g}\bar{n}_{\rm t}(i)$ photons in $\Delta t$ seconds, or a power of $p_{\rm r}(i)=g_{\rm g}p_{\rm t}(i)$ Watts. Consequently, if we choose to transmit at a rate of $\bar{n}_{\rm t}(i)$ photons per $\Delta t$ seconds, the number of detected photons will fluctuate around the mean $\bar{n}_{\rm r}(i)=g_{\rm g}\bar{n}_{\rm t}(i)$. But how exactly is this fluctuation described? 

Let us focus on a specific interval $i=1$ and drop the time index for readability, and let us divide the corresponding interval $[0,\Delta t)$ into $m$ intervals $\mathcal{I}_j=[(j-1)\frac{\Delta t}{m},j\frac{\Delta t}{m})$, $j=1,\ldots,m$. % so that at most one photon is received in each time step. 
If $m$ is large enough, the detector will receive no more than one photon per interval $\mathcal{I}_j$. Since $\bar{n}_{\rm r}$ is the average number of photons received in $[0,\Delta t)$, the probability of receiving a photon in $\mathcal{I}_j$ is $\epsilon=\frac{\bar{n}_{\rm r}}{m}$. Assuming that photon detection is independent through $j=1,\ldots,m$, the probability of receiving $n_{\rm r}$ photons in $[0,\Delta t]$ is given by 
\begin{align}
\mathbb{P}_{N_{\rm r}}(n_{\rm r})=\binom{m}{n_{\rm r}}\epsilon^{n_{\rm r}}(1-\epsilon)^{m-n_{\rm r}},
\end{align}
which is the probability of receiving a photon in $n_{\rm r}$ intervals $\mathcal{I}_j$, and no photon in the remaining $m-n_{\rm r}$ intervals. Letting $m\to\infty$, we obtain 
\begin{align}
\label{PoissonOpticalNoiseless}
\mathbb{P}_{N_{\rm r}}(n_{\rm r})\to \frac{\bar{n}_{\rm r}^{n_{\rm r}}e^{-\bar{n}_{\rm r}}}{n_{\rm r}!} \ \text{as } m\to\infty.
\end{align}
Therefore, the number of received photons in $[0,\Delta t)$ is Poisson distributed with mean $\bar{n}_{\rm r}=g_{\rm g}\bar{n}_{\rm t}$. 

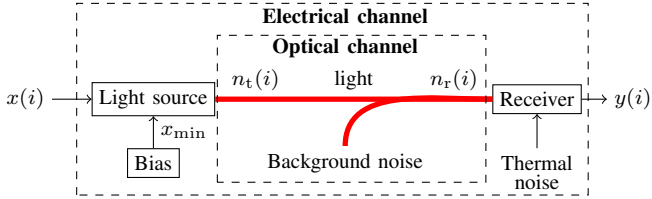
\begin{figure}
\centering
\begin{tikzpicture}[scale = .85]
\footnotesize
\node (s) at (0,0) [draw] {Light source};
\node (dc) at (0,-1) [draw] {Bias};
\draw[->] (dc) to node[right] {$x_{\rm min}$} (s);
\node (d) at (6,0) [draw] {Receiver};
\draw[line width=2,red] (s.east) to node[above] {\textcolor{black}{light}} (d.west);
\node (on) at (3,-1) {Background noise};
\draw[line width=2,red] (on.north) to[out=90,in=180] (d.west);
\node (x) at (-2,0) {$x(i)$};
\node (pt) at (1.6,.3) {$n_{\rm t}(i)$};
\node (pr) at (4.7,.3) {$n_{\rm r}(i)$};
\node (y) at (7.5,0) {$y(i)$};
\draw[->] (x) to (s);
\draw[->] (d) to (y);
\draw[dashed] (1,-1.3) rectangle (5.2,1);
\node at (3,.8) {\bf Optical channel};
\draw[dashed] (-1.2,-1.5) rectangle (6.8,1.5);
\node at (3,1.3) {\bf Electrical channel};
\node (tn) at (6,-1) {Thermal};
\node at (6,-1.3) {noise};
\draw[->] (tn.north) to (d.south);
\end{tikzpicture}
\caption{The optical channel can be described as a channel with input $n_{\rm t}(i)$ and output $n_{\rm r}(i)$ which represent the number of emitted photons and received photons, respectively. The electrical channel can be described as a channel with input $x(i)$ representing the modulating current, and output $y(i)$ representing the received current after processing and sampling.}
\label{Fig:Channels}
\end{figure}

But this is not the only effect that takes place in a practical system. In addition to this, some unwanted photons from background radiation (ambient light e.g.) will be received by the detector. This {\it optical noise} increases the average number of received photons to $$\bar{n}_{\rm r}=g_{\rm g}\bar{n}_{\rm t}+\bar{n}_{\rm bg},$$ where $\bar{n}_{\rm bg}$ is the average number of background radiation photons received in $\Delta t$ seconds. Thus, the number of received photons is distributed as
\begin{align}
\mathbb{P}_{N_{\rm r}}(n_{\rm r})=\frac{(g_{\rm g}\bar{n}_{\rm t}+\bar{n}_{\rm bg})^{n_{\rm r}}e^{-(g_{\rm g}\bar{n}_{\rm t}+\bar{n}_{\rm bg})}}{n_{\rm r}!}.
\end{align}
This describes the optical channel as shown in Fig. \ref{Fig:Channels}. Now we are ready to define the Poisson channel.

\begin{definition}[Poisson Channel]
\label{Def:Poisson}
The optical channel corresponding to an IM/DD OWC system can be modelled as a Poisson channel with input $\bar{N}_{\rm t}\geq 0$ and output $N_{\rm r}$ where the distribution of $N_{\rm r}$ conditioned on $\bar{N}_{\rm t}$ is given by
\begin{align}
\label{PoissonOptical}
\mathbb{P}_{N_{\rm r}|\bar{N}_{\rm t}}(n_{\rm r}|\bar{n}_{\rm t})=\frac{(g_{\rm g}\bar{n}_{\rm t}+\bar{n}_{\rm bg})^{n_{\rm r}}e^{-(g_{\rm g}\bar{n}_{\rm t}+\bar{n}_{\rm bg})}}{n_{\rm r}!}.
\end{align}
\end{definition}

%Fig. \ref{Fig:PoissonNoiseless_and_Noisy} shows the distribution of $N_{\rm r}$ for different values of $\bar{n}_{\rm bg}$.\footnote{We plot this as a continuous distribution instead of a discrete distribution for illustrative purposes.} It can be seen that the Poisson distribution in \eqref{PoissonOpticalNoiseless} can already compromise the infinite amplitude resolution we have seen in Example \ref{Eg:Noiseless} especially if $p_{\rm t}$ is low (Fig. \ref{Fig:Poisson_Noiseless}). This effect is more prominent when background radiation is present (Fig. \ref{Fig:Poisson_Noisy}). 

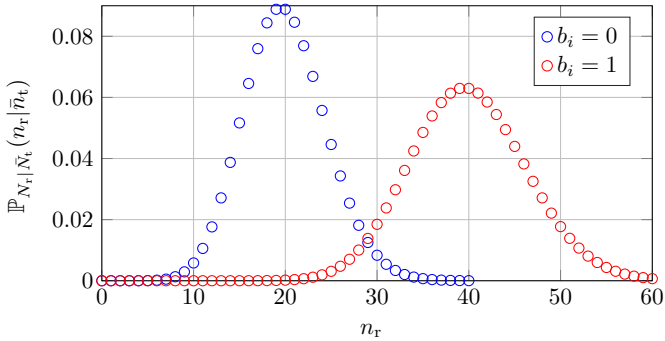
\begin{figure}[t]
\centering
\tikzset{every picture/.style={scale=.98}, every node/.style={scale=.9}}
\definecolor{mycolor1}{rgb}{0.00000,0.44700,0.74100}%
\definecolor{mycolor2}{rgb}{0.92900,0.69400,0.12500}%
\begin{tikzpicture}

\begin{axis}[%
width=3in,
height=1.5in,
scale only axis,
xmin=0,
xmax=60,
xlabel={$n_{\rm r}$},
xmajorgrids,
ymin=0,
ymax=0.09,
ylabel={$\mathbb{P}_{N_{\rm r}|\bar{N}_{\rm t}}(n_{\rm r}|\bar{n}_{\rm t})$},
ymajorgrids,
legend style={legend cell align=left,align=left,draw=white!15!black},
scaled y ticks = false,
y tick label style={/pgf/number format/fixed,/pgf/number format/1000 sep = \thinspace % Optional if you want to replace comma as the 1000 separator 
}
]
\addplot [color=blue,solid,only marks, mark = o]
  table[row sep=crcr]{%
0	2.06115362243856e-09\\
1	4.12230724487712e-08\\
2	4.12230724487712e-07\\
3	2.74820482991808e-06\\
4	1.37410241495904e-05\\
5	5.49640965983615e-05\\
6	0.000183213655327872\\
7	0.000523467586651062\\
8	0.00130866896662766\\
9	0.00290815325917257\\
10	0.00581630651834514\\
11	0.0105751027606275\\
12	0.0176251712677125\\
13	0.0271156481041731\\
14	0.0387366401488188\\
15	0.0516488535317583\\
16	0.0645610669146979\\
17	0.0759541963702329\\
18	0.084393551522481\\
19	0.0888353173920852\\
20	0.0888353173920852\\
21	0.0846050641829383\\
22	0.0769136947117621\\
23	0.0668814736624018\\
24	0.0557345613853348\\
25	0.0445876491082679\\
26	0.0342981916217445\\
27	0.0254060678679589\\
28	0.0181471913342564\\
29	0.0125153043684527\\
30	0.00834353624563511\\
31	0.00538292661008717\\
32	0.00336432913130448\\
33	0.00203898735230575\\
34	0.00119940432488573\\
35	0.000685373899934704\\
36	0.000380763277741502\\
37	0.00020581798796838\\
38	0.000108325256825463\\
39	5.55514137566477e-05\\
40	2.77757068783238e-05\\
};
\addlegendentry{$b_i=0$};

\addplot [color=red,solid,only marks, mark = o]
  table[row sep=crcr]{%
0	4.24835425529159e-18\\
1	1.69934170211664e-16\\
2	3.39868340423327e-15\\
3	4.53157787231103e-14\\
4	4.53157787231103e-13\\
5	3.62526229784882e-12\\
6	2.41684153189921e-11\\
7	1.38105230394241e-10\\
8	6.90526151971204e-10\\
9	3.06900511987202e-09\\
10	1.22760204794881e-08\\
11	4.46400744708657e-08\\
12	1.48800248236219e-07\\
13	4.57846917649905e-07\\
14	1.3081340504283e-06\\
15	3.4883574678088e-06\\
16	8.720893669522e-06\\
17	2.051974981064e-05\\
18	4.55994440236444e-05\\
19	9.5998829523462e-05\\
20	0.000191997659046924\\
21	0.000365709826756046\\
22	0.000664926957738265\\
23	0.00115639470911003\\
24	0.00192732451518338\\
25	0.0030837192242934\\
26	0.00474418342198985\\
27	0.00702841988442941\\
28	0.0100405998348992\\
29	0.0138491032205506\\
30	0.0184654709607341\\
31	0.0238264141428827\\
32	0.0297830176786033\\
33	0.0361006274892162\\
34	0.0424713264579014\\
35	0.0485386588090301\\
36	0.0539318431211446\\
37	0.0583046952661023\\
38	0.0613733634380024\\
39	0.0629470394235922\\
40	0.0629470394235922\\
41	0.0614117457791143\\
42	0.0584873769324898\\
43	0.0544068622627813\\
44	0.0494607838752557\\
45	0.0439651412224495\\
46	0.0382305575847387\\
47	0.0325366447529691\\
48	0.0271138706274743\\
49	0.0221337719407953\\
50	0.0177070175526362\\
51	0.0138878569040284\\
52	0.0106829668492526\\
53	0.00806261649000199\\
54	0.00597230851111259\\
55	0.00434349709899097\\
56	0.00310249792785069\\
57	0.00217719152831628\\
58	0.00150151139883881\\
59	0.00101797382972123\\
60	0.000678649219814152\\
};
\addlegendentry{$b_i=1$};

\end{axis}
\end{tikzpicture}%
\caption{The {conditional} distribution of $N_{\rm r}$ given $\bar{N}_{\rm t}$ in \eqref{PoissonOptical} when $p_{\rm t}=0$ and $p_{\rm t}=2$~mW, corresponding $b_t=0$ and $b_t=1$ in Example \ref{Eg:Noiseless}, and $\bar{n}_{\rm bg}=20$ photons per symbol duration which is equivalent to a background noise power of $20$~$\mu$W according to the parameters of Example \ref{Eg:Noiseless}.}
\label{Fig:PoissonNoiseless_and_Noisy}
\end{figure}

Fig. \ref{Fig:PoissonNoiseless_and_Noisy} shows the distribution of $N_{\rm r}$ for a given value of $\bar{n}_{\rm bg}$. It can be seen that the Poisson distribution in \eqref{PoissonOpticalNoiseless} can already compromise the infinite amplitude resolution we have seen in Example \ref{Eg:Noiseless} even for weak background noise.

The capacity of the discrete-time Poisson channel has been studied in \cite{PiercePosnerRodemich,Shamai_DT_Poisson,LapidothMoser_Poisson,
LapidothShapiro_Poisson,Martinez,CaoHranilovicChen1} {and considering secrecy constraints in \cite{SoltaniRezki_ISIT2019}}, and that of the continuous-time Poisson channel ($\Delta t\to0$) has been studied in \cite{PiercePosnerRodemich,Wyner_Poisson,Wyner_Poisson_II,Kabanov,Davis,HaasShapiro,
HaasShapiro_MIMOPoissonCap} {and considering a secrecy constraint in \cite{LaourineWagner}}. The capacity of multi-user Poisson channels has been studied in \cite{LapidothMoser_Poisson,LapidothShapiro_Poisson,
LapidothShamaiPoissonMAC,KimNachmanElGamal_Journal}. 

{Note that apart from the continuous time case, i.e., the unconstrained bandwidth scenario ($\Delta t\to0$), where capacity results are sharp, in the most practical case of constrained bandwidth, capacity results are generally not exact and only bounds and/or asymptotic results have been established. To provide additional insight into the capacity of the constrained bandwidth case, it is helpful to simplify the Poisson model to models which are relatively easier to study. One such simplification is the Gaussian model, which has been used frequently in the literature as an approximation under some practical assumptions. With this in mind, we are now ready to transit to defining Gaussian channel models.}

\subsection{The Input-Dependent Gaussian Noise Channel}
We will split the discussion here into three parts. First, we will describe the input-dependent Gaussian noise channel model in the optical domain. Then, we will describe a continuous-time input-dependent Gaussian noise channel model in the electrical domain (in terms of electric currents). Finally, we will describe its discrete-time counterpart.

\subsubsection{Optical Domain}
Observing Fig. \ref{Fig:PoissonNoiseless_and_Noisy}, we see that \ac{the number of received photons} $N_{\rm r}$ follows a distribution which is `nearly' Gaussian (clipped at zero). The reason is that a Poisson distribution with mean $\mu$ approaches a Gaussian distribution with mean and variance $\mu$ as $\mu$ increases. The mean of $N_{\rm r}$ for a given \ac{mean number of transmitted photons} $\bar{N}_{\rm t}=\bar{n}_{\rm t}$ is 
\begin{align}
\label{Nrbar}
\bar{n}_{\rm r}&=g_{\rm g}\bar{n}_{\rm t}+\bar{n}_{\rm bg}\\
&=g_{\rm g}\bar{n}_{\rm t}+\frac{p_{\rm bg}\Delta t}{{\sf e}_{\lambda}},
\end{align}
where $p_{\rm bg}$ is the power of the background radiation. Thus, the mean is large if one or more of $\bar{n}_{\rm t}$, $\Delta t$, and $p_{\rm bg}$ is relatively large. The following example shows that it is not a stretch to assume a large $\bar{n}_{\rm r}$. 

\begin{example}
Suppose that $\Delta t=23$~ps and $\lambda=850$~nm, and that we want to send $b_i=0$ in Example \ref{Eg:Noiseless}. Thus, $\bar{n}_{\rm t}=0$. %If we accept a Gaussian approximation for the Poisson distribution when $\bar{n}_{\rm r}>100$ e.g., then we need $p_{\rm bg}$ to be larger than $\frac{100{\sf e}_{\lambda}}{\Delta t}=100\mu$W. 
The solar radiation at $\lambda=850$~nm is in the order of hundreds of W/m$^2$ \cite{NASA_SolarRadiation} which for the $1$cm$^2$ detector we assumed in this example induces a received noise power of tens of mW. %Thus, $\bar{n}_{\rm r}\gg100$ if $b_i=0$, and even larger if $b_i=1$, and a Gaussian approximation is acceptable. 
This produces $\bar{n}_{\rm r}$ in the order of $10^5$ for $b_i=0$, and an even larger $\bar{n}_{\rm r}$ for $b_i=1$.
\end{example}

This example demonstrates that background radiation can be large enough, and a Gaussian approximation is acceptable. Given that the solar background radiation power received by the detector is tens of mW at $850$~nm, the approximation will be acceptable even for smaller values of $\Delta t$ (femtoseconds), and more so when $\Delta t$ is larger which is likely the case in practice given the current technology. 

The previous discussion suggests that a Poisson model is important when the number of received photons per symbol duration is low (inter-satellite and ground-satellite communication, scattering non-line-of-sight UV communication), and a Gaussian model is sufficient for most other cases. For most terrestrial systems, the Gaussian approximation is acceptable.

Let us describe the number of received photons based on this model. Since $N_{\rm r}$ is Poisson distributed with mean $\bar{n}_{\rm r}$, then for large $\bar{n}_{\rm r}$, we can use the approximation $N_{\rm r}\sim\mathcal{N}(\bar{n}_{\rm r},\bar{n}_{\rm r})$. Thus $N_{\rm r}$ can be written as
\begin{align}
N_{\rm r}&=\bar{n}_{\rm r}+\sqrt{\bar{n}_{\rm r}}\tilde{Z},
\end{align}
where $\tilde{Z}\sim\mathcal{N}(0,1)$. Substituting $\bar{n}_{\rm r}$ from \eqref{Nrbar} leads to
\begin{align}
N_{\rm r}&=g_{\rm g}\bar{n}_{\rm t}+\bar{n}_{\rm bg}+\sqrt{g_{\rm g}\bar{n}_{\rm t}+\bar{n}_{\rm bg}}\tilde{Z}.
\end{align}
Note that the receiver can subtract $\bar{n}_{\rm bg}$ because this is a constant, and that $\sqrt{g_{\rm g}\bar{n}_{\rm t}+\bar{n}_{\rm bg}}\tilde{Z}$ can be written as the sum of two independent Gaussian random variables with zero mean and variances $g_{\rm g}\bar{n}_{\rm t}$ and $\bar{n}_{\rm bg}$. {This leads to the following signal at the receiver}
\begin{align}
\label{InOutGaussianDependentOptical}
\tilde{N}_{\rm r}&=g_{\rm g}\bar{n}_{\rm t}+\sqrt{g_{\rm g}\bar{n}_{\rm t}}\tilde{Z}_1+\sqrt{\bar{n}_{\rm bg}}\tilde{Z}_2,
\end{align}
where both $\tilde{Z}_1$ and $\tilde{Z}_2$ are $\mathcal{N}(0,1)$. The term $\tilde{Z}_1$ can be interpreted as noise induced by the fluctuation of the number of photons that reach the receiver from the transmitter due to the randomness of photon arrivals. The term $\tilde{Z}_2$ can be interpreted as shot-noise induced by photons from background radiation. This allows us to write the following definition.

\begin{definition}[Optical Input-Dependent Gaussian Channel]
\label{Def:GaussianOptical}
The optical channel corresponding to IM/DD OWC can be modelled, in the high received-photon-rate regime, as a channel with input $\bar{N}_{\rm t}\geq 0$ and output $\tilde{N}_{\rm r}$ given by
\begin{align}
\label{GaussianOptical}
\tilde{N}_{\rm r}&=g_{\rm g}\bar{N}_{\rm t}+\sqrt{g_{\rm g}\bar{N}_{\rm t}}\tilde{Z}_1+\sqrt{\bar{n}_{\rm bg}}\tilde{Z}_2,
\end{align}
where $\tilde{Z}_1,\tilde{Z}_2$ are independent $\mathcal{N}(0,1)$.
\end{definition}

\subsubsection{Electrical Domain -- Continuous-Time}
So far, we have been discussing optical aspects of the transmitter-receiver system. The distribution in \eqref{PoissonOptical} and the relation in \eqref{InOutGaussianDependentOptical} describe a discrete-time optical-input optical-output channel. 

\ac{But in IM/DD systems, the modulating signal is an electric signal, used as an input to an LED e.g., or as an input to an electro-optical modulator \cite{FengJiangWu_CLEO}. Also, receiver processing is done in the electrical domain after converting the received optical signal to an electric signal. Thus, it is important to represent} the system in the electrical domain.

\begin{figure}
\centering
\begin{tikzpicture}
\footnotesize
\node (f) at (0,0) {\includegraphics[width=8cm]{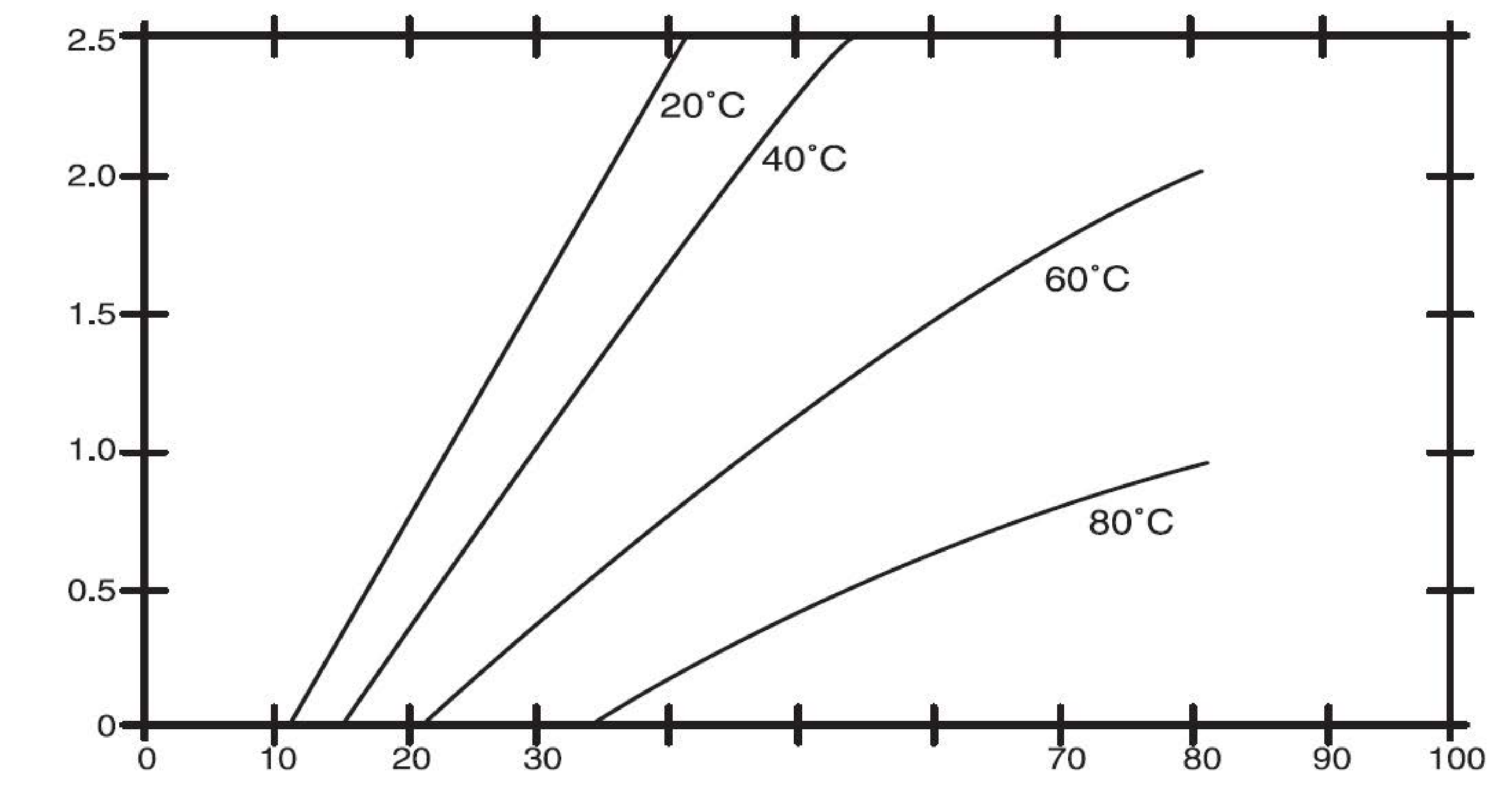}};
\node at ($(f)+(.3,-2)$) {Current (mA)};
\node at ($(f)+(-3.9,1.4)$) [left,rotate=90] {Light output (mW)};
\end{tikzpicture}
\caption{Typical L-I characteristics of a laser diode \cite{ILX_LD}. The linear relation between current and light power is measured by the electrical-optical conversion efficiency in mW/mA.}
\label{Fig:LI}
\end{figure}

Modulation of $\bar{n}_{\rm r}(i)$ is achieved by modulating the emitted optical power $p_{\rm t}(i)$. \ac{This in turn is achieved by using an electrical signal, so that the relation between the electrical signal and the optical power is linear. For instance,} the optical power--current characteristics of a solid-state light source has a linear range (cf. Fig. \ref{Fig:LI}) which is used for modulation in practice. An electro-optical modulation is modulated similarly in a linear range \cite{newport_EO_mod}. Thus, we will write the emitted optical power as $p_{\rm t}(i)=\eta_{\rm eo}x(i)$ where $\eta_{\rm eo}$ is the electrical-optical conversion efficiency (in Watts per Ampere) which is the slope in Fig. \ref{Fig:LI}.

Photodetectors have a power-current characteristics similar to Fig. \ref{Fig:LI} \cite{New_port_PDSensors}. Thus, the current generated by the detector follows a linear relation with the received optical power with slope $\eta_{\rm oe}$ (in Amperes per Watt).

Based on this description of electrical--optical conversion, we describe the following IM/DD OWC system. Let the modulating current be $q_{\rm t}(t)\geq0$ where $t\in\mathbb{R}_+$. Consider an instance of $q_{\rm t}(t)$ given by $q_{\rm t}(t_0)=q_{\rm t}(t)\delta(t-t_0)$ where $\delta(t)$ is the Dirac delta function. This can be written as 
\begin{align}
q_{\rm t}(t_0)=\lim_{\Delta t\to0}\frac{1}{\Delta t}q_{\rm t}(t)r_{\Delta t}(t-t_0),
\end{align} 
where $r_\tau(t)$ is a rectangular function with $r_\tau(t)=1$ for $t\in[0,\tau)$ and $0$ elsewhere. Let us focus on an interval $[t_0,t_0+\Delta t)$, i.e., where $\dot{q}_{\rm t}(t_0)=q_{\rm t}(t)r_{\Delta t}(t-t_0)$ is nonzero. During this interval, the transmitted optical power is approximately $p_{\rm t}(t_0)=\eta_{\rm eo}\dot{q}_{\rm t}(t_0)$, where the approximation becomes accurate as $\Delta t\to0$. The average number of emitted photons is $\bar{n}_{\rm t}(t_0)=\frac{\Delta t}{{\sf e}_{\lambda}}\eta_{\rm eo}\dot{q}_{\rm t}(t_0)$. According to \eqref{InOutGaussianDependentOptical}, the detector will receive a number of photons in this interval given by
\begin{align}
n_{\rm r}(t_0)=g_{\rm g}\bar{n}_{\rm t}(t_0)+\sqrt{g_{\rm g}\bar{n}_{\rm t}(t_0)}\tilde{z}_{1,t_0}+\sqrt{\bar{n}_{\rm bg}}\tilde{z}_{2,t_0},
\end{align}
where $\tilde{z}_{1,t_0}$ and $\tilde{z}_{2,t_0}$ are realizations of independent $\mathcal{N}(0,1)$ noises, i.i.d. through time $t_0$. The received power in the same interval will be $p_{\rm r}(t_0)=\frac{{\sf e}_{\lambda}}{\Delta t}n_{\rm r}(t_0)$, and the current generated by the detector will be
\begin{align}
\dot{q}_{\rm r}(t_0)=\frac{{\sf e}_{\lambda}}{\Delta t}\eta_{\rm oe}n_{\rm r}(t_0).
\end{align}
Therefore, we have 
\begin{align}
\dot{q}_{\rm r}(t_0)%\nonumber\\
&=g_{\rm g}\eta_{\rm eo}\eta_{\rm oe}\dot{q}_{\rm t}(t_0)+\sqrt{g_{\rm g}\eta_{\rm eo}\eta_{\rm oe}^2\frac{{\sf e}_{\lambda}}{\Delta t}\dot{q}_{\rm t}(t_0)}\tilde{z}_{1,t_0}\nonumber\\&\quad 
+\sqrt{\eta_{\rm oe}^2\frac{{\sf e}_{\lambda}}{\Delta t}p_{\rm bg}}\tilde{z}_{2,t_0}.
\end{align}

Letting $\Delta t\to0$, and denoting $\lim_{\Delta t\to0}\dot{q}_{\rm r}(t_0)$ by $q_{\rm r}(t_0)$, we obtain
\begin{align}
q_{\rm r}(t_0)&=g_{\rm g}\eta_{\rm eo}\eta_{\rm oe}q_{\rm t}(t_0)+\sqrt{g_{\rm g}\eta_{\rm eo}\eta_{\rm oe}^2{\sf e}_{\lambda}q_{\rm t}(t_0)\delta(t-t_0)}\tilde{z}_{1,t_0}\nonumber\\&\quad +\sqrt{\eta_{\rm oe}^2{\sf e}_{\lambda}\,p_{\rm bg}\delta(t-t_0)}\tilde{z}_{2,t_0}.
\end{align}
This follows by replacing $\frac{1}{\Delta t}$ by $\frac{r_{\Delta t}(t-t_0)}{\Delta t}$ which does not change the expression, and since $\lim_{\Delta\to0}\frac{r_{\Delta t}(t-t_0)}{\Delta t}=\delta(t-t_0)$ (in s$^{-1}$). Note that $\sqrt{\delta(t-t_0)}\tilde{z}_{j,t_0}$, $j=1,2,$ is a sample of a white Gaussian noise process $\tilde{z}_j(t)$ with zero mean and unit power spectral density. Thus, we can write
\begin{align}
q_{\rm r}(t_0)&=g_{\rm g}\eta_{\rm eo}\eta_{\rm oe}q_{\rm t}(t_0)+\sqrt{g_{\rm g}\eta_{\rm eo}\eta_{\rm oe}^2{\sf e}_{\lambda}q_{\rm t}(t_0)}\tilde{z}_1(t_0)\nonumber\\&\quad+\sqrt{\eta_{\rm oe}^2{\sf e}_{\lambda}p_{\rm bg}}\tilde{z}_2(t_0).
\end{align}

We still need to account for thermal noise generated by the receiver electronics. This can be combined with the background noise. Assuming that thermal noise is white Gaussian with power spectral density $s_{\rm th}$ (in A$^2$s), we can replace $\sqrt{\eta_{\rm oe}^2{\sf e}_{\lambda}p_{\rm bg}}$ above with $\sqrt{\eta_{\rm oe}^2{\sf e}_{\lambda}p_{\rm bg}+s_{\rm th}}$ to obtain
\begin{align}
\label{Cont_time_Gaussian_channel}
q_{\rm r}(t_0)=g_{\rm g}q_{\rm t}(t_0)+\sqrt{\tilde{a}_1q_{\rm t}(t_0)}\tilde{z}_1(t_0)+\sqrt{\tilde{a}_2}\tilde{z}_2(t_0),
\end{align}
where we divided by $\eta_{\rm eo}\eta_{\rm oe}$, and where $\tilde{a}_1=\frac{g_{\rm g}{\sf e}_{\lambda}}{\eta_{\rm eo}}$  (in A\,s) and $\tilde{a}_2=\frac{{\sf e}_{\lambda}p_{\rm bg}}{\eta_{\rm eo}^2}+\frac{s_{\rm th}}{\eta_{\rm eo}^2\eta_{\rm oe}^2}$ (in A$^2$s). This leads to the following continuous-time input-dependent Gaussian noise channel model.

\begin{definition}[Continuous-Time Electrical Input-Dependent Gaussian Channel]
\label{Def:GaussianContinuous}
An IM/DD OWC system can be modelled, in the high received optical power regime, as a continuous-time input-dependent Gaussian noise electrical channel with input current $q_{\rm t}(t)\geq0$ and output current $q_{\rm r}(t)$ described as in \eqref{Cont_time_Gaussian_channel}.
\end{definition}

Now we are ready to describe the discrete-time input-dependent Gaussian noise channel model.

\subsubsection{Electrical Domain -- Discrete-Time}
Here, the channel described in \eqref{Def:GaussianContinuous} is used for digital transmission of $\kappa$ symbols $x_1,\ldots,x_\kappa$ for some $\kappa\in\mathbb{N}$, where $x_k\in\mathcal{X}$ is the $k$th transmit symbol chosen from an alphabet $\mathcal{X}\subset\mathbb{R}_+$. The transmitter modulates the current $q_{\rm t}(t)$ as
\begin{align}
\label{qt_digital_transmission}
q_{\rm t}(t)=\sum_{k=1}^{\kappa} x_k \gamma(t-kt_{\rm s})+q_{\rm dc},
\end{align}
where $\gamma(\cdot)$ is a pulse-shaping function which has finite energy, $t_{\rm s}$ is the symbol duration (inverse of symbol rate), and $q_{\rm dc}$ is a constant DC-offset applied, if necessary, to make the signal positive.\footnote{such as when optical OFDM schemes are used \cite{ZhouZhang}.} If we use optical intensity pulses, which are positive pulses,\footnote{For more detail on optical intensity pulses, the reader is referred to \cite{Hranilovic_Pulses}.} then a DC-offset will not be necessary, i.e., $q_{\rm dc}=0$.

The received signal $q_{\rm r}(t)$ is given by
\begin{align}
q_{\rm r}(t)&=g_{\rm g}\sum_{k=1}^{\kappa} x_k \gamma(t-kt_{\rm s})+g_{\rm g}q_{\rm dc}\\&\quad+\sqrt{\tilde{a}_1\sum_{k=1}^{\kappa} x_k \gamma(t-kt_{\rm s})+\tilde{a}_1q_{\rm dc}}\tilde{z}_1(t)+\sqrt{\tilde{a}_2}\tilde{z}_2(t).\nonumber
\end{align}
The receiver subtracts $g_{\rm g}q_{\rm dc}$, filters $q_{\rm r}(t)$ using a filter with impulse response $\tilde{\gamma}(t)$, and then samples at $i t_{\rm s}$. Let 
\begin{align}
\zeta(t)&=\tilde{\gamma}(t)\circledast\gamma(t),
\end{align}
where $\circledast$ denotes convolution. The filtered signal $\tilde{q}_{\rm r}(t)$ can thus be written as 
\begin{align}
\tilde{q}_{\rm r}(t)&=g_{\rm g}\sum_{k=1}^{\kappa} x_k \zeta(t-kt_{\rm s})+\breve{z}_{1}(t)+\breve{z}_2(t),
\end{align}
where the filtered noises $\breve{z}_{1}(t)$ and $\breve{z}_2(t)$ are given by 
\begin{align}
\breve{z}_{1}(t)&=\tilde{\gamma}(t)\circledast\sqrt{\tilde{a}_1\sum_{k=1}^{\kappa} x_k \gamma(t-kt_{\rm s})+\tilde{a}_1q_{\rm dc}}\tilde{z}_1(t)\\
\breve{z}_2(t)&=\tilde{\gamma}(t)\circledast \sqrt{\tilde{a}_2}\tilde{z}_2(t).
\end{align}

To avoid inter-symbol interference, we require that
\begin{align}
\label{Nyquist}
\zeta(it_{\rm s})&=\begin{cases}g_{\zeta}, & i=0,\\
0, & \text{otherwise,}\end{cases}
\end{align}
which is the first Nyquist criterion \cite{ProakisSelehi}. Under this condition, the sampled filtered signal $\tilde{q}_{\rm r}(i t_{\rm s})$ can be written as 
\begin{align}
\label{SampledFilteredOutput}
\tilde{q}_{\rm r}(it_{\rm s})&=g_{\rm g}g_{\zeta}x_i +\breve{z}_{1}(i t_{\rm s})+\breve{z}_2(it_{\rm s}).
\end{align}

It remains to characterize noise. Since $\breve{z}_2(t)$ is obtained by filtering the white Gaussian noise process $\tilde{z}_2(t)$ with zero mean and unit power spectral density using a filter $\sqrt{\tilde{a}_2}\tilde{\gamma}(t)$, then $\breve{z}_2(t)$ is a zero mean Gaussian noise process with power spectral density $|\mathcal{F}\{\sqrt{\tilde{a}_2}\tilde{\gamma}(t)\}|^2$ where $\mathcal{F}\{\cdot\}$ is the Fourier transform. This implies that $\breve{z}_2(it_{\rm s})$ is i.i.d. (with respect to $i$) Gaussian with zero mean and variance 
\begin{align}
\breve{\sigma}^2_2=\tilde{a}_2\int_{-\infty}^{\infty}\tilde{\gamma}^2(t){\rm d}t.
\end{align}

On the other hand, we have
\begin{align}
&\breve{z}_{1}(it_{\rm s})\nonumber\\
&=\left[\tilde{\gamma}(t)\circledast\sqrt{\tilde{a}_1\sum_{k=1}^{\kappa} x_k \gamma(t-kt_{\rm s})+\tilde{a}_1q_{\rm dc}}\tilde{z}_1(t)\right]_{t=it_{\rm s}}\\
&=\int_{-\infty}^\infty\hspace{-1.5mm} \tilde{\gamma}(\tau)\sqrt{\tilde{a}_1\hspace{-1.0mm}\sum_{k=1}^{\kappa}\hspace{-1.0mm} x_k \gamma((i-k)t_{\rm s}\hspace{-.5mm}-\hspace{-.5mm}\tau)\hspace{-.5mm}+\hspace{-.5mm}\tilde{a}_1q_{\rm dc}}\tilde{z}_1(it_{\rm s}\hspace{-.5mm}-\hspace{-.5mm}\tau){\rm d}\tau\nonumber\\
&=\left[\left(\hspace{-.5mm}\tilde{\gamma}(t)\sqrt{\hspace{-.5mm}\tilde{a}_1\hspace{-.5mm}\sum_{k=1}^{\kappa} x_k \gamma((i\hspace{-.5mm}-\hspace{-.5mm}k)t_{\rm s}\hspace{-.5mm}-\hspace{-.5mm}t)\hspace{-.5mm}+\hspace{-.5mm}\tilde{a}_1q_{\rm dc}}\right)\hspace{-.5mm}\circledast\hspace{-.5mm}\tilde{z}_1(t)\right]_{t=it_{\rm s}}\hspace{-1.5mm}.\nonumber
\end{align}
Thus, similar to $\breve{z}_2(it_{\rm s})$, we have that $\breve{z}_1(it_{\rm s})$ is a sample of filtered white Gaussian noise with filter $\tilde{\gamma}(t)\sqrt{\tilde{a}_1\sum_{k=1}^{\kappa} x_k \gamma((i-k)t_{\rm s}-t)+\tilde{a}_1q_{\rm dc}}$, and hence $\breve{z}_1(it_{\rm s})$ is i.i.d. Gaussian with zero mean and variance 
\begin{align}
\breve{\sigma}_1^2
&=\int_{-\infty}^{\infty} \tilde{\gamma}^2(t)\left(\tilde{a}_1\sum_{k=1}^{\kappa} x_k \gamma((i-k)t_{\rm s}-t)+\tilde{a}_1q_{\rm dc}\right) {\rm d}t\nonumber\\
&=\tilde{a}_1\sum_{k=1}^{\kappa} x_k \int_{-\infty}^{\infty} \tilde{\gamma}^2(t)\gamma((i-k)t_{\rm s}-t){\rm d}t\nonumber\\
&\qquad +\tilde{a}_1q_{\rm dc}\int_{-\infty}^{\infty} \tilde{\gamma}^2(t){\rm d}t\\
\label{eq:brevesigma1s}
&=x_i \underbrace{\tilde{a}_1\int_{-\infty}^{\infty} \tilde{\gamma}^2(t)\gamma(-t){\rm d}t}_{\sigma_{0}^2}\nonumber\\
&\qquad+\sum_{k=1,k\neq i}^{\kappa} x_k \underbrace{\tilde{a}_1\int_{-\infty}^{\infty} \tilde{\gamma}^2(t)\gamma((i-k)t_{\rm s}-t){\rm d}t}_{\breve{\sigma}_{0,ik}^2}\nonumber\\
&\qquad +\underbrace{\tilde{a}_1q_{\rm dc}\int_{-\infty}^{\infty} \tilde{\gamma}^2(t){\rm d}t}_{\breve{\sigma}_{12}^2}.
\end{align}
We assume that 
\begin{align}
\label{IDNoiseCondition}
\breve{\sigma}^2_{0,ik} &\ll \sigma_0^2,\ \forall k\neq i.
\end{align} 
This assumption holds when the energy of $\tilde{\gamma}(t)$ and $\gamma(-t)$ is `concentrated' within an interval around $t=0$ of width $t_{\rm s}$. This takes place for instance if we use time-disjoint pulses, which is common in practice due to the simplicity of implementation \cite{FaridHranilovic_SelectedAreas,NuwanpriyaHoZhang}.
%Thus, when we shift $\gamma(-t)$ by nonzero multiples of $t_{\rm s}$, we get `weak' noise. 
As a result, if we write $\breve{z}_1(it_{\rm s})$ as $\sqrt{x_i}z_{0}(it_{\rm s})+\breve{z}_{12}(it_{\rm s})$ with variances $\sigma_{0}^2$ and $\breve{\sigma}_{12}^2$, respectively (corresponding to the first and last terms in \eqref{eq:brevesigma1s}), we obtain
\begin{align}
\tilde{q}_{\rm r}(it_{\rm s})&=g_{\rm g}g_{\zeta}x_i +\sqrt{x_i}z_{0}(i t_{\rm s})+\breve{z}_{12}(i t_{\rm s})+\breve{z}_2(it_{\rm s})\\
\label{eq:qrtildeits}
&=g x_i +\sqrt{x_i }z_{0,i}+z_{i},
\end{align}
where $g=g_{\rm g}g_{\zeta}$, $z_{0,i}$ is an i.i.d. Gaussian noise with zero mean and variance $\sigma_0^2$ and $z_i$ is an i.i.d. Gaussian noise with zero mean and variance $\sigma^2=\breve{\sigma}_{12}^2+\breve{\sigma}_2^2$, which combines contributions from $\breve{z}_{12}(it_{\rm s})$ and $\breve{z}_{2}(it_{\rm s})$. Now, we can write the following definition.

\begin{definition}[Discrete-Time Electrical Input-Dependent Gaussian Channel]
\label{Def:GaussianInputDep}
Under conditions \eqref{Nyquist} and \eqref{IDNoiseCondition}, the discrete-time electrical channel corresponding to an IM/DD OWC system can be modelled as a channel with input $X$ and output $Y=gX+\sqrt{X}Z_0+Z$, where $Z_0\sim\mathcal{N}(0,\sigma_0^2)$ and $Z\sim\mathcal{N}(0,\sigma^2)$.
\end{definition}

In general, condition \eqref{IDNoiseCondition} is satisfied for any time-disjoint pulse-shaping scheme. The following example illustrates this statement.

\begin{example}
Suppose that we choose $\gamma(t)=\tilde{\gamma}(-t)=r_{t_{\rm s}}(t)$. In this case, $q_{\rm dc}=0$, $\zeta(0)=g_{\zeta}=t_{\rm s}$, $\zeta(it_{\rm s})=0$ for all $i\neq 0$, $\sigma^2=\tilde{a}_2t_{\rm s}$, $\breve{\sigma}_{0,ik}^2=0$ for all $k\neq i$ and $\sigma_0^2=\tilde{a}_1t_{\rm s}$. This satisfies the Nyquist criterion \eqref{Nyquist} and condition \eqref{IDNoiseCondition} and the channel can be described as in Definition \ref{Def:GaussianInputDep}.
\end{example}

%Since noise in the model in Definition \ref{Def:GaussianInputDep} has a conditionally Gaussian distribution, we call this channel a {\bf conditionally-Gaussian channel}. 

The capacity of this channel has been studied in \cite{Moser,WangYangWangChen,WangWangChenWang,
AminianGhourchianGohariMirmohseniNasiri-Kenari,GhourchianAminianGohariMirmohseniKenari,
DadamahallehHodtani,SoltaniRezki,SoltaniRezkiChaaban}. In general, for channels with input-dependent noise and with peak-constrained inputs, \cite{ElmoslimanyDuman} showed that the capacity-achieving distribution is discrete with a finite alphabet.  {However, when only an average intensity constraint is considered, the capacity achieving distribution turns out to have a countably infinite support set \cite{SoltaniRezki_CL2020}.}

\subsection{The Input-Independent Gaussian Noise Channel}
Finally, we arrive at what is perhaps the most common model in the recent OWC literature, which is the discrete-time Gaussian channel model where noise is input-independent. This can be obtained by imposing an additional constraint on the discrete-time input-dependent Gaussian channel model in Definition \ref{Def:GaussianInputDep}. In particular, we require that 
\begin{align}
\label{IIndNoiseCondition}
\breve{\sigma}^2_{0,ik} &\ll \sigma^2,\ \forall k\neq i,\text{ and } \sigma_0^2\ll \sigma^2,
\end{align} 
instead of \eqref{IDNoiseCondition}. In other words, we require that all input-dependent noise components in \eqref{SampledFilteredOutput} are negligible with respect to the background-plus-thermal noise $z_i$ in \eqref{eq:qrtildeits}. One way this can take place is if $\tilde{a}_1\ll\tilde{a}_2$. The following example illustrates this possibility.

\begin{example}
Consider the parameters used in Example \ref{Eg:Noiseless}, in addition to $\eta_{\rm eo}=0.5$~A/W, $\eta_{\rm eo}=0.5$~W/A, $p_{\rm bg}=50$~mW, and $s_{\rm th}=10^{-20}$A$^2$s. In this scenario, we have $\tilde{a}_1=\frac{g_{\rm g}{\sf e}_{\lambda}}{\eta_{\rm eo}}\approx 4.6\times10^{-21}$, while $\tilde{a}_2=\frac{{\sf e}_{\lambda}p_{\rm bg}}{\eta_{\rm eo}^2}+\frac{s_{\rm th}}{\eta_{\rm eo}^2\eta_{\rm oe}^2}\approx2\times 10^{-19}$ which is 2 orders of magnitude larger than $\tilde{a}_1$.
\end{example}

Generally, the background-plus-thermal noise is stronger than the input dependent noise in most terrestrial OWC applications. This forms the basis of the above approximation, leading the following model.

\begin{definition}[Input-Independent Gaussian Channel]
\label{Def:GaussianInputInDep}
Under conditions \eqref{Nyquist} and \eqref{IIndNoiseCondition}, the discrete-time electrical channel corresponding to an IM/DD OWC system can be modelled as a channel with input $X$ and output $Y=gX+Z$, where $Z\sim\mathcal{N}(0,\sigma^2)$.
\end{definition}

For brevity, we call this channel model an {\it IM/DD Gaussian channel} henceforth. Its capacity has been studied in \cite{LapidothMoserWigger,FaridHranilovic,FaridHranilovic_SelectedAreas,
ChaabanMorvanAlouini,WangHuWang_Dimmable,XuYuZhuSun} among others. 

\begin{remark}
\ac{The approximation given in Def. \ref{Def:GaussianInputInDep} looses its accuracy if the mean of the received number of photons is small, or if the power of input-dependent noise due to fluctuations of the number of photons that reach the receiver from the transmitter is large relative to background noise and thermal noise. While the former is uncommon in terrestrial applications, the latter may take place in some scenarios, and hence care must be taken when using this approximation.}
\end{remark}

Now that we have arrived at the final channel model, we are ready to introduce the input constraints.

\section{Optical Transmission Constraints}
\label{Sec:Constraints}
We talked about the amplitude resolution of the system and how this is impaired by noise. Despite noise, the channel capacity can still be infinite if we do not have constraints on the transmit signal. However, this can not happen in practice due to practical constraints. So what type of constraints apply in IM/DD OWC?

\subsection{Intensity Constraints}
We focus on the constraints of the IM/DD Gaussian channel model in Def. \ref{Def:GaussianInputInDep} since this is the main focus of this tutorial. For the above discussion, we have the following relation between the electric current and the optical power
\begin{align}
%\bar{n}_{\rm t}(i)&=\frac{p_{\rm t}(i)}{{\sf e}}\Delta t,\quad\text{Definition \ref{Def:Poisson} and \ref{Def:GaussianOptical},}\\
%q_{\rm t}(t)&=\frac{p_{\rm t}(t)}{\eta_{\rm eo}},\qquad \text{Definition \ref{Def:GaussianContinuous},}\\
\label{currentOpticalPowerRelation}
\sum_{k=1}^{\kappa} x_k\gamma(t-kt_{\rm s})+q_{\rm dc}&=\frac{p_{\rm t}(t)}{\eta_{\rm eo}}.%,\qquad \text{Definition \ref{Def:GaussianInputDep} and \ref{Def:GaussianInputInDep}.}
\end{align}
The optical power signal $p_{\rm t}(t)$ has to satisfy two types of constraints due to practical considerations and safety standards: Average and/or peak constraints. We start by deriving the average constraint.

\subsubsection{Average Intensity Constraint}
Eye safety limitations are generally expressed in terms of exposure duration at a specific optical power \cite[Table 6.1]{Sliney1993}. Illumination constraints which are relevant in VLC are expressed in terms of Lumens per square meter (Lux) (see \cite{osti_10125246} e.g.) which translates to a constraint on the optical power per unit area. Both constraints can be satisfied if we constrain the energy emitted over a given transmission duration, i.e.,
\begin{align}
\int_{0}^{\tau} p_{\rm t}(t){\rm d}t &\leq \tau \bar{p}_{\rm t,max},
\end{align} 
for some average power $\bar{p}_{\rm t,max}$, and $\tau>0$. Combining this with \eqref{currentOpticalPowerRelation} leads to
\begin{align}
\int_{0}^{\tau}\sum_{k=1}^{\kappa} x_k\gamma(t-kt_{\rm s}){\rm d}t+\tau q_{\rm dc}&\leq\frac{\tau \bar{p}_{\rm t,max}}{\eta_{\rm eo}}.
\end{align}

We shall assume that $\tau=\kappa t_{\rm s}$ in the following arguments by choosing $\kappa$ accordingly. Note that $\kappa$ can still be large since the symbol duration $t_{\rm s}$ can be made very small in nowadays' technologies. Then we have
\begin{align}
\frac{1}{\kappa}\int_{0}^{\tau}\sum_{k=1}^{\kappa} x_k\gamma(t-kt_{\rm s}){\rm d}t&\leq \frac{t_{\rm s} \bar{p}_{\rm t,max}}{\eta_{\rm eo}}-t_{\rm s}q_{\rm dc}.
\end{align}
Note that
\begin{align}
\int_{0}^{\tau}\sum_{k=1}^{\kappa} x_k\gamma(t-kt_{\rm s}){\rm d}t
&\leq \int_{-\infty}^{\infty}\sum_{k=1}^{\kappa} x_k\gamma(t-kt_{\rm s}){\rm d}t\\
&= \sum_{k=1}^{\kappa} x_k {\sf e}_{\gamma},
%&\approx \kappa \mathbb{E}[X] {\sf e}_{\gamma}
\end{align}
where ${\sf e}_{\rm \gamma}=\int_{-\infty}^{\infty}\gamma(t){\rm d}t$, and where the inequality is fairly tight for large $\tau$ and for functions $\gamma(t)$ whose energy is concentrated around $t=0$ (which is common in pulse-shaping). %, and where $X$ is a random variable which represents $x_k$. The approximation follows by the law of large numbers and is precise for large $\kappa$. 
Consequently, the average constraint becomes
\begin{align}
\label{AverageConst0}
%\mathbb{E}[X]\leq \frac{t_{\rm s} \left(\frac{\bar{p}_{\rm t,max}}{\eta_{\rm eo}}-q_{\rm dc}\right)}{{\sf e}_{\gamma}}\triangleq \mathcal{E}.
\frac{1}{\kappa}\sum_{k=1}^{\kappa}x_k\leq \frac{t_{\rm s} \left(\frac{\bar{p}_{\rm t,max}}{\eta_{\rm eo}}-q_{\rm dc}\right)}{{\sf e}_{\gamma}}\triangleq \mathcal{E}.
\end{align}

%If we represent $x_k$ by a random variable $X$, and choose $\kappa$ large, then we have
%\begin{align}
%\label{AverageConst}
%\mathbb{E}[X]\leq\mathcal{E},
%\end{align}
%which is the average constraint that is common in IM/DD OWC systems. We shall use the representations \eqref{AverageConst0} and \eqref{AverageConst} interchangeably in the sequel.

\subsubsection{Peak Intensity Constraint}
A peak constraint also arises due to safety constraints \cite{Boucouvalas_Safety} which induce a constraint of the type 
\begin{align}
p_{\rm t}(t)\leq p_{\rm t,max}, \ \forall t.
\end{align} 
Practical operation also induces peak constraints. Namely, the assumed linear optical power--current relationship (Fig. \ref{Fig:LI}) holds true in an interval of driving currents $[q_{\rm min},q_{\rm max}]$ where $q_{\rm min}$ is the threshold current beyond which light emission starts,\footnote{The threshold current $q_{\rm min}$ can be ignored from a communications perspective because this is a constant bias that has to be applied to operate the device.} and $q_{\rm max}$ is the maximum current beyond which the optical power nearly saturates (or the device burns). This $q_{\rm max}$ sets a peak constraints on $q_{\rm t}(t)$ and hence also $p_{\rm t}(t)$ for all $t$. 

To satisfy $p_{\rm t}(t)\leq p_{\rm t,max}$ for all $t$, the symbols $x_k$ have to be bounded. %Suppose that $x_k\leq\mathcal{A}$ for all $k$ and let us calculate $\mathcal{A}$ for which the constraint $p_{\rm t}(t)\leq p_{\rm t,max}$ is satisfied with equality. 
The maximum emitted optical power is given by 
\begin{align}
\max_t p_{\rm t}(t)&= \eta_{\rm eo}\max_{t,x_k}\sum_{k=1}^{\kappa} x_k\gamma(t-kt_{\rm s})+\eta_{\rm eo}q_{\rm dc}\\
&=\eta_{\rm eo}x_{\rm max}\max_t\sum_{k=1}^{\kappa}\gamma(t-k t_{\rm s})+\eta_{\rm eo}q_{\rm dc},
\end{align} 
where $x_{\rm max}$ is the largest value $x_k$ can take. Denoting $\max_t\sum_{k=1}^{\kappa}\gamma(t-k t_{\rm s})$ by $\gamma_{\Sigma, \rm max}$, and since $p_{\rm t}(t)\leq p_{\rm t,max}$ implies $\max_t p_{\rm t}(t)\leq p_{\rm t,max}$, we obtain
\begin{align}
x_{\rm max}\leq \frac{\frac{p_{\rm t, max}}{\eta_{\rm eo}}-q_{\rm dc}}{\gamma_{\Sigma,\rm max}}.
\end{align}
This leads to the peak constraint
\begin{align}
\label{PeakConst}
x_k&\leq \frac{\frac{p_{\rm t, max}}{\eta_{\rm eo}}-q_{\rm dc}}{\gamma_{\Sigma,\rm max}}\triangleq \mathcal{A},\ \forall k.
\end{align}

As a result, we have the following definition.

\begin{definition}[Average and Peak Constraints]
The input $(x_1,x_2,\ldots,x_\kappa)$ of the IM/DD channels in Definitions \ref{Def:GaussianInputDep} and \ref{Def:GaussianInputInDep}, where $x_k$ is the transmit symbol at time $k$, is constrained by an average and/or a peak constraint given by 
\begin{align}
\label{AverageConst2}
\frac{1}{\kappa}\sum_{k=1}^{\kappa}x_k&\leq\mathcal{E},\\
%\mathbb{E}[X]&\leq \mathcal{E},\\
x_k&\leq \mathcal{A},\ \forall k\in\{1,\ldots,\kappa\},
\end{align}
where $\mathcal{E}$ and $\mathcal{A}$ are as defined in \eqref{AverageConst0} and \eqref{PeakConst}, respectively.
\end{definition}

\begin{remark}
We shall see in Sec. \ref{Sec:SISO_Cap} that constraint \eqref{AverageConst2} is equivalent to 
\begin{align}
\label{AverageConst}
\mathbb{E}[X]\leq\mathcal{E}
\end{align}
from a capacity perspective, where $X$ is a random variable representing the input. 
\end{remark}

At this point, the model of the static discrete-time Gaussian channel is complete. Static here means that the channel gain $g$ is constant. We will discuss time-variations of $g$ in subsection \ref{Sec:Fading}. Next, we will discuss the transmission bandwidth.

\subsection{Transmission Bandwidth}
In the above derivations of the discrete-time channel models, we have referred to a symbol duration $t_{\rm s}$. The capacity of these discrete-time channels in terms of bits/symbol is finite as long as the constraints $\mathcal{E}$ and $\mathcal{A}$ and the noise variances $\sigma^2$ and $\sigma_0^2$ are finite. Thus, they have finite `amplitude resolution'. However, we can still approach infinite capacity in bits/second by decreasing $t_{\rm s}$ to an arbitrarily small value. This infinite `temporal resolution' is impossible in practice because devices set a limitation on $t_{\rm s}$. 

Photonic devices have a limited bandwidth. Ideally, a photonic device can be seen as a filter with an optical/electrical input and an electrical/optical output. This filter has a bandwidth, know as its {\it modulation bandwidth}, which typically depends on the device and can be as large as tens or hundreds of MHz, and can even be  above 1GHz for laser diodes \cite{WangHaldarLiMendis}. Transmitter and receiver circuits also have limited bandwidth. The overall bandwidth of the channel is defined by the interplay of the transfer functions of its components. This channel bandwidth sets a minimum $t_{\rm s}$ that can be supported by the transmitter-receiver system. 

For the purpose of this tutorial, since we deal with discrete-time systems, we will measure the bandwidth in terms of $t_{\rm s}$ as $b=\frac{1}{t_{\rm s}}$. In other words, a system with bandwidth $b$ can send at most $\frac{1}{b}$ symbols per second (Baud rate).

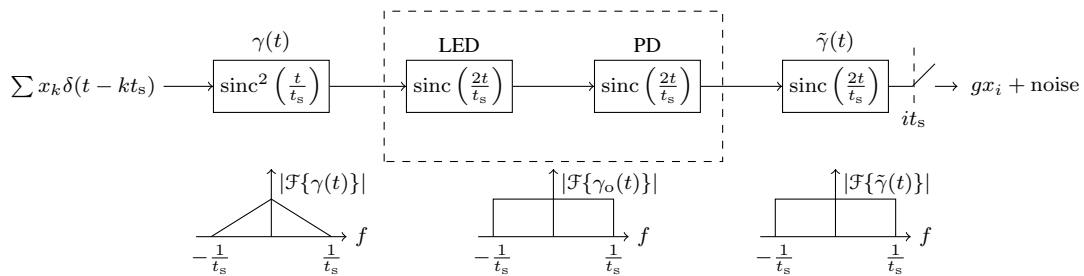
\begin{figure*}[t]
\centering
\begin{tikzpicture}
\footnotesize
\node (s) at (-.5,0) [draw] {${\rm sinc}^2\left(\frac{t}{t_{\rm s}}\right)$};
\node (led) at (2,0) [draw] {${\rm sinc}\left(\frac{2t}{t_{\rm s}}\right)$};
\node (pd) at (4.5,0) [draw] {${\rm sinc}\left(\frac{2t}{t_{\rm s}}\right)$};
\node (d) at (7,0) [draw] {${\rm sinc}\left(\frac{2t}{t_{\rm s}}\right)$};
\node (x) at (-3,0) {$\sum x_k\delta(t-kt_{\rm s})$};
\node (y) at (8.5,0) [anchor=west] {\ \ $gx_i+{\rm noise}$};
\node at ($(y.south east)-(2.3,0)$) [below] {$it_{\rm s}$};
\draw[dashed] ($(y.south east)-(2.3,0)$) to ($(y.north east)+(-2.3,.3)$);
\draw[->] (x) to (s);
\draw[->] (s) to (led);
\draw[->] (led) to (pd);
\draw[->] (pd) to (d);
\draw[-] (d) to ($(y.center)-(1.4,0)$) to ($(y.center)+(-1.4+.3,.3)$);
\draw[->] ($(y.center)+(-1.4+.3,0)$) to ($(y.center)+(-1.4+.6,0)$);
\node at (s.north) [above] {$\gamma(t)$};
\node at (led.north) [above] {LED};
\node at (pd.north) [above] {PD};
\node at (d.north) [above] {$\tilde{\gamma}(t)$};

\node (fs) at ($(s)-(0,2)$) {};
\draw[->] ($(fs)-(1,0)$) to ($(fs)-(-1,0)$);
\draw[->] ($(fs)-(0,0)$) to ($(fs)-(0,-.9)$);
\draw ($(fs)-(.8,0)$) to ($(fs)+(0,.5)$) to ($(fs)-(-.8,0)$);
\node at ($(fs)-(.8,.3)$) {$-\frac{1}{t_{\rm s}}$};
\node at ($(fs)-(-.8,.3)$) {$\frac{1}{t_{\rm s}}$};
\node at ($(fs)-(-1.2,0)$) {$f$};
\node at ($(fs)+(.0,.7)$) [anchor=west] {$|\mathcal{F}\{\gamma(t)\}|$};

\node (fl) at ($.5*(led)+.5*(pd)-(0,2)$) {};
\draw[->] ($(fl)-(1,0)$) to ($(fl)-(-1,0)$);
\draw[->] ($(fl)-(0,0)$) to ($(fl)-(0,-.9)$);
\draw ($(fl)-(.8,0)$) to ($(fl)-(.8,-.5)$) to ($(fl)+(.8,.5)$) to ($(fl)+(.8,0)$);
\node at ($(fl)-(.8,.3)$) {$-\frac{1}{t_{\rm s}}$};
\node at ($(fl)-(-.8,.3)$) {$\frac{1}{t_{\rm s}}$};
\node at ($(fl)-(-1.2,0)$) {$f$};
\node at ($(fl)+(.0,.7)$) [anchor=west] {$|\mathcal{F}\{\gamma_{\rm o}(t)\}|$};

\node (fd) at ($(d)-(0,2)$) {};
\draw[->] ($(fd)-(1,0)$) to ($(fd)-(-1,0)$);
\draw[->] ($(fd)-(0,0)$) to ($(fd)-(0,-.9)$);
\draw ($(fd)-(.8,0)$) to ($(fd)-(.8,-.5)$) to ($(fd)+(.8,.5)$) to ($(fd)+(.8,0)$);
\node at ($(fd)-(.8,.3)$) {$-\frac{1}{t_{\rm s}}$};
\node at ($(fd)-(-.8,.3)$) {$\frac{1}{t_{\rm s}}$};
\node at ($(fd)-(-1.2,0)$) {$f$};
\node at ($(fd)+(.0,.7)$) [anchor=west] {$|\mathcal{F}\{\tilde{\gamma}(t)\}|$};

\draw[dashed] ($(led)-(1,1)$) rectangle ($(pd)+(1,1)$);
\end{tikzpicture}
\caption{An exemplary IM/DD OWC system (Example \ref{Eg:Bandwidth}) with ${\rm sinc}^2$ pulse shaping, an LED, photodetector (PD), and receiver filter with ideal low-pass response ($\gamma_{\rm o}(t)$ is the optical channel response). In this system, we can send $b=\frac{1}{t_{\rm s}}$ symbols per second.}
\label{Fig:z}
\end{figure*}

\begin{example}
\label{Eg:Bandwidth}
Suppose that the light emitter and light detector have a response which can be assumed as an ideal low-pass filter with cut-off frequency $\frac{1}{t_{\rm s}}$ (or larger), and the receiver uses an ideal low-pass filter with impulse response $\tilde{\gamma}(t)={\rm sinc}\left(\frac{2t}{t_{\rm s}}\right)$. In this case, the minimum-bandwidth optical intensity Nyquist pulse (satisfies \eqref{Nyquist}) is $\gamma(t)={\rm sinc}^2\left(\frac{t}{t_{\rm s}}\right)$ where ${\rm sinc}(x)=\frac{\sin(\pi x)}{\pi x}$ \cite{Hranilovic_Pulses} (Fig. \ref{Fig:z}). This system can send $\frac{1}{t_{\rm s}}$ symbols per second, and hence its bandwidth is $b=\frac{1}{t_{\rm s}}$.
\end{example}

\begin{remark}
Note that this model accommodates all unipolar transmission schemes that send $b$ symbols per second, including unipolar PAM sending $b$ symbols per second \cite{LapidothMoserWigger,FaridHranilovic,FaridHranilovic_SelectedAreas,
ChaabanMorvanAlouini}, unipolar OFDM schemes with a sampling rate of $b$ samples per second \cite{LiVucicJungnickelArmstrong, DissanayakeArmstrong, ArmstrongLowery, FernandoHongViterbo, LamWilsonElgalaLittle}, and PPM with an `on' pulse duration of $\frac{1}{b}$ seconds \cite{ElmirghaniCryan,QuirkGin,LetzepisFabregas,KazemiMostaaniUysalGhassemlooy,
WilsonBrandtPearceCaoLeveque,WilsonBrandtPearceCaoBaedke}.
\end{remark}

\subsection{Constraints due to Channel Variations}
\label{Sec:Fading}
We have assumed at the beginning of our analysis that the system is static. We have then described the optical channel response by a constant $g_{\rm g}$ (Definition \ref{Def:GaussianContinuous}). These assumptions do not hold in general. The channel is generally time varying, and its variation can be represented using several statistical models. We will not delve deep into this topic, since this has been discussed thoroughly in \cite{KhalighiUysal}, but we will focus on its impact on the channel model and on coding.

Firstly, the optical channel response is in general a function of time $\gamma_{\rm o}(t)$ which includes reflections due to multipath propagation \cite{BarryKahnKrause}. However, reflections are normally weak, and the channels delay spread (nanoseconds) is significantly smaller than the symbol duration (microseconds). Thus, the response $\gamma_{\rm o}(t)$ can be assumed equal to $\delta(t)$, and $\mathcal{F}\{\gamma_{\rm o}(t)\}$ is flat in the regime of operation \cite{KahnBarry}. This allows abstracting the channel response as a constant $g_{\rm g}$.\footnote{Note that if reflections are significant, then, assuming perfect equalization, the overall channel from $x_i$ to output after equalization can still be modelled as a constant $g_{\rm g}$.}

Secondly, the number of received photons will be affected by physical propagation phenomena such as scattering, absorption, and refraction. These effects are generally combined under an {\it atmospheric turbulence} coefficient $g_{\rm a}$, which varies with time. In addition to this, the received number of photons will also depend on the alignment of the transmitter and receiver, or the lack thereof. This effect is captured by a pointing error term $g_{\rm p}$. Thus, instead of the channel gain $g_{\rm g}$, we would have an effective channel gain of $g=g_{\rm g}g_{\rm a}g_{\rm p}$.

The variables $g_{\rm a}$ and $g_{\rm p}$ are random. Several distributions have been used to model their statistics. For instance, the atmospheric turbulence coefficient $g_{\rm a}$ can follow a log-normal distribution under weak turbulence \cite{KhalighiUysal}, a Gamma-Gamma distribution under moderate-to-strong turbulence\cite{GarciaZambranaCastilloVazquez}, an exponential distribution or $K-$distribution under strong turbulence \cite{TsiftsisSandalidisKaragiannidisUysal}, or a M{\'a}laga distribution in general \cite{Navas_Malaga,AnsariYilmazAlouini}. The pointing error follows a Rayleigh distribution \cite{Arnon03,Liu} or more generally a Hoyt distribution \cite{GappmairHranilovicLeitgeb,AlQuwaieeYangAlouini}. 

Although the effective channel gain varies with time, this time variation is much slower than the symbol rate. For instance, while the symbol duration can be in the range of microseconds, the coherence time of the channel can be in the range of milliseconds \cite{KhalighiUysal}. Thus, it is reasonable to assume that the channel remains constant throughout a transmission block \cite{FaridHranilovic_Outage}. Consequently, to encode over this constant channel, the codewords have to be shorter than the coherence time of the channel. However, this is not a crucial constraint since the coherence time is orders of magnitude larger than the symbol duration.

Another constraint that arises due to channel variations is related to channel state information (CSI). CSI can be obtained at the receiver using a channel estimation mechanism to estimate $g$. The CSI at the transmitter (CSIT) can be available or unavailable. CSIT can be acquired using a feedback channel from the receiver to the transmitter \cite{Djordjevic,KhalighiUysal}. Since the channel varies slowly, the CSIT acquisition overhead can be neglected in performance evaluation. In the absence of a feedback channel, the system is said to have no CSIT, in which case the performance is studied in terms of outage probability {or the compound channel capacity.}

In what follows, we assume the channel $g$ to be static and known globally. Next, we discuss the capacity of the single-user point-to-point (P2P) IM/DD Gaussian channel.

\begin{figure}[t]
\centering
\begin{tikzpicture}
\node (tx) at (-1,0) [draw,rectangle,minimum height=1.5cm,minimum width=1.5cm,fill=gray!50!white] {};
\node at ($(tx)-(1.2,0)$) {Tx};
\node (rx) at (4.7,0) [draw,rectangle,minimum height=1.6cm,minimum width=1.5cm,fill=gray!50!white] {};
\node at ($(rx)+(1.2,0)$) {Rx};

\draw[-] ($(tx.east)+(0,0)$) to ($(tx.east)+(0.2,.2)$) to ($(tx.east)+(0.2,-.2)$) to ($(tx.east)+(0,0)$);

\draw[-] ($(rx.west)+(0,0)$) to ($(rx.west)+(-0.2,.2)$) to ($(rx.west)+(-0.2,-.2)$) to ($(rx.west)+(0,0)$);

\draw[-,dashed,fill=yellow,opacity=.3] ($(tx.east)+(0.2,.2)$) %to ($(tx.east)+(1.5,1)$) 
to ($(rx.west)+(-0.2,+1)$)  
to ($(rx.west)+(-0.2,-1)$)  to ($(tx.east)+(0.2,-.2)$) ;

\node (x1) at (0-1,0) {\footnotesize $X$};
\node (z1) at (4.4,.5+.1) {\footnotesize $Z$};
\node (y1) at (5.2,0) {\footnotesize $Y$};
\node (p1) at (4.4,0) {};
\node at (p1) {$\oplus$};

\draw[->] (x1) to ($(tx.east)+(0,0)$);
\draw[->] ($(tx.east)+(0.2,0)$) to node {\scriptsize \contour{yellow!30!white}{$g$}} ($(rx.west)+(-0.2,0)$);%(p1.west);
\draw[->] ($(z1.south)+(0,.1)$) to (p1.north);
\draw[->] (p1.east) to (y1.west);
\draw[->] ($(rx.west)+(0,0)$) to (p1.west);

\end{tikzpicture}
\caption{An IM/DD Gaussian channel.}
\label{Fig:SISOChannel}
\end{figure}
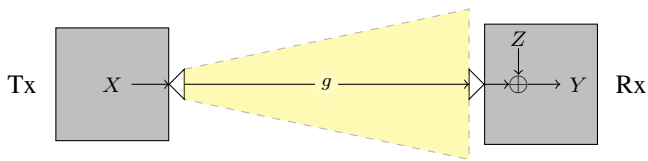

\section{Capacity of the Single User IM/DD Gaussian Channel}
\label{Sec:SISO_Cap}
\ac{We focus on the IM/DD Gaussian channel described in Def.~\ref{Def:GaussianInputInDep}, whose input-output relation is given by (see Fig. \ref{Fig:SISOChannel})}
\begin{align}
Y_i=gX_i+Z_i,
\end{align}
where at time $i$, $Y_i$ is the output, $Z_i\sim\mathcal{N}(0,1)$ is i.i.d. noise, $g\geq0$ is the channel coefficient which is fixed and known at both nodes, and $X_i$ is the input. We choose the noise variance $\sigma^2=1$ without loss of generality since this can be absorbed into $g$ by normalization. For an input signal $(x_1,\ldots,x_n)$ where $n$ is the number of transmissions, the following constraint must be satisfied
\begin{align}
0\leq x_i & \leq\mathcal{A},\ \forall i\in\{1,\ldots,n\}\\
\frac{1}{n}\sum_{i=1}^{n}x_i&\leq \mathcal{E}.
\end{align}
\ac{Before we discuss the capacity of this channel, we present some preliminaries}

\subsection{Information-Theoretic Preliminaries}
\ac{This section relies heavily on some information-theoretic quantities which are introduced here. For a more comprehensive overview, the reader is referred to \cite{CoverThomas}.} 

\ac{Consider a discrete random variable $X\in\mathcal{X}$ with distribution $\mathbb{P}_X(x)$. The entropy $H(X)$ of this random variable is defined as
\begin{align}
H(X) = -\sum_{x\in\mathcal{X}}\mathbb{P}_X(x)\log(\mathbb{P}_X(x)).
\end{align}
This quantity measures the uncertainty in $X$, and is maximum if $X$ is uniform over $\mathcal{X}$.}

\begin{example}
\ac{Given $X\sim{\rm Bern}(\nicefrac{1}{2})$ and $Y\sim{\rm Bern}(\nicefrac{1}{10})$, then $H(X)=\log(2)=0.6931$ nats ($1$ bit) whereas $H(Y)=0.3251$ nats ($0.469$ bits). Thus, $X$ is more uncertain that $Y$.}
\end{example}

\ac{Given another discrete random variable $Y\in\mathcal{Y}$, so that $(X,Y)$ is distributed according to $\mathbb{P}_{X,Y}(x,y)=\mathbb{P}_X(x)\mathbb{P}_{Y|X}(y|x)$, the uncertainty of $Y$ given $X$ is measured by the conditional entropy
\begin{align}
H(Y|X)=-\sum_{x\in\mathcal{X}}\sum_{y\in\mathcal{Y}}\mathbb{P}_{X,Y}(x,y)\log(\mathbb{P}_{Y|X}(y|x)),
\end{align}
and the uncertainty of the pair $(X,Y)$ is given by the joint entropy
\begin{align}
H(X,Y)=-\sum_{x\in\mathcal{X}}\sum_{y\in\mathcal{Y}}\mathbb{P}_{X,Y}(x,y)\log(\mathbb{P}_{X,Y}(x,y)).
\end{align}
Also, the amount of information that $Y$ contains about $X$ can be measured by the mutual information defined as
\begin{align}
I(X;Y)=H(Y)-H(Y|X)=H(X)-H(X|Y).
\end{align}
This is the amount of reduction of uncertainty in $Y$ when $X$ is observed, or vice versa, the amount of reduction of uncertainty in $X$ when $Y$ is observed.}

\begin{example}
\ac{Given $X\sim{\rm Bern}(\nicefrac{1}{2})$, $Y_1=X+Z\bmod2$ where $Z\sim{\rm Bern}(\nicefrac{1}{10})$, then 
$Y_1\sim{\rm Bern}(\nicefrac{1}{2})$ and $Y_1$ given $X$ is ${\rm Bern}(\nicefrac{1}{10})$, thus $I(X;Y_1)=H(Y_1)-H(Y_1|X)=\log(2) - 0.3251=0.368$ nats. Thus, $Y_1$ `reveals' $0.368$ nats of uncertainty about $X$. If on the other hand $Y_2=X$, then $I(X;Y_2)=\log(2)$ since $H(Y_2|X)=0$,\footnote{\ac{We use the convention $0\log(0)=0$.}} i.e., $Y_2$ `reveals' all uncertainty about $X$. If $Y_3$ is independent of $X$, then $I(X;Y_3)=0$ because $H(Y_3|X)=H(Y_3)$, i.e., $Y_3$ does not tell us anything about $X$.}
\end{example}

\ac{Analogous quantities for continuous random variables $X$ and $Y$ are defined as the differential entropy, differential joint entropy, differential conditional entropy, and mutual information, given respectively as
\begin{align}
h(X) &= -\int_{\mathcal{X}}\mathbb{P}_X(x)\log(\mathbb{P}_X(x)){\rm d}x,\\
h(Y|X)&=-\int_{\mathcal{X}\times\mathcal{Y}}\mathbb{P}_{X,Y}(x,y)\log(\mathbb{P}_{Y|X}(y|x)){\rm d}x{\rm d}y,\\
h(X,Y)&=-\int_{\mathcal{X}\times\mathcal{Y}}\mathbb{P}_{X,Y}(x,y)\log(\mathbb{P}_{X,Y}(x,y)){\rm d}x{\rm d}y,\\
I(X;Y)&=h(Y)-h(Y|X)=h(X)-h(X|Y),
\end{align}
assuming the integrals exist.}

\ac{The entropy and mutual information satisfy the following properties that will be needed in the sequel.
\begin{itemize}
\item P1: $h(Y|X)=h(Y-f(X)|X)$ \acc{for any deterministic mapping $f(\cdot)$} \cite[Thm. 8.6.3]{CoverThomas};
\item P2: Conditioning does not increase entropy: $h(X_1|X_2)\leq h(X_1)$ with equality if and only if $X_1$ and $X_2$ are independent \cite[p. 253]{CoverThomas};
\item P3: Chain rule: $h(\X)=\sum_{i=1}^nh(X_i|X_1,\ldots,X_{i-1})$ \cite[Thm. 8.6.2]{CoverThomas};
\item P4: Concavity of mutual information in the input distribution: $I(X;Y)$ is concave in $\mathbb{P}_{X}$ for a given $\mathbb{P}_{Y|X}$ \cite[Thm. 2.7.4]{CoverThomas}.
\end{itemize}
Next, we describe how communication is realized over the IM/DD Gaussian channel.}

\subsection{The Communication Problem}
\ac{After modelling an IM/DD OWC system (in the infra-red, visible-light, or ultra-violet ranges) as an IM/DD Gaussian channel, we describe communication over this channel.} Communication over the IM/DD Gaussian channel  can be generally described as follows. The transmitter wants to send a message of $m$ bits to the receiver. The set of all possible messages has size $2^m$. Without loss of generality, the messages can be labelled by integers from $1$ to $2^m$, and the set of all messages can be denoted $\mathcal{W}=\{1,\ldots,2^m\}$. Thus, the message can be represented by a random variable $W\in\mathcal{W}$. The transmitter desires to `load' $W$ with the largest number of information bits. This is achieved when $W$ is uniformly distributed on $\mathcal{W}$, since a uniformly distributed random variable has the largest `information content' measured by its entropy. For a uniform random variable $W$ distributed on $\mathcal{W}$, the information content is $H(W)=m$. Thus, we can assume that the transmitter picks a message $w$ uniformly at random from $\mathcal{W}$.

To send a message $w$, the transmitter assigns a codeword $\x(w)$ from a codebook of $2^m$ codewords, where $\x(w)=(x_1(w),\ldots,x_n(w))$ and $n$ is the code length (Table \ref{Tab:Codebook}). For all $w\in\mathcal{W}$, the codewords satisfy $x_i(w)\in[0,\mathcal{A}]$ and $\frac{1}{n}\sum_{i=1}^{n}x_i(w)\leq\mathcal{E}$, which satisfies the intensity constraints. Then, to send a message $w$, the transmitter sends $\x(w)$ through the channel in $n$ transmissions (i.e., uses $\x(w)$ to modulate the light source).\footnote{\ac{In a VLC system where a desired lighting level is required, modulating the light source using $\x$ has to occur at a frequency higher than the eye's `flicker fusion threshold' (10s of Hz). This is ensured in practice since the modulation frequency is much larger (in MHz \cite{HAAS_PhyRev}).}}

In this construction, the transmitter sends $m$ bits in $n$ transmissions, for a rate of $\frac{m}{n}$ bits per transmission. To send at rate $r$ bits per transmission, we require that $\mathcal{W}$ has $2^{nr}$ elements, and hence $2^{nr}$ must be an integer. We will assume that this is true henceforth, by proper choice of $n$ and $r$. 

The receiver records the received signal $y_i$ over $n$ transmissions to obtain $\y$. Then it uses a decoder to decide that message $\hat{w}$ has been sent. This process incurs an error probability ${\sf p}_{{\rm e},n}=\mathbb{P}\{w\neq\hat{w}\}$, which is required to be sufficiently small. 

This description of message set $\mathcal{W}=\{1,\ldots,2^{nr}\}$, encoder and decoder defines a channel code, which we denote as a $(2^{nr},n)$ code. We say that a rate $r$ is achievable if there exists a sequence of $(2^{nr},n)$ codes so that ${\sf p}_{{\rm e},n}\to0$ as $n\to\infty$. Indeed, we seek to maximize the achievable rate $r$. The maximum $r$ is the channel capacity denoted $c_g(\mathcal{A},\mathcal{E})$, to indicate its dependence on $g$, $\mathcal{A}$ and $\mathcal{E}$ explicitly.

One can ask several question at this point: First, if the input is constrained by $\frac{1}{n}\sum_{i=1}^{n}x_i\leq \mathcal{E}$, then why does \cite{LapidothMoserWigger,FaridHranilovic,FaridHranilovic_SelectedAreas,ChaabanMorvanAlouini} and many other works study the channel with a constraint $\mathbb{E}[X]\leq\mathcal{E}$? Second, {\it what is the value of $c_g(\mathcal{A},\mathcal{E})$}? Third, {\it can we express $c_g(\mathcal{A},\mathcal{E})$ in a simple form}? We discuss these questions next. 

\begin{table}
\centering
\begin{tabular}{c||c|c|c|c}
Message & \multicolumn{4}{c}{codeword}\\\hline\hline
1 & $x_1(1)$&$x_2(1)$&$\ldots$&$x_n(1)$\\\hline
2 & $x_1(2)$&$x_2(2)$&$\ldots$&$x_n(2)$\\\hline
$\vdots$ & $\vdots$&$\vdots$&$\vdots$&$\vdots$\\\hline
$2^m$ & $x_1(2^m)$&$x_2(2^m)$&$\ldots$&$x_n(2^m)$\\\hline
\end{tabular}
\caption{Codebook consisting of $2^m$ codewords, each of length $n$ symbols. To send message $w$, the transmitter sends the symbols of the codeword $\boldsymbol{x}(w)$ over $n$ transmissions.}
\label{Tab:Codebook}
\end{table}

\subsection{Channel Capacity as Mutual Information Maximization}
\label{Sec:MutualInfoMax}
In his 1948 seminal paper \cite{Shannon}, Shannon derived the capacity of a memoryless channel. \ac{The theory asserts that as long as $r$ is smaller than the channel capacity, no matter how noisy the channel is,  one can send information at a rate of $r$ bits per transmission {\it `error free'}.\footnote{\ac{More specifically, an error rate that vanishes as the code length grows.}} The key is to insert enough redundancy in the codewords so as to be able to correct errors at the receiver. Specifically, Shannon showed that there exists codes which allows `error-free' transmission as long as the rate $r$ is smaller than capacity, and some such codes have been discovered over the past decades such as LDPC codes and Polar codes \cite{Gallager_LDPC,Arikan}. This capacity is given} in the following lemma.

\begin{lemma}[Capacity of a memoryless channel \cite{Shannon}]
\label{Lem:ShannonCapacity}
For a memoryless channel with input $X\in\mathcal{X}$, output $Y\in\mathcal{Y}$, and channel law $\mathbb{P}_{Y|X}$, the channel capacity is given by $\max_{\mathbb{P}_X\in\mathcal{P}_X}I(X;Y)$.
\end{lemma}

Here, $\mathcal{P}_X$ is the set of all $\mathbb{P}_{X}$ defined on $\mathcal{X}$. \ac{Thus, the capacity is the maximum (with respect to $\mathbb{P}_X$) of the amount of information that $Y$ tells us about $X$. This capacity is achievable by a random i.i.d. code, where code symbols are chosen independently at random according to $\mathbb{P}_X$, i.e., the error probability vanishes as $n\to\infty$.}

\begin{example}[BSC Capacity]
\ac{To provide a simple (and rather crude) interpretation of Lemma \ref{Lem:ShannonCapacity}, let us consider a binary-symmetric channel with input $X\in\{0,1\}$, and output $Y=X+Z\mod 2$ where $Z\sim{\rm Bern}(\epsilon)$. Here $\epsilon$ can be interpreted as the probability that the channel flips a bit. Let us consider coding over $n$ transmission. A transmitted codeword $\X(w)=(X_1,\ldots,X_n)$ will be corrupted by $n$ instances of noise $\Z=(Z_1,\ldots,Z_n)$. Since $n$ is large, $\Z$ will likely have around $n\epsilon$ ones and $n(1-\epsilon)$ zeros. Such $\Z$ is said to be `typical' \cite{ElgamalKim}, and the number of typical $\Z$ is around $e^{nH(Z)}$.\footnote{\ac{This is when $H(Z)$ is in nats. If we use bits, then the number of typical $\Z$ is around $2^{nH(Z)}$.}} Thus, $\X(w)$ will be received as one of around $e^{nH(Z)}$ possible corrupted versions of itself. On the other hand, the received signal $\Y=(Y_1,\ldots,Y_n)$ can take any of $2^n=e^{n\log(2)}$ possible values. To make sure the decoder does not confuse corrupted versions of $\X(w)$ with corrupted versions of another codeword $\X(w')$, we should send less than $\frac{e^{n\log(2)}}{e^{nH(Z)}}=e^{n(\log(2)-H(z))}$ codewords. Thus, the rate must be less than $\frac{1}{n}\log\left(e^{n(\log(2)-H(z))}\right)=\log(2)-H(Z)$ bits per transmission, which is exactly $I(X;Y)$ when $X\sim{\rm Bern}(\nicefrac{1}{2})$. This can be achieved by using a random i.i.d. binary code, or a rather more structured LDPC or Polar code \cite{Gallager,Arikan}.}
\end{example}

Lemma \ref{Lem:ShannonCapacity} presents the channel capacity in a `single-letter' form where the time index is obsolete. However, in the channel under consideration, the time index is not obsolete since the constraint $\frac{1}{n}\sum_{i=1}^nx_i\leq\mathcal{E}$ is a time average. In this case, instead of finding the best $\mathbb{P}_X$, we have to find the best $\mathbb{P}_{\X}$ where $\X=(X_1,\ldots,X_n)$. This seems to prevent using Lemma \ref{Lem:ShannonCapacity} directly to express the capacity of the channel under consideration. Nevertheless, we can still apply this lemma in a `multi-letter' form as follows.

Consider the $n$-symbol extended channel $\Y=g\X+\Z$ where $\X=(X_1,\ldots,X_n)$, $Y_i=gX_i+Z_i$, and $Z_i$ is i.i.d. $\mathcal{N}(0,1)$. The input alphabet of this channel is the set $\mathcal{X}^{[n]}=\{\X\in[0,\mathcal{A}]^n| \|\X\|_1\leq n\mathcal{E}\}$, its output alphabet is $\mathbb{R}^n$, and its channel law $\mathbb{P}_{\Y|\X}=\prod_{i=1}^n\mathbb{P}_{Y_i|X_i}(y_i|x_i)=\prod_{i=1}^n\frac{1}{\sqrt{2\pi}}e^{-\frac{(y_i-gx_i)^2}{2}}$. This is a memoryless channel whose capacity is
\begin{align}
c^{[n]}_g(\mathcal{A},\mathcal{E})=\max_{\mathbb{P}_{\X}\in\mathcal{P}_{\X}} I(\X;\Y)
\end{align}
by Lemma \ref{Lem:ShannonCapacity}. Here, $\mathbb{P}_{\X}$ is the distribution of $\X$, and $\mathcal{P}_{\X}$ is the collection of all $\mathbb{P}_{\X}$ defined on $\mathcal{X}^{[n]}$. Since the resulting vector channel is defined as $n$-uses of the scalar channel, its capacity is also equal to $c^{[n]}_g(\mathcal{A},\mathcal{E})=nc_g(\mathcal{A},\mathcal{E})$.\footnote{{This can be shown using standard steps as in \cite[Ch. 7]{CoverThomas}.}} Thus, we have
\begin{align}
\label{P2PCapNuses}
nc_g(\mathcal{A},\mathcal{E})=\max_{\mathbb{P}_{\X}\in\mathcal{P}_{\X}} I(\X;\Y).
\end{align}
The maximization with respect to $\mathbb{P}_{\X}$ is required to determine the distribution of codeword symbols from the alphabet $\mathcal{X}^{[n]}$.

To simplify this \acc{expression} into a single letter form, we note the following. While the constraint $\|\X\|_1\leq n\mathcal{E}$ permits $\mathbb{P}_{\X}$ to be non i.i.d. in general \ac{(such as $X_i=n\mathcal{E}$ and $X_j=0$ $\forall j\neq i$)}, property P4 forces the optimal $\mathbb{}P_{\X}$ to be i.i.d., \acc{as shown next in Theorem \ref{Thm:P2P_IMDD_Capacity}}, \ac{which allows us to replace the constraint} $\|\X\|_1\leq n\mathcal{E}$ with $\mathbb{E}[X_i]\leq\mathcal{E}$ for all $i$.

\begin{theorem}
\label{Thm:P2P_IMDD_Capacity}
The capacity of a channel with input $\X=(X_1,\ldots,X_n)$ satisfying $X_i\in[0,\mathcal{A}]$ for all $i$ and $\|\X\|_1\leq n\mathcal{E}$, and output $Y_i=gX_i+Z_i$, $i=1,\ldots,n$, where $g$ is a constant and $Z_i$ is i.i.d. $\mathcal{N}(0,1)$ is given by
\begin{align}
c_g(\mathcal{A},\mathcal{E})=\max_{\mathbb{P}_{X}\in\mathcal{P}_{X}} I(X;gX+Z),
\end{align}
{when $n$ is large enough,} where $\mathcal{P}_{X}$ is the collection of all $\mathbb{P}_{X}$ defined on $[0,\mathcal{A}]$ with $\mathbb{E}[X]\leq \mathcal{E}$.
\end{theorem}
\begin{IEEEproof}
Starting from \eqref{P2PCapNuses}, we have 
\begin{align}
I(\X;\Y)
&=h(\Y)-h(\Y|\X)\\
&\stackrel{\rm P1}{=} h(\Y)-h(\Z|\X)\\
&\stackrel{\rm P2}{=} h(\Y)-h(\Z)\\
&\stackrel{\rm P3}{=} \sum_{i=1}^nh(Y_i|Y_1,\ldots,Y_{i-1})-h(Z_i|Z_1,\ldots,Z_{i-1})\nonumber\\
&\stackrel{\rm P2}{\leq} \sum_{i=1}^nh(Y_i)-h(Z_i)\\
&= \sum_{i=1}^nI(X_i;Y_i),
\end{align}
\ac{for any distribution $\mathbb{P}_{\X}$ with marginals $\mathbb{P}_{X_i}$.} This leads to
\begin{align}
c
&\leq \max_{\mathbb{P}_{\X}\in\mathcal{P}_{\X}} \frac{1}{n}\sum_{i=1}^n I(X_i;gX_i+Z_i).
\end{align}
\ac{Let the support of $\mathbb{P}_{X_i}$ be $\mathcal{X}_i$ and its mean $\mathbb{E}[X_i]=\mathcal{E}_i$.} %Note that we can write $\mathbb{P}_{X_i}(x)=\mathbb{E}_{\X}[\frac{1}{2\epsilon}\mathbb{I}_{X_i\in[x-\epsilon,x+\epsilon]}]$ with $\epsilon>0$ arbitrarily small. 
Then, using property P4, we have
\begin{align}
\frac{1}{n}\sum_{i=1}^n I(X_i;gX_i+Z_i)
&\stackrel{P4}{\leq} I(\bar{X};g\bar{X}+Z),
\end{align}
where $Z\sim\mathcal{N}(0,1)$, and $\bar{X}$ is a random variable defined on $\bar{\mathcal{X}}=\bigcup_{i=1}^n\mathcal{X}_i$ with distribution $\mathbb{P}_{\bar{X}}(x)=\frac{1}{n}\sum_{i=1}^n\mathbb{P}_{X_i}(x)$. Therefore, we obtain
\begin{align}
c_g(\mathcal{A},\mathcal{E})&\leq \max_{\mathbb{P}_{\X}\in\mathcal{P}_{\X}} I(\bar{X};g\bar{X}+Z).
\end{align}

Note that $\bar{\mathcal{X}}\subseteq[0,\mathcal{A}]$ since $\mathcal{X}_i\subseteq[0,\mathcal{A}]$. Moreover, we have 
\begin{align}
\mathbb{E}_{\bar{X}}[\bar{X}]
&=\int_{\bar{\mathcal{X}}}x\,\frac{1}{n}\sum_{i=1}^n\mathbb{P}_{X_i}(x){\rm d}{x}\\
&=\frac{1}{n}\sum_{i=1}^n \int_{\bar{\mathcal{X}}}x\,\mathbb{P}_{X_i}(x){\rm d}{x}\\
&=\frac{1}{n}\sum_{i=1}^n\mathcal{E}_i \leq \mathcal{E},
\end{align} 
where the last step follows since $\|\X\|_1\leq n\mathcal{E}$ implies that $\mathbb{E}[\|\X\|_1]\leq n\mathcal{E}$ and hence $\frac{1}{n}\sum_{i=1}^n\mathcal{E}_i\leq\mathcal{E}$. 
%Note that $\bar{\mathcal{X}}\subseteq[0,\mathcal{A}]$ since $\mathcal{X}_i\subseteq[0,\mathcal{A}]$. Moreover, we have 
%\begin{align}
%\mathbb{E}_{\bar{X}}[\bar{X}]
%%&=\int_{\bar{\mathcal{X}}}{x}\mathbb{P}_{\bar{X}}({x}){\rm d}{x}\\
%%&=\frac{1}{n}\int_{\bar{\mathcal{X}}}{x}\sum_{i=1}^n\mathbb{P}_{X_i}({x}){\rm d}{x}\\
%&=\frac{1}{n}\int_{\bar{\mathcal{X}}}{x}\sum_{i=1}^n\mathbb{E}_{\X}\left[\frac{1}{2\epsilon}\mathbb{I}_{X_i\in[x-\epsilon,x+\epsilon]}\right]{\rm d}{x}\\
%%&=\frac{1}{n}\int_{\bar{\mathcal{X}}}\bar{x}\mathbb{E}_{\X}\left[\sum_{i=1}^n\mathbb{I}(X_i=\bar{x})\right]{\rm d}\bar{x}\\
%%&=\frac{1}{n}\mathbb{E}_{\X}\left[\int_{\bar{\mathcal{X}}}\bar{x}\sum_{i=1}^n\mathbb{I}(X_i=\bar{x}){\rm d}\bar{x}\right]\\
%&=\frac{1}{n}\mathbb{E}_{\X}\left[\sum_{i=1}^n\int_{\bar{\mathcal{X}}}\frac{x}{2\epsilon}\mathbb{I}_{X_i\in[x-\epsilon,x+\epsilon]}{\rm d}{x}\right]\\
%&=\frac{1}{n}\mathbb{E}_{\X}\left[\sum_{i=1}^nX_i\right]\\
%&\leq \mathcal{E},
%\end{align} 
%where the last step follows since $\|\X\|_1\leq n\mathcal{E}$. 
Let $\mathcal{P}_{\bar{X}}$ be the set of all $\mathbb{P}_{\bar{X}}$ satisfying $\bar{X}\in[0,\mathcal{A}]$ and $\mathbb{E}_{\bar{X}}[\bar{X}]\leq\mathcal{E}$. Then, $\mathbb{P}_{\X}\in\mathcal{P}_{\X}\Rightarrow \mathbb{P}_{\bar{X}}\in\mathcal{P}_{\bar{X}}$. Hence,
\begin{align}
c_g(\mathcal{A},\mathcal{E})&\leq \max_{\mathbb{P}_{\bar{X}}\in\mathcal{P}_{\bar{X}}}  I(\bar{X};g\bar{X}+Z).
\end{align}
 
This upper bound is achievable as $n\to\infty$ using i.i.d. $\X\sim\prod_{i=1}^n\mathbb{P}_{\bar{X}}(x_i)$. The resulting codebook satisfies the constraints for large $n$, since for each codeword $(x_1,\ldots,x_n)$ we have $x_i\in[0,\mathcal{A}]$ for all $i$, and $\lim_{n\to\infty}\frac{1}{n}\sum_{i=1}^n x_i=\mathbb{E}[X]\leq\mathcal{E}$ by the law of large numbers. This concludes the proof.
\end{IEEEproof}

This demonstrates that the constraint $\frac{1}{n}\sum_{i=1}^nx_i\leq\mathcal{E}$ is equivalent to $\mathbb{E}[X]\leq\mathcal{E}$ from a capacity perspective, and $c_g(\mathcal{A},\mathcal{E})=\max_{\mathbb{P}_{X}\in\mathcal{P}_{X}} I(X;gX+Z)$. 

Now some remarks about the constraints are due. One can show that the channel capacity under the constraint $\mathbb{E}[X]=\mathcal{E}$ is equal to that under the constraint $\mathbb{E}[X]=\mathcal{A}-\mathcal{E}$ by symmetry of the Gaussian noise distribution \cite{LapidothMoserWigger,ChaabanMorvanAlouini}. By property P4, one can also show that capacity increases as $\mathbb{E}[X]$ increases from $0$ to $\frac{\mathcal{A}}{2}$ \cite{ChaabanMorvanAlouini}. Hence, the optimal input distribution satisfies $\mathbb{E}[X]=\mathcal{E}$ if $\mathcal{E}\leq \frac{\mathcal{A}}{2}$ and satisfies $\mathbb{E}[X]=\frac{\mathcal{A}}{2}$ otherwise \cite{LapidothMoserWigger,ChaabanMorvanAlouini}. In the latter case, we can show using P4 that an input distribution symmetric with respect to $\frac{\mathcal{A}}{2}$ is optimal.

Due to this, it suffices to study the capacity when $\mathcal{E}\leq\frac{\mathcal{A}}{2}$, with the understanding that a channel with a peak constraint only has the same capacity as a channel with an average and a peak constraint with $\mathcal{E}=\frac{\mathcal{A}}{2}$.

To evaluate capacity, it remains to compute 
\begin{align}
\label{CapMax0}
c_g(\mathcal{A},\mathcal{E})=\max_{\mathbb{P}_{X}\in\mathcal{P}_{X}'} I(X;gX+Z),
\end{align}
where $\mathcal{P}_{X}'$ is the set of distributions $\mathbb{P}_X$ on $[0,\mathcal{A}]$ satisfying $\mathbb{E}[X]=\mathcal{E}\leq\frac{\mathcal{A}}{2}$. \ac{Methods for evaluating this capacity are discussed next.}

\section{Numerical Evaluation of the Capacity of a Single User IM/DD Gaussian channel}
\label{Sec:NumCapacity}
While the objective of the maximization in \eqref{CapMax0} is the same as that in the standard AWGN channel \cite[Ch. 9]{CoverThomas}, the constraint set $\mathcal{P}_{X}'$ is different. This alone makes this problem elusive contrary to the standard AWGN channel for which the optimal input is known to be Gaussian.

However, numerical methods can be used to solve this problem, due to the following result from \cite{ChanHranilovicKschischang}.
\begin{theorem}[Discreteness of Optimal Distribution \cite{ChanHranilovicKschischang}]
\label{Thm:Discreteness}
The capacity-achieving input distribution $\mathbb{P}_{X}$ for the IM/DD Gaussian channel with a peak constraint is a discrete distribution with a finite number of probability mass points.
\end{theorem}

Thus, one can restrict attention to distributions of the form
\begin{align}
\mathbb{P}_{X}^{\rm d}(x)=\sum_{i=1}^{k} a_i\delta(x-x_i)
\end{align}
for some $k$, $x_i\in[0,\mathcal{A}]$, and $a_i>0$ such that $\sum_ia_i=1$, satisfying the constraints on $X$. Thus, to find capacity, we have to solve
\begin{align}\label{CapOpt}
\begin{split}
\max_{k,x_i,a_i} \quad& I(X;gX+Z)|_{X\sim\mathbb{P}_{X}^{\rm d}}\\
{\rm subject\ to} \quad & k\geq2;\quad \sum_{i=1}^ka_i=1; \quad \sum_{i=1}^k a_ix_i\leq\mathcal{E}\\ & x_i\in[0,\mathcal{A}]\quad\forall i=1,\ldots,k,
\end{split}
\end{align}
where
\begin{align}
&I(X;gX+Z)\nonumber\\
&=-\int_{-\infty}^{\infty}\left[\left(\sum_{i=1}^k\frac{a_i}{\sqrt{2\pi}}e^{-\frac{(y-gx_i)^2}{2}}\right)\right.\\
&\qquad\qquad\left.\log\left(\sum_{i=1}^k\frac{a_i}{\sqrt{2\pi}}e^{-\frac{(y-gx_i)^2}{2}}\right)dy\right]-\frac{1}{2}\log(2\pi e).\nonumber
\end{align}
This gives the capacity in nats/transmission. To convert to bits/transmission, we divide by $\log(2)$, or we replace $\log(\cdot)$ with $\log_2(\cdot)$. 

The problem of finding the optimal $\mathbb{P}_X$ involves the following three problems: (i) Finding the optimal $k$, (ii) finding the optimal $x_i$, and (iii) finding the optimal $a_i$. The overall problem is nonlinear, but can be solved numerically. With currently existing numerical solvers, it is common to solve for $a_i$ and $x_i$ jointly for a given $k$ \cite{ChanHranilovicKschischang}. We shall describe this later, but first, let us study a method for finding the optimal $a_i$ for a given $x_i$ and $k$ that dates back to 1972.

\subsection{The Blahut-Arimoto Algorithm: Optimal $a_i$ given $k$ and $x_i$}
If $k$ and $x_i$ are given, then the maximization with respect to $a_i$ can be solved numerically due to property P4, i.e., the concavity of $I(X;gX+Z)$ in $\mathbb{P}_X$. A famous elegant algorithm that solves this maximization is the Blahut-Arimoto algorithm \cite{Blahut,Arimoto}. It is based on a rewriting of the mutual information as follows
\begin{align}
I(X;Y)&= \sum_{x\in\mathcal{X}}\int_{\mathbb{R}}\left[\mathbb{P}_{X,Y}(x,y)\log\left( \frac{\mathbb{P}_{X,Y}(x,y)}{ \mathbb{P}_{X}(x)\mathbb{P}_{Y}(y) } \right){\rm d}y\right],\nonumber\\
&=\sum_{x\in\mathcal{X}}\int_{\mathbb{R}}\mathbb{P}_{X}(x)\mathbb{P}_{Y|X}(y|x)\log\left( \frac{\mathbb{P}_{X|Y}(x|y)}{ \mathbb{P}_{X}(x) } \right){\rm d}y,\nonumber
\end{align}
and recasting the maximization problem as 
\begin{align}
\max_{\mathbb{P}_{X|Y}} \max_{\mathbb{P}_{X}} \sum_{x\in\mathcal{X}}\int_{\mathbb{R}}\mathbb{P}_{X}(x)\mathbb{P}_{Y|X}(y|x)\log\left( \frac{\mathbb{P}_{X|Y}(x|y)}{ \mathbb{P}_{X}(x) } \right){\rm d}y.
\end{align}
{Solving this double-maximization leads to the capacity achieving distribution as shown in \cite[Ch. 10]{CoverThomas}.} This problem is convex in $\mathbb{P}_X$ given $\mathbb{P}_{X|Y}$, and also in $\mathbb{P}_{X|Y}$ given $\mathbb{P}_X$, and hence can be solved by alternating maximization with respect to $\mathbb{P}_X$ and $\mathbb{P}_{X|Y}$. 

To solve this maximization numerically, we discretize the interval $[y_{\rm l},y_{\rm u}]$ where $y_{\rm l}<0$ and $y_{\rm u}> g\mathcal{A}$ are chosen so that $\mathbb{P}\{y<y_{\rm l}\}$ and $\mathbb{P}\{y>y_{\rm u}\}$ are arbitrarily small for any $\mathbb{P}_{X}$ (typically $y_{\rm l}<-3\sigma$ and $y_{\rm u}>g\mathcal{A}+3\sigma$, recall that $\sigma=1$). We define $\mathcal{Y}=\{y_{\rm l}+j\delta|j=0,1,\ldots,\ell\}$ for $0<\delta\ll1$ and $\ell=\left\lceil\frac{y_{\rm u}-y_{\rm l}}{\delta}\right\rceil$. This way, we can discretize $Y$ and describe $\mathbb{P}_{Y|X}$ as a matrix $(p_{j,i})$, $i\in\{1,\ldots,k\}$ and $j\in\{1,\ldots,\ell\}$, where 
\begin{align}
\label{BAA_pji}
p_{j,i}=\mathbb{P}\{Y\in[y_{\rm l}+(j-1)\delta,y_{\rm l}+j\delta) |\text{$x_i$ was sent}\}.
\end{align}
This can be calculated numerically from the channel law $\mathbb{P}_{Y|X}$. Similarly, define 
\begin{align}
\label{BAA_pji}
q_{i,j}=\mathbb{P}\{ \text{$x_i$ was sent}|Y\in[y_{\rm l}+(j-1)\delta,y_{\rm l}+j\delta)\}.
\end{align}

Now we can recast the optimization as 
\begin{align}
\max_{q_{i,j}} \max_{a_i} \sum_{i=1}^k\sum_{j=1}^\ell a_ip_{j,i}\log\left( \frac{q_{i,j}}{ a_i } \right).
\end{align}
Fixing $a_i$ and consequently also $p_{j,i}$, we need to solve
\begin{align}
\max_{q_{i,j}} & \sum_{i=1}^k\sum_{j=1}^\ell a_i p_{j,i} \log\left( \frac{q_{i,j}}{ a_i } \right)\\
\nonumber
{\rm s.t.} & \sum_{i=1}^k q_{i,j}=1,\ \forall j\in\{1,\ldots,\ell\}.
\end{align}
The problem is convex, its Lagrangian is $\mathcal{L}(q_{i,j},\boldsymbol{\psi})=-\sum_{i,j} a_i p_{j,i} \log\left( \frac{q_{i,j}}{ a_i } \right)+\sum_j \psi_j(\sum_iq_{i,j}-1)$, and the optimal solution satisfies $\frac{\partial\mathcal{L}(q_{i,j},\boldsymbol{\psi})}{\partial q_{i,j}}=0$. Combined with $\sum_{i=1}^k q_{i,j}=1$, this yields
\begin{align}
\label{BAA_qij}
q_{i,j}=\frac{a_ip_{j,i}}{\sum_{i'} a_{i'}p_{j,i'}}.
\end{align}
Note that this solution preserves the mean of $X$, i.e., if $\sum_i x_ia_i=\mu$ for some $\mu$, then we also have $\sum_ix_i\sum_j\tilde{p}_jq_{i,j}=\mu$ where $\tilde{p}_j=\sum_i a_i p_{j,i}$.

On the other hand, fixing $q_{i,j}$ and $p_{j,i}$, we need to solve
\begin{align}
\max_{a_i} & \sum_{i,j} a_i p_{j,i} \log\left( \frac{q_{i,j}}{ a_i } \right)\\
\nonumber
{\rm s.t.} & \sum_{i=1}^k a_i=1,\ \sum_{i=1}^k x_ia_i=\mathcal{E}.
\end{align}
The sum in the objective function is with respect to $i,j$ for which $q_{i,j}>0$ and $p_{j,i}>0$. This problem is also convex, its Lagrangian is $\mathcal{L}(a_i,\psi,\nu)=-\sum_{i,j} a_i p_{j,i} \log\left( \frac{q_{i,j}}{ a_i } \right)+\psi(\sum_{i} a_i-1)+\nu(\sum_{i} x_ia_i-\mathcal{E})$, and the optimal solution satisfies $\frac{\partial\mathcal{L}(a_i,\psi,\nu)}{\partial a_i}=0$. Combined with the two constraints, this yields
\begin{align}
\label{BAA_ri}
a_i=\frac{e^{-\nu x_i}\prod_jq_{i,j}^{p_{j,i}}}{\sum_{i'} e^{-\nu x_{i'}}\prod_j q_{i',j}^{p_{j,i'}}},
\end{align}
where the product is with respect to $j$ for which $p_{j,i}>0$ and $q_{i,j}>0$, and $\nu$ is the solution of 
\begin{align}
\label{BAA_lambda}
\mathcal{E}=\frac{\sum_i x_ie^{-\nu x_i}\prod_jq_{i,j}^{p_{j,i}}}{\sum_{i'} e^{-\nu x_{i'}}\prod_j q_{i',j}^{p_{j,i'}}}.
\end{align}

\begin{algorithm}[t]
\caption{Blahut-Arimoto Algorithm for the IM/DD Gaussian channel}
\begin{algorithmic}[1]
\State {\bf Inputs:} $\mathcal{A}$; $\mathcal{E}$; $g$; $(x_1,\ldots,x_k)$; $\delta$; $y_{\rm l}$; $y_{\rm u}$;
\State $a_i\leftarrow\frac{1}{k}$; $\forall i\in\{1,\ldots,k\}$; $\ell\leftarrow\frac{y_{\rm u}-y_{\rm l}}{\delta}$; $r\leftarrow0$
%\Do %{$\frac{r-r'}{r'}>\epsilon$}
\Repeat
\State $p_{j,i}\leftarrow \mathbb{P}\{Y\in[y_{\rm l}+(j-1)\delta,y_{\rm l}+j\delta)|\text{$x_i$ was sent}\}$
%\State Calculate $p_{j,i}$ from \eqref{BAA_pji}
\State $q_{i,j}\leftarrow a_ip_{j,i} \left(\sum_{i'} a_{i'}p_{j,i'}\right)^{-1}$
%\State Calculate $q_{i,j}$ from \eqref{BAA_qij}
\State $\nu\leftarrow$ Solution of \eqref{BAA_lambda}%$\mathcal{E}=\left(\sum_i x_ie^{-\nu x_i}\prod_jq_{i,j}^{p_{j,i}}\right)\left(\sum_{i'} e^{-\nu x_{i'}}\prod_j q_{i',j}^{p_{j,i'}}\right)^{-1}$ 
%\State Solve \eqref{BAA_lambda} for $\nu$
\State $a_i\leftarrow \left(e^{-\nu x_i}\prod_jq_{i,j}^{p_{j,i}}\right)\left(\sum_{i'} e^{-\nu x_{i'}}\prod_j q_{i',j}^{p_{j,i'}}\right)^{-1}$
\State %$r'\leftarrow r$; 
$r\leftarrow \sum_{i,j} a_i p_{j,i} \log\left( \frac{q_{i,j}}{ a_i }\right)$
%\State Calculate $r_i$ from \eqref{BAA_ri} 
%\If{$\frac{r-r'}{r'}>\epsilon$}{ go to \ref{Steppji}}
%\EndIf
%\doWhile{$\frac{r-r'}{r'}>\epsilon$}
\Until{$r$ converges}
\State {\bf Outputs:} $r$; $a_i$
\end{algorithmic}
\label{Alg:BAA}
\end{algorithm}

Based on this, the optimal $a_i$ for given $k$ and $x_i$  can be computed using the Blahut-Arimoto algorithm as described in the Algorithm \ref{Alg:BAA}. By discretizing the interval $[y_{\rm l},y_{\rm u}]$ into infinitesimally small intervals ($\delta\ll1$) and choosing $-y_{\rm l}$ and $y_{\rm u}-g\mathcal{A}$ large enough,\footnote{\ac{so that the interval $[y_{\rm l},y_{\rm u}]$ contains a large enough margin around $[0,g\mathcal{A}]$ to accommodate a large subset of the support of $Y$.}} the rate $r$ in Algorithm \ref{Alg:BAA} converges to $\max_{a_i}I(X;Y)$ where $X\in\{x_1,\ldots,x_k\}$.

\begin{example}
\label{Eg:BAA1}
Consider an IM/DD Gaussian channel with $g=1$, $\mathcal{A}=5$, and $\mathcal{E}=1.25$. Let $k=3$, $(x_1,x_2,x_3)=(0,2,5)$, $y_{\rm l}=-10$, $y_{\rm u}=g\mathcal{A}+10$, and $\delta=10^{-3}$. %, and $\epsilon=10^{-4}$. 
With these parameters, the solution of $\max_{a_i}I(X;Y)$ using Algorithm \ref{Alg:BAA} is $0.61$ nats ($0.88$ bits) per transmission achieved when $(a_1,a_2,a_3)=(0.638,    0.1866,    0.1753)$. Table \ref{Tab:BAA_eg} shows the evolution of $(a_i)$ for this channel until convergence. 
\end{example}

\begin{table}[t]
\centering
\begin{tabular}{c||c|c}
Iteration & $(a_1,a_2,a_{3})$ & $r$ (nats/transmission)\\\hline\hline
1		  & $(0.5912,    0.2647,    0.1441)$ & $0.5679$\\
2		  & $(0.6150,    0.2250,    0.1600)$ & $0.6053$\\
3		  & $(0.6264,    0.2060,    0.1676)$ & $0.6087$\\
4	      & $(0.6323,    0.1962,    0.1715)$ & $0.6096$\\
$\vdots$  & $\vdots$						 & $\vdots$\\
7		  & $(0.6380,    0.1866,    0.1753)$ & $0.6100$
\end{tabular}
\caption{Evolution of $(a_i)$ for the channel in Example \ref{Eg:BAA1} until convergence. The solution of $\max_{a_i}I(X;Y)$ for this example using exhaustive (grid) search is $0.61$ nats per transmission which matches the result of Algorithm \ref{Alg:BAA}.}
\label{Tab:BAA_eg}
\end{table}

Now that we have a method for finding $(a_i)$ for a given $k$ and $(x_i)$, we move to the maximization with respect to $(x_i)$.

\subsection{Optimal $a_i$ and $x_i$ given $k$}
To find the optimal $(x_i)$ for a given $k$, we rely on a statement that dates back to 1971 \cite{SMITH1971203}. The statement provides a necessary {and sufficient} condition for a distribution $\mathbb{P}_{X}$ to be optimal for a Gaussian channel with a peak constrained input without and with a power constraint. The statement has been generalized later on to different types of channels such as Poisson \cite{Shamai_DT_Poisson}, quadrature Gaussian \cite{ShamaiBarDavid}, Rayleigh-fading \cite{AbouFaycalTrottShamai}, and conditionally Gaussian channels \cite{ChanHranilovicKschischang}. The statement for the channel under consideration is given in \cite{ChanHranilovicKschischang}, and is repeated next.

\begin{theorem}[Optimality Condition \cite{ChanHranilovicKschischang}]
\label{Thm:DiscreteInputConditions}
Let $\mathbb{P}_{X}$ be the capacity achieving distribution of the IM/DD Gaussian channel with support $\mathcal{X}=\{x_1,\ldots,x_k\}$ for some $k$, and define 
\begin{align}
{\sf I}(\mathbb{P}_X)&=\left.I(X;Y)\right|_{X\sim\mathbb{P}_X}\\
Q(x,\mathbb{P}_X)&=-\int_{\mathbb{R}}\mathbb{P}_{Y|X}(y|x)\log(\mathbb{P}_{Y}(y)){\rm d}y\\
\label{lambda_Def_Opt}
\psi&=\frac{1}{\mathcal{E}}\left( {\sf I}(\mathbb{P}_X)+\frac{1}{2}\log(2\pi e)-Q(0,\mathbb{P}_X)\right)\\
\label{JxP_Def}
J(x,\mathbb{P}_X)&={\sf I}(\mathbb{P}_X)-Q(x,\mathbb{P}_X)+\frac{1}{2}\log(2\pi e)+\psi(x-\mathcal{E}).
\end{align}
Then the following statements hold:
\begin{enumerate}
\item $0\in \mathcal{X}$ and $\mathbb{P}_{X}(0)>0$;
\item $\psi>0$; and 
\item $J(x,\mathbb{P}_X)\geq 0$ for all $x\in[0,\mathcal{A}]$ with equality if $x\in\mathcal{X}$.
\end{enumerate}
\end{theorem}

This statement {provides a necessary and sufficient condition for the optimality of an input distribution and} can be used to check if an input distribution for a given $k$ is optimal. We apply this to the channel in Example \ref{Eg:BAA1}.

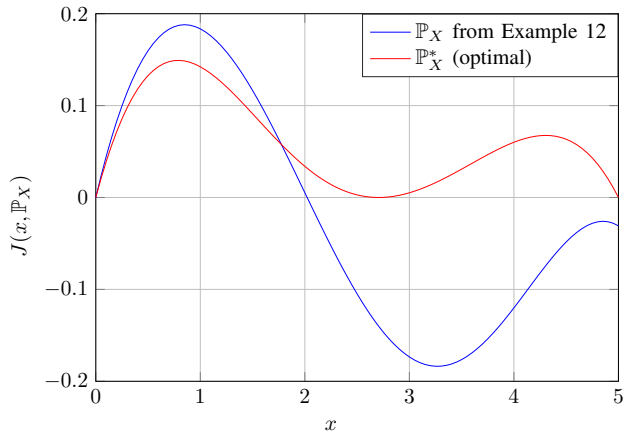
\begin{figure}[t]
\centering
\tikzset{every picture/.style={scale=.9}, every node/.style={scale=.9}}
%\input{matlab/JxP_1}
% Minimal pgfplots version: 1.3
%  http://www.mathworks.com/matlabcentral/fileexchange/22022-matlab2tikz
\definecolor{mycolor1}{rgb}{0.00000,0.44700,0.74100}%
\begin{tikzpicture}

\begin{axis}[%
width=4in,
height=3in,
xmin=0,
xmax=5,
xlabel={$x$},
xmajorgrids,
ymin=-.2,
ymax=.2,
ylabel={$J(x,\mathbb{P}_X)$},
ymajorgrids,
legend style={at = {(axis cs: 5,.2)}, anchor = north east, legend cell align=left,align=left,draw=white!15!black}
]
\addplot [color=blue,solid]
  table[row sep=crcr]{%
0	1.66533453693773e-16\\
0.05	0.0232927927457282\\
0.1	0.0448547891643662\\
0.15	0.0647235654414329\\
0.2	0.0829375916628723\\
0.25	0.0995361924800229\\
0.3	0.114559504977583\\
0.35	0.128048434069335\\
0.4	0.140044605782412\\
0.45	0.150590318822686\\
0.5	0.159728494840523\\
0.55	0.167502627838207\\
0.6	0.173956733175105\\
0.65	0.179135296634908\\
0.7	0.183083224019408\\
0.75	0.185845791725966\\
0.8	0.187468598748581\\
0.85	0.187997520517641\\
0.9	0.187478664958444\\
0.95	0.185958331105272\\
1	0.183482970555044\\
1.05	0.180099151983794\\
1.1	0.175853528880384\\
1.15	0.170792810575792\\
1.2	0.164963736564132\\
1.25	0.158413054025078\\
1.3	0.151187498365556\\
1.35	0.143333776506742\\
1.4	0.134898552548523\\
1.45	0.125928435351289\\
1.5	0.116469967485984\\
1.55	0.106569614919674\\
1.6	0.0962737567259212\\
1.65	0.0856286740415691\\
1.7	0.074680537434002\\
1.75	0.0634753917967825\\
1.8	0.0520591378621233\\
1.85	0.0404775094015348\\
1.9	0.028776045187993\\
1.95	0.0170000548119778\\
2	0.00519457748048185\\
2.05	-0.00659566701533421\\
2.1	-0.0183263359034004\\
2.15	-0.0299535302513519\\
2.2	-0.0414338715397024\\
2.25	-0.0527245883951671\\
2.3	-0.0637836137488847\\
2.35	-0.0745696925278499\\
2.4	-0.0850424998208714\\
2.45	-0.0951627692833558\\
2.5	-0.104892431363367\\
2.55	-0.114194760745936\\
2.6	-0.123034532226616\\
2.65	-0.131378184043477\\
2.7	-0.139193987520556\\
2.75	-0.146452221709466\\
2.8	-0.153125351563514\\
2.85	-0.159188208040254\\
2.9	-0.164618168411244\\
2.95	-0.169395334960295\\
3	-0.173502710178257\\
3.05	-0.176926366514778\\
3.1	-0.179655608727568\\
3.15	-0.181683126875818\\
3.2	-0.183005138041438\\
3.25	-0.183621514926501\\
3.3	-0.183535899566762\\
3.35	-0.182755800522121\\
3.4	-0.181292672049185\\
3.45	-0.179161973930933\\
3.5	-0.17638321082716\\
3.55	-0.172979950219931\\
3.6	-0.168979818249536\\
3.65	-0.164414472974829\\
3.7	-0.159319554833176\\
3.75	-0.15373461432614\\
3.8	-0.147703017204456\\
3.85	-0.141271827672968\\
3.9	-0.13449167037499\\
3.95	-0.12741657214506\\
4	-0.120103784732961\\
4.05	-0.112613589901567\\
4.1	-0.105009088478014\\
4.15	-0.0973559750954553\\
4.2	-0.089722300494025\\
4.25	-0.0821782233571329\\
4.3	-0.0747957537389199\\
4.35	-0.0676484901919959\\
4.4	-0.0608113527298025\\
4.45	-0.0543603137566058\\
4.5	-0.0483721290701505\\
4.55	-0.0429240709890942\\
4.6	-0.038093665581167\\
4.65	-0.0339584358692691\\
4.7	-0.0305956527752328\\
4.75	-0.0280820954265102\\
4.8	-0.0264938223009814\\
4.85	-0.0259059545247947\\
4.9	-0.0263924724670064\\
4.95	-0.0280260265993966\\
5	-0.0308777634089213\\
};
\addlegendentry{$\mathbb{P}_X$ from Example \ref{Eg:Discrete1}};
\addplot [color=red,solid]
  table[row sep=crcr]{%
0	1.84187005092262e-08\\
0.05	0.0204664537528466\\
0.1	0.0391938653589182\\
0.15	0.0562313087975236\\
0.2	0.0716296387093887\\
0.25	0.0854414519581717\\
0.3	0.097721018692715\\
0.35	0.108524201171359\\
0.4	0.117908360302824\\
0.45	0.125932249977561\\
0.5	0.132655899389442\\
0.55	0.138140483677299\\
0.6	0.142448183351561\\
0.65	0.14564203310421\\
0.7	0.147785760737158\\
0.75	0.14894361707575\\
0.8	0.149180197862547\\
0.85	0.148560258748012\\
0.9	0.147148524607621\\
0.95	0.14500949451697\\
1	0.14220724380575\\
1.05	0.138805224686021\\
1.1	0.134866067008416\\
1.15	0.130451380740054\\
1.2	0.12562156177889\\
1.25	0.120435602719829\\
1.3	0.114950910167531\\
1.35	0.109223130148972\\
1.4	0.103305983115495\\
1.45	0.0972511099395837\\
1.5	0.0911079302065439\\
1.55	0.08492351397711\\
1.6	0.0787424680537685\\
1.65	0.0726068376259042\\
1.7	0.0665560239954348\\
1.75	0.0606267188993896\\
1.8	0.0548528557530191\\
1.85	0.0492655779361323\\
1.9	0.0438932240421203\\
1.95	0.0387613298054666\\
2	0.0338926462229264\\
2.05	0.0293071731891661\\
2.1	0.0250222077825648\\
2.15	0.0210524061641591\\
2.2	0.0174098578951166\\
2.25	0.0141041713382596\\
2.3	0.0111425686893532\\
2.35	0.00852998908606406\\
2.4	0.00626919816828353\\
2.45	0.00436090241409803\\
2.5	0.00280386655189707\\
2.55	0.00159503235092057\\
2.6	0.000729637121095805\\
2.65	0.000201330304809844\\
2.7	2.28662170076444e-06\\
2.75	0.000123314326766377\\
2.8	0.000553957263215443\\
2.85	0.00128258953110444\\
2.9	0.00229650174873852\\
2.95	0.00358197805322114\\
3	0.00512436316657872\\
3.05	0.00690811904129479\\
3.1	0.0089168707907516\\
3.15	0.0111334418030416\\
3.2	0.0135398781261001\\
3.25	0.0161174623987902\\
3.3	0.0188467177777191\\
3.35	0.0217074024760939\\
3.4	0.0246784956825277\\
3.45	0.0277381757637152\\
3.5	0.0308637917729442\\
3.55	0.0340318293850096\\
3.6	0.0372178724559938\\
3.65	0.0403965614627437\\
3.7	0.0435415501113193\\
3.75	0.0466254614161282\\
3.8	0.049619844542212\\
3.85	0.0524951336730568\\
3.9	0.0552206101162082\\
3.95	0.0577643687904156\\
4	0.0600932901525864\\
4.05	0.062173018522385\\
4.1	0.0639679476489237\\
4.15	0.065441214240012\\
4.2	0.0665547000420146\\
4.25	0.0672690429201597\\
4.3	0.0675436572472018\\
4.35	0.0673367637665093\\
4.4	0.066605428950646\\
4.45	0.0653056137451059\\
4.5	0.0633922314458516\\
4.55	0.060819214340831\\
4.6	0.0575395886262726\\
4.65	0.0535055570040388\\
4.7	0.0486685882726106\\
4.75	0.0429795131434583\\
4.8	0.0363886254473196\\
4.85	0.0288457878416207\\
4.9	0.0203005410912535\\
4.95	0.0107022159700642\\
5	0\\
};
\addlegendentry{$\mathbb{P}_X^*$ (optimal)};
\end{axis}
\end{tikzpicture}%
\caption{Plot of $J(x,\mathbb{P}_X)$ defined in Theorem \ref{Thm:DiscreteInputConditions} as a function of $x$ for Examples \ref{Eg:Discrete1} and \ref{Eg:Discrete2}.}
\label{Fig:JxP_1}
\end{figure}

\begin{example}
\label{Eg:Discrete1}
For the IM/DD Gaussian channel in Example \ref{Eg:BAA1}, we have $\mathbb{P}_{X}=\sum_{i=1}^3 a_i\delta(x-x_i)$ where $(x_1,x_2,x_3)=(0,2,5)$ and $(a_1,a_2,a_3)=(0.638,    0.1866,    0.1753)$, and ${\sf I}(\mathbb{P}_{X})=0.61$ nats per transmission. Since $x_1=0$ and $a_1>0$, then the first condition in Theorem \ref{Thm:DiscreteInputConditions} is satisfied. To check the second and third conditions, we evaluate $\psi$ and $J(x,\mathbb{P}_X)$ numerically. We obtain $\psi=0.2528$ which satisfies the second condition. However, $J(x,\mathbb{P}_X)$ does not satisfy the third condition as shown in Fig. \ref{Fig:JxP_1}. Thus, this input distribution is not optimal.
\end{example}

To find out whether a distribution with $k$ mass points is optimal, one can fix $x_1=0$, vary $x_2,\ldots,x_k$, find $a_1,\ldots,a_k$ using Algorithm \ref{Alg:BAA}, and repeat until either the conditions in Theorem \ref{Thm:DiscreteInputConditions} are satisfied in which case we have the optimal distribution, or all (discretized) values of $x_2,\ldots,x_k$ are exhausted in which case $k$ is too small to achieve capacity. %This is summarized in Algorithm \ref{Alg:DiscOptCond}. 
%\begin{algorithm}[t]
%\caption{Searching for an optimal input for an IM-DD channel with $k$ mass points}
%\begin{algorithmic}[1]
%\State {\bf Inputs:} $\mathcal{A}$, $\mathcal{E}$; $g$; $\sigma$; $k$
%\State $r_{\rm max}\leftarrow 0$; $\mathbb{P}_X^*=\emptyset$
%\For{$0=x_1<x_2<\cdots<x_k\leq\mathcal{A}$}
%\State $r,(a_1,\ldots,a_k)\leftarrow$ Outputs of Algorithm \ref{Alg:BAA} for $(x_1,\ldots,x_k)$ 
%\If{$r>r_{\rm max}$}
%\State $r_{\rm max}\leftarrow r$; $\mathbb{P}_X\leftarrow\sum_{i=1}^ka_i\delta(x-x_i)$
%\State Compute $\lambda$ and $J(x,\mathbb{P}_X)$ using \eqref{lambda_Def_Opt} and \eqref{JxP_Def}
%\If{$\lambda>0$; $J(x,\mathbb{P}_X)\geq 0$ $\forall x\in[0,\mathcal{A}]$ and $J(x_i,\mathbb{P}_X)=0$ $\forall i\in\{1,\ldots,k\}$}
%\State $\mathbb{P}_X^*=\mathbb{P}_X$; {\bf break}
%%\State Break;
%\EndIf
%\EndIf
%\EndFor
%\State {\bf Output:} $\mathbb{P}_X^*$
%\end{algorithmic}
%\label{Alg:DiscOptCond}
%\end{algorithm}
Alternatively, one can rely on the concavity of $I(X;Y)$ in $\mathbb{P}_X$, and maximize jointly with respect to $(x_1,\ldots,x_k)$ and $(a_1,\ldots,a_k)$ using numerical solvers for a given $k$. Checking whether the obtained distribution satisfies the conditions in Theorem \ref{Thm:DiscreteInputConditions} reveals if this distribution is optimal or not (in which case we conclude that $k$ is too small) \cite{AbouFaycalTrottShamai,ChanHranilovicKschischang}. This is summarized in Algorithm \ref{Alg:DiscOptCond2}.

\begin{algorithm}[t]
\caption{Combined search for optimal $x_1,\ldots,x_k$ and $a_1,\ldots,a_k$ for the IM/DD Gaussian  channel}
\begin{algorithmic}[1]
\State {\bf Inputs:} $\mathcal{A}$, $\mathcal{E}$; $g$; $k$
\State $\mathbb{P}_X^*=\emptyset$
\State $\mathbb{P}_X\leftarrow$ Solution of $\max_{\mathbb{P}_X: k \text{ mass points}} I(X;Y)$ using numerical solvers
\State Compute $\psi$ and $J(x,\mathbb{P}_X)$ using \eqref{lambda_Def_Opt} and \eqref{JxP_Def}
\If{$\mathbb{P}_X(0)>0$, $\psi>0$; $J(x,\mathbb{P}_X)\geq 0$ $\forall x\in[0,\mathcal{A}]$ with equality if $\mathbb{P}_X(x)>0$}
\State $\mathbb{P}_X^*=\mathbb{P}_X$
\EndIf
\State {\bf Output:} $\mathbb{P}_X^*$
\end{algorithmic}
\label{Alg:DiscOptCond2}
\end{algorithm}

To test this algorithm, we use the parameters in example \ref{Eg:BAA1}.

\begin{example}\label{Eg:Discrete2}
For the channel in Example \ref{Eg:BAA1}, Algorithm \ref{Alg:DiscOptCond2} yields $\mathbb{P}_X^*=\sum_{i=1}^3a_i\delta(x-x_i)$ where $(x_1,x_2,x_3)=(0,2.7058,5)$ and $(a_1,a_2,a_3)=(0.6643,0.1869,0.1489)$, and the rate is $r=0.626$ nats ($0.9031$ bits) per transmission. Moreover, $\mathbb{P}_X(0)>0$, $\psi=0.2501>0$, and $J(x,\mathbb{P}_X^*)\geq0$ for all $x$ with equality when $x=a_i$ as shown in Fig. \ref{Fig:JxP_1}, which satisfies the conditions in Theorem~\ref{Thm:DiscreteInputConditions}. Hence, $\mathbb{P}_X^*$ is optimal and the capacity of this channel is $c_1(5,1.25)=0.626$ nats/transmission. 
\end{example}

If Algorithm \ref{Alg:DiscOptCond2} outputs $\mathbb{P}_X^*=\emptyset$, i.e., the conditions in Theorem \ref{Thm:DiscreteInputConditions} are not satisfied for any $\mathbb{P}_X$ with $k$ mass points. This indicates that the optimal input distribution has more than $k$ points. It remains to find the optimal $k$.

\subsection{Optimal $k$}
To find the optimal $k$, we also use the necessary conditions in Theorem \ref{Thm:DiscreteInputConditions}. In particular, for a given channel, we start by setting $k=2$ and using Algorithm \ref{Alg:DiscOptCond2} to find the optimal $\mathbb{P}_X$. If the output is $\mathbb{P}_X^*=\emptyset$, then we increment $k$. This is repeated until an optimal $\mathbb{P}_X$ is found as summarized in Algorithm \ref{Alg:OptInputDist}.

\begin{algorithm}[t]
\caption{Search for the optimal $\mathbb{P}_X$ for an IM/DD Gaussian channel}
\begin{algorithmic}[1]
\State {\bf Inputs:} $\mathcal{A}$, $\mathcal{E}$; $g$; 
\State $r_{\rm max}\leftarrow 0$; $\mathbb{P}_X^*=\emptyset$; $k\leftarrow 1$
\While{$\mathbb{P}_X^*=\emptyset$}
\State $k\leftarrow k+1$
\State $\mathbb{P}_X^*\leftarrow$ Output of Algorithm \ref{Alg:DiscOptCond2} given $k$ 
\EndWhile
\State {\bf Output:} $\mathbb{P}_X^*$
\end{algorithmic}
\label{Alg:OptInputDist}
\end{algorithm}

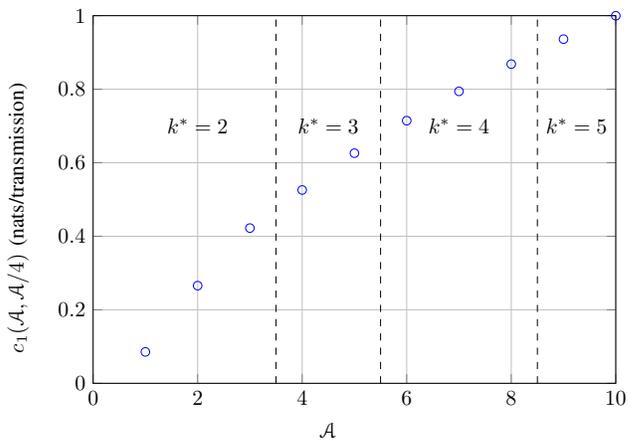
\begin{figure}[t]
\centering
\tikzset{every picture/.style={scale=.9}, every node/.style={scale=.9}}
%\input{fig/Capacity_vs_A_alpha_p25_s_1_g_1}
% Minimal pgfplots version: 1.3
%  http://www.mathworks.com/matlabcentral/fileexchange/22022-matlab2tikz
\definecolor{mycolor1}{rgb}{0.00000,0.44700,0.74100}%
\begin{tikzpicture}

\begin{axis}[%
width=4in,
height=3in,
xmin=0,
xmax=10,
xlabel={$\mathcal{A}$},
xmajorgrids,
ymin=0,
ymax=1,
ylabel={$c_1(\mathcal{A},\mathcal{A}/4)$ (nats/transmission)},
ymajorgrids,
legend style={legend cell align=left,align=left,draw=white!15!black}
]
\addplot [color=blue,only marks,mark=o,mark options={solid},forget plot]
  table[row sep=crcr]{%
1	0.0854482126852514\\
2	0.265653273432171\\
3	0.422224167909863\\
4	0.525909389774591\\
5	0.625975976279065\\
6	0.714411356895874\\
7	0.794232011975669\\
8	0.868266145925912\\
9	0.936148081645106\\
10	0.999761867641674\\
};
\draw[dashed] (axis cs: 3.5,0) to (axis cs: 3.5,1);
\node at (axis cs: 2,0.7) {$k^*=2$};

\draw[dashed] (axis cs: 5.5,0) to (axis cs: 5.5,1);
\node at (axis cs: 4.5,0.7) {$k^*=3$};

\draw[dashed] (axis cs: 8.5,0) to (axis cs: 8.5,1);
\node at (axis cs: 7,0.7) {$k^*=4$};

\node at (axis cs: 9.25,0.7) {$k^*=5$};

\end{axis}
\end{tikzpicture}%
\caption{Capacity versus $\mathcal{A}$ for an IM/DD Gaussian channel with $\mathcal{E}=\mathcal{A}/4$, and $g=1$, indicating the optimal number of mass points $k^*$.}
\label{Fig:CapvsA}
\end{figure}

Using Algorithm \ref{Alg:OptInputDist}, we can find the optimal input distribution for the single-user IM/DD Gaussian channel. Fig. \ref{Fig:CapvsA} shows the channel capacity obtained using Algorithm \ref{Alg:OptInputDist} for a channel with $\mathcal{E}=\frac{\mathcal{A}}{4}$ as a function of $\mathcal{A}$ when $g=1$. 

While Fig. \ref{Fig:CapvsA} shows the behaviour of capacity as a function of the peak intensity, it does not provide an explicit relation between the two. Thus, the following question arises: Can we express capacity in a simple form? The following section discusses this issue.

\section{Capacity Bounds and Asymptotics}
\label{Sec:CapacityBounds}
The capacity of the single-user IM/DD Gaussian channel can be obtained by solving problem \eqref{CapOpt} using Algorithm \ref{Alg:OptInputDist}. While this does not have the elegance of $\log\left(1+{\rm SNR}\right)$ (the capacity of a Gaussian memoryless channel with a power constraint, see Remark \ref{Rem:AWGN_Capacity} below), this is still useful for evaluating the capacity of a static channel. {Furthermore, the optimality of discrete inputs with a finite number of mass points as asserted by Theorem \ref{Thm:Discreteness} is appealing from an engineering point of view.}

\begin{remark}
\label{Rem:AWGN_Capacity}
For a Gaussian memoryless channel with input $X\in\mathbb{C}$ and output $Y=gX+Z$ where $g\in\mathbb{C}$, $\mathbb{E}[|X|^2]\leq p$, and $Z\sim\mathcal{CN}(0,\sigma^2)$, the capacity equals $\log\left(1+{\rm SNR}\right)$ where ${\rm SNR}$ is the signal-to-noise ratio ${\rm SNR}=\frac{|g|^2p}{\sigma^2}$. This is achieved by choosing $X\sim\mathcal{CN}(0,p)$.
\end{remark}

The simplicity of $\log(1+{\rm SNR})$ enabled a large body of literature on wireless communications focusing on capacity and power allocation for parallel channels, MIMO channels, and multi-user channels, in addition to ergodic and outage capacities of time-varying channels \cite{TseViswanath,Goldsmith}. However, the capacity of the IM/DD Gaussian channel obtained using Algorithm \ref{Alg:OptInputDist} is not amenable to similar analysis. This makes it important to derive capacity bounds and approximations that enable further analysis of IM/DD Gaussian channels based on information-theoretic fundamentals. {Moreover, capacity bounds are important in cases where the capacity is not numerically computable, such as for the channel with an average intensity constraint only, whose capacity achieving distribution has an infinite number of mass points.}

This section presents such results.

\subsection{Capacity Lower Bounds -- Achievable Rates}
Two methods have been used in the literature to derive capacity lower bounds. One method focuses on continuous input distributions, and the other focuses on discrete input distributions. The common factor between the two is that they both rely on the entropy-power inequality (EPI) stated next.

\begin{lemma}[{EPI \cite[Theorem 17.7.3]{CoverThomas}}]\label{Lem:EPI}
If $\X$ and $\Y$ are $n$-dimensional independent random vectors with densities, then $e^{\frac{2}{n}h(\X+\Y)}\geq e^{\frac{2}{n}h(\X)}+e^{\frac{2}{n}h(\Y)}$.
\end{lemma}

To derive a capacity lower bound using this lemma, we write
\begin{align}
I(X;gX+Z)&=h(gX+Z)-h(gX+Z|X)\\
&\stackrel{P1}{=}h(gX+Z)-h(Z)\\
&\stackrel{\hspace{-.2cm}\rm EPI}{\geq} \frac{1}{2}\log\left( e^{2h(gX)}+e^{2h(Z)} \right)-h(Z)\\
&= \frac{1}{2}\log\left( \frac{g^2e^{2h(X)}}{2\pi e}+1 \right),
\end{align}
where in the last step we used $h(Z)=\frac{1}{2}\log(2\pi e)$ and $h(gX)=h(X)+\log(g)$ \cite[(8.71)]{CoverThomas}. Let $\tilde{\mathbb{P}}_X$ be a selected input distribution, then it holds that
\begin{align}
c_g(\mathcal{A},\mathcal{E})&=\max_{\mathbb{P}_{X}\in\mathcal{P}_{X}} I(X;gX+Z)\\
&\geq \left.I(X;gX+Z)\right|_{X\sim\tilde{\mathbb{P}}_X}\\
&\geq \frac{1}{2}\log\left( \frac{g^2e^{2\left.h(X)\right|_{X\sim\tilde{\mathbb{P}}_X}}}{2\pi e}+1 \right).
\end{align}
Thus, the problem of deriving a capacity lower bound boils down to choosing a `good' $\tilde{\mathbb{P}}_X$, i.e., one which maximizes $h(X)$. Next, we apply this to a channel with average and a peak constraints, followed by one with an average constraint only.

\subsubsection{Average and Peak Constraints} 
We distinguish between bounds using continuous input distributions and ones using discrete input distributions.

\paragraph{Continuous Input Distributions} 
\label{Sec:LowerBoundsBothConstraintsContDist}
The solution of $\max_{\mathbb{P}_X} h(X)$ where $\mathbb{P}_X$ is a continuous distribution satisfying $X\in[0,\mathcal{A}]$ and $\mathbb{E}[X]\leq\mathcal{E}$ was given in \cite{LapidothMoserWigger}. The results rely on finding the max-entropic distribution using \cite[Thm. 12.1.1]{CoverThomas} which states the following.

\begin{lemma}[{\cite[Thm. 12.1.1]{CoverThomas}}]
\label{Lem:MaxEnt}
The solution of 
\begin{align} 
\max_{\mathbb{P}_X}\quad & h(X)\\
{\rm s.t.}\quad & X\in\mathcal{S},\quad \int_{\mathcal{S}}\mathbb{P}_X(x){\rm d}x=1\\
&\mathbb{E}[r_i(X)]=m_i,\ \forall i=1,\ldots,n
\end{align}
for some measurable functions $r_i:\mathcal{S} \to\mathbb{R}$, is given by $\mathbb{P}^*_X(x)=e^{a_0+\sum_{i=1}^n a_ir_i(x)}$, where $a_0,\ldots,a_n$ are chosen to satisfy the constraints.
\end{lemma}

\begin{figure}[t]
\centering
\begin{subfigure}[]{.22\textwidth}
\tikzset{every picture/.style={scale=.92}, every node/.style={scale=.9}}
%\input{matlab/TruncExp}
% Minimal pgfplots version: 1.3
%  http://www.mathworks.com/matlabcentral/fileexchange/22022-matlab2tikz
\definecolor{mycolor1}{rgb}{0.00000,0.44700,0.74100}%
\begin{tikzpicture}

\begin{axis}[%
width=2in,
height=3in,
xmin=-1,
xmax=5,
xlabel={$x$},
xmajorgrids,
ymin=0,
ymax=1,
ylabel={$\mathbb{P}_X(x)$},
ylabel style = {at = {(axis cs: -.7,.5)}},
ymajorgrids,
legend style={legend cell align=left,align=left,draw=white!15!black}
]
\addplot [color=mycolor1,solid,forget plot]
  table[row sep=crcr]{%
-3	0\\
-2.99	0\\
-2.98	0\\
-2.97	0\\
-2.96	0\\
-2.95	0\\
-2.94	0\\
-2.93	0\\
-2.92	0\\
-2.91	0\\
-2.9	0\\
-2.89	0\\
-2.88	0\\
-2.87	0\\
-2.86	0\\
-2.85	0\\
-2.84	0\\
-2.83	0\\
-2.82	0\\
-2.81	0\\
-2.8	0\\
-2.79	0\\
-2.78	0\\
-2.77	0\\
-2.76	0\\
-2.75	0\\
-2.74	0\\
-2.73	0\\
-2.72	0\\
-2.71	0\\
-2.7	0\\
-2.69	0\\
-2.68	0\\
-2.67	0\\
-2.66	0\\
-2.65	0\\
-2.64	0\\
-2.63	0\\
-2.62	0\\
-2.61	0\\
-2.6	0\\
-2.59	0\\
-2.58	0\\
-2.57	0\\
-2.56	0\\
-2.55	0\\
-2.54	0\\
-2.53	0\\
-2.52	0\\
-2.51	0\\
-2.5	0\\
-2.49	0\\
-2.48	0\\
-2.47	0\\
-2.46	0\\
-2.45	0\\
-2.44	0\\
-2.43	0\\
-2.42	0\\
-2.41	0\\
-2.4	0\\
-2.39	0\\
-2.38	0\\
-2.37	0\\
-2.36	0\\
-2.35	0\\
-2.34	0\\
-2.33	0\\
-2.32	0\\
-2.31	0\\
-2.3	0\\
-2.29	0\\
-2.28	0\\
-2.27	0\\
-2.26	0\\
-2.25	0\\
-2.24	0\\
-2.23	0\\
-2.22	0\\
-2.21	0\\
-2.2	0\\
-2.19	0\\
-2.18	0\\
-2.17	0\\
-2.16	0\\
-2.15	0\\
-2.14	0\\
-2.13	0\\
-2.12	0\\
-2.11	0\\
-2.1	0\\
-2.09	0\\
-2.08	0\\
-2.07	0\\
-2.06	0\\
-2.05	0\\
-2.04	0\\
-2.03	0\\
-2.02	0\\
-2.01	0\\
-2	0\\
-1.99	0\\
-1.98	0\\
-1.97	0\\
-1.96	0\\
-1.95	0\\
-1.94	0\\
-1.93	0\\
-1.92	0\\
-1.91	0\\
-1.9	0\\
-1.89	0\\
-1.88	0\\
-1.87	0\\
-1.86	0\\
-1.85	0\\
-1.84	0\\
-1.83	0\\
-1.82	0\\
-1.81	0\\
-1.8	0\\
-1.79	0\\
-1.78	0\\
-1.77	0\\
-1.76	0\\
-1.75	0\\
-1.74	0\\
-1.73	0\\
-1.72	0\\
-1.71	0\\
-1.7	0\\
-1.69	0\\
-1.68	0\\
-1.67	0\\
-1.66	0\\
-1.65	0\\
-1.64	0\\
-1.63	0\\
-1.62	0\\
-1.61	0\\
-1.6	0\\
-1.59	0\\
-1.58	0\\
-1.57	0\\
-1.56	0\\
-1.55	0\\
-1.54	0\\
-1.53	0\\
-1.52	0\\
-1.51	0\\
-1.5	0\\
-1.49	0\\
-1.48	0\\
-1.47	0\\
-1.46	0\\
-1.45	0\\
-1.44	0\\
-1.43	0\\
-1.42	0\\
-1.41	0\\
-1.4	0\\
-1.39	0\\
-1.38	0\\
-1.37	0\\
-1.36	0\\
-1.35	0\\
-1.34	0\\
-1.33	0\\
-1.32	0\\
-1.31	0\\
-1.3	0\\
-1.29	0\\
-1.28	0\\
-1.27	0\\
-1.26	0\\
-1.25	0\\
-1.24	0\\
-1.23	0\\
-1.22	0\\
-1.21	0\\
-1.2	0\\
-1.19	0\\
-1.18	0\\
-1.17	0\\
-1.16	0\\
-1.15	0\\
-1.14	0\\
-1.13	0\\
-1.12	0\\
-1.11	0\\
-1.1	0\\
-1.09	0\\
-1.08	0\\
-1.07	0\\
-1.06	0\\
-1.05	0\\
-1.04	0\\
-1.03	0\\
-1.02	0\\
-1.01	0\\
-1	0\\
-0.99	0\\
-0.98	0\\
-0.97	0\\
-0.96	0\\
-0.95	0\\
-0.94	0\\
-0.93	0\\
-0.92	0\\
-0.91	0\\
-0.9	0\\
-0.89	0\\
-0.88	0\\
-0.87	0\\
-0.86	0\\
-0.85	0\\
-0.84	0\\
-0.83	0\\
-0.82	0\\
-0.81	0\\
-0.8	0\\
-0.79	0\\
-0.78	0\\
-0.77	0\\
-0.76	0\\
-0.75	0\\
-0.74	0\\
-0.73	0\\
-0.72	0\\
-0.71	0\\
-0.7	0\\
-0.69	0\\
-0.68	0\\
-0.67	0\\
-0.66	0\\
-0.65	0\\
-0.64	0\\
-0.63	0\\
-0.62	0\\
-0.61	0\\
-0.6	0\\
-0.59	0\\
-0.58	0\\
-0.57	0\\
-0.56	0\\
-0.55	0\\
-0.54	0\\
-0.53	0\\
-0.52	0\\
-0.51	0\\
-0.5	0\\
-0.49	0\\
-0.48	0\\
-0.47	0\\
-0.46	0\\
-0.45	0\\
-0.44	0\\
-0.43	0\\
-0.42	0\\
-0.41	0\\
-0.4	0\\
-0.39	0\\
-0.38	0\\
-0.37	0\\
-0.36	0\\
-0.35	0\\
-0.34	0\\
-0.33	0\\
-0.32	0\\
-0.31	0\\
-0.3	0\\
-0.29	0\\
-0.28	0\\
-0.27	0\\
-0.26	0\\
-0.25	0\\
-0.24	0\\
-0.23	0\\
-0.22	0\\
-0.21	0\\
-0.2	0\\
-0.19	0\\
-0.18	0\\
-0.17	0\\
-0.16	0\\
-0.15	0\\
-0.14	0\\
-0.13	0\\
-0.12	0\\
-0.11	0\\
-0.1	0\\
-0.0899999999999999	0\\
-0.0800000000000001	0\\
-0.0699999999999998	0\\
-0.0600000000000001	0\\
-0.0499999999999998	0\\
-0.04	0\\
-0.0299999999999998	0\\
-0.02	0\\
-0.00999999999999979	0\\
0	0.923783494271392\\
0.0100000000000002	0.915521593776606\\
0.02	0.907333583971803\\
0.0300000000000002	0.89921880401217\\
0.04	0.891176598963196\\
0.0500000000000003	0.883206319747802\\
0.0600000000000001	0.875307323093962\\
0.0700000000000003	0.867478971482783\\
0.0800000000000001	0.859720633097053\\
0.0899999999999999	0.852031681770244\\
0.1	0.844411496935979\\
0.11	0.836859463577946\\
0.12	0.829374972180258\\
0.13	0.821957418678262\\
0.14	0.814606204409785\\
0.15	0.807320736066819\\
0.16	0.800100425647631\\
0.17	0.792944690409311\\
0.18	0.785852952820735\\
0.19	0.778824640515957\\
0.2	0.771859186248012\\
0.21	0.764956027843133\\
0.22	0.758114608155384\\
0.23	0.751334375021685\\
0.24	0.744614781217254\\
0.25	0.73795528441144\\
0.26	0.731355347123951\\
0.27	0.724814436681474\\
0.28	0.718332025174685\\
0.29	0.711907589415642\\
0.3	0.705540610895557\\
0.31	0.699230575742951\\
0.32	0.692976974682178\\
0.33	0.686779302992323\\
0.34	0.680637060466463\\
0.35	0.674549751371303\\
0.36	0.668516884407158\\
0.37	0.662537972668306\\
0.38	0.656612533603691\\
0.39	0.650740088977971\\
0.4	0.644920164832926\\
0.41	0.639152291449203\\
0.42	0.633436003308406\\
0.43	0.627770839055524\\
0.44	0.622156341461696\\
0.45	0.616592057387308\\
0.46	0.611077537745423\\
0.47	0.605612337465531\\
0.48	0.600196015457633\\
0.49	0.594828134576638\\
0.5	0.589508261587083\\
0.51	0.584235967128166\\
0.52	0.579010825679092\\
0.53	0.573832415524733\\
0.54	0.568700318721588\\
0.55	0.563614121064055\\
0.56	0.558573412050998\\
0.57	0.553577784852616\\
0.58	0.548626836277611\\
0.59	0.543720166740643\\
0.6	0.538857380230085\\
0.61	0.534038084276055\\
0.62	0.529261889918745\\
0.63	0.524528411677028\\
0.64	0.519837267517344\\
0.65	0.515188078822869\\
0.66	0.510580470362954\\
0.67	0.506014070262845\\
0.68	0.501488509973667\\
0.69	0.497003424242677\\
0.7	0.492558451083788\\
0.71	0.488153231748353\\
0.72	0.483787410696209\\
0.73	0.479460635566984\\
0.74	0.475172557151656\\
0.75	0.470922829364372\\
0.76	0.466711109214511\\
0.77	0.462537056779008\\
0.78	0.458400335174912\\
0.79	0.454300610532204\\
0.8	0.450237551966844\\
0.81	0.44621083155407\\
0.82	0.442220124301931\\
0.83	0.438265108125055\\
0.84	0.434345463818657\\
0.85	0.430460875032773\\
0.86	0.426611028246731\\
0.87	0.422795612743845\\
0.88	0.41901432058634\\
0.89	0.415266846590493\\
0.9	0.411552888302009\\
0.91	0.407872145971605\\
0.92	0.404224322530822\\
0.93	0.400609123568043\\
0.94	0.397026257304739\\
0.95	0.393475434571912\\
0.96	0.389956368786763\\
0.97	0.386468775929558\\
0.98	0.383012374520709\\
0.99	0.37958688559805\\
1	0.376192032694329\\
1.01	0.372827541814891\\
1.02	0.369493141415564\\
1.03	0.366188562380744\\
1.04	0.362913538001678\\
1.05	0.35966780395493\\
1.06	0.356451098281056\\
1.07	0.353263161363458\\
1.08	0.350103735907431\\
1.09	0.346972566919398\\
1.1	0.343869401686326\\
1.11	0.340793989755336\\
1.12	0.337746082913484\\
1.13	0.334725435167731\\
1.14	0.331731802725087\\
1.15	0.328764943972937\\
1.16	0.32582461945954\\
1.17	0.3229105918747\\
1.18	0.32002262603062\\
1.19	0.317160488842912\\
1.2	0.314323949311791\\
1.21	0.311512778503429\\
1.22	0.308726749531479\\
1.23	0.305965637538761\\
1.24	0.303229219679116\\
1.25	0.300517275099422\\
1.26	0.297829584921764\\
1.27	0.295165932225774\\
1.28	0.292526102031121\\
1.29	0.289909881280159\\
1.3	0.287317058820736\\
1.31	0.284747425389145\\
1.32	0.282200773593242\\
1.33	0.279676897895703\\
1.34	0.277175594597438\\
1.35	0.274696661821145\\
1.36	0.272239899495027\\
1.37	0.269805109336633\\
1.38	0.267392094836864\\
1.39	0.265000661244108\\
1.4	0.262630615548522\\
1.41	0.260281766466457\\
1.42	0.257953924425017\\
1.43	0.255646901546762\\
1.44	0.253360511634539\\
1.45	0.251094570156462\\
1.46	0.248848894231012\\
1.47	0.246623302612279\\
1.48	0.244417615675336\\
1.49	0.242231655401738\\
1.5	0.240065245365158\\
1.51	0.237918210717145\\
1.52	0.235790378173013\\
1.53	0.233681575997857\\
1.54	0.231591633992688\\
1.55	0.229520383480704\\
1.56	0.227467657293668\\
1.57	0.225433289758421\\
1.58	0.223417116683511\\
1.59	0.221418975345938\\
1.6	0.219438704478024\\
1.61	0.217476144254395\\
1.62	0.215531136279083\\
1.63	0.213603523572742\\
1.64	0.211693150559977\\
1.65	0.20979986305679\\
1.66	0.207923508258135\\
1.67	0.206063934725583\\
1.68	0.204220992375103\\
1.69	0.202394532464948\\
1.7	0.200584407583649\\
1.71	0.198790471638118\\
1.72	0.197012579841858\\
1.73	0.195250588703276\\
1.74	0.193504356014103\\
1.75	0.191773740837917\\
1.76	0.190058603498767\\
1.77	0.188358805569899\\
1.78	0.186674209862587\\
1.79	0.185004680415058\\
1.8	0.183350082481519\\
1.81	0.181710282521283\\
1.82	0.180085148187989\\
1.83	0.178474548318924\\
1.84	0.176878352924433\\
1.85	0.17529643317743\\
1.86	0.173728661403001\\
1.87	0.172174911068096\\
1.88	0.170635056771321\\
1.89	0.169108974232816\\
1.9	0.167596540284222\\
1.91	0.166097632858743\\
1.92	0.164612130981292\\
1.93	0.163139914758729\\
1.94	0.161680865370184\\
1.95	0.160234865057467\\
1.96	0.158801797115563\\
1.97	0.157381545883215\\
1.98	0.155973996733586\\
1.99	0.154579036065012\\
2	0.153196551291828\\
2.01	0.151826430835285\\
2.02	0.150468564114546\\
2.03	0.149122841537755\\
2.04	0.147789154493199\\
2.05	0.146467395340537\\
2.06	0.145157457402115\\
2.07	0.143859234954356\\
2.08	0.142572623219226\\
2.09	0.141297518355779\\
2.1	0.140033817451774\\
2.11	0.138781418515371\\
2.12	0.1375402204669\\
2.13	0.136310123130701\\
2.14	0.135091027227038\\
2.15	0.133882834364089\\
2.16	0.132685447030006\\
2.17	0.131498768585039\\
2.18	0.130322703253742\\
2.19	0.129157156117243\\
2.2	0.128002033105576\\
2.21	0.126857240990101\\
2.22	0.125722687375966\\
2.23	0.124598280694662\\
2.24	0.123483930196624\\
2.25	0.122379545943911\\
2.26	0.121285038802946\\
2.27	0.120200320437322\\
2.28	0.119125303300674\\
2.29	0.11805990062961\\
2.3	0.117004026436712\\
2.31	0.115957595503595\\
2.32	0.114920523374025\\
2.33	0.11389272634711\\
2.34	0.11287412147054\\
2.35	0.111864626533892\\
2.36	0.110864160061996\\
2.37	0.10987264130836\\
2.38	0.108889990248649\\
2.39	0.107916127574233\\
2.4	0.10695097468578\\
2.41	0.105994453686917\\
2.42	0.105046487377937\\
2.43	0.104106999249577\\
2.44	0.103175913476834\\
2.45	0.102253154912852\\
2.46	0.101338649082853\\
2.47	0.100432322178129\\
2.48	0.0995341010500813\\
2.49	0.0986439132043216\\
2.5	0.0977616867948173\\
2.51	0.0968873506180941\\
2.52	0.0960208341074896\\
2.53	0.0951620673274573\\
2.54	0.0943109809679227\\
2.55	0.0934675063386892\\
2.56	0.0926315753638941\\
2.57	0.091803120576514\\
2.58	0.0909820751129207\\
2.59	0.0901683727074832\\
2.6	0.0893619476872207\\
2.61	0.0885627349665015\\
2.62	0.0877706700417905\\
2.63	0.0869856889864428\\
2.64	0.0862077284455444\\
2.65	0.085436725630799\\
2.66	0.0846726183154602\\
2.67	0.0839153448293096\\
2.68	0.0831648440536791\\
2.69	0.0824210554165182\\
2.7	0.0816839188875053\\
2.71	0.0809533749732032\\
2.72	0.0802293647122565\\
2.73	0.0795118296706337\\
2.74	0.0788007119369105\\
2.75	0.0780959541175964\\
2.76	0.0773974993325022\\
2.77	0.0767052912101491\\
2.78	0.0760192738832194\\
2.79	0.0753393919840473\\
2.8	0.0746655906401504\\
2.81	0.0739978154698007\\
2.82	0.0733360125776358\\
2.83	0.0726801285503089\\
2.84	0.0720301104521782\\
2.85	0.0713859058210339\\
2.86	0.0707474626638645\\
2.87	0.0701147294526607\\
2.88	0.069487655120256\\
2.89	0.0688661890562057\\
2.9	0.0682502811027016\\
2.91	0.0676398815505247\\
2.92	0.0670349411350323\\
2.93	0.0664354110321826\\
2.94	0.0658412428545935\\
2.95	0.0652523886476381\\
2.96	0.0646688008855737\\
2.97	0.0640904324677063\\
2.98	0.0635172367145893\\
2.99	0.0629491673642556\\
3	0.0623861785684845\\
3.01	0.0618282248891007\\
3.02	0.0612752612943074\\
3.03	0.0607272431550519\\
3.04	0.0601841262414233\\
3.05	0.0596458667190832\\
3.06	0.0591124211457273\\
3.07	0.0585837464675798\\
3.08	0.0580598000159183\\
3.09	0.0575405395036302\\
3.1	0.0570259230217994\\
3.11	0.0565159090363244\\
3.12	0.0560104563845656\\
3.13	0.0555095242720236\\
3.14	0.0550130722690463\\
3.15	0.0545210603075663\\
3.16	0.0540334486778667\\
3.17	0.0535501980253761\\
3.18	0.0530712693474929\\
3.19	0.0525966239904368\\
3.2	0.0521262236461295\\
3.21	0.0516600303491028\\
3.22	0.0511980064734342\\
3.23	0.0507401147297107\\
3.24	0.0502863181620187\\
3.25	0.0498365801449616\\
3.26	0.0493908643807036\\
3.27	0.0489491348960404\\
3.28	0.0485113560394958\\
3.29	0.048077492478444\\
3.3	0.0476475091962584\\
3.31	0.047221371489485\\
3.32	0.0467990449650421\\
3.33	0.0463804955374436\\
3.34	0.045965689426049\\
3.35	0.0455545931523363\\
3.36	0.0451471735372003\\
3.37	0.0447433976982746\\
3.38	0.0443432330472778\\
3.39	0.0439466472873832\\
3.4	0.0435536084106124\\
3.41	0.0431640846952518\\
3.42	0.0427780447032925\\
3.43	0.0423954572778927\\
3.44	0.0420162915408634\\
3.45	0.0416405168901762\\
3.46	0.041268102997493\\
3.47	0.0408990198057189\\
3.48	0.0405332375265761\\
3.49	0.0401707266381994\\
3.5	0.0398114578827538\\
3.51	0.0394554022640733\\
3.52	0.0391025310453202\\
3.53	0.0387528157466662\\
3.54	0.0384062281429937\\
3.55	0.0380627402616177\\
3.56	0.037722324380028\\
3.57	0.0373849530236523\\
3.58	0.0370505989636382\\
3.59	0.0367192352146559\\
3.6	0.0363908350327199\\
3.61	0.0360653719130312\\
3.62	0.0357428195878374\\
3.63	0.035423152024313\\
3.64	0.0351063434224583\\
3.65	0.0347923682130172\\
3.66	0.0344812010554132\\
3.67	0.0341728168357047\\
3.68	0.0338671906645575\\
3.69	0.0335642978752365\\
3.7	0.0332641140216148\\
3.71	0.0329666148762003\\
3.72	0.0326717764281807\\
3.73	0.0323795748814856\\
3.74	0.0320899866528657\\
3.75	0.0318029883699898\\
3.76	0.0315185568695581\\
3.77	0.0312366691954328\\
3.78	0.0309573025967854\\
3.79	0.0306804345262605\\
3.8	0.0304060426381562\\
3.81	0.0301341047866201\\
3.82	0.0298645990238625\\
3.83	0.0295975035983846\\
3.84	0.0293327969532234\\
3.85	0.0290704577242112\\
3.86	0.0288104647382522\\
3.87	0.0285527970116127\\
3.88	0.0282974337482284\\
3.89	0.0280443543380252\\
3.9	0.0277935383552564\\
3.91	0.0275449655568538\\
3.92	0.0272986158807941\\
3.93	0.0270544694444796\\
3.94	0.0268125065431336\\
3.95	0.0265727076482097\\
3.96	0.0263350534058164\\
3.97	0.0260995246351542\\
3.98	0.0258661023269683\\
3.99	0.0256347676420139\\
4	0.0254055019095359\\
4.01	0\\
4.02	0\\
4.03	0\\
4.04	0\\
4.05	0\\
4.06	0\\
4.07	0\\
4.08	0\\
4.09	0\\
4.1	0\\
4.11	0\\
4.12	0\\
4.13	0\\
4.14	0\\
4.15	0\\
4.16	0\\
4.17	0\\
4.18	0\\
4.19	0\\
4.2	0\\
4.21	0\\
4.22	0\\
4.23	0\\
4.24	0\\
4.25	0\\
4.26	0\\
4.27	0\\
4.28	0\\
4.29	0\\
4.3	0\\
4.31	0\\
4.32	0\\
4.33	0\\
4.34	0\\
4.35	0\\
4.36	0\\
4.37	0\\
4.38	0\\
4.39	0\\
4.4	0\\
4.41	0\\
4.42	0\\
4.43	0\\
4.44	0\\
4.45	0\\
4.46	0\\
4.47	0\\
4.48	0\\
4.49	0\\
4.5	0\\
4.51	0\\
4.52	0\\
4.53	0\\
4.54	0\\
4.55	0\\
4.56	0\\
4.57	0\\
4.58	0\\
4.59	0\\
4.6	0\\
4.61	0\\
4.62	0\\
4.63	0\\
4.64	0\\
4.65	0\\
4.66	0\\
4.67	0\\
4.68	0\\
4.69	0\\
4.7	0\\
4.71	0\\
4.72	0\\
4.73	0\\
4.74	0\\
4.75	0\\
4.76	0\\
4.77	0\\
4.78	0\\
4.79	0\\
4.8	0\\
4.81	0\\
4.82	0\\
4.83	0\\
4.84	0\\
4.85	0\\
4.86	0\\
4.87	0\\
4.88	0\\
4.89	0\\
4.9	0\\
4.91	0\\
4.92	0\\
4.93	0\\
4.94	0\\
4.95	0\\
4.96	0\\
4.97	0\\
4.98	0\\
4.99	0\\
5	0\\
5.01	0\\
5.02	0\\
5.03	0\\
5.04	0\\
5.05	0\\
5.06	0\\
5.07	0\\
5.08	0\\
5.09	0\\
5.1	0\\
5.11	0\\
5.12	0\\
5.13	0\\
5.14	0\\
5.15	0\\
5.16	0\\
5.17	0\\
5.18	0\\
5.19	0\\
5.2	0\\
5.21	0\\
5.22	0\\
5.23	0\\
5.24	0\\
5.25	0\\
5.26	0\\
5.27	0\\
5.28	0\\
5.29	0\\
5.3	0\\
5.31	0\\
5.32	0\\
5.33	0\\
5.34	0\\
5.35	0\\
5.36	0\\
5.37	0\\
5.38	0\\
5.39	0\\
5.4	0\\
5.41	0\\
5.42	0\\
5.43	0\\
5.44	0\\
5.45	0\\
5.46	0\\
5.47	0\\
5.48	0\\
5.49	0\\
5.5	0\\
5.51	0\\
5.52	0\\
5.53	0\\
5.54	0\\
5.55	0\\
5.56	0\\
5.57	0\\
5.58	0\\
5.59	0\\
5.6	0\\
5.61	0\\
5.62	0\\
5.63	0\\
5.64	0\\
5.65	0\\
5.66	0\\
5.67	0\\
5.68	0\\
5.69	0\\
5.7	0\\
5.71	0\\
5.72	0\\
5.73	0\\
5.74	0\\
5.75	0\\
5.76	0\\
5.77	0\\
5.78	0\\
5.79	0\\
5.8	0\\
5.81	0\\
5.82	0\\
5.83	0\\
5.84	0\\
5.85	0\\
5.86	0\\
5.87	0\\
5.88	0\\
5.89	0\\
5.9	0\\
5.91	0\\
5.92	0\\
5.93	0\\
5.94	0\\
5.95	0\\
5.96	0\\
5.97	0\\
5.98	0\\
5.99	0\\
6	0\\
6.01	0\\
6.02	0\\
6.03	0\\
6.04	0\\
6.05	0\\
6.06	0\\
6.07	0\\
6.08	0\\
6.09	0\\
6.1	0\\
6.11	0\\
6.12	0\\
6.13	0\\
6.14	0\\
6.15	0\\
6.16	0\\
6.17	0\\
6.18	0\\
6.19	0\\
6.2	0\\
6.21	0\\
6.22	0\\
6.23	0\\
6.24	0\\
6.25	0\\
6.26	0\\
6.27	0\\
6.28	0\\
6.29	0\\
6.3	0\\
6.31	0\\
6.32	0\\
6.33	0\\
6.34	0\\
6.35	0\\
6.36	0\\
6.37	0\\
6.38	0\\
6.39	0\\
6.4	0\\
6.41	0\\
6.42	0\\
6.43	0\\
6.44	0\\
6.45	0\\
6.46	0\\
6.47	0\\
6.48	0\\
6.49	0\\
6.5	0\\
6.51	0\\
6.52	0\\
6.53	0\\
6.54	0\\
6.55	0\\
6.56	0\\
6.57	0\\
6.58	0\\
6.59	0\\
6.6	0\\
6.61	0\\
6.62	0\\
6.63	0\\
6.64	0\\
6.65	0\\
6.66	0\\
6.67	0\\
6.68	0\\
6.69	0\\
6.7	0\\
6.71	0\\
6.72	0\\
6.73	0\\
6.74	0\\
6.75	0\\
6.76	0\\
6.77	0\\
6.78	0\\
6.79	0\\
6.8	0\\
6.81	0\\
6.82	0\\
6.83	0\\
6.84	0\\
6.85	0\\
6.86	0\\
6.87	0\\
6.88	0\\
6.89	0\\
6.9	0\\
6.91	0\\
6.92	0\\
6.93	0\\
6.94	0\\
6.95	0\\
6.96	0\\
6.97	0\\
6.98	0\\
6.99	0\\
7	0\\
};
\end{axis}
\end{tikzpicture}%
%\vspace{-.4cm}
\caption{Truncated-exponential.}
\label{Fig:TExp}
\end{subfigure}\hspace{.3cm}
\begin{subfigure}[]{.22\textwidth}
\tikzset{every picture/.style={scale=.92}, every node/.style={scale=.9}}
%\input{matlab/TruncGeo}
%\vspace{-.4cm}
% Minimal pgfplots version: 1.3
%  http://www.mathworks.com/matlabcentral/fileexchange/22022-matlab2tikz
\definecolor{mycolor1}{rgb}{0.00000,0.44700,0.74100}%
\begin{tikzpicture}

\begin{axis}[%
width=2in,
height=3in,
xmin=-1,
xmax=5,
xlabel={$x$},
xmajorgrids,
ymin=0,
ymax=0.5,
ylabel={$\mathbb{P}_X(x)$},
ylabel style = {at = {(axis cs: -.7,.25)}},
ymajorgrids,
legend style={legend cell align=left,align=left,draw=white!15!black}
]
\addplot[ycomb,color=mycolor1,solid,mark=o,mark options={solid}] plot table[row sep=crcr] {%
0	0.390341906973409\\
0.4	0.238655041923542\\
0.8	0.145913692632054\\
1.2	0.0892116316752376\\
1.6	0.0545439916062403\\
2	0.0333481964680553\\
2.4	0.020389087320569\\
2.8	0.0124658879877959\\
3.2	0.00762164391573841\\
3.6	0.00465987309008247\\
4	0.00284904640727642\\
};
\end{axis}
\end{tikzpicture}%
\caption{Truncated-geometric.}
\label{Fig:TGeo}
\end{subfigure}
\caption{Truncated-exponential distribution \eqref{TEDistribution} and truncated-geometric distribution \eqref{InputDistKMassPoints} for $\mathcal{A}=4$ and $\mathcal{E}=1$.}
\label{Fig:TruncatedDist}
\end{figure}
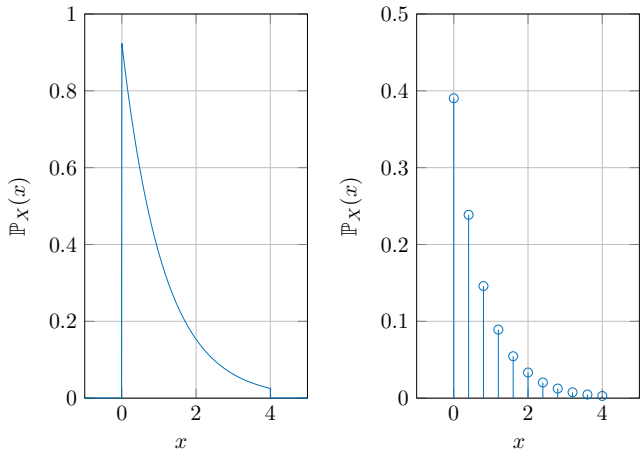

Using Lemma \ref{Lem:MaxEnt} with $\mathcal{S}=[0,\mathcal{A}]$, $n=1$, $r_1(x)=x$, and $m_1=\mathcal{E}$ leads to the following solution \cite{LapidothMoserWigger}
\begin{align}
\label{TEDistribution}
\mathbb{P}_{X}(x)=\begin{cases}
\frac{1}{\mathcal{A}}\frac{\mu^*}{1-e^{-\mu^*}}e^{-\frac{\mu^*x}{\mathcal{A}}},&\text{if  } \frac{\mathcal{E}}{\mathcal{A}}<\frac{1}{2}\\
\frac{1}{\mathcal{A}},&\text{if  } \frac{\mathcal{E}}{\mathcal{A}}=\frac{1}{2}.\end{cases}
\end{align}
for $x\in[0,\mathcal{A}]$, where $\mu^*>0$ satisfies $\frac{1}{\mu^*}-\frac{e^{-\mu^*}}{1-e^{-\mu^*}}=\frac{\mathcal{E}}{\mathcal{A}}$. These distributions are the `truncated-exponential' distribution (Fig. \ref{Fig:TExp}) and uniform distribution, respectively. The resulting maximum entropy is 
\begin{align*}
h(X)=\begin{cases}
\frac{1}{2}\log\left( \mathcal{A}^2 e^{2\mu^* \frac{\mathcal{E}}{\mathcal{A}}}\left( \frac{1-e^{-\mu^*}}{\mu^*}\right)^2\right),&\text{if  } \frac{\mathcal{E}}{\mathcal{A}}<\frac{1}{2}\\
\frac{1}{2}\log(\mathcal{A}^2),&\text{if  } \frac{\mathcal{E}}{\mathcal{A}}=\frac{1}{2}.\end{cases}
\end{align*}
This leads to the following statement.

\begin{theorem}[\cite{LapidothMoserWigger}]
\label{Thm:LowerBoundLMW}
The capacity of the IM/DD Gaussian channel with a peak constraint $\mathcal{A}$ and an average constraint $\mathcal{E}\leq\frac{\mathcal{A}}{2}$ satisfies $c_g(\mathcal{A},\mathcal{E}) \geq r_g^{\rm lmw}(\mathcal{A},\mathcal{E})$ where
\begin{align}
&r_g^{\rm lmw}(\mathcal{A},\mathcal{E})\\
&=\begin{cases} \frac{1}{2}\log\left( 1+ \frac{g^2\mathcal{A}^2}{2\pi e} e^{2\mu^* \frac{\mathcal{E}}{\mathcal{A}}}\left( \frac{1-e^{-\mu^*}}{\mu^*}\right)^2\right)&\text{if }\frac{\mathcal{E}}{\mathcal{A}}<\frac{1}{2},\\
\frac{1}{2}\log\left( 1+ \frac{g^2\mathcal{A}^2}{2\pi e}\right)&\text{if }\frac{\mathcal{E}}{\mathcal{A}}=\frac{1}{2}.\end{cases}\nonumber
\end{align}
\end{theorem}

In addition to this result, it is interesting to calculate the achievable rate using a `truncated-Gaussian' distribution, especially since a Gaussian distribution achieves the capacity of the standard AWGN channel (Remark \ref{Rem:AWGN_Capacity}). This has been derived in \cite{ChaabanMorvanAlouini}. Let the Gaussian distribution function with mean $\mu$ and variance $\nu^2$ be denoted $\mathbb{P}_{\mu,\nu}^{\rm G}(x)=\frac{1}{\sqrt{2\pi}\nu}e^{-\frac{(x-\mu)^2}{2\nu^2}}$. We construct a truncated Gaussian distribution as 
\begin{align}
\label{TGDist}
\tilde{\mathbb{P}}^{\rm G}_{\mu,\nu}(x)=\eta \mathbb{P}^{\rm G}_{\mu,\nu}(x),\quad x\in[0,\mathcal{A}],
\end{align}
where 
\begin{align}
\label{EtaTG}
\eta=\left(\mathbb{F}^{\rm G}_{\mu,\nu}(\mathcal{A})-\mathbb{F}^{\rm G}_{\mu,\nu}(0)\right)^{-1},
\end{align}
and $\mathbb{F}^{\rm G}_{\mu,\nu}(x)$ is the cumulative distribution function corresponding to $\mathbb{P}^{\rm G}_{\mu,\nu}(x)$. The mean of $\tilde{\mathbb{P}}^{\rm G}_{\mu,\nu}(x)$ is 
\begin{align}
\label{MuTilde}
\tilde{\mu}=\mu+\nu^2\eta(\mathbb{P}^{\rm G}_{\mu,\nu}(0)-\mathbb{P}^{\rm G}_{\mu,\nu}(\mathcal{A})),
\end{align} 
and its variance is 
\begin{align}
\label{NuTilde}
\tilde{\nu}^2=\nu^2\left(1-\mathcal{A}\tilde{\mathbb{P}}^{\rm G}_{\mu,\nu}(\mathcal{A})-\tilde{\mu}(\tilde{\mathbb{P}}^{\rm G}_{\mu,\nu}(0)-\tilde{\mathbb{P}}^{\rm G}_{\mu,\nu}(\mathcal{A}))\right).
\end{align} 

By choosing $\mu\in\mathbb{R}$ and $\nu>0$ so that $\tilde{\mu}\leq\mathcal{E}$, we can derive a feasible truncated-Gaussian input distribution. This leads to the following statement.\footnote{We only give a simplified version of the achievable rate of the Truncated-Gaussian distribution here. The achievable rate of this distribution is larger than that in Theorem \ref{Thm:TGLowerBound}. The reader is referred to \cite{ChaabanMorvanAlouini} for details.}

\begin{theorem}[\cite{ChaabanMorvanAlouini}]
\label{Thm:TGLowerBound}
The capacity of the IM/DD Gaussian channel with a peak constraint $\mathcal{A}$ and an average constraint $\mathcal{E}\leq\frac{\mathcal{A}}{2}$ satisfies $c_g(\mathcal{A},\mathcal{E})\geq r_g^{\rm cma}(\mathcal{A},\mathcal{E})$ where
\begin{align}
r_g^{\rm cma}(\mathcal{A},\mathcal{E})&=\max_{\mu,\nu} \frac{1}{2}\log\left(1+g^2\nu^2\right)-\log(\eta)\\
&\quad -\frac{\eta g^2\nu^2}{2(g^2\nu^2+1)}\left( (\mathcal{A}-\mu)\mathbb{P}^{\rm G}_{\mu,\nu}(\mathcal{A})+\mu \mathbb{P}^{\rm G}_{\mu,\nu}(0)\right),\nonumber
\end{align}
where $\mu\in\mathbb{R}$ and $\nu>0$ are chosen such that $\tilde{\mu}\leq\mathcal{E}$. 
\end{theorem}

In \cite{ChaabanMorvanAlouini}, specific choices of $\mu$ and $\nu$ are given that make $r_g^{\rm cma}$ approach capacity within a gap $<0.164$ nats as $\mathcal{A}$ and $\mathcal{E}$ increase (high SNR). It is also shown numerically that the gap nearly vanishes at asymptotically high SNR if we optimize $r_g^{\rm cma}$ with respect to $\mu$ and $\nu$. Note that the simple choice of $\nu=\frac{\mu}{3}$ and $\mu$ slightly lower than $\mathcal{E}$ so that $\tilde{\mu}=\mathcal{E}$ leads to the achievable rate $\frac{1}{2}\log\left(1+g^2\mathcal{E}^2\right)$ which is a simple expression but is suboptimal.

\paragraph{Discrete Input Distributions} 
We know from \cite{ChanHranilovicKschischang} (see Theorem \ref{Thm:Discreteness}) that the optimal input distribution is discrete. Thus, it is relevant to derive a lower bound using a simple discrete input distribution. To this end, \cite{FaridHranilovic_SelectedAreas} maximizes $H(X)$ while imposing the constraint that $\mathbb{P}_X$ has the form 
\begin{align}
\label{InputDistKMassPoints}
\tilde{\mathbb{P}}_X^{[k]}(x)=\sum_{i=0}^k a_i\delta\left(x-i\frac{\mathcal{A}}{k}\right),
\end{align}
i.e., a discrete distribution with support $\mathcal{X}=\{0,\frac{\mathcal{A}}{k},\frac{2\mathcal{A}}{k},\ldots,\mathcal{A}\}$. The optimization problem becomes $\max_{a_i} \left.H(X)\right|_{X\sim\tilde{\mathbb{P}}^{[k]}_X}$ subject to $\mathbb{E}[X]\leq\mathcal{E}$. This can be solved using Lagrangian duality \cite{Boyd} to obtain the following.

\begin{theorem}[{\cite{FaridHranilovic_SelectedAreas}}]
\label{Thm:FaridHranilovicLB}
The solution of the optimization $\max_{a_i} H(X)$ subject to $X\in[0,\mathcal{A}]$, $X\sim\tilde{\mathbb{P}}_X^{[k]}$ given in \eqref{InputDistKMassPoints}, and $\mathbb{E}[X]\leq\mathcal{E}$ is 
\begin{align}
a_i=\begin{cases}
\frac{t_0^i}{1+t_0+\cdots+t_0^k}, & \text{if } \frac{\mathcal{E}}{\mathcal{A}}<\frac{1}{2},\\
\frac{1}{k+1} & \text{if } \frac{\mathcal{E}}{\mathcal{A}}=\frac{1}{2},
\end{cases}
\end{align} 
where $t_0$ is the unique positive root of $\sum_{i=0}^k\left(1-\frac{i\mathcal{A}}{k\mathcal{E}}\right)t^i$. The resulting achievable rate for the IM/DD Gaussian channel is $r_g^{\rm fh}(\mathcal{A},\mathcal{E})=\max_{k} I(X;Y)$ with $X$ distributed according to the optimal solutions above.
\end{theorem}

These distributions are respectively a `truncated-geometric' distribution (Fig. \ref{Fig:TGeo}) and a discrete uniform distribution, and have been shown to be capacity-approaching in \cite{FaridHranilovic_SelectedAreas}. However, contrary to the lower bounds in Theorems \ref{Thm:LowerBoundLMW} and \ref{Thm:TGLowerBound}, the achievable rate in this case is evaluated numerically, which leads to a numerical capacity lower bound instead of an analytical expression.

\begin{example}
Consider an IM/DD Gaussian channel with $\mathcal{A}=5$, $\mathcal{E}=1.25$ and $g=1$. The lower bounds in Theorems \ref{Thm:LowerBoundLMW}-\ref{Thm:FaridHranilovicLB} yield $r^{\rm lmw}_1(5,1.25)=0.3493$, $r^{\rm cma}_1(5,1.25)=0.1242$, and $r^{\rm fh}_1(5,1.25)=0.6134$ in nats/transmission. Recall from Example \ref{Eg:Discrete2} that the capacity of this channel is $c_1(5,1.25)=0.626$ nats/transmission which is very close to $r^{\rm fh}_1(5,1.25)$ but far from $r_1^{\rm lmw}(5,1.25)$ and $r_1^{\rm cma}(5,1.25)$.
\end{example}

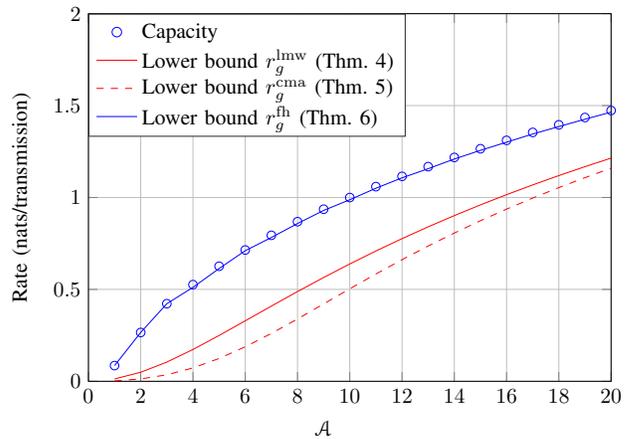
\begin{figure}[t]
\centering
\tikzset{every picture/.style={scale=.9}, every node/.style={scale=.9}}
%\input{fig/Capacity_vs_A_alpha_p25_s_1_g_1_and_LBs}
% Minimal pgfplots version: 1.3
%  http://www.mathworks.com/matlabcentral/fileexchange/22022-matlab2tikz
\begin{tikzpicture}

\begin{axis}[%
width=4in,
height=3in,
xmin=0,
xmax=20,
xlabel={$\mathcal{A}$},
xmajorgrids,
ymin=0,
ymax=2,
ylabel={Rate (nats/transmission)},
ylabel style = {at = {(axis cs: .3,1)}},
ymajorgrids,
legend style={at = {(axis cs: 0,2)}, anchor = north west, legend cell align=left,align=left,draw=white!15!black}
]
\addplot [color=blue,only marks,mark=o,mark options={solid}]
  table[row sep=crcr]{%
1	0.0854482126852514\\
2	0.265653273432171\\
3	0.422224167909863\\
4	0.525909389774591\\
5	0.625975976279065\\
6	0.714411356895874\\
7	0.794232011975669\\
8	0.868266145925912\\
9	0.936148081645106\\
10	0.999761867641674\\
11	1.05912386075089\\
    12      1.11508375525223\\
      13    1.16785448054832\\
        14  1.21789764896208\\
15          1.26528823040629\\
  16        1.31072776564075\\
    17      1.35398530185399\\
      18     1.3953914147047\\
        19  1.43510341058738\\
          20 1.47324945988358\\
};
\addlegendentry{Capacity};

\addplot [color=red,solid]
  table[row sep=crcr]{%
1	0.0127644225629829\\
2	0.049211749346658\\
3	0.10461052941365\\
4	0.173112246106981\\
5	0.249306907354123\\
6	0.328985309745506\\
7	0.409232511241468\\
8	0.488199689786092\\
9	0.564806262571931\\
10	0.638483784426669\\
11 0.708988210709817\\
   12      0.776272859750962\\
     13    0.840406222792222\\
       14  0.901520247515704\\
15         0.959778451113847\\
  16        1.01535663458169\\
    17      1.06843149496692\\
      18     1.1191741417655\\
        19  1.16774663294667\\
20          1.21430035093709\\
};
\addlegendentry{Lower bound $r^{\rm lmw}_g$ (Thm. \ref{Thm:LowerBoundLMW})};

\addplot [color=red,dashed]
  table[row sep=crcr]{%
1	0.00235890953422199\\
2	0.0127938348913469\\
3	0.0354813563389772\\
4	0.0727251702057689\\
5	0.124223692835128\\
6	0.187805704143861\\
7	0.260509813059802\\
8	0.339226324670374\\
9	0.420991379428624\\
10	0.50320506378959\\
11 0.583868878132628\\
   12      0.661699546667754\\
     13    0.736034591212084\\
       14  0.806640167581324\\
15         0.873537310305114\\
  16       0.936882765179455\\
    17     0.996894815449553\\
      18    1.05381551730894\\
        19  1.10788521092517\\
20          1.15933309314181\\
};
\addlegendentry{Lower bound $r^{\rm cma}_g$ (Thm. \ref{Thm:TGLowerBound})};

\addplot [color=blue,solid]
  table[row sep=crcr]{%
1	0.08544941268263\\
2	0.265653333497137\\
3	0.422225367906943\\
4	0.511059758855946\\
5	0.613469718138603\\
6	0.710515671121357\\
7	0.782437627787918\\
8	0.858338907090682\\
9	0.929570910582446\\
10	0.987764759201703\\
11 1.05033899612093\\
   12       1.10698853630084\\
     13     1.15661188696772\\
       14   1.20929517922524\\
15          1.25647124603967\\
  16         1.3008655300832\\
    17      1.34533812183841\\
      18    1.38582556949851\\
        19  1.42593093495513\\
20          1.46444527396744\\
};
\addlegendentry{Lower bound $r^{\rm fh}_g$ (Thm. \ref{Thm:FaridHranilovicLB})};

\end{axis}
\end{tikzpicture}%
\caption{Achievable rates of Theorems \ref{Thm:LowerBoundLMW}-\ref{Thm:FaridHranilovicLB} versus $\mathcal{A}$ for a channel with $\mathcal{E}=\mathcal{A}/4$, and $g=1$.}
\label{Fig:RatesvsA}
\end{figure}

This example suggests that $r_g^{\rm fh}(\mathcal{A},\mathcal{E})$ is a good lower bound. While $r_g^{\rm lmw}(\mathcal{A},\mathcal{E})$ and $r_g^{\rm cma}(\mathcal{A},\mathcal{E})$ are away from capacity in this example, they both become closer to capacity as $\mathcal{A}$ increases with $\mathcal{E}$ held proportional to $\mathcal{A}$ as we shall see later. Fig. \ref{Fig:RatesvsA} shows these bounds graphically.

\begin{figure*}[!b]
\hrule
\begin{align}
\label{bglmw1}
b_g^{\rm lmw}(\nu,\mu,\mathcal{A},\mathcal{E})&=
\left(1-Q\left(\nu+g\mathcal{E}\right)-Q\left(\nu+g(\mathcal{A}-\mathcal{E})\right)\right)\log\left(g\mathcal{A} \frac{e^{\frac{\mu\nu}{g\mathcal{A}}}-e^{-\mu(1+\frac{\nu}{g\mathcal{A}})}}{\sqrt{2\pi}\mu (1-2Q(\nu))} \right)-\frac{1}{2}\nonumber\\
&\quad +Q\left(\nu\right)+\frac{\nu e^{-\frac{\nu^2}{2}}}{\sqrt{2\pi}}+\frac{\mu}{\nu\sqrt{2\pi}}\left( e^{-\frac{\nu^2}{2}}-e^{-\frac{(g\mathcal{A}+\nu)^2}{2}} \right)+\mu\frac{\mathcal{E}}{\mathcal{A}}\left( 1-2Q\left(\nu+\frac{g\mathcal{A}}{2}\right) \right),\\
\label{bglmw2}
\tilde{b}_g^{\rm lmw}(\nu,\mathcal{A})&=
\left(1-2Q\left(\nu+\frac{g\mathcal{A}}{2}\right)\right)\log\left(\frac{g\mathcal{A} + 2\nu}{\sqrt{2\pi}\mu (1-2Q(\nu))} \right)-\frac{1}{2} +Q\left(\nu\right)+\frac{\nu e^{-\frac{\nu^2}{2}}}{\sqrt{2\pi}}.
\end{align}
\end{figure*}

\subsubsection{Average Constraint Only} 
\label{Sec:CapBoundsAvgConstOnly}
Using the EPI to lower bound the channel capacity in this case requires finding an input distribution on $[0,\infty)$ satisfying $\mathbb{E}[X]\leq\mathcal{E}$ which maximizes $h(X)$ or $H(X)$. We start with continuous distributions.

\paragraph{Continuous Input Distributions}
The max-entropic continous distribution which maximizes $h(X)$ in this case is the exponential distribution $\mathbb{P}_{X}(x)=\frac{1}{\mathcal{E}}e^{-\frac{x}{\mathcal{E}}}$ for $x\geq0$ \cite[Example 12.2.5]{CoverThomas}. This leads to the following lower bound given in \cite{LapidothMoserWigger}.

\begin{theorem}[{\cite{LapidothMoserWigger}}]
\label{Thm:LB_Exp}
The capacity of the IM/DD Gaussian channel with only an average constraint $\mathcal{E}$ satisfies 
\begin{align}
c_g(\infty,\mathcal{E})& \geq r_g^{\rm lmw,a}(\mathcal{E})=\frac{1}{2}\log\left( 1+ \frac{eg^2\mathcal{E}^2}{2\pi} \right).
\end{align}
\end{theorem}

The truncated-Gaussian lower bound given in Theorem \ref{Thm:TGLowerBound} can be specialized for this case by setting $\mathcal{A}=\infty$ leading to the following statement.

\begin{corollary}
For $\mathcal{A}=\infty$, the truncated-Gaussian distribution capacity lower bound in Theorem \ref{Thm:TGLowerBound} becomes
\begin{align}
r_g^{\rm cma,a}(\mathcal{E})&=\max_{\mu,\nu}\frac{1}{2}\log\left(1+g^2\nu^2\right)-\log(\eta)\nonumber\\
&\qquad-\frac{\eta g^2\nu^2}{2(g^2\nu^2+1)}\mu \mathbb{P}^{\rm G}_{\mu,\nu}(0),
\end{align}
where $\eta=(1-\mathbb{F}^{\rm G}_{\mu,\nu}(0))^{-1}$, and where $\mu\in\mathbb{R}$ and $\nu>0$ are chosen such $\mu+\nu^2\eta \mathbb{P}^{\rm G}_{\mu,\nu}(0)\leq\mathcal{E}$. This is an achievable rate for an IM/DD Gaussian channel with only an average constraint.
\end{corollary}

Next, we consider discrete input distributions.

\paragraph{Discrete Input Distributions}
We want to find the discrete input distribution which satisfies the constraints and maximizes $H(X)$. To simplify the search, we restrict our attention to distributions of the form
\begin{align}
\label{InputDistInfMassPoints}
\hat{\mathbb{P}}_X^{[\ell]}(x)=\sum_{i=0}^\infty a_i\delta (x-i\ell)
\end{align}
for some $\ell>0$. The max-entropic $\hat{\mathbb{P}}_X^{[\ell]}$ was derived in \cite{FaridHranilovic}, as stated next.

\begin{theorem}[{\cite{FaridHranilovic}}]
\label{Thm:LB_Geom}
The solution of the optimization $\max_{a_i} H(X)$ subject to $X\geq0$, $X\sim\hat{\mathbb{P}}_X^{[\ell]}$ given in \eqref{InputDistInfMassPoints}, and $\mathbb{E}[X]\leq\mathcal{E}$ is the geometric distribution $a_i=\frac{\ell}{\ell+\mathcal{E}}\left(\frac{\mathcal{E}}{\ell+\mathcal{E}}\right)^i$. The resulting achievable rate in an IM/DD Gaussian channel is $r_g^{\rm fh,a}(\mathcal{E})=\max_{\ell}I(X;Y)$ with $X$ distributed according to the geometric distribution above.
\end{theorem}

Generally, the achievable rate of this distribution is higher than $r_g^{\rm lmw,a}$ and $r_g^{\rm cma,a}$, but lacks an analytical expression. We shall see that analytical lower bounds are very useful for deriving asymptotic capacity results in Sec. \ref{Sec:AsymptoticCapacity}. Next, we discuss capacity upper bounds.

\subsection{Capacity Upper Bounds}
\label{Sec:CapacityUpperBounds}
Capacity upper bounds for the IM/DD Gaussian channel have been derived using one of three methods: Duality, sphere packing, or constraint relaxation. Duality bounds are derived using a dual expression of the channel capacity that has been given in  \cite{Moser_Dissertation,LapidothMoser_Duality} as follows.

\begin{lemma}[{\cite{LapidothMoserWigger}}]
\label{Lem:Duality}
For a channel with input $X\in\mathcal{X}$ and output $Y\in\mathcal{Y}$ described by the transition probability $\mathbb{P}_{Y|X}$, the capacity is upper bounded by $\sup_{\mathbb{P}_X} \mathbb{E}_X\left[ D( \mathbb{P}_{Y|X}\| \mathbb{P}_Y) \right]$ for any distribution $\mathbb{P}_Y$.
\end{lemma}

The sphere packing approach has been used earlier for the standard AWGN channel in \cite{ShannonNoise} and will be detailed in Sec. \ref{Sec:UpperBoundsSpherePacking}. Constraint relaxations refers to replacing the constraints of the capacity maximization problem by ones which simplify the maximization problem as we shall see in Sec. \ref{Sec:UpperBoundsBothConstraints}. Upper bounds on the capacity of the IM/DD Gaussian channel which use these methods are discussed next.

\subsubsection{Average and Peak Constraints} 
\label{Sec:UpperBoundsBothConstraints}
In this case, we have bounds based on the duality approach and bounds using constraint relaxation.

\paragraph{Duality Upper Bound}
\label{Sec:UpperBoundsBothConstraintsDuality}
To apply Lemma \ref{Lem:Duality} for the IM/DD Gaussian channel in this case, we restrict $\mathbb{P}_X$ to satisfy $\mathcal{X}=[0,\mathcal{A}]$ and $\mathbb{E}[X]\leq\mathcal{E}$, and we fix $\mathbb{P}_{Y|X}$ to be the IM/DD channel law, i.e., Gaussian. The main difficulties in deriving an upper bound based on Lemma \ref{Lem:Duality} are finding a good $\mathbb{P}_Y$, calculating the expectation $\mathbb{E}_X$, and maximizing with respect to $\mathbb{P}_X$. The last two challenges can be simplified by upper bounding $D( \mathbb{P}_{Y|X}\| \mathbb{P}_Y)$ using bounds that simplify the expectation $\mathbb{E}_X$. Determining a good $\mathbb{P}_Y$ requires some intuition. Let us focus on high SNR first, and let us study an equivalent normalized channel where $\tilde{Y}=X+\tilde{Z}$ where $\tilde{Z}\sim\mathcal{N}(0,\tilde{\sigma}^2)$ and $\tilde{\sigma}=\frac{1}{g}$. At high SNR, one expects the output distribution $\mathbb{P}_{\tilde{Y}}$ to be `similar to' the input distribution $\mathbb{P}_X$ since the noise variance is negligible at high SNR relative to $\mathcal{A}$. Thus, the maximum $h(\tilde{Y})$ should be close to the maximum $h(X)$, which is achieved by the distribution in \eqref{TEDistribution}. The maximum $h(\tilde{Y})$ determines the channel capacity since $I(X;\tilde{Y})=h(\tilde{Y})-h(\tilde{Z})$. Thus, one expects that choosing $\mathbb{P}_{\tilde{Y}}$ as given in \eqref{TEDistribution} is a good choice. To generalize this insight to any SNR, we choose $\mathbb{P}_{\tilde{Y}}$ to be `similar to' $\mathbb{P}_X$ between $0$ and $\mathcal{A}$, and to have a Gaussian roll-off outside this interval. To quantify this statement, we can use a parameter $\delta>0$, and choose $\mathbb{P}_{\tilde{Y}}(\tilde{y})$ to be similar to $\mathbb{P}_X(x)$ for $\tilde{y}\in[-\delta,\mathcal{A}+\delta]$ and to have a Gaussian roll-off for $\tilde{y}<-\delta$ and $\tilde{y}>\mathcal{A}+\delta$. With appropriate normalization, this leads to the following distribution \cite{LapidothMoserWigger}
\begin{align*}
\mathbb{P}_{\tilde{Y}}(\tilde{y})=
\begin{cases}
\frac{1}{\sqrt{2\pi}\tilde{\sigma}}e^{-\frac{\tilde{y}^2}{2\tilde{\sigma}^2}},& \tilde{y}<-\delta\\
\frac{1}{\mathcal{A}}\frac{\mu (1-2Q(\frac{\delta}{\tilde{\sigma}}))}{e^{\frac{\mu\delta}{\mathcal{A}}}-e^{-\mu(1+\frac{\delta}{\mathcal{A}})}}e^{-\frac{\mu \tilde{y}}{\mathcal{A}}}, & -\delta\leq \tilde{y}\leq \mathcal{A}+\delta\\
\frac{1}{\sqrt{2\pi}\tilde{\sigma}}e^{-\frac{(\tilde{y}-\mathcal{A})^2}{2\tilde{\sigma}^2}},&\tilde{y}>\mathcal{A}+\delta.\end{cases}
\end{align*}
Plugging this distribution in Lemma \ref{Lem:Duality}, it remains to maximize the expectation with respect to $\mathbb{P}_X$ on $[0,\mathcal{A}]$ with $\mathbb{E}[X]\leq\mathcal{E}$. This has been bounded in \cite{LapidothMoserWigger}, leading to the following.

\begin{theorem}[{\cite{LapidothMoserWigger}}]
\label{Thm:DualityBnd1}
The capacity of the IM/DD Gaussian channel with a peak constraint $\mathcal{A}$ and an average constraint $\mathcal{E}\leq\frac{\mathcal{A}}{2}$ satisfies $c_g(\mathcal{A},\mathcal{E})\leq \overline{r}_g^{\rm lmw}(\mathcal{A},\mathcal{E})$ where 
\begin{align*}
\overline{r}_g^{\rm lmw}(\mathcal{A},\mathcal{E})&=\begin{cases}
\min_{\nu>0,\,\mu>0} b_g^{\rm lmw}(\nu,\mu,\mathcal{A},\mathcal{E}),& \mathcal{E}<\frac{\mathcal{A}}{2},\\
\min_{\nu>0} \tilde{b}_g^{\rm lmw}(\nu,\mathcal{A}),& \mathcal{E}=\frac{\mathcal{A}}{2},
\end{cases}
\end{align*}
and $b_g^{\rm lmw}$ and $\tilde{b}_g^{\rm lmw}$ are given in \eqref{bglmw1} and \eqref{bglmw2} given at the bottom of the page.
\end{theorem}

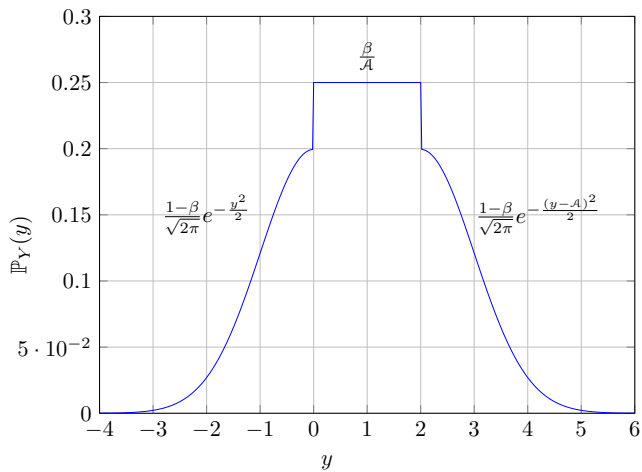
\begin{figure}[t]
\centering
\tikzset{every picture/.style={scale=.85}, every node/.style={scale=1}}
%\input{matlab/McKellips.tex}
% Minimal pgfplots version: 1.3
%  http://www.mathworks.com/matlabcentral/fileexchange/22022-matlab2tikz
\definecolor{mycolor1}{rgb}{0.00000,0.44700,0.74100}%
\begin{tikzpicture}

\begin{axis}[%
width=4.5in,
height=3.5in,
xmin=-4,
xmax=6,
xlabel={$y$},
xmajorgrids,
ymin=0,
ymax=0.3,
ylabel={$\mathbb{P}_Y(y)$},
ymajorgrids,
legend style={legend cell align=left,align=left,draw=white!15!black}
]
\addplot [color=blue,solid,forget plot]
  table[row sep=crcr]{%
-4	6.69151128824427e-05\\
-3.98	7.24737802119455e-05\\
-3.96	7.84628170327661e-05\\
-3.94	8.49127997146718e-05\\
-3.92	9.18562490012286e-05\\
-3.9	9.93277356963864e-05\\
-3.88	0.000107363990750184\\
-3.86	0.000116004019828471\\
-3.84	0.00012528922244543\\
-3.82	0.000135263515730761\\
-3.8	0.00014597346289573\\
-3.78	0.000157468406453761\\
-3.76	0.000169800606241827\\
-3.74	0.000183025382278667\\
-3.72	0.000197201262484578\\
-3.7	0.000212390135275376\\
-3.68	0.000228657407029928\\
-3.66	0.000246072164416447\\
-3.64	0.000264707341547467\\
-3.62	0.000284639891917126\\
-3.6	0.000305950965056886\\
-3.58	0.000328726087827338\\
-3.56	0.000353055350244018\\
-3.54	0.000379033595714355\\
-3.52	0.000406760615540904\\
-3.5	0.00043634134752288\\
-3.48	0.00046788607846374\\
-3.46	0.000501510650367119\\
-3.44	0.000537336670076868\\
-3.42	0.000575491722089242\\
-3.4	0.00061610958423651\\
-3.38	0.000659330445911371\\
-3.36	0.000705301128470692\\
-3.34	0.000754175307425154\\
-3.32	0.000806113735988562\\
-3.3	0.000861284469526841\\
-3.28	0.000919863090412139\\
-3.26	0.000982032932752188\\
-3.24	0.00104798530642897\\
-3.22	0.00111791971984427\\
-3.2	0.00119204410073242\\
-3.18	0.00127057501436326\\
-3.16	0.00135373787842035\\
-3.14	0.00144176717380172\\
-3.12	0.00153490665055237\\
-3.1	0.00163340952809996\\
-3.08	0.00173753868892747\\
-3.06	0.00184756686477952\\
-3.04	0.00196377681446239\\
-3.02	0.00208646149226198\\
-3	0.002215924205969\\
-2.98	0.00235247876346699\\
-2.96	0.00249644960680619\\
-2.94	0.00264817193265551\\
-2.92	0.00280799179799548\\
-2.9	0.00297626620988793\\
-2.88	0.00315336319813296\\
-2.86	0.00333966186960131\\
-2.84	0.00353555244300972\\
-2.82	0.00374143626289028\\
-2.8	0.00395772579148998\\
-2.78	0.00418484457732651\\
-2.76	0.00442322719911861\\
-2.74	0.00467331918380614\\
-2.72	0.00493557689737557\\
-2.7	0.0052104674072113\\
-2.68	0.00549846831470279\\
-2.66	0.00580006755685128\\
-2.64	0.00611576317563899\\
-2.62	0.00644606305394765\\
-2.6	0.00679148461684282\\
-2.58	0.00715255449707484\\
-2.56	0.00752980816368872\\
-2.54	0.00792378951268041\\
-2.52	0.00833505041869053\\
-2.5	0.00876415024678427\\
-2.48	0.00921165532343102\\
-2.46	0.00967813836586848\\
-2.44	0.0101641778691129\\
-2.42	0.0106703574499614\\
-2.4	0.0111972651474214\\
-2.38	0.0117454926791007\\
-2.36	0.0123156346531913\\
-2.34	0.0129082877357938\\
-2.32	0.0135240497734409\\
-2.3	0.0141635188708006\\
-2.28	0.0148272924236706\\
-2.26	0.0155159661075041\\
-2.24	0.0162301328218487\\
-2.22	0.0169703815912246\\
-2.2	0.0177372964231157\\
-2.18	0.0185314551239033\\
-2.16	0.0193534280737278\\
-2.14	0.0202037769614302\\
-2.12	0.0210830534808852\\
-2.1	0.0219917979902136\\
-2.08	0.0229305381355274\\
-2.06	0.0238997874410385\\
-2.04	0.0249000438675354\\
-2.02	0.0259317883414103\\
-2	0.026995483256594\\
-1.98	0.028091570951934\\
-1.96	0.0292204721667257\\
-1.94	0.0303825844772824\\
-1.92	0.0315782807175993\\
-1.9	0.0328079073873383\\
-1.88	0.0340717830505223\\
-1.86	0.0353701967284917\\
-1.84	0.0367034062908285\\
-1.82	0.0380716368481037\\
-1.8	0.0394750791504471\\
-1.78	0.0409138879960714\\
-1.76	0.0423881806540111\\
-1.74	0.0438980353054528\\
-1.72	0.0454434895081414\\
-1.7	0.0470245386884435\\
-1.68	0.0486411346657337\\
-1.66	0.0502931842138453\\
-1.64	0.0519805476643821\\
-1.62	0.0537030375567419\\
-1.6	0.0554604173397278\\
-1.58	0.0572524001296462\\
-1.56	0.0590786475297911\\
-1.54	0.0609387685162009\\
-1.52	0.0628323183945441\\
-1.5	0.0647587978329459\\
-1.48	0.0667176519755012\\
-1.46	0.0687082696411409\\
-1.44	0.0707299826124194\\
-1.42	0.0727820650186738\\
-1.4	0.0748637328178724\\
-1.38	0.0769741433813169\\
-1.36	0.0791123951851915\\
-1.34	0.0812775276127671\\
-1.32	0.0834685208708569\\
-1.3	0.0856842960239037\\
-1.28	0.0879237151488312\\
-1.26	0.0901855816135402\\
-1.24	0.0924686404816527\\
-1.22	0.0947715790458201\\
-1.2	0.0970930274916065\\
-1.18	0.0994315596936379\\
-1.16	0.10178569414538\\
-1.14	0.104153895023554\\
-1.12	0.106534573387859\\
-1.1	0.108926088516275\\
-1.08	0.111326749375881\\
-1.06	0.113734816228693\\
-1.04	0.116148502371683\\
-1.02	0.11856597600969\\
-1	0.120985362259572\\
-0.98	0.123404745283521\\
-0.96	0.125822170549059\\
-0.94	0.12823564721281\\
-0.92	0.130643150624777\\
-0.9	0.133042624949377\\
-0.88	0.135431985899169\\
-0.86	0.137809123576728\\
-0.84	0.14017190541981\\
-0.82	0.142518179244504\\
-0.8	0.144845776380741\\
-0.78	0.147152514894163\\
-0.76	0.149436202887976\\
-0.74	0.15169464187815\\
-0.72	0.153925630234927\\
-0.7	0.156126966683381\\
-0.68	0.158296453855446\\
-0.66	0.160431901885586\\
-0.64	0.162531132042041\\
-0.62	0.164591980385382\\
-0.6	0.1666123014459\\
-0.58	0.16858997191119\\
-0.56	0.170522894315176\\
-0.54	0.172409000719667\\
-0.52	0.174246256379487\\
-0.5	0.17603266338215\\
-0.48	0.177766264252999\\
-0.46	0.179445145516772\\
-0.44	0.181067441206546\\
-0.42	0.182631336311077\\
-0.4	0.184135070151662\\
-0.38	0.185576939679733\\
-0.36	0.186955302686564\\
-0.34	0.188268580916627\\
-0.32	0.189515263076351\\
-0.3	0.190693907730262\\
-0.28	0.191803146076739\\
-0.26	0.192841684595908\\
-0.24	0.193808307562507\\
-0.22	0.194701879416895\\
-0.2	0.195521346987728\\
-0.18	0.196265741560214\\
-0.16	0.19693418078427\\
-0.14	0.197525870417306\\
-0.12	0.198040105896828\\
-0.1	0.198476273738506\\
-0.0800000000000001	0.198833852755804\\
-0.0600000000000001	0.199112415097803\\
-0.04	0.199311627102303\\
-0.02	0.199431249961833\\
0	0.25\\
0.0200000000000005	0.25\\
0.04	0.25\\
0.0600000000000005	0.25\\
0.0800000000000001	0.25\\
0.0999999999999996	0.25\\
0.12	0.25\\
0.14	0.25\\
0.16	0.25\\
0.18	0.25\\
0.2	0.25\\
0.22	0.25\\
0.24	0.25\\
0.26	0.25\\
0.28	0.25\\
0.3	0.25\\
0.32	0.25\\
0.34	0.25\\
0.36	0.25\\
0.38	0.25\\
0.4	0.25\\
0.42	0.25\\
0.44	0.25\\
0.46	0.25\\
0.48	0.25\\
0.5	0.25\\
0.52	0.25\\
0.54	0.25\\
0.56	0.25\\
0.58	0.25\\
0.600000000000001	0.25\\
0.62	0.25\\
0.64	0.25\\
0.66	0.25\\
0.68	0.25\\
0.7	0.25\\
0.72	0.25\\
0.74	0.25\\
0.76	0.25\\
0.78	0.25\\
0.8	0.25\\
0.82	0.25\\
0.84	0.25\\
0.86	0.25\\
0.88	0.25\\
0.9	0.25\\
0.92	0.25\\
0.94	0.25\\
0.96	0.25\\
0.98	0.25\\
1	0.25\\
1.02	0.25\\
1.04	0.25\\
1.06	0.25\\
1.08	0.25\\
1.1	0.25\\
1.12	0.25\\
1.14	0.25\\
1.16	0.25\\
1.18	0.25\\
1.2	0.25\\
1.22	0.25\\
1.24	0.25\\
1.26	0.25\\
1.28	0.25\\
1.3	0.25\\
1.32	0.25\\
1.34	0.25\\
1.36	0.25\\
1.38	0.25\\
1.4	0.25\\
1.42	0.25\\
1.44	0.25\\
1.46	0.25\\
1.48	0.25\\
1.5	0.25\\
1.52	0.25\\
1.54	0.25\\
1.56	0.25\\
1.58	0.25\\
1.6	0.25\\
1.62	0.25\\
1.64	0.25\\
1.66	0.25\\
1.68	0.25\\
1.7	0.25\\
1.72	0.25\\
1.74	0.25\\
1.76	0.25\\
1.78	0.25\\
1.8	0.25\\
1.82	0.25\\
1.84	0.25\\
1.86	0.25\\
1.88	0.25\\
1.9	0.25\\
1.92	0.25\\
1.94	0.25\\
1.96	0.25\\
1.98	0.25\\
2	0.25\\
2.02	0.199431249961833\\
2.04	0.199311627102303\\
2.06	0.199112415097803\\
2.08	0.198833852755804\\
2.1	0.198476273738506\\
2.12	0.198040105896828\\
2.14	0.197525870417306\\
2.16	0.19693418078427\\
2.18	0.196265741560214\\
2.2	0.195521346987728\\
2.22	0.194701879416895\\
2.24	0.193808307562507\\
2.26	0.192841684595908\\
2.28	0.191803146076739\\
2.3	0.190693907730262\\
2.32	0.189515263076351\\
2.34	0.188268580916627\\
2.36	0.186955302686564\\
2.38	0.185576939679733\\
2.4	0.184135070151662\\
2.42	0.182631336311077\\
2.44	0.181067441206546\\
2.46	0.179445145516772\\
2.48	0.177766264252999\\
2.5	0.17603266338215\\
2.52	0.174246256379487\\
2.54	0.172409000719667\\
2.56	0.170522894315176\\
2.58	0.16858997191119\\
2.6	0.1666123014459\\
2.62	0.164591980385382\\
2.64	0.162531132042041\\
2.66	0.160431901885586\\
2.68	0.158296453855446\\
2.7	0.156126966683381\\
2.72	0.153925630234927\\
2.74	0.15169464187815\\
2.76	0.149436202887976\\
2.78	0.147152514894163\\
2.8	0.144845776380741\\
2.82	0.142518179244504\\
2.84	0.14017190541981\\
2.86	0.137809123576728\\
2.88	0.135431985899169\\
2.9	0.133042624949377\\
2.92	0.130643150624777\\
2.94	0.12823564721281\\
2.96	0.125822170549059\\
2.98	0.123404745283521\\
3	0.120985362259572\\
3.02	0.11856597600969\\
3.04	0.116148502371683\\
3.06	0.113734816228693\\
3.08	0.111326749375881\\
3.1	0.108926088516275\\
3.12	0.106534573387859\\
3.14	0.104153895023554\\
3.16	0.10178569414538\\
3.18	0.0994315596936379\\
3.2	0.0970930274916065\\
3.22	0.0947715790458201\\
3.24	0.0924686404816527\\
3.26	0.0901855816135402\\
3.28	0.0879237151488312\\
3.3	0.0856842960239037\\
3.32	0.0834685208708569\\
3.34	0.0812775276127671\\
3.36	0.0791123951851915\\
3.38	0.0769741433813169\\
3.4	0.0748637328178724\\
3.42	0.0727820650186738\\
3.44	0.0707299826124194\\
3.46	0.0687082696411409\\
3.48	0.0667176519755012\\
3.5	0.0647587978329459\\
3.52	0.0628323183945441\\
3.54	0.0609387685162009\\
3.56	0.0590786475297911\\
3.58	0.0572524001296462\\
3.6	0.0554604173397278\\
3.62	0.0537030375567419\\
3.64	0.0519805476643821\\
3.66	0.0502931842138453\\
3.68	0.0486411346657337\\
3.7	0.0470245386884435\\
3.72	0.0454434895081414\\
3.74	0.0438980353054528\\
3.76	0.0423881806540111\\
3.78	0.0409138879960714\\
3.8	0.0394750791504471\\
3.82	0.0380716368481037\\
3.84	0.0367034062908285\\
3.86	0.0353701967284917\\
3.88	0.0340717830505223\\
3.9	0.0328079073873383\\
3.92	0.0315782807175993\\
3.94	0.0303825844772824\\
3.96	0.0292204721667257\\
3.98	0.028091570951934\\
4	0.026995483256594\\
4.02	0.0259317883414103\\
4.04	0.0249000438675354\\
4.06	0.0238997874410385\\
4.08	0.0229305381355274\\
4.1	0.0219917979902136\\
4.12	0.0210830534808852\\
4.14	0.0202037769614302\\
4.16	0.0193534280737278\\
4.18	0.0185314551239033\\
4.2	0.0177372964231157\\
4.22	0.0169703815912246\\
4.24	0.0162301328218487\\
4.26	0.0155159661075041\\
4.28	0.0148272924236706\\
4.3	0.0141635188708006\\
4.32	0.0135240497734409\\
4.34	0.0129082877357938\\
4.36	0.0123156346531913\\
4.38	0.0117454926791007\\
4.4	0.0111972651474214\\
4.42	0.0106703574499614\\
4.44	0.0101641778691129\\
4.46	0.00967813836586848\\
4.48	0.00921165532343102\\
4.5	0.00876415024678427\\
4.52	0.00833505041869054\\
4.54	0.00792378951268041\\
4.56	0.00752980816368872\\
4.58	0.00715255449707484\\
4.6	0.00679148461684282\\
4.62	0.00644606305394765\\
4.64	0.00611576317563899\\
4.66	0.00580006755685128\\
4.68	0.00549846831470279\\
4.7	0.0052104674072113\\
4.72	0.00493557689737557\\
4.74	0.00467331918380614\\
4.76	0.00442322719911861\\
4.78	0.00418484457732651\\
4.8	0.00395772579148998\\
4.82	0.00374143626289028\\
4.84	0.00353555244300972\\
4.86	0.00333966186960131\\
4.88	0.00315336319813296\\
4.9	0.00297626620988792\\
4.92	0.00280799179799548\\
4.94	0.00264817193265551\\
4.96	0.00249644960680619\\
4.98	0.00235247876346699\\
5	0.002215924205969\\
5.02	0.00208646149226198\\
5.04	0.00196377681446239\\
5.06	0.00184756686477952\\
5.08	0.00173753868892747\\
5.1	0.00163340952809996\\
5.12	0.00153490665055237\\
5.14	0.00144176717380172\\
5.16	0.00135373787842035\\
5.18	0.00127057501436326\\
5.2	0.00119204410073242\\
5.22	0.00111791971984427\\
5.24	0.00104798530642897\\
5.26	0.000982032932752188\\
5.28	0.000919863090412139\\
5.3	0.000861284469526841\\
5.32	0.000806113735988562\\
5.34	0.000754175307425154\\
5.36	0.000705301128470691\\
5.38	0.000659330445911371\\
5.4	0.000616109584236509\\
5.42	0.000575491722089242\\
5.44	0.000537336670076868\\
5.46	0.000501510650367119\\
5.48	0.000467886078463739\\
5.5	0.00043634134752288\\
5.52	0.000406760615540905\\
5.54	0.000379033595714355\\
5.56	0.000353055350244019\\
5.58	0.000328726087827338\\
5.6	0.000305950965056886\\
5.62	0.000284639891917126\\
5.64	0.000264707341547468\\
5.66	0.000246072164416447\\
5.68	0.000228657407029929\\
5.7	0.000212390135275376\\
5.72	0.000197201262484578\\
5.74	0.000183025382278667\\
5.76	0.000169800606241827\\
5.78	0.000157468406453761\\
5.8	0.00014597346289573\\
5.82	0.00013526351573076\\
5.84	0.00012528922244543\\
5.86	0.000116004019828471\\
5.88	0.000107363990750184\\
5.9	9.93277356963862e-05\\
5.92	9.18562490012286e-05\\
5.94	8.49127997146716e-05\\
5.96	7.84628170327661e-05\\
5.98	7.24737802119454e-05\\
6	6.69151128824427e-05\\
};

\node at (axis cs: -2, .15) {$\frac{1-\beta}{\sqrt{2\pi}}e^{-\frac{y^2}{2}}$};
\node at (axis cs: 4.2, .15) {$\frac{1-\beta}{\sqrt{2\pi}}e^{-\frac{(y-\mathcal{A})^2}{2}}$};
\node at (axis cs: 1, .27) {$\frac{\beta}{\mathcal{A}}$};

\end{axis}
\end{tikzpicture}%
\caption{Distribution $\mathbb{P}_Y$ for deriving the McKellips bound in Theorem \ref{Thm:McKellipsBound} \cite{ThangarajKramerBoecherer}. Here, $\mathcal{A}=2$.}
\label{Fig:McKellips}
\end{figure}

Note that in \cite{McKellips}, McKellips derived a capacity upper bound for the additive Gaussian noise channel with peak constraints $|X|\leq \sqrt{P}$. In \cite{ThangarajKramerBoecherer}, it was shown that McKellips' bound can be obtained as a special case of the Duality bound in Lemma \ref{Lem:Duality} with the distribution shown in Fig. \ref{Fig:McKellips}. This bound can be easily modified to obtain a bound for the IM/DD Gaussian channel with a constraint $0\leq X\leq\mathcal{A}$, and also holds under a redundant average constraint $\mathcal{E}=\frac{\mathcal{A}}{2}$, as stated next.

\begin{theorem}[{\cite{McKellips}}]
\label{Thm:McKellipsBound}
The capacity of the IM/DD Gaussian channel with a peak constraint $\mathcal{A}$ satisfies $c_g(\mathcal{A},\frac{\mathcal{A}}{2})\leq \overline{r}_g^{\rm m}(\mathcal{A})\triangleq \log\left(1+\frac{g\mathcal{A}}{\sqrt{2\pi e}}\right)$.
\end{theorem}

The upper bounds in Theorems \ref{Thm:DualityBnd1} and \ref{Thm:McKellipsBound} are both asymptomatically tight at high SNR for the respective cases where they hold, as we shall see later. An upper bound which is tight at low SNR can be derived using the constraint relaxation methods and also the Duality method, as discussed next.

\paragraph{Constraint-Relaxation Upper Bound}
In this approach, the input constraints are relaxed into a variance constraint. Namely, for a random variable $X\in[0,\mathcal{A}]$ with $\mathbb{E}[X]\leq\mathcal{E}\leq\frac{\mathcal{A}}{2}$, the maximum variance is $\mathcal{E}(\mathcal{A}-\mathcal{E})$ achieved by the binary distribution $\mathbb{P}_X(x)=\left(1-\frac{\mathcal{E}}{\mathcal{A}}\right)\delta(x)+\frac{\mathcal{E}}{\mathcal{A}}\delta(x-\mathcal{A})$ \cite{ChaabanMorvanAlouini}. Therefore, we can upper bound the maximization $\max_{\mathbb{P}_X}I(X;Y)$ where $\mathbb{P}_X$ is defined on $[0,\mathcal{A}]$ with $\mathbb{E}[X]\leq\mathcal{E}$ by enlarging the feasible set to the set of all $\mathbb{P}_X$ on $\mathbb{R}$ with $\mathbb{E}[(X-\mathbb{E}[X])^2]\leq\mathcal{E}(\mathcal{A}-\mathcal{E})$. The solution of this maximization is known to be achieved by the Gaussian distribution leading to the following statement.

\begin{theorem}{\cite{LapidothMoserWigger,ChaabanMorvanAlouini}}
\label{Thm:UBRelaxation}
The capacity of the IM/DD Gaussian channel with a peak constraint $\mathcal{A}$ and an average constraint $\mathcal{E}\in[0,\frac{\mathcal{A}}{2}]$ satisfies 
\begin{align}
c_g(\mathcal{A},\mathcal{E})\leq \overline{r}_g^{\rm 0}(\mathcal{A},\mathcal{E})=\frac{1}{2}\log\left(1+g^2\mathcal{E}(\mathcal{A}-\mathcal{E})\right).
\end{align}
\end{theorem}

Note that this upper bound was derived in \cite{LapidothMoserWigger} using the duality approach. Next, we turn our attention to cases with either an average or a peak constraint, where sphere packing bounds have been derived.

\begin{figure*}[!ht]
\centering
\begin{subfigure}[t]{.3\textwidth}
\begin{center}
\begin{tikzpicture}[scale=0.9, every node/.style={scale=0.9}]
%\node (o) at (0,0) [below] {$0$};
\draw (0,0,0) circle (.1cm);
\draw[->,line width=1pt] (0,0) to (3.2,0);
\draw[->,line width=1pt] (0,0) to (0,3.1);
\draw[->,line width=1pt] (0,0,0) to (0,0,3.5);
\node at (2.4,-.25) {$3\mathcal{E}$};
\node at (-.5,2.4) {$3\mathcal{E}$};
\node at (-.5,-.1,2.15) {$3\mathcal{E}$};
\node at (3.1,-.3) {$x_1$};
\node at (.3,2.9) {$x_2$};
\node at (.1,-.4,2.8) {$x_3$};
\draw (2.4,0,0) to (0,2.4,0) to (0,0,2.4) to (2.4,0,0);
\node at (2.4,0,0) [smalldot] {};
\node at ({2.4-cos(45)},0,{sin(45)}) [smalldot] {};
\node at ({2.4-2*cos(45)},0,{2*sin(45)}) [smalldot] {};
\node at ({2.4-3*cos(45)},0,{3*sin(45)}) [smalldot] {};
\def\x1{1.693};
\def\y1{.707};
\def\z1{0};
\node at (\x1,\y1,\z1) [smalldot] {};
\node at ({\x1-cos(45)},\y1,{\z1+sin(45)}) [smalldot] {};
\node at ({\x1-2*cos(45)},\y1,{\z1+2*sin(45)}) [smalldot] {};
\draw[dotted] (\x1,\y1,\z1) to (\z1,\y1,\x1);

\def\x1{.986};
\def\y1{1.414};
\def\z1{0};
\node at (\x1,\y1,\z1) [smalldot] {};
\node at ({\x1-cos(45)},\y1,{\z1+sin(45)}) [smalldot] {};
\draw[dotted] (\x1,\y1,\z1) to (\z1,\y1,\x1);

\def\x1{.279};
\def\y1{2.121};
\def\z1{0};
\node at (\x1,\y1,\z1) [smalldot] {};
\draw[dotted] (\x1,\y1,\z1) to (\z1,\y1,\x1);

\draw[->] (1.9,1.9) to (1.806,0.8607);
\draw[->] (1.9,1.9) to (1.039,1.507);
\node at (2,2.1) {Codewords};

\end{tikzpicture}
\end{center}
\vspace{-.3cm}
\caption{A $2$-simplex $\mathcal{S}_\mathcal{E}^2$ defined by $\sum_{i=1}^3X_i=3\mathcal{E}$. Codewords lie on this simplex.}
\label{Fig:SimplexCounting0}
\end{subfigure}
\hspace{.5cm}
\begin{subfigure}[t]{.3\textwidth}
\begin{center}
\begin{tikzpicture}[scale=0.9, every node/.style={scale=0.9}]
%\node (o) at (0,0) [below] {$0$};
\draw (0,0,0) circle (.1cm);
\draw[->,line width=1pt] (0,0) to (3.2,0);
\draw[->,line width=1pt] (0,0) to (0,3.1);
\draw[->,line width=1pt] (0,0,0) to (0,0,3.5);
\node at (2.4,-.25) {$3\mathcal{E}$};
\node at (-.5,2.4) {$3\mathcal{E}$};
\node at (-.5,-.1,2.15) {$3\mathcal{E}$};
\node at (3.1,-.3) {$X_1$};
\node at (.3,2.9) {$X_2$};
\node at (.1,-.4,2.8) {$X_3$};
\draw (2.4,0,0) to (0,2.4,0) to (0,0,2.4) to (2.4,0,0);

\draw (2.4,0,0) circle (.5cm);
\shade[ball color=blue!10!white,opacity=0.20] (2.4,0,0) circle (.5cm);
\node at (2.4,0,0) [smalldot] {};
\draw ({2.4-cos(45)},0,{sin(45)}) circle (.5cm);
\shade[ball color=blue!10!white,opacity=0.20] ({2.4-cos(45)},0,{sin(45)}) circle (.5cm);
\node at ({2.4-cos(45)},0,{sin(45)}) [smalldot] {};
\draw ({2.4-2*cos(45)},0,{2*sin(45)}) circle (.5cm);
\shade[ball color=blue!10!white,opacity=0.20] ({2.4-2*cos(45)},0,{2*sin(45)}) circle (.5cm);
\node at ({2.4-2*cos(45)},0,{2*sin(45)}) [smalldot] {};
\draw ({2.4-3*cos(45)},0,{3*sin(45)}) circle (.5cm);
\shade[ball color=blue!10!white,opacity=0.20] ({2.4-3*cos(45)},0,{3*sin(45)}) circle (.5cm);
\node at ({2.4-3*cos(45)},0,{3*sin(45)}) [smalldot] {};

\def\x1{1.693};
\def\y1{.707};
\def\z1{0};
\draw (\x1,\y1,\z1) circle (.5cm);
\shade[ball color=blue!10!white,opacity=0.20] (\x1,\y1,\z1) circle (.5cm);
\draw ({\x1-cos(45)},\y1,{\z1+sin(45)}) circle (.5cm);
\shade[ball color=blue!10!white,opacity=0.20] ({\x1-cos(45)},\y1,{\z1+sin(45)}) circle (.5cm);
\draw ({\x1-2*cos(45)},\y1,{\z1+2*sin(45)}) circle (.5cm);
\shade[ball color=blue!10!white,opacity=0.20] ({\x1-2*cos(45)},\y1,{\z1+2*sin(45)}) circle (.5cm);
\draw[dotted] (\x1,\y1,\z1) to (\z1,\y1,\x1);
\node at (\x1,\y1,\z1) [smalldot] {};
\node at ({\x1-cos(45)},\y1,{\z1+sin(45)}) [smalldot] {};
\node at ({\x1-2*cos(45)},\y1,{\z1+2*sin(45)}) [smalldot] {};

\def\x1{.986};
\def\y1{1.414};
\def\z1{0};
\draw (\x1,\y1,\z1) circle (.5cm);
\shade[ball color=blue!10!white,opacity=0.20] (\x1,\y1,\z1) circle (.5cm);
\draw ({\x1-cos(45)},\y1,{\z1+sin(45)}) circle (.5cm);
\shade[ball color=blue!10!white,opacity=0.20] ({\x1-cos(45)},\y1,{\z1+sin(45)}) circle (.5cm);
\draw[dotted] (\x1,\y1,\z1) to (\z1,\y1,\x1);
\node at (\x1,\y1,\z1) [smalldot] {};
\node at ({\x1-cos(45)},\y1,{\z1+sin(45)}) [smalldot] {};

\def\x1{.279};
\def\y1{2.121};
\def\z1{0};
\draw (\x1,\y1,\z1) circle (.5cm);
\shade[ball color=blue!10!white,opacity=0.20] (\x1,\y1,\z1) circle (.5cm);
\draw[dotted] (\x1,\y1,\z1) to (\z1,\y1,\x1);
\node at (\x1,\y1,\z1) [smalldot] {};

\draw[->] (1.9,1.9) to (2.046,1.0607);
\draw[->] (1.9,1.9) to (1.339,1.767);
\node at (2,2.1) {noise spheres};
\draw[<->] (1.69,0.707) to (2.043,1.0605);
\node at (1.97,0.73) {$r_{\rm b}$};
\end{tikzpicture}
\end{center}
\vspace{-.3cm}
\caption{Noise-perturbed codewords form spheres centered on the $2$-simplex $\mathcal{S}_\mathcal{E}^2$.}
\label{Fig:SimplexCounting1}
\end{subfigure}
\hspace{.5cm}
\begin{subfigure}[t]{.3\textwidth}
\begin{center}
\begin{tikzpicture}[scale=0.9, every node/.style={scale=0.9}]
\draw (0,0) to (-3.39,0) to (-1.697,2.939) to (0,0);
\draw (0,0) circle (.5cm);
\draw (-1,0) circle (.5cm);
\draw (-2,0) circle (.5cm);
\draw (-3,0) circle (.5cm);
\draw (-.5,.866) circle (.5cm);
\draw (-1.5,.866) circle (.5cm);
\draw (-2.5,.866) circle (.5cm);
\draw (-1,2*.866) circle (.5cm);
\draw (-2,2*.866) circle (.5cm);
\draw (-1.5,3*.866) circle (.5cm);
\draw[<->] (-.5,0.866) to node[above] {$r_{\rm b}$} (0,.866);
\draw[<->] (0,-1) to node[below] {$3\mathcal{E}\sqrt{2}$} (-3.39,-1);
\draw[dotted] (0,0) to (0,-1);
\draw[dotted] (-3.39,0) to (-3.39,-1);
\end{tikzpicture}
\end{center}
\vspace{-.3cm}
\caption{A projection of the spheres in Fig. \ref{Fig:SimplexCounting1} onto the plane containing the $2$-simplex $\mathcal{S}_\mathcal{E}^2$.}
\label{Fig:SimplexCounting2}
\end{subfigure}
\caption{A graphical illustration of sphere packing in a simplex (from \cite{ChaabanMorvanAlouini}).}
\label{Fig:2DSimplex}
\end{figure*}
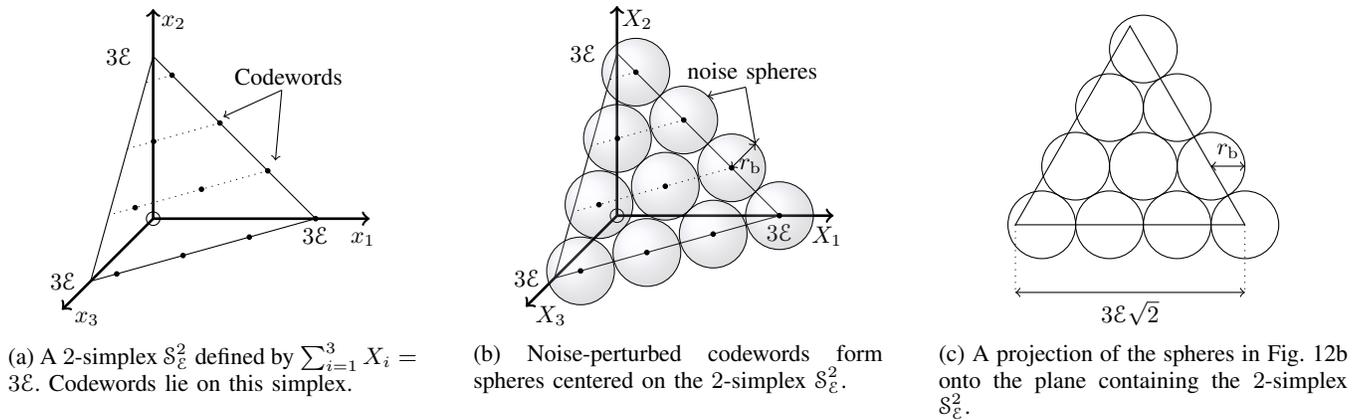

\subsubsection{Average or Peak Constraints} 
\label{Sec:UpperBoundsSpherePacking}
Capacity upper bounds for the IM/DD Gaussian channel with either an average or a peak constraint have been derived in \cite{LapidothMoserWigger,FaridHranilovic,ChaabanMorvanAlouini}. To complement the approaches presented above, we focus on the sphere-packing bounding approach which was studied in \cite{FaridHranilovic,ChaabanMorvanAlouini}. 

\paragraph{Sphere packing under an Average Constraint}
Again, we consider the equivalent channel $\tilde{Y}=X+\tilde{Z}$ where $\tilde{Z}\sim\mathcal{N}(0,\tilde{\sigma}^2)$ and $\tilde{\sigma}=\frac{1}{g}$. For a channel with an average constraint only, the capacity achieving distribution satisfies the average constraint with equality \cite{ChaabanMorvanAlouini}. A codeword $(x_1,x_2,\cdots,x_n)$ generated using $\mathbb{P}_X$ satisfying $\mathbb{E}[X]=\mathcal{E}$ almost certainly satisfies $\sum_{i=1}^nx_i=n\mathcal{E}$ for large $n$ by the law of large numbers. Moreover, $x_i\geq 0$. This confines the codewords to a regular $(n-1)$-simplex defined by the set $\mathcal{S}^{n-1}_{\mathcal{E}}=\left\{\x\in\mathbb{R}^n_+\left|\sum_{i=1}^nx_i= n\mathcal{E}\right.\right\}$ with side-length $n\mathcal{E}\sqrt{2}$ (Fig. \ref{Fig:SimplexCounting0} shows an example with $n=3$ taken from \cite{ChaabanMorvanAlouini}). 

On the other hand, noise $(\tilde{z}_1,\ldots,\tilde{z}_n)$ is i.i.d. $\mathcal{N}(0,\tilde{\sigma}^2)$. For large $n$, this noise tuple will be confined almost certainly to points near the surface of an $n$-dimensional ball of radius $r_{\rm b}=\sqrt{n\tilde{\sigma}^2}$ by the sphere hardening effect \cite{WozencraftJacobs,ShannonNoise}. Thus, the noise-perturbed codeword $(y_1,\ldots,y_n)$, where $y_i=x_i+\tilde{z}_i$, lies almost surely near the surface of a ball with radius $r_{\rm b}$ about $(x_1,\ldots,x_n)$ as shown in Fig. \ref{Fig:SimplexCounting1}. This is called ``decoding sphere'' in \cite{CoverThomas}. An upper bound for the IM/DD channel capacity can be obtained by computing or upper bounding the maximum number of disjoint $n$-dimensional balls that can be packed centered in $\mathcal{S}^{n-1}_\mathcal{E}$, in the limit as $n\to\infty$. Equivalently, we can bound the number of $n-1$ dimensional balls with radius $r_{\rm b}$ that can be packed centered in $\mathcal{S}^{n-1}_\mathcal{E}$ (Fig. \ref{Fig:SimplexCounting2}).\footnote{See \cite[Chapter 5]{WozencraftJacobs} and \cite[Appendix B]{ChaabanMorvanAlouini_TechReport} for a justification of this sphere-packing bounding approach.}  The following example shows a `back-of-the-envelope' calculation which explains this idea.

\begin{example}\label{Eg:SimplexPacking}
Consider a channel with an average constraint $\mathcal{E}$ over $n$ transmissions. Codewords are confined in a simplex $\mathcal{S}^{n-1}_\mathcal{E}$ whose volume is given by $\frac{\sqrt{n}(n\mathcal{E})^{n-1}}{(n-1)!}$ \cite{Rabinowitz}. On the other hand, noise is confined in an $n$-dimensional ball of radius $\sqrt{n\tilde{\sigma}}$, whose intersection with the simplex is an $(n-1)$-dimensional ball of volume $\frac{(\pi (n-1) \tilde{\sigma}^2)^{\frac{n-1}{2}}}{\Gamma(1+\frac{n-1}{2})}$ where $\Gamma(\cdot)$ is the Gamma function. These balls will be centered in the simplex and, for $\mathcal{E}\gg\tilde{\sigma}$, will not enlarge the simplex by much (realtive to the volume of the simplex). Thus, we can approximate the number of balls that can be packed in the simplex by the ratio of volumes, i.e., $n_{\rm balls}\approx\frac{\sqrt{n}(n\mathcal{E})^{n-1}}{(n-1)!}\left(\frac{(\pi (n-1) \tilde{\sigma}^2)^{\frac{n-1}{2}}}{\Gamma(1+\frac{n-1}{2})}\right)^{-1}\approx \frac{\sqrt{n}(e\mathcal{E}^2)^{\frac{n-1}{2}}}{\sqrt{2}(2\pi\tilde{\sigma}^2)^{\frac{n-1}{2}}}$ for $n$ large, where we used Stirling's approximation \cite{AbramowitzStegun}. With this number of balls, we can send at a rate of $\frac{1}{n}\log(n_{\rm balls})\approx\frac{1}{2}\log\left(\frac{eg^2\mathcal{E}^2}{2\pi}\right)$ nats/transmission, for $n$ large since $\tilde{\sigma}=\frac{1}{g}$.
\end{example}

We shall see that the rate calculated in this example is in fact the high-SNR capacity of this channel ($\mathcal{E}\to\infty$). For a more careful calculation of a bound on the the number of spheres that can be packed in the simplex, two approaches can be used. One can bound the `Minkowski sum' of the simplex and the ball using the Steiner-Minkowski theorem for polytopes \cite[Proposition 12.3.6]{BergerGeometryII}\cite{Morvan}. This approach has been used in \cite{FaridHranilovic}. The second approach bounds the volume of portions of spheres inside the simplex and the volume of portions outside the simplex in a recursive manner and has been used in \cite{ChaabanMorvanAlouini}. The second approach leads to a tighter bound given next.

\begin{theorem}{\cite{ChaabanMorvanAlouini}}
\label{Thm:SpherePackingSimplex}
The capacity of the IM/DD Gaussian channel with only an average constraint $\mathcal{E}$  satisfies $c_g(\infty,\mathcal{E})\leq \overline{r}_g^{\rm cma,a}(\mathcal{E})$ where
\begin{align*}
\overline{r}_g^{\rm cma,a}(\mathcal{E})=\sup_{\mu\in[0,1]} \mu\log\left(\frac{\sqrt{e}g\mathcal{E}}{\sqrt{2\pi}}\right)
-\log\left((1-\mu)^{1-\mu}\mu^{\frac{3\mu}{2}}\right).
\end{align*}
\end{theorem}

This bound is tight at high SNR. Under a peak constraint, the problem becomes one of sphere-packing in a cube as discussed next.

\paragraph{Sphere Packing under a Peak Constraint}
For a channel with a peak constraint only, a codeword $(x_1,x_2,\cdots,x_n)$ satisfying $x_i\in[0,\mathcal{A}]$ lives in an $n$-dimensional cube with side-length $\mathcal{A}$. In this case, a capacity upper bound can be derived by computing the maximum number of disjoint decoding spheres that can be packed centered in this cube, in the limit as $n\to\infty$. Again, let us start with a `back-of-the-envelope' calculation.

\begin{example}\label{Eg:CubePacking}
Consider a channel with a peak constraint $\mathcal{A}$ over $n$ transmissions. Codewords are confined in a cube whose volume is $\mathcal{A}^n$, while noise is confined in an $n$-dimensional ball of volume $\frac{(\pi n \tilde{\sigma}^2)^{\frac{n}{2}}}{\Gamma(1+\frac{n}{2})}$. Noise balls will not enlarge the cube by much when $\mathcal{A}$ is large. Thus, we can approximate the number of balls that can be packed in the cube by the ratio of volumes, i.e., $n_{\rm balls}\approx\mathcal{A}^{n}\left(\frac{(\pi n \tilde{\sigma}^2)^{\frac{n}{2}}}{\Gamma(1+\frac{n}{2})}\right)^{-1}\approx \frac{\sqrt{n\pi}(\mathcal{A}^2)^{\frac{n}{2}}}{(2\pi e\tilde{\sigma}^2)^{\frac{n}{2}}}$ for $n$ large, using Stirling's approximation \cite{AbramowitzStegun}. Thus, the rate is $\frac{1}{n}\log(n_{\rm balls})\approx\frac{1}{2}\log\left(\frac{g^2\mathcal{A}^2}{2\pi e}\right)$ nats/transmission for $n$ large since $\tilde{\sigma}=\frac{1}{g}$.
\end{example}

Again, the result of this example is exactly the high SNR capacity ($\mathcal{A}\to\infty$) under a peak constraint only, and also under a peak constraint and an average constraint $\mathcal{E}=\alpha\mathcal{A}$ with $\alpha\geq\frac{1}{2}$. For a more careful analysis, we can bound the number of balls using either the Steiner-Minkowski theorem for polytopes \cite[Proposition 12.3.6]{BergerGeometryII}\cite{Morvan} or the recursive approach in \cite{ChaabanMorvanAlouini}. Both bounds have been derived in \cite{ChaabanMorvanAlouini} and are summarized next.

\begin{theorem}{\cite{ChaabanMorvanAlouini}}
\label{Thm:SpherePackingCube}
The capacity of the IM/DD Gaussian channel with only a peak constraint $\mathcal{A}$ satisfies $c_g(\mathcal{A},\infty)\leq \overline{r}_g^{\rm cma}(\mathcal{A})=\min\{\sup_{\mu\in[0,1]}b_g^{\rm cma,1}(\mu,\mathcal{A}),\sup_{\mu\in[0,1]}b_g^{\rm cma,2}(\mu,\mathcal{A})\}$ where 
\begin{align*}
b_g^{\rm cma,1}(\mu,\mathcal{A})&=\mu\log\left(\frac{g\mathcal{A}}{\sqrt{2\pi e}}\right)-\log\left(\mu^\mu(1-\mu)^{\frac{3(1-\mu)}{2}}\right),\\
b_g^{\rm cma,2}(\mu,\mathcal{A})&=\mu\log\left(\frac{g\mathcal{A}}{\sqrt{2\pi e}}\right)-\log\left(\mu^{\frac{\mu}{2}}(1-\mu)^{1-\mu}2^{\mu-1}\right).
\end{align*}
\end{theorem}
Here, the bound $b_g^{\rm cma,1}$ is obtained using the Steiner-Minkowski theorem, and $b_g^{\rm cma,2}$ is obtained using the recursive approach of \cite{ChaabanMorvanAlouini}. None of these two bounds is tighter than the other over the whole range of SNR, but they are both asymptotically tight at high SNR.

Note that both bounds $\overline{r}_g^{\rm cma,a}$ and $\overline{r}_g^{\rm cma}$ (Theorems \ref{Thm:SpherePackingSimplex} and \ref{Thm:SpherePackingCube}) are also upper bounds on the capacity of a channel with both average and peak constraints, since omitting a constraint can only increase capacity.  

\begin{example}
Consider an IM/DD Gaussian channel with $g=1$, $\mathcal{A}=5$, and $\mathcal{E}=1.25$. For this channel, the upper bounds in Theorems \ref{Thm:DualityBnd1}-\ref{Thm:SpherePackingCube} evaluate to $\overline{r}_1^{\rm lmw}(5,1.25)=0.8394$, $\overline{r}_1^{\rm 0}(5,1.25)=0.8691$, $\overline{r}_1^{\rm cma,a}(1.25)=0.7806$, $\overline{r}_1^{\rm cma}(5)=0.9734$, in nats/transmission. Recall from Example \ref{Eg:Discrete2} that the capacity of this channel is $c_1(5,1.25)=0.626$ nats/transmission, which shows that $\overline{r}_1^{\rm cma,a}(1.25)$ is the tightest in this example.
\end{example}

We shall see later that the bounds $\overline{r}_g^{\rm lmw}(\mathcal{A},\mathcal{E})$ and $\overline{r}_g^{\rm 0}(\mathcal{A},\mathcal{E})$ are in fact is tight at high and low SNR, i.e., $\mathcal{A}\to\infty$ and $\mathcal{A}\to0$, respectively, with $\mathcal{E}$ proportional to $\mathcal{A}$. Moreover, the bounds $\overline{r}_g^{\rm cma,a}(\mathcal{A},\mathcal{E})$ and $\overline{r}_g^{\rm cma}(\mathcal{A},\mathcal{E})$ improve at high SNR. 

At this point, we are ready to compare the bounds and develop asymptotic capacity results.

\subsection{Asymptotic Capacity Results}
\label{Sec:AsymptoticCapacity}
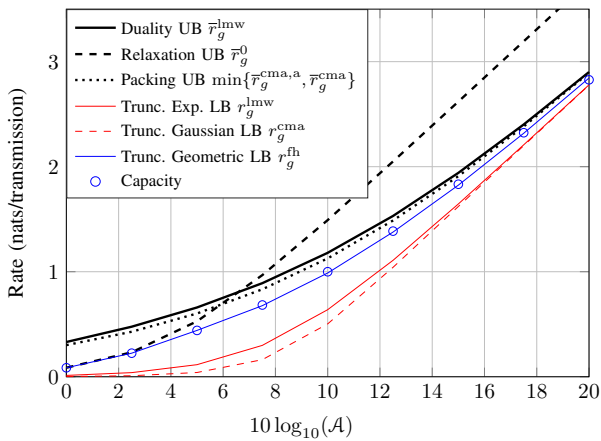
\begin{figure}[t]
\centering
\tikzset{every picture/.style={scale=.9}, every node/.style={scale=.9}}
%\input{fig/Bounds_vs_A_alpha_p25_s_1_g_1}
% Minimal pgfplots version: 1.3
%  http://www.mathworks.com/matlabcentral/fileexchange/22022-matlab2tikz
\begin{tikzpicture}

\begin{axis}[%
width=4in,
height=3in,
xmin=0,
xmax=20,
xlabel={$10\log_{10}(\mathcal{A})$},
xmajorgrids,
ymin=0,
ymax=3.5,
ylabel={Rate (nats/transmission)},
ymajorgrids,
ylabel style={at = {(axis cs: 1,1.75)}},
legend style={at = {(axis cs: 0,3.5)}, anchor = north west, legend cell align=left,align=left,draw=white!15!black}
]
\addplot [color=black,solid,line width=1]
  table[row sep=crcr]{%
0	0.330802307075241\\
2.5	0.476533356989043\\
5	0.659623771146775\\
7.5	0.891002672011962\\
10	1.1799191291609\\
12.5	1.53090631026757\\
15	1.94105983525619\\
17.5	2.40162329452223\\
20	2.90105709727509\\
};
\addlegendentry{\footnotesize Duality UB $\overline{r}_g^{\rm lmw}$};

\addplot [color=black,dashed,line width=1]
  table[row sep=crcr]{%
0	0.0859251284633296\\
2.5	0.232786621431218\\
5	0.528026337124657\\
7.5	0.967877278493314\\
10	1.49157674567357\\
12.5	2.04960556834243\\
15	2.61954900344403\\
17.5	3.19337825968388\\
20	3.76844856478308\\
};
\addlegendentry{\footnotesize Relaxation UB $\overline{r}_g^{\rm 0}$};

\addplot [color=black,dotted,line width=1]
  table[row sep=crcr]{%
0	0.301031885188911\\
2.5	0.428491096711509\\
5	0.602294948640875\\
7.5	0.832616173560003\\
10	1.12701075176893\\
12.5	1.48764728620224\\
15	1.90975480576041\\
17.5	2.38272131749117\\
20	2.89343561975787\\
};
\addlegendentry{\footnotesize Packing UB $\min\{\overline{r}_g^{\rm cma,a},\overline{r}_g^{\rm cma}\}$};

\addplot [color=red,solid]
  table[row sep=crcr]{%
0	0.0127644225629829\\
2.5	0.0392986400412301\\
5	0.114990051702486\\
7.5	0.298782089369128\\
10	0.638483784426669\\
12.5	1.10834246508465\\
15	1.6452724968086\\
17.5	2.20802428806107\\
20	2.77952277807944\\
};
\addlegendentry{\footnotesize Trunc. Exp. LB $r_g^{\rm lmw}$};

\addplot [color=red,dashed]
  table[row sep=crcr]{%
0	0.00235890953422199\\
2.5	0.00955095489169926\\
5	0.0405130604054697\\
7.5	0.162623583839935\\
10	0.50320506378959\\
12.5	1.04170255576796\\
15	1.62107473134447\\
17.5	2.2000258995632\\
20	2.77683493471029\\
};
\addlegendentry{\footnotesize Trunc. Gaussian LB $r_g^{\rm cma}$};

\addplot [color=blue,solid]
  table[row sep=crcr]{%
0	0.08544941268263\\
2.5	0.224446345877416\\
5	0.441274133062622\\
7.5	0.676942001269714\\
10	0.987764759201703\\
12.5	1.37736821372002\\
15	1.8243035067447\\
17.5	2.31807664598834\\
20	2.8441585291182\\
};
\addlegendentry{\footnotesize Trunc. Geometric LB $r_g^{\rm fh}$};

\addplot [color=blue,only marks,mark=o,mark options={solid}]
  table[row sep=crcr]{%
0	0.0854482126766123\\
2.5	0.224446285877423\\
5	0.441274073062633\\
7.5	0.683093777438541\\
10	0.999762778159381\\
12.5	1.38654728338782\\
15	1.83293969565236\\
17.5	2.32326143506224\\
20	2.82705963198911\\
};
\addlegendentry{\footnotesize Capacity};

\end{axis}
\end{tikzpicture}%
\caption{Capacity bounds for an IM/DD Gaussian channel with $g=1$ and $\mathcal{E}=\frac{\mathcal{A}}{4}$.}
\label{Fig:Cap_A_p25}
\end{figure}

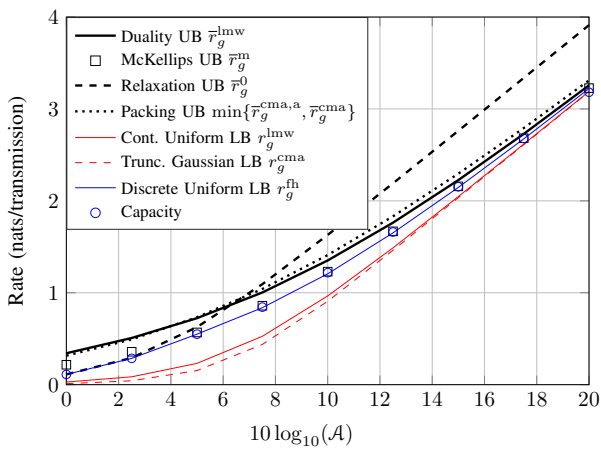
\begin{figure}[t]
\centering
\tikzset{every picture/.style={scale=.9}, every node/.style={scale=.9}}
%\input{fig/Bounds_vs_A_alpha_p5_s_1_g_1}
% Minimal pgfplots version: 1.3
%  http://www.mathworks.com/matlabcentral/fileexchange/22022-matlab2tikz
\begin{tikzpicture}

\begin{axis}[%
width=4in,
height=3in,
xmin=0,
xmax=20,
xlabel={$10\log_{10}(\mathcal{A})$},
xmajorgrids,
ymin=0,
ymax=4,
ylabel={Rate (nats/transmission)},
ymajorgrids,
ylabel style={at = {(axis cs: 1,2)}},
legend style={at = {(axis cs: 0,4)}, anchor = north west, legend cell align=left,align=left,draw=white!15!black}
]

\addplot [color=black,solid,line width=1]
  table[row sep=crcr]{%
0	0.34173561052554\\
2.5	0.507902787797933\\
5	0.723554227999454\\
7.5	1.00306145599636\\
10	1.35136782079401\\
12.5	1.76273685572604\\
15	2.22593989227945\\
17.5	2.72801600703755\\
20	3.25717932724879\\
};
\addlegendentry{\footnotesize Duality UB $\overline{r}_g^{\rm lmw}$};

\addplot [color=black,only marks,mark=square,mark options={solid}]
  table[row sep=crcr]{%
0	0.2167    \\
2.5	0.3579    \\
5	0.5683    \\
7.5	0.8590    \\
10	1.2296    \\
12.5	1.6683   \\ 
15	2.1578    \\
17.5	2.6815   \\ 
20	3.2267\\
};
\addlegendentry{\footnotesize McKellips UB $\overline{r}_g^{\rm m}$};

\addplot [color=black,dashed,line width=1]
  table[row sep=crcr]{%
0	0.111571775657105\\
2.5	0.2912668391144\\
5	0.626381484247684\\
7.5	1.09334543217023\\
10	1.62904826901074\\
12.5	2.19136907514461\\
15	2.76272646956589\\
17.5	3.33700878804866\\
20	3.91222296543881\\
};
\addlegendentry{\footnotesize Relaxation UB $\overline{r}_g^{\rm 0}$};

\addplot [color=black,dotted,line width=1]
  table[row sep=crcr]{%
0	0.315588757992238\\
2.5	0.492652639972576\\
5	0.733919135170184\\
7.5	1.04135662095803\\
10	1.4103504917176\\
12.5	1.83196118297515\\
15	2.29569988378843\\
17.5	2.79157807098175\\
20	3.31114075197476\\
};
\addlegendentry{\footnotesize Packing UB $\min\{\overline{r}_g^{\rm cma,a},\overline{r}_g^{\rm cma}\}$};

\addplot [color=red,solid]
  table[row sep=crcr]{%
0	0.0286336629150533\\
2.5	0.0854537577839386\\
5	0.231674688963071\\
7.5	0.526076491769635\\
10	0.965317424358291\\
12.5	1.48873583490436\\
15	2.04666247354223\\
17.5	2.61657204779347\\
20	3.19039043354759\\
};
\addlegendentry{\footnotesize Cont. Uniform LB $r_g^{\rm lmw}$};

\addplot [color=red,dashed]
  table[row sep=crcr]{%
0	0.0106640521088969\\
2.5	0.042231873996437\\
5	0.15350969471446\\
7.5	0.437525574126678\\
10	0.907623627768855\\
12.5	1.46241126600725\\
15	2.03528227258944\\
17.5	2.61059215659934\\
20	3.18622362750736\\
};
\addlegendentry{\footnotesize Trunc. Gaussian LB $r_g^{\rm cma}$};

\addplot [color=blue,solid]
  table[row sep=crcr]{%
0	0.111421508064465\\
2.5	0.286059848402752\\
5	0.549604364097013\\
7.5	0.837416203204684\\
10	1.20618361100664\\
12.5	1.65187968797312\\
15	2.14651489745927\\
17.5	2.67380096979743\\
20	3.22172557285844\\
};
\addlegendentry{\footnotesize Discrete Uniform LB $r_g^{\rm fh}$};

\addplot [color=blue,only marks,mark=o,mark options={solid}]
  table[row sep=crcr]{%
0	0.111420693778014\\
2.5	0.286059034123771\\
5	0.549603337280622\\
7.5	0.843767739647681\\
10	1.21883021721387\\
12.5	1.6613299785767\\
15	2.15352937559112\\
17.5	2.67844071888978\\
20	3.18185570945203\\
};
\addlegendentry{\footnotesize Capacity};

\end{axis}
\end{tikzpicture}%
\caption{Capacity bounds for an IM/DD Gaussian channel with $g=1$ and $\mathcal{E}=\frac{\mathcal{A}}{2}$.}
\label{Fig:Cap_A_p5}
\end{figure}

We start by plotting the bounds presented so far. Fig. \ref{Fig:Cap_A_p25} shows the bounds for a channel with $\mathcal{E}=\frac{\mathcal{A}}{4}$. It shows that the lower bound (LB) $r_g^{\rm lmw}$  (Thm. \ref{Thm:LowerBoundLMW}) and the upper bound (UB) $\overline{r}_g^{\rm lmw}$  (Thm. \ref{Thm:DualityBnd1}) converge at high SNR (large $\mathcal{A}$). It also shows that the lower bound $r_g^{\rm fh}$ (Thm. \ref{Thm:FaridHranilovicLB}) and the upper bound $\overline{r}_g^{\rm 0}$  (Thm. \ref{Thm:UBRelaxation}) converge at low SNR. Fig. \ref{Fig:Cap_A_p5} shows a similar plot for a channel with $\mathcal{E}=\frac{\mathcal{A}}{2}$, where similar observations hold. The figures also show that the truncated-Geometric and discrete uniform distributions approaches capacity over the whole SNR range (LB $r_g^{\rm fh}$), and that the truncated-Gaussian distribution is close to optimal at high SNR (LB $r_g^{\rm cma}$, Thm. \ref{Thm:TGLowerBound}). Moreover, the sphere-packing bounds $\overline{r}_g^{\rm cma,a}$ and $\overline{r}_g^{\rm cma}$ (Thm. \ref{Thm:SpherePackingSimplex} \& \ref{Thm:SpherePackingCube}) are fairly tight at high SNR, and close to the duality bound $\overline{r}_g^{\rm lmw}$ over the whole SNR range.

\begin{remark}
Fig. \ref{Fig:Cap_A_p5} shows that a combinations of bounds $\overline{r}_g^{\rm 0}$ and $\overline{r}_g^{\rm m}$ provides a fairly tight capacity approximation since their minimum nearly meets $r_g^{\rm fh}$ over the whole range of $\mathcal{A}$. Thus, one can use $\min\{\overline{r}_g^{\rm 0},\overline{r}_g^{\rm m}\}$ as a capacity approximation for a peak constrained channel or one with $\mathcal{E}=\frac{\mathcal{A}}{2}$. 
\end{remark}

Using these bounds, we can characterize the asymptotic capacity of the channel. The asymptotic capacity results are given next.

\subsubsection{High-SNR Asymptotic Capacity}
The asymptotic capacity of the IM/DD Gaussian channel at high SNR was characterized in \cite{LapidothMoserWigger}, when $X$ is subject to both average and peak constraints, and also when $X$ is subject to an average constraint only. Note that the high SNR regime is of interest because it is the regime of operation of many OWC systems (cf. \cite{KazemiHaas}~e.g.). The following theorem presents asymptotic capacity results for the IM/DD Gaussian channel at high SNR.

\begin{theorem}[\cite{LapidothMoserWigger}]
\label{Thm:HighSNR}
The capacity of the IM/DD Gaussian channel with a peak constraint $\mathcal{A}$ and an average constraint $\mathcal{E}=\alpha\mathcal{A}$ satisfies 
\begin{align}\label{HighSNRCap1}
\lim_{\mathcal{A}\to\infty}\left(c_g(\mathcal{A},\alpha\mathcal{A})-\frac{1}{2}\log\left(\frac{g^2\mathcal{A}^2}{2\pi e}\right)\right)=0,
\end{align}
for $\alpha=\frac{1}{2}$, and 
\begin{align}\label{HighSNRCap2}
\lim_{\mathcal{A}\to\infty}\left(c_g(\mathcal{A},\alpha\mathcal{A})-\frac{1}{2}\log\left(\frac{g^2\mathcal{A}^2e^{2\alpha\mu}(1-e^{-\mu})^2}{2\pi e\mu^2}\right)\right)=0,
\end{align}
for $\alpha<\frac{1}{2}$, where $\mu$ is the unique solution of $\alpha=\frac{1}{\mu}-\frac{e^{-\mu}}{1-e^{-\mu}}$.
\end{theorem}

This theorem proves that the truncated-exponential and the uniform input distributions are optimal at high SNR. Recall that a channel with a peak constraint only has the same capacity as one with both a peak and an average constraint with $\alpha=\frac{1}{2}$. Hence, the statement of Theorem \ref{Thm:HighSNR} applies for this case as well. Note also that the statement for $\alpha=\frac{1}{2}$ also applies for $\alpha\geq\frac{1}{2}$ (since the average constraint can be replaced with $\frac{\mathcal{A}}{2}$ in this case, cf. Sec. \ref{Sec:MutualInfoMax}), and coincides with the calculation in Example \ref{Eg:CubePacking}.

A simplified asymptotic capacity expression was given in \cite{ChaabanMorvanAlouini} by observing that upper bounds $\overline{r}_g^{\rm cma,a}(\mathcal{E})$ and $\overline{r}_g^{\rm cma}(\mathcal{A})$ (Theorems \ref{Thm:SpherePackingSimplex} and \ref{Thm:SpherePackingCube}) are fairly tight at high SNR, with a gap to capacity $<0.1$ nats/transmission. Thus, the high-SNR asymptotic capacity can be well approximated as $\min\left\{\frac{1}{2}\log\left(\frac{e\alpha^2\mathcal{A}^2}{2\pi}\right),\frac{1}{2}\log\left(\frac{\mathcal{A}^2}{2\pi e}\right)\right\}$, where these two expressions are the high-SNR asymptotes of $\overline{r}_g^{\rm cma,a}(\mathcal{E})$ and $\overline{r}_g^{\rm cma}(\mathcal{A})$, respectively. 

If the channel is subject to an average constraint only, then its high-SNR capacity is given as follows.

\begin{theorem}[\cite{LapidothMoserWigger}]
\label{Thm:HighSNR_average_only}
The capacity of the IM/DD Gaussian channel with only an average constraint $\mathcal{E}$ satisfies 
\begin{align}\label{HighSNRCap1}
\lim_{\mathcal{E}\to\infty}\left(c_g(\infty,\mathcal{E})-\frac{1}{2}\log\left(\frac{eg^2\mathcal{E}^2}{2\pi}\right)\right)=0.
\end{align}
\end{theorem}

Thus, the high-SNR capacity in this case is $\frac{1}{2}\log\left(\frac{eg^2\mathcal{E}^2}{2\pi}\right)$, which coincides with the high-SNR asymptote of $\overline{r}_g^{\rm cma,a}(\mathcal{E})$ (Theorem \ref{Thm:SpherePackingSimplex}) and $r_g^{\rm cma}(\mathcal{E})$ (Theorem \ref{Thm:LB_Exp}), and also with the calculation in Example \ref{Eg:SimplexPacking}. An exponential input  distribution is optimal at high SNR in this case.

\subsubsection{Low-SNR Asymptotic Capacity}
At low SNR, the upper bound $\overline{r}_g^{\rm 0}(\mathcal{A},\mathcal{E})$ obtained using constraint relaxation in Theorem \ref{Thm:UBRelaxation} is tight if $X$ is subject to both average and peak constraints. It matches the achievable rate $r_g^{\rm fh}(\mathcal{A},\mathcal{E})$ (Theorem \ref{Thm:FaridHranilovicLB}) in this regime. This is proved by using a result by Prelov and van der Meulen \cite{PrelovVanDerMeulen} dating back to 1993, which provides an asymptotic expression for the mutual information under weak input signals. The result states that, under some technical conditions which are all satisfied by the IM/DD Gaussian channel, $I(X;Y)$ in a peak constrained channel $Y=X+Z$ can be written as
\begin{align}\label{IvsVar}
I(X;Y)=\frac{{\rm Var}(X)}{2\sigma^2}+o(\mathcal{A}^2),
\end{align}
where $\sigma^2$ is the noise variance, the term $o(\mathcal{A}^2)$ satisfies $\lim_{\mathcal{A}\to0}\frac{o(\mathcal{A}^2)}{\mathcal{A}^2}=0$, and ${\rm Var}(X)$ is the variance of $X$. The significant term in this expression, i.e., $\frac{{\rm Var}(X)}{2\sigma^2}$ is achieved at low SNR ($\mathcal{A}\to0$) using coded OOK, which is a special case of the truncated-geometric and discrete uniform input distributions with 2 mass points. 

Using \eqref{IvsVar} to express the low-SNR asymptotic behaviour of the lower bound $r_g^{\rm fh}(\mathcal{A},\mathcal{E})$ in Theorem \ref{Thm:FaridHranilovicLB} when $\mathbb{P}_X$ has two mass points only leads to the following result. 

\begin{theorem}[\cite{LapidothMoserWigger}]
\label{Thm:LowSNR}
The capacity of the IM/DD Gaussian channel with a peak constraint $\mathcal{A}$ and an average costraint $\mathcal{E}=\alpha\mathcal{A}$ satisfies 
\begin{align}\label{HighSNRCap1}
\lim_{\frac{\mathcal{A}}{\sigma}\to0}\frac{c_g(\mathcal{A},\alpha\mathcal{A})}{\alpha'(1-\alpha')\frac{\mathcal{A}^2}{2}}=1,
\end{align}
where $\alpha'=\min\{\alpha,\frac{1}{2}\}$.
\end{theorem}

Thus, the low-SNR asymptotic capacity is $\alpha'(1-\alpha')\frac{\mathcal{A}^2}{2}$. This shows that coded OOK is optimal in this regime. Unfortunately, such an expression for a channel with an average constraint only does not exit to-date. Bounds on the low-SNR asymptotic capacity for this case were given in \cite{LapidothMoserWigger}.

The results discussed in this section have been used to study multi-aperture \ac{IM/DD OWC systems} (MIMO) and multi-user \ac{IM/DD OWC systems}. Results on the capacity of these systems are reviewed in the following two sections.

\section{Multi-Aperture Systems}
\label{Sec:MIMO}

\begin{figure}[t]
\centering
\begin{tikzpicture}
\node (tx) at (-1,0) [draw,rectangle,minimum height=2.6cm,minimum width=1.5cm,fill=gray!50!white] {};
\node at ($(tx)-(1.2,0)$) {Tx};

\node (rx) at (4.7,0) [draw,rectangle,minimum height=2.6cm,minimum width=1.5cm,fill=gray!50!white] {};
\node at ($(rx)+(1.2,0)$) {Rx};

\draw[-] ($(tx.east)+(0,.5)$) to ($(tx.east)+(0.2,.7)$) to ($(tx.east)+(0.2,.3)$) to ($(tx.east)+(0,.5)$);
\draw[-] ($(tx.east)+(0,-.5)$) to ($(tx.east)+(0.2,-.7)$) to ($(tx.east)+(0.2,-.3)$) to ($(tx.east)+(0,-.5)$);

\draw[-] ($(rx.west)+(0,.5)$) to ($(rx.west)+(-0.2,.7)$) to ($(rx.west)+(-0.2,.3)$) to ($(rx.west)+(0,.5)$);
\draw[-] ($(rx.west)+(0,-.5)$) to ($(rx.west)+(-0.2,-.7)$) to ($(rx.west)+(-0.2,-.3)$) to ($(rx.west)+(0,-.5)$);

\draw[-,dashed,fill=yellow,opacity=.3] ($(tx.east)+(0.2,.7)$) to ($(tx.east)+(1.5,1)$) to ($(rx.west)+(-0.2,+1)$)  
to ($(rx.west)+(-0.2,-1)$)  to ($(tx.east)+(0.2,.3)$) ;
\draw[-,dashed,fill=yellow,opacity=.3] ($(tx.east)+(0.2,-.7)$) to ($(tx.east)+(1.5,-1)$) to ($(rx.west)+(-0.2,-1)$)  to ($(rx.west)+(-0.2,1)$)  to ($(tx.east)+(0.2,-.3)$) ;

\node (x1) at (0-1,0+.5) {\footnotesize $X_1$};
\node (z1) at (4.4,1+.1) {\footnotesize $Z_1$};
\node (y1) at (5.2,0+.5) {\footnotesize $Y_1$};
\node (p1) at (4.4,0+.5) {};
\node at (p1) {$\oplus$};

\node (x2) at (0-1,0-.5) {\footnotesize $X_2$};
\node (z2) at (4.4,-1-.1) {\footnotesize $Z_2$};
\node (y2) at (5.2,0-.5) {\footnotesize $Y_2$};
\node (p2) at (4.4,0-.5) {};
\node at (p2) {$\oplus$};

\draw[->] (x1) to ($(tx.east)+(0,.5)$);
\draw[->] ($(tx.east)+(0.2,.5)$) to node {\scriptsize \contour{yellow!30!white}{$g_{1,1}$}} ($(rx.west)+(-0.2,.5)$);%(p1.west);
\draw[->] ($(tx.east)+(0.2,.5)$) to ($(rx.west)+(-0.2,-.5)$);%(p2.north west);
\node at ($(x1.east)+(1.5,-.3)$) {\scriptsize \contour{yellow!30!white}{$g_{2,1}$}};
\draw[->] ($(z1.south)+(0,.1)$) to (p1.north);
\draw[->] (p1.east) to (y1.west);
\draw[->] ($(rx.west)+(0,.5)$) to (p1.west);

\draw[->] (x2) to ($(tx.east)+(0,-.5)$);
\draw[->] ($(tx.east)+(0.2,-.5)$) to node {\scriptsize \contour{yellow!30!white}{$g_{2,2}$}} ($(rx.west)+(-0.2,-.5)$);%(p2.west);
\draw[->] ($(tx.east)+(0.2,-.5)$) to ($(rx.west)+(-0.2,.5)$);%(p1.south west);
\node at ($(x2.east)+(1.5,.32)$) {\scriptsize \contour{yellow!30!white}{$g_{1,2}$}};
\draw[->] ($(z2.north)-(0,.1)$) to (p2.south);
\draw[->] (p2.east) to (y2.west);
\draw[->] ($(rx.west)+(0,-.5)$) to (p2.west);

%
%\draw[->] ($(tx.east)+(0,.5)$) to (x1.west);
%\draw[->] ($(tx.east)+(0,-.5)$) to (x2.west);
%\draw[<-] ($(rx.west)+(0,.5)$) to (y1.east);
%\draw[<-] ($(rx.west)+(0,-.5)$) to (y2.east);
\end{tikzpicture}
\caption{A MIMO IM/DD Gaussian channel \ac{with crosstalk}. In a parallel channel, $g_{1,2}=g_{2,1}=0$ \ac{(no crosstalk)}.}
\label{Fig:MIMOChannel}
\end{figure}
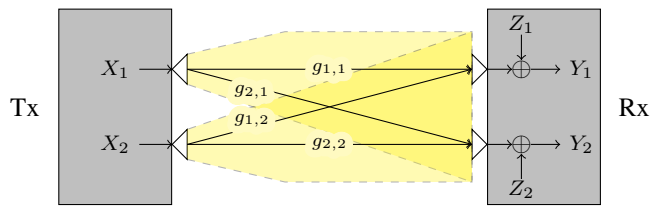

For a MIMO system with $\nt$ transmit apertures (LEDs or LED groups) and $\nr$ receive apertures (photodetectors or photodetector groups), the transmit signal $X$ becomes a vector $\X$ of dimension $\nt$,\footnote{With some notational abuse, we reuse $\X$ (which we used in Sec. \ref{Sec:MutualInfoMax} to denote $n$ scalar transmissions $(X_1,\ldots,X_n)$) here to denote a single vector transmission $\X=(X_1,\ldots,X_{\nt})$. We also reuse $\nt$ and $\nr$ (which we used in Sec. \ref{Model} to denote a number of photons) here to denote the number of transmit and receive apertures, respectively.} and the received signal becomes (Fig. \ref{Fig:MIMOChannel})
\begin{align}
\Y=\G\X+\Z,
\end{align}
where $\G$ is an $\nr\times\nt$ channel matrix whose component $g_{i,j}$ represents the channel gain from transmit aperture $j$ to receive aperture $i$,\footnote{The matrix $\G$ is assumed to be known at the transmitter through estimation and feedback e.g., which can be achieved in OWC without major impact on performance since the coherence time of OWC is typically much larger than the symbol duration \cite{FaridHranilovic_Outage,KhalighiUysal}.} and $\Z$ is $\nr$-dimensional noise with i.i.d. $\mathcal{N}(0,1)$ components. The transmit signal is subject to the constraints
\begin{align}
\label{MIMO_Constraints}
\X\in[0,\mathcal{A}]^{\nt}\text{ \ and \ }\sum_{i=1}^{\nt} \mathbb{E}[X_i]\leq \mathcal{E}.
\end{align}

This channel model appears in various OWC applications including indoors VLC, color-multiplexing, or multi-user OWC systems employing TDMA \cite{FathHaas,ButalaElgalaLittle,KazemiMostaaniUysalGhassemlooy,
FaridHranilovic_OutageMIMO,TsiftsisSandalidisKaragiannidisUysal,RiedigerSchoberLampe,
ZengObrienFaulkner,GaoWangXuHua,MonteiroHranilovic,AbdelhadyAminChaabanShihadaAlouini}. The capacity of this channel can be written as
\begin{align}
c_{\G}^{\rm mimo}(\mathcal{A},\mathcal{E})=\max_{\mathbb{P}_{\X}\in\mathcal{P}_{\X}} I(\X,\Y),
\end{align}
where $\mathcal{P}_{\X}$ is the set of all distributions of $\X$ that satisfy \eqref{MIMO_Constraints}. This maximization is achieved by a discrete input distribution {(in the form of a sum of multi-dimensional Dirac delta functions)}, and can be solved using the algorithm in \cite{ChanHranilovicKschischang}. For analytical results, this problem has been studied in \cite{ChaabanRezkiAlouini_ParallelOWC_ICC,
ChaabanRezkiAlouini_ParallelOWC_TCOM,ChaabanRezkiAlouini_LowSNR_CommL,ChaabanRezkiAlouini_MIMOOWC_ICC,MoserMylonakisWangWigger,
MoserWangWigger,
ChaabanRezkiAlouini_MIMOOWC_TWC,ChaabanRezkiAlouini_MIMOOWC_LowSNR_ISIT,
ChaabanRezkiAlouini_MIMO_IMDD_Low_SNR_TCOM,MoserWangWigger_IT,
LiMoserWangWigger,LiMoserWangWigger_Fading_Globecom}. Next, we review some main results and discuss them.

\subsection{Parallel IM/DD Gaussian Channels}
\label{Sec:ParallelChannels}
In this case, $\nt=\nr$ and there is no cross-talk between transmit aperture $i$ and receive aperture $j\neq i$, i.e., the channel matrix $\G$ is a diagonal matrix, with diagonal $\boldsymbol{g}=(g_1,\ldots,g_{\nt})$ (with $g_i=g_{i,i}$). The received signal becomes 
\begin{align}
Y_i=g_iX_i+Z_i,
\end{align}
and hence, the channel decomposes into a set of single-aperture (SISO) IM/DD Gaussian channels. This model arises in MIMO channels with little or no cross-talk, such as some RGB-multiplexing systems, and in MIMO channels with a channel inversion receiver.\footnote{\ac{Here, we assume that the MIMO channel matrix is invertible, which requires proper spacing between the transmitters and receivers. Generally, a smaller spacing is required in FSO due to the narrow beamwidth, whereas a larger spacing is required in VLC.}}

The capacity of this model was studied in \cite{ChaabanRezkiAlouini_ParallelOWC_TCOM} under an average constraint only, and in \cite{ChaabanRezkiAlouini_MIMO_IMDD_Low_SNR_TCOM} under both average and peak constraints. Generally, the capacity of this channel can be written as the sum of the capacities of the individual channels. The optimal input distribution is a product distribution, i.e., $\mathbb{P}_{\X}^*=\prod_{i=1}^{\nt} \mathbb{P}_{X_i}^*$. Thus, the parallel IM/DD Gaussian channel capacity can be expressed as 
\begin{align}
c_{\boldsymbol{g}}^{\rm parallel}(\mathcal{A},\mathcal{E})=\sum_{i=1}^{\nt} c_{g_i}(\mathcal{A},\mathcal{E}_i),
\end{align}
where $c_{g_i}(\mathcal{A},\mathcal{E}_i)=\max_{\mathbb{P}_{X_i}}I(X_i,Y_i)$ {and $\mathcal{E}_i=\mathbb{E}[X_i]$}. One can bound this capacity using the bounds in Sec. \ref{Sec:CapacityBounds}. Since these bounds are functions of the average constraint, one has to maximize them with respect to the allocation $\mathcal{E}_i$ where $\mathcal{E}_i=\mathbb{E}[X_i]$ is the average constraint allocated to the $i$-th channel with $\sum_{i=1}^{\nt}\mathcal{E}_i\leq\mathcal{E}$. For example, for a channel with an average constraint only, this leads to the following statement. 

\begin{theorem}[\cite{ChaabanRezkiAlouini_ParallelOWC_TCOM}]
The capacity of parallel IM/DD Gaussian channels with only an average constraint $\mathcal{E}$ satisfies 
\begin{align}
\label{LBParallel}
c^{\rm parallel}_{\boldsymbol{g}}&\geq \max_{\mathcal{E}_i} \sum_{i=1}^{\nt} r_{g_{i}}^{\ell,\rm a}(\mathcal{E}_i) ,\quad \ell\in\{\rm lmw,cma,fh\},\\
\label{UBParallel}
c^{\rm parallel}_{\boldsymbol{g}}&\leq \max_{\mathcal{E}_i} \sum_{i=1}^{\nt} \overline{r}_{g_{i}}^{\rm cma,a}(\mathcal{E}_i),
\end{align}
where $r_{g_{i}}^{\ell,\rm a}(\mathcal{E}_i)$ and $\overline{r}_{g_{i}}^{\rm cma,a}(\mathcal{E}_i)$ are given in Sec. \ref{Sec:CapBoundsAvgConstOnly}. 
\end{theorem}

In practice, it is relevant to obtain a close-to-optimal intensity allocation $\mathcal{E}_i$. To achieve this, \cite{ChaabanRezkiAlouini_ParallelOWC_TCOM} relies on the lower bound $r_{g_{i}}^{\rm lmw,a}(\mathcal{E}_i)=\frac{1}{2}\log\left(1+\frac{eg_{i,i}^2\mathcal{E}_i^2}{2\pi}\right)$ as a surrogate. The maximization problem becomes 
\begin{align}
\max_{\mathcal{E}_i\geq0} &\quad \sum_{i=1}^{\nt} \frac{1}{2}\log\left(1+\frac{eg_{i}^2\mathcal{E}_i^2}{2\pi}\right)\\
{\rm s.t.} & \quad \sum_{i=1}^{\nt}\mathcal{E}_i\leq\mathcal{E}.
\end{align}

This problem is nonconvex and its solution differs from the standard water-filling solution \cite{Haas_Dissertation,TseViswanath}. By describing its solution using the KKT conditions \cite{Boyd}, one can describe the optimal solution of this problem. Based on this, \cite{ChaabanRezkiAlouini_ParallelOWC_TCOM} devised a simple algorithm which can approach this optimal solution. The obtained solution can then be used in the lower bounds $r_{g_{i}}^{\rm lmw,a}$, $r_{g_{i}}^{\rm cma,a}$ and $r_{g_{i}}^{\rm fh,a}$. Note that this can not be used for the upper bound $\overline{r}_{g_{i}}^{\rm cma,a}$, which only remains an upper bound if we find the optimal solution of $\max_{\mathcal{E}_i} \sum_{i=1}^{\nt} \overline{r}_{g_{i}}^{\rm cma,a}(\mathcal{E}_i)$. This can be found numerically for benchmarking purposes. Following this approach, the following high-SNR asymptotic capacity can be derived.

\begin{theorem}[\cite{ChaabanRezkiAlouini_ParallelOWC_TCOM}]
The capacity of parallel IM/DD Gaussian channels with only an average constraint $\mathcal{E}$ satisfies 
\begin{align}
\lim_{\mathcal{E}\to\infty} \left(c_{\boldsymbol{g}}(\infty,\mathcal{E}) - \sum_{i=1}^{\nt} \frac{1}{2}\log\left(\frac{eg_{i}^2\mathcal{E}^2}{2\pi \nt^2}\right)\right)=0.
\end{align} 
\end{theorem}

This statement indicates that the optimal solution at high SNR is to allocate $\mathcal{E}$ equally over the channels, as expected, while this is not necessarily true for any SNR. The optimal input distribution for each channel at high SNR is the exponential distribution.

Under both average and peak constraints, \cite{ChaabanRezkiAlouini_ParallelOWC_TCOM} also provides capacity bounds in terms of $r_{g_i}^{\rm lmw}$, $r_{g_i}^{\rm cma}$, and $\overline{r}_{g_i}^{\rm cma}$, in addition to the following asymptotic capacity characterizations. 

\begin{theorem}[\cite{ChaabanRezkiAlouini_ParallelOWC_TCOM}]
The capacity of parallel IM/DD Gaussian channels with a peak constraint $\mathcal{A}$ and an average constraint $\mathcal{E}=\alpha\mathcal{A}$ satisfies 
\begin{align}
\lim_{\mathcal{A}\to\infty} \left(c_{\boldsymbol{g}}(\mathcal{A},\alpha\mathcal{A}) - \sum_{i=1}^{\nt} \frac{1}{2}\log\left(\frac{g_{i}^2\mathcal{A}^2}{2\pi e}\right)\right)=0,
\end{align} 
if $\alpha>\frac{\nt}{2}$, and
\begin{align}
\lim_{\mathcal{A}\to\infty} \left(c_{\boldsymbol{g}}(\mathcal{A},\alpha\mathcal{A}) - \sum_{i=1}^{\nt} \frac{1}{2}\log\left(\frac{eg_{i}^2\mathcal{E}^2}{2\pi \nt^2}\right)\right)\leq 0.1 \nt,
\end{align} 
if $\alpha\leq\frac{\nt}{2}$.
\end{theorem}

This is shown using a continuous uniform input distribution which is optimal at high SNR if $\alpha>\frac{\nt}{2}$, and a truncated-Gaussian distribution which is close-to-optimal at high SNR if $\alpha\leq\frac{\nt}{2}$. Note that the last asymptotic expression can be refined using bounds $r_{g_i}^{\rm lmw}$ and $\overline{r}_{g_i}^{\rm lmw}$ in Sec. \ref{Sec:LowerBoundsBothConstraintsContDist} and \ref{Sec:UpperBoundsBothConstraintsDuality}, respectively.

The low-SNR asymptotic capacity was studied in \cite{ChaabanRezkiAlouini_LowSNR_CommL} under both average and peak constraints. An optimal intensity allocation $\mathcal{E}_i$ was derived, and was described as {\it inverted water-filling}. The following statement was proved in \cite{ChaabanRezkiAlouini_LowSNR_CommL}.

\begin{theorem}[\cite{ChaabanRezkiAlouini_LowSNR_CommL}]
\label{Thm:LowSNRParallel}
The capacity of parallel IM/DD Gaussian channels with a peak constraint $\mathcal{A}$ and an average constraint $\mathcal{E}$ satisfies 
\begin{align}
\label{LBParallel_low}
c^{\rm parallel}_{\boldsymbol{g}}&\geq \sum_{i=1}^{\nt} r_{g_{i}}^{\rm fh}(\mathcal{A},\mathcal{E}_i^*),\\
\label{UBParallel_low}
c^{\rm parallel}_{\boldsymbol{g}}&\leq \max_{\mathcal{E}_i} \sum_{i=1}^{\nt} \overline{r}_{g_{i}}^{\rm 0}(\mathcal{A},\mathcal{E}_i),
\end{align}
and
\begin{align}
\label{AsympParallel_low}
\lim_{\mathcal{E}\to0} \frac{c_{\boldsymbol{g}}^{\rm parallel}(\mathcal{A},\mathcal{E})}{\sum_{i=1}^{\nt} \frac{1}{2}g_{i}^2\mathcal{E}_i^*(\mathcal{A}-\mathcal{E}_i^*)}&=1,
\end{align} 
where $r_{g_{i}}^{\rm fh}$ and $\overline{r}_{g_{i}}^{\rm 0}$ are defined in Theorems \ref{Thm:FaridHranilovicLB} and \ref{Thm:UBRelaxation}, $\mathcal{E}_i^*=\max\left\{0,\frac{\mathcal{A}}{2}-\frac{\mu}{g_{i}^2}\right\}$, and $\mu$ is chosen so that $\sum_{i=1}^{\nt}\mathcal{E}_i=\min\{\mathcal{E},\frac{\nt \mathcal{A}}{2}\}$. 
\end{theorem}

Note that the lower bound $\overline{r}_{g_{i}}^{\rm 0}$ will be the largest if only $2$ mass points are used are low SNR, in which case the transmission scheme can be described as coded OOK. The asymptotic capacity statement of Theorem \ref{Thm:LowSNRParallel} applies both when $\mathcal{A}$ is held fixed, and when $\mathcal{A}$ vanishes proportional to $\mathcal{E}$, i.e., $\mathcal{E}=\alpha\mathcal{A}$. In the former case, as $\mathcal{E}\to0$, only the strongest channel (largest $g_{i}$) will be activated, and will be allocated the full $\mathcal{E}$. The optimal scheme in this case is coded OOK over the strongest channel. For the latter case, interestingly, multiple channels may remain active as $\mathcal{E}\to0$, and coded OOK over multiple channels (not only the strongest channel) is optimal.

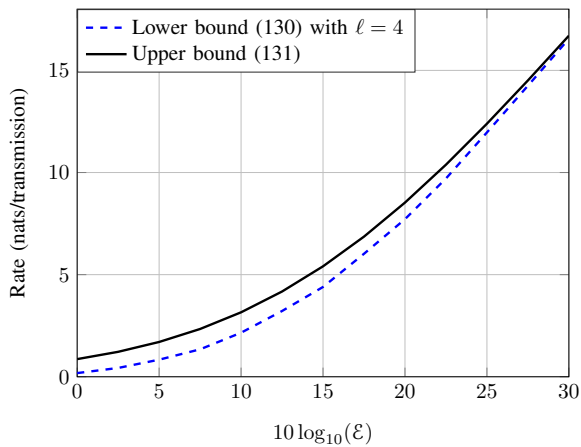
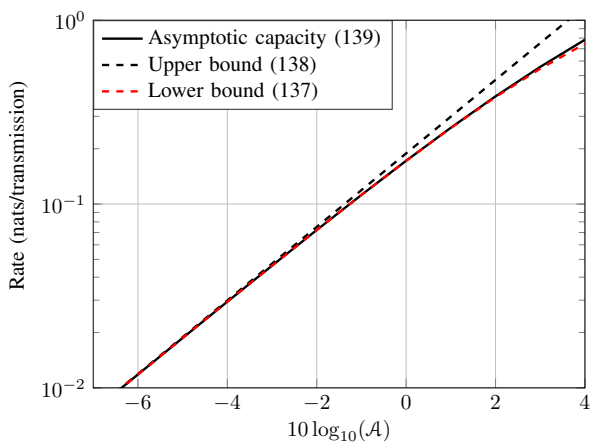
\begin{figure}[t]
%\centering
\begin{subfigure}[]{.46\textwidth}
\centering
\tikzset{every picture/.style={scale=.9}, every node/.style={scale=.9}}
%\input{matlab/rate_parallel_channels_1p7p3p1}
% Minimal pgfplots version: 1.3
%  http://www.mathworks.com/matlabcentral/fileexchange/22022-matlab2tikz
\begin{tikzpicture}

\begin{axis}[%
width=3.8in,
height=3in,
xmin=0,
xmax=30,
xlabel={$10\log_{10}(\mathcal{E})$},
xmajorgrids,
x label style={at={(axis cs:15,-.3)},anchor=north},
ymin=0,
ymax=18,
ylabel={Rate (nats/transmission)},
ymajorgrids,
y label style={at={(axis cs:0,9)},anchor=north},
legend style={at = {(axis cs: 0,18)}, anchor = north west, legend cell align=left,align=left,draw=white!15!black}
]
\addplot [color=blue,dashed,line width=1.0pt]
  table[row sep=crcr]{%
0	0.17975525635616\\
2.5	0.431041825009864\\
5	0.836326519494487\\
7.5	1.34327360569902\\
10	2.15701444678124\\
12.5	3.21922612587096\\
15	4.3963935262973\\
17.5	6.02180338321605\\
20	7.71679445659379\\
22.5	9.70732274191327\\
25	11.9629251328379\\
27.5	14.2502800189442\\
30	16.5489434623582\\
};
\addlegendentry{Lower bound \eqref{LBParallel} with $\ell=4$};

\addplot [color=black,solid,line width=1.0pt]
  table[row sep=crcr]{%
0	0.864161309588369\\
2.5	1.22112390101414\\
5	1.70336670755989\\
7.5	2.33928212082165\\
10	3.1555031258572\\
12.5	4.17362036028227\\
15	5.4075502418166\\
17.5	6.86174693375786\\
20	8.52828140750773\\
22.5	10.3835768131981\\
25	12.3906708993728\\
27.5	14.5081226405879\\
30	16.6992316249257\\
};
\addlegendentry{Upper bound \eqref{UBParallel}};

\end{axis}
\end{tikzpicture}%
\caption{Channel with only an average constraint $\mathcal{E}$.}
\label{Fig:ParallelHigh}
\end{subfigure}
%\hspace{.4cm}
\begin{subfigure}[]{.46\textwidth}
\vspace{.4cm}
\centering
\tikzset{every picture/.style={scale=.9}, every node/.style={scale=.9}}
%\input{matlab/rate_parallel_channels_1p7p3p1_low_SNR}
%\vspace{-.4cm}
% Minimal pgfplots version: 1.3
%  http://www.mathworks.com/matlabcentral/fileexchange/22022-matlab2tikz
\definecolor{mycolor1}{rgb}{1.00000,0.00000,1.00000}%
\begin{tikzpicture}

\begin{axis}[%
width=3.8in,
height=3in,
xmin=-7,
xmax=4,
xlabel={$10\log_{10}(\mathcal{A})$},
xmajorgrids,
x label style={at={(axis cs:-1.5,1.1e-2)},anchor=north},
ymin=1e-2,
ymax=1,
ylabel={Rate (nats/transmission)},
ymajorgrids,
ymode=log,
legend style={at = {(axis cs: -7,1)}, anchor = north west, legend cell align=left,align=left,draw=white!15!black}
]
\addplot [color=black,solid,line width=1.0pt]
  table[row sep=crcr]{%
%-8	0.00472711129027954\\
-7	0.00748091164078211\\
-6	0.0118288325706152\\
-5	0.0186787214089977\\
-4	0.0294335622690098\\
-3	0.0462308099070658\\
-2	0.0722556545642408\\
-1	0.112097454504202\\
0	0.172044838334764\\
1	0.26010779126231\\
2	0.385491492936531\\
3	0.557390424056022\\
4	0.78341588868209\\
};
\addlegendentry{Asymptotic capacity \eqref{AsympParallel_low}};

\addplot [color=black,dashed,line width=1.0pt]
  table[row sep=crcr]{%
%-8	0.00473910781323511\\
-7	0.00751097972089393\\
-6	0.0119041006377166\\
-5	0.0188667268084451\\
-4	0.029901748847395\\
-3	0.0473910779723506\\
-2	0.0751097970489396\\
-1	0.119041004937165\\
0	0.188667280564474\\
1	0.299017489049951\\
2	0.473910782843516\\
3	0.751097973289397\\
4	1.19041005737164\\
};
\addlegendentry{Upper bound \eqref{UBParallel_low}};

\addplot [color=red,dashed,line width=1.0pt]
  table[row sep=crcr]{%
%-8	0.00472711020519667\\
-7	0.0074809074402975\\
-6	0.0118288159764888\\
-5	0.0186786131328525\\
-4	0.029433083111023\\
-3	0.0462286161608927\\
-2	0.072245418648321\\
-1	0.112050050357402\\
0	0.171831707146785\\
1	0.259202741976387\\
2	0.381958116985947\\
3	0.545005329949031\\
4	0.745176657119805\\
};
\addlegendentry{Lower bound \eqref{LBParallel_low}};

\end{axis}
\end{tikzpicture}%
\caption{Channel with a peak constraint $\mathcal{A}$ and an average constraint $\mathcal{E}=\alpha\mathcal{A}$ with $\alpha=1$.}
\label{Fig:ParallelLow}
\end{subfigure}
\caption{capacity bounds for an parallel IM/DD Gaussian channel with $\boldsymbol{g}=(1,0.7,0.3,0.1)$.}
\label{Fig:Parallel}
\end{figure}

Fig. \ref{Fig:Parallel} shows achievable rates and upper bounds for an exemplary parallel channel showing the asymptotic capacity. In Fig. \ref{Fig:ParallelHigh}, only one channel is active when $10\log_{10}(\mathcal{E})$ is $7.5$ and lower, two channels are active when $10\log_{10}(\mathcal{E})$ is between $10$ and $12.5$, three channels are active when $10\log_{10}(\mathcal{E})$ is between $15$ and $20$, and all channels are active when $10\log_{10}(\mathcal{E})$ is larger than $22.5$. In Fig. \ref{Fig:ParallelLow}, all channels are always active.

\subsection{MIMO IM/DD Gaussian Channels}
When there is cross-talk between the transmit and receiver apertures, the problem of finding the capacity becomes more complicated. Indeed, in this case one has to find the optimal $\mathbb{P}_{\X}$ which is not a product distribution any more. The element of transmitter cooperation {(by multiplexing)} and receiver combining comes into play. The capacity in this case has been bounded  and the asymptotic capacity derived in the literature. We split the discussion into three parts, where we discuss the SIMO, MISO, and MIMO channels.

\subsubsection{SIMO IM/DD Gaussian Channels}
The SIMO channel has $\nt=1$ and $\nr>1$, and hence $\G$ is a column vector. The capacity of the SIMO channel can be easily expressed in terms of the capacity of the SISO channel. 

Let $\U$ be an orthogonal matrix defined as $\U=\left[\frac{\G}{\|\G\|_2},\ \G^{\perp}\right]$ where $\G^{\perp}$ is an $\nr\times(\nt-1)$ matrix orthogonal to $\G$ with orthogonal columns of unit norm. This matrix is invertible, and hence the channel 
$$\U^T\Y=\left[\|\G\|_2X + \frac{\G^T}{\|\G\|_2}\Z,\G^{\perp^T}\Z\right]^T$$ 
has the same capacity as the original SIMO channel since this transformation from $\Y$ to $\U^T\Y$ is information-lossless. Since the noise $\G^{\perp^T}\Z$ is uncorrelated with $\frac{\G^T}{\|\G\|_2}\Z$ and independent of $\X$, it can be ignored. Thus, the capacity of the SIMO IM/DD Gaussian channel is equal to the capacity of the channel with input $X$ and output 
$\|\G\|_2X + \frac{\G^T}{\|\G\|_2}\Z$. Note that $\|\G\|_2$ is a scalar, and that $\frac{\G^T}{\|\G\|_2}\Z$ is $\mathcal{N}(0,1)$. This channel is a SISO IM/DD Gaussian channel, with channel gain $\|\G\|_2$. Thus, we can state the following.

\begin{theorem}
\label{Thm:SIMOCapacity}
The capacity of the SIMO IM/DD Gaussian channel satisfies 
\begin{align}
c^{\rm simo}_{\G}(\mathcal{A},\mathcal{E})=c_{\|\G\|_2}(\mathcal{A},\mathcal{E}).
\end{align} 
\end{theorem}

This capacity can be characterized and bounded using the statements in Sec. \ref{Sec:SISO_Cap}--\ref{Sec:CapacityBounds}. 

The multiplication of $\Y$ by $\U$ is known as maximum-ratio combining (MRC). Thus, MRC is the optimal receiver combining in a SIMO IM/DD Gaussian channel. The choice of the input distribution then determines the achievable rate.

\subsubsection{MISO IM/DD Gaussian Channels}
In this case, $\nt>1$ and $\nr=1$, and hence $\G$ is a row vector, which we denote for convenience $(g_1,\ldots,g_{\nt})$ (not to be confused with the diagonal channel in Sec. \ref{Sec:ParallelChannels}). We assume without loss of generality that $g_1\geq g_2\geq\cdots\geq g_{n_{\rm t}}$, i.e., the first channel is the strongest channel, followed by the second channel, and so on. 

Achievable rates for the MISO channel have been derived in \cite{ChaabanRezkiAlouini_MIMOOWC_TWC,MoserWangWigger_IT,
ChaabanRezkiAlouini_MIMO_IMDD_Low_SNR_TCOM}. In \cite{MoserWangWigger_IT}, the problem of finding the capacity of the MISO channel was approached from the perspective of finding the optimal distribution $\G\X$, which is a scalar in this case. The capacity of the MISO channel can be expressed as 
\begin{align}
c^{\rm miso}_{\G}(\mathcal{A},\mathcal{E}) = \max_{\mathbb{P}_{\G\X}} I(\G\X;Y),
\end{align}
where the maximization is with respect to feasible distributions of $\G\X$ given $\G$ and the constraints on $\X$. This enables expressing the capacity in terms of the capacity of an SISO IM/DD Gaussian channel. In \cite{ChaabanRezkiAlouini_MIMOOWC_TWC}, the problem was approached from a pre-/post-coding perspective, i.e., construction of $\X$ as $\mathbb{V}\tilde{\X}$ where $\tilde{\X}$ has independent components, and decoding from $\mathbb{U}\Y$ where $\mathbb{V}$ and $\mathbb{U}$ are pre- and post-coding matrices, respectively. The work in \cite{ChaabanRezkiAlouini_MIMO_IMDD_Low_SNR_TCOM} focuses on low SNR.

\paragraph{Average Constraint Only}
Under an average constraint only, it was shown that it is optimal to transmit through the strongest channel $g_{1}$, i.e., activate the first transmit aperture only. The intuition is that since the capacity of the SISO channel increases with the average of the input (under an average constraint only), then one needs to maximize the average of $\G\X$. This is achieved when $X_1>0$ with $\mathbb{E}[X_1]=\mathcal{E}$, and $X_2=X_3=\cdots=X_{\nt}=0$. Thus the optimal pre-coder in this case is $(1,0,0,\ldots,0)^T$, i.e., {\it best aperture selection}, since the transmit signal can be written as $\X=(1,0,0,\ldots,0)^TX$ where $X\sim\mathbb{P}_X$ and satisfies the average constraint. This leads to the following statement.

\begin{theorem}[\cite{MoserWangWigger_IT}]
\label{Thm:MISOAverageOnly}
The capacity of MISO IM/DD Gaussian channels with only an average constraint $\mathcal{E}$ satisfies 
\begin{align}
c^{\rm miso}_{\G}(\infty,\mathcal{E})=c_{g_1}(\infty,\mathcal{E}).
\end{align} 
\end{theorem}

Generally, one can use statements in Sec. \ref{Sec:CapacityBounds} to bound this capacity expression or derive the asymptotic capacity at high SNR. This is stated next, and is easy to prove using Theorem \ref{Thm:MISOAverageOnly} and Theorem \ref{Thm:HighSNR_average_only}.

\begin{theorem}[\cite{ChaabanRezkiAlouini_MIMOOWC_TWC}]
The capacity of the MISO IM/DD Gaussian channel with only an average constraint $\mathcal{E}$ satisfies
\begin{align}
\lim_{\mathcal{E}\to\infty}\left( c^{\rm miso}_{\G}(\infty,\mathcal{E})-\frac{1}{2}\log\left(\frac{eg_1^2\mathcal{E}^2}{2\pi}\right)\right)=0.
\end{align}
\end{theorem}

\paragraph{Peak and Average Constraints}
Under both peak and average constraints with an average constraint $\mathcal{E}\geq\frac{\nt\mathcal{A}}{2}$, $\G\X$ is subject to a peak constraint $\|\G\|_1\mathcal{A}$ and an average constraint $\mathbb{E}[\G\X]=\sum_{i=1}^{\nt}g_i\mathcal{E}_i\leq\mathcal{E}$ where $\mathcal{E}_i=\mathbb{E}[X_i]$. Recall from Sec. \ref{Sec:MutualInfoMax} that for a peak and average constrained channel, capacity is maximum if the average is equal to half the peak constraint. This is feasible in this case, since one can choose $\mathcal{E}_i=\frac{\mathcal{E}}{\nt}$ leading to and average $\sum_{i=1}^{\nt}g_i\mathcal{E}_i=\|\G\|_1\frac{\mathcal{E}}{\nt}\geq\frac{\|\G\|_1\mathcal{A}}{2}$, where the last inequality follows due to the assumption $\mathcal{E}\geq\frac{\nt\mathcal{A}}{2}$. Hence, in this case, the channel $Y=\G\X+Z$ has the same capacity as SISO IM/DD Gaussian channel with channel gain $1$, a peak constraint $\|\G\|_1\mathcal{A}$, and an average constraint $\frac{\|\G\|_1\mathcal{A}}{2}$. This in turn has the same capacity as a SISO IM/DD Gaussian channel with channel gain $\|\G\|_1$, a peak constraint $\mathcal{A}$, and an average constraint $\frac{\mathcal{A}}{2}$. The same applies if the channel is only subject to a peak constraint, leading to the following statement.

\begin{theorem}[\cite{MoserWangWigger_IT}]
\label{Thm:MISOPeakOnly}
The capacity of MISO IM/DD Gaussian channels with a peak and an average constraint $\mathcal{E}\geq\frac{\nt\mathcal{A}}{2}$ (or $\mathcal{E}=\infty$) satisfies 
\begin{align}
c^{\rm miso}_{\G}(\mathcal{A},\mathcal{E})=c_{\|\G\|_1}\left(\mathcal{A},\frac{\mathcal{A}}{2}\right).
\end{align} 
\end{theorem}

Note that this statement implies that it is optimal to use a {\it repetition code }in this case. By sending $\X=(1,1,\ldots,1)^TX$, where $X\in[0,\mathcal{A}]$ and $\mathbb{E}[X]=\frac{\mathcal{A}}{2}$ over this MISO channel, we obtain an effective SISO IM/DD Gaussian channel with channel gain $\|\G\|_1$, peak constraint $\mathcal{A}$, and average constraint $\frac{\mathcal{A}}{2}$, which has the capacity in Theorem \ref{Thm:MISOPeakOnly}. 

One can develop capacity bounds and asymptotic capacity expressions for this case using statements in Sec. \ref{Sec:CapacityBounds}.

For $\mathcal{E}<\frac{\nt\mathcal{A}}{2}$, the situation is more difficult. In this case, \cite{MoserWangWigger_IT} derives some properties of the capacity achieving input distributions which aid in the analysis. Namely, the optimal distribution activates aperture $i$ only if apertures $1,\ldots,i-1$ (which have stronger channels) are transmitting at peak intensity. Formally, this means that  $X_i>0\Rightarrow X_j=\mathcal{A}\,\forall j<i$. This can be understood as a two-layer modulation, where the first layer modulates the number of active apertures $i\in\{0,\ldots,\nt\}$, and the second layer modulates the intensity of $i$-th aperture according to some probability distribution. A similar behaviour has been identified in \cite{ChaabanRezkiAlouini_MIMO_IMDD_Low_SNR_TCOM} for the MIMO channel at low SNR as described later. 

Using the properties derived in \cite{MoserWangWigger_IT}, capacity lower bounds for the MISO IM/DD Gaussian channel with $\mathcal{E}<\frac{\nt\mathcal{A}}{2}$ can be derived using the EPI. Recall that this requires finding the solution of $\max_{\mathbb{P}_{\G\X}}h(\G\X)$ over the set of feasible distributions subject to the input constraints. This problem was solved in \cite{MoserWangWigger_IT} which led to a capacity lower bound which is tight at high SNR. Capacity upper bounds were also derived in \cite{MoserWangWigger_IT} using the peak-only constrained channel, constraint relaxation (maximum variance) and using the dual capacity expression in Lemma \ref{Lem:Duality}. These bounds have rather sophisticated expressions, and are not repeated here. Instead, we focus on asymptotic capacity expressions. We start with high SNR.

\begin{theorem}{\cite{MoserWangWigger_IT}}
\label{Thm:HighSNRAvgPeakMISO}
The capacity of the MISO IM/DD Gaussian channel with a peak constraint $\mathcal{A}$ and an average constraint $\mathcal{E}=\alpha\mathcal{A}$ satisfies
\begin{align}
\lim_{\mathcal{A}\to\infty}\left( c^{\rm miso}_{\G}(\mathcal{A},\alpha\mathcal{A})-\frac{1}{2}\log\left( \frac{\|\G\|_1^2\mathcal{A}^2}{2\pi e}\right)\right)=0
\end{align}
if $\alpha>\alpha_{\rm th}$, and
\begin{align}
\lim_{\mathcal{A}\to\infty}\left( c^{\rm miso}_{\G}(\mathcal{A},\alpha\mathcal{A})-\frac{1}{2}\log\left(\frac{\|\G\|_1^2\mathcal{A}^2}{2\pi e }\right)-\nu\right)=0
\end{align}
if $\alpha\leq\alpha_{\rm th}$, where $\alpha_{\rm th}	=\frac{1}{2}+\frac{1}{\|\G\|_1}\sum_{i=1}^{\nt}g_i(i-1)$, 
\begin{align}
\nu&=\hspace{-.1cm}\sup_{ \omega \in \left( \max\{0,\frac{1}{2}+\alpha-\alpha_{\rm th}\},\min\{\frac{1}{2},\alpha\} \right)} \hspace{-.1cm} \left(1-\log\left(\frac{\mu(\omega)}{1-e^{-\mu(\omega)}}\right)\right.\nonumber\\
&\qquad\qquad -\left.\frac{\mu(\omega)e^{-\mu(\omega)}}{1-e^{-\mu(\omega)}}-D\left(\boldsymbol{p}\left\|\frac{\G}{\|\G\|_1}\right.\right)\right),
\end{align}
$\mu(\omega)$ is the unique positive solution of $\frac{1}{\mu}-\frac{e^{-\mu}}{1-e^{-\mu}}=\omega$, $\boldsymbol{p}=(p_1,\ldots,p_{\nt})$, $p_i=\frac{g_ia^i}{\sum_{k=1}^{\nt}g_ka^k}$, and $a$ is the unique positive solution of $\frac{\sum_{i=1}^{\nt}ig_ia^i}{\sum_{k=1}^{\nt}g_ka^k}=\alpha-\omega+1$.
\end{theorem}

Thus, the asymptotic high-SNR capacity is $\frac{1}{2}\log\left(\frac{\|\G\|_1^2\mathcal{A}^2}{2\pi e }\right)$ if $\alpha>\alpha_{\rm th}$ which is also the high-SNR capacity when there is no average constraint. The asymptotic high SNR capacity is $\frac{1}{2}\log\left(\frac{\|\G\|_1^2\mathcal{A}^2}{2\pi e }\right)+\nu$ when $\alpha\leq\alpha_{\rm th}$. 

At low SNR, the asymptotic capacity was given in \cite{ChaabanRezkiAlouini_MIMO_IMDD_Low_SNR_TCOM,MoserWangWigger_IT} as follows.

\begin{theorem}[\cite{ChaabanRezkiAlouini_MIMO_IMDD_Low_SNR_TCOM,MoserWangWigger_IT}]
\label{Thm:LowSNRMISOAvgPeak}
The capacity of the MISO IM/DD Gaussian channel with a peak constraint $\mathcal{A}$ and an average constraint $\mathcal{E}=\alpha\mathcal{A}$ satisfies
\begin{align}
\lim_{\mathcal{A}\to0}\frac{c^{\rm miso}_{\G}(\mathcal{A},\alpha\mathcal{A})}{\frac{\gamma\mathcal{A}^2}{2}}=1,
\end{align}
where 
\begin{align*}
\gamma &= 
\max_{a_i: \sum_{i=1}^{\nt}a_i\leq \alpha} \sum_{i=1}^{\nt}\sum_{j=1}^{\nt} g_ig_j\min\{a_i,a_j\}(1-\max\{a_i,a_j\}).
\end{align*}
\end{theorem}

Here, the expression $\gamma\mathcal{A}^2$ is in fact the maximum variance of $\G\X$ when $\X\in[0,\mathcal{A}]^{\nt}$ and $\sum_{i=1}^{\nt}\mathbb{E}[X_i]\leq \mathcal{E}$. It is achieved when $\X$ follows a maximally-correlated $\nt$-dimensional binary distribution as shown in \cite{ChaabanRezkiAlouini_MIMO_IMDD_Low_SNR_TCOM}. In this case, $\X$ is distributed on $\{0,\mathcal{A}\}^{\nt}$ with $\mathbb{P}_{X_i}(\mathcal{A})=a_i$ for some $a_i$ that satisfies $\sum_{i=1}^{\nt}a_i=\frac{\mathcal{E}}{\mathcal{A}}$ in order to satisfy the average constraint, with maximum correlation. Maximum correlation is achieved using the structure in Table \ref{Tab:ProbX}. The proof of this statement is based on the mutual-information expression for weak signals given in \cite{PrelovVanDerMeulen}.

\begin{table*}[t]
\caption{The maximally-correlated $\nt$-dimensional distribution of $\X\in\{0,\mathcal{A}\}^{\nt}$ with $\mathbb{P}_{X_i}(\mathcal{A})=a_i$ and $a_1\geq a_2\geq\cdots\geq a_{\nt}$. When optimized with respect to $a_i$ with $\sum_{i=1}^{\nt}a_i\leq\frac{\mathcal{E}}{\mathcal{A}}$, this distribution achieves the low-SNR capacity of the MISO and MIMO IM/DD Gaussian channel.}
\label{Tab:ProbX}
\centering
\setlength\extrarowheight{4pt}
\begin{tabular}{|c||c|c|c|c|c|c|}
\hline
$\x^T$    & $(0,0,0,\ldots,0,0)$ & $(\mathcal{A},0,0,\ldots,0,0)$ & $(\mathcal{A},\mathcal{A},0,\ldots,0,0)$ & $\cdots$ & $(\mathcal{A},\mathcal{A},\mathcal{A},\ldots,\mathcal{A},0)$ & $(\mathcal{A},\mathcal{A},\mathcal{A},\ldots,\mathcal{A},\mathcal{A})$\\\hline
$\mathbb{P}_{\X}(\x)$ & $1-a_1$ & $a_1-a_2$ & $a_2-a_3$ & $\cdots$ & $a_{\nt-1}-a_{\nt}$ & $a_{\nt}$\\\hline
\end{tabular}
%\caption{Capacity achieving distribution at low SNR for $\mathcal{E}_1\geq\mathcal{E}_2\geq\cdots\geq\mathcal{E}_M$. Realizations of $\x$ with zero probability are not shown.}
%\label{Tab:ProbX}
\end{table*}

\subsubsection{MIMO IM/DD Gaussian Channels}
In the MIMO channel, we have that both $\nt>1$ and $\nr>1$. The channel gain matrix in this case is $\nr\times\nt$ denoted $\G=(\boldsymbol{g}_1,\ldots,\boldsymbol{g}_{n_{\rm t}})$ where $\boldsymbol{g}_i$ is a column vector representing the channel gains from transmit aperture $i$ to all receive apertures.

Recall that capacity in this case is given by
\begin{align}
c_{\G}^{\rm mimo}(\mathcal{A},\mathcal{E})=\max_{\mathbb{P}_{\X}}I(\X,\Y).
\end{align}
This problem was studied in \cite{ChaabanRezkiAlouini_MIMOOWC_ICC,MoserMylonakisWangWigger,
ChaabanRezkiAlouini_MIMOOWC_TWC,ChaabanRezkiAlouini_MIMOOWC_LowSNR_ISIT,
ChaabanRezkiAlouini_MIMO_IMDD_Low_SNR_TCOM,
LiMoserWangWigger}, where capacity bounds and asymptotic capacity characterizations were given. 

The work in \cite{ChaabanRezkiAlouini_MIMOOWC_TWC} focused on pre-coding and post-coding to convert the MIMO channel into a set of parallel channels. For the scenario with $\nr\geq\nt$, three schemes were compared: Channel inversion post-coding; DC-biased singular-value decomposition pre-/post-coding; and QR-decomposition post-coding. These schemes allow converting the MIMO channel into a set of parallel channel where the intensity allocation in \cite{ChaabanRezkiAlouini_ParallelOWC_TCOM} is applied. It is shown that the QR-decomposition scheme outperforms the rest, so we describe it here. 

Assume that $\G$ has full column rank, and $\nr\geq\nt$. In a QR-decomposition scheme, the transmitter sends $\X$ consisting of independently coded symbols $X_1,\ldots,X_{\nt}$, each of which is from an independent stream. The receiver computes the QR-decomposition of $\G$, i.e., a decomposition $\G=\mathbb{Q}\mathbb{U}$ (we use $\mathbb{U}$ instead of $\mathbb{R}$ for convenience), where $\mathbb{Q}$ is an $\nr\times\nr$ orthogonal matrix, and $\mathbb{U}$ is an $\nr\times\nt$ upper triangular matrix. The receiver multiplies $\Y$ with $\mathbb{Q}$ to obtain the signal $\tilde{\Y}=\mathbb{U}\X+\tilde{\Z}$, where $\tilde{\Z}$ is i.i.d. $\mathcal{N}(0,1)$. Then the receiver starts decoding from $\tilde{Y}_{\nt}$ which has an interference free observation of $X_{\nt}$ due to the upper triangular structure of $\mathbb{U}$, i.e., $Y_{\nt}=u_{\nt,\nt}X_{\nt}+\tilde{Z}_{\nt}$. Then, the receiver decodes the stream sent over $X_{\nt}$ (by considering $n$ observations of $\tilde{Y}_{\nt}$, where $n$ is the code-length), subtracts its contribution from $\tilde{Y}_{\nt-1}$ and proceeds by decoding $X_{\nt-1}$ interference free. {The receiver can effectively do this if the rate of $X_{n_t}$ is smaller than $I(X_{n_t};Y_{n_t})$, i.e., information is encoded in $X_{n_t}$ at any achievable rate described earlier.} The receiver proceeds this way until all streams are decoded. The resulting achievable rate is in the form
\begin{align}
\label{AchRateMIMOAvgOnly}
\max_{\mathcal{E}_i}\sum_{i=1}^{\nt} r^{\ell,\rm a}_{u_{i,i}}(\mathcal{E}_i),\quad \ell\in\{\rm lma,cma,fh\},
\end{align}
where $\mathcal{E}_i$ satisfies $\sum_{i=1}^{\nt}\mathcal{E}_i\leq\mathcal{E}$, and $r^{\ell,\rm a}_{u_{i,i}}(\mathcal{E}_i)$ is as defined in Theorems \ref{Thm:LB_Exp}-\ref{Thm:LB_Geom}. The intensity allocation $\mathcal{E}_i$ is done using the algorithm in \cite{ChaabanRezkiAlouini_ParallelOWC_TCOM}. In addition to achievable rates, \cite{ChaabanRezkiAlouini_MIMOOWC_TWC} also derives capacity upper bounds and asymptotic capacity results at high SNR. Asymptotic capacity results at low SNR are given in \cite{ChaabanRezkiAlouini_MIMO_IMDD_Low_SNR_TCOM}. The case $\nr<\nt$ is also discussed.

The work in \cite{LiMoserWangWigger} focused on a novel approach, wherein the capacity is rewritten as
\begin{align}
c^{\rm mimo}_{\G}(\mathcal{A},\mathcal{E})=\max_{\mathbb{P}_{\G\X}\in\mathcal{P}_{\G\X}}I(\G\X,\Y),
\end{align}
where $\mathcal{P}_{\G\X}$ is the set of all distributions $\mathbb{P}_{\G\X}$ over the {\it zenotope} $\mathcal{R}(\G)=\{\sum_{i=1}^{\nr}a_i\boldsymbol{g}_i| (a_1,\ldots,a_{\nt})\in[0,\mathcal{A}]^{\nt}\}$ which satisfy the power constraints on $\X$. This zenotope is the image of the hypercube $[0,\mathcal{A}]^{\nt}$ after multiplying by $\G$. Using this formulation, \cite{LiMoserWangWigger} derived a minimum energy signalling scheme, capacity bounds, and asymptotic capacity results. 

We discuss some of the results of \cite{ChaabanRezkiAlouini_ParallelOWC_TCOM,
ChaabanRezkiAlouini_MIMO_IMDD_Low_SNR_TCOM,LiMoserWangWigger} in what follows.

\paragraph{Average Constraint Only}
Capacity bounds for the average constrained MIMO channel were derived in \cite{ChaabanRezkiAlouini_MIMOOWC_TWC} when $\nr\geq\nt$. Using the derived bounds, it is shown that the QR-scheme achieves the high-SNR asymptotic capacity for a MIMO IM/DD Gaussian channel with only an average constraint when $\nr\geq\nt$, given as follows.

\begin{theorem}[\cite{ChaabanRezkiAlouini_MIMOOWC_TWC}]
\label{Thm:HighSNRAvgMIMO}
The capacity of a MIMO IM/DD Gaussian channel with $\nr\geq\nt$, with only an average constraint $\mathcal{E}$, satisfies
\begin{align}
\lim_{\mathcal{E}\to\infty}\left( c^{\rm mimo}_{\G}(\infty,\mathcal{E})-\frac{1}{2}\log\left|\frac{e\mathcal{E}^2}{2\pi\nt^2}\G^T\G\right|\right)=0.
\end{align}
\end{theorem}

If $\nr<\nt$, then \cite{ChaabanRezkiAlouini_MIMOOWC_TWC} shows that the capacity pre-log is $\nr$, i.e., capacity scales as $\nr\log(\mathcal{E})$ at high SNR ($\mathcal{E}\to\infty$).

\paragraph{Peak and Average Constraints}
The work in \cite{ChaabanRezkiAlouini_MIMOOWC_TWC,LiMoserWangWigger} studies the MIMO channel with both average and peak constraints. Achievable rates, capacity upper bounds, and asymptotic high-SNR capacity expressions/approximations were given using different approaches. We only state asymptotic capacities next.

In \cite{LiMoserWangWigger}, the asymptotic high-SNR capacity was characterized for a MIMO channel with both peak and average constraints, under $\nt>\nr$. The asymptotic capacity expression is a function of the volume of the zenotope $\mathcal{R}(\G)$ and related parameters, and is not reviewed here due to its complicated nature. The reader is referred to \cite{LiMoserWangWigger} for further details. For $\nr\geq \nt$, we review some asymptotic high SNR results from \cite{ChaabanRezkiAlouini_MIMOOWC_TWC} which have simpler expressions.

\begin{theorem}[\cite{ChaabanRezkiAlouini_MIMOOWC_TWC}]
\label{Thm:HighSNRAvgPeakMIMO}
The capacity of a MIMO IM/DD Gaussian channel with $\nr\geq\nt$, with a peak constraint $\mathcal{A}$ and average constraint $\mathcal{E}=\alpha\mathcal{A}$  satisfies
\begin{align}
\lim_{\mathcal{A}\to\infty}\left( c^{\rm mimo}_{\G}(\mathcal{A},\alpha\mathcal{A})-\frac{1}{2}\log\left|\frac{\mathcal{A}^2}{2\pi e}\G^T\G\right|\right)=0
\end{align}
if $\alpha\geq\frac{\nt}{2}$, and 
\begin{align}
&\lim_{\mathcal{A}\to\infty}\left( c^{\rm mimo}_{\G}(\mathcal{A},\alpha\mathcal{A})-\frac{1}{2}\log\left|\frac{e\min\left\{\frac{\alpha^2\mathcal{A}^2}{\nt^2},\frac{\mathcal{A}^2}{e^2}\right\}}{2\pi}\G^T\G\right|\right)\nonumber\\
&\qquad\qquad \leq 0.1\nt.
\end{align}
if $\alpha<\frac{\nt}{2}$.
\end{theorem}

The asymptotic high-SNR results in Theorems \ref{Thm:HighSNRAvgMIMO} and \ref{Thm:HighSNRAvgPeakMIMO} are achievable using the QR-decomposition scheme in combination with exponential or truncated-Gaussian distributions on each $X_i$. 

Achievable rates and upper bounds for the case $\nr<\nt$ are also given in \cite{ChaabanRezkiAlouini_MIMOOWC_TWC}. While the high-SNR capacity is characterized/approximated in \cite{ChaabanRezkiAlouini_MIMOOWC_TWC} for the case $\nr\geq\nt$, only the high-SNR capacity per-log is given for $\nr<\nt$. Namely, it is shown that capacity scales as $\nr\log(\mathcal{A})$ in this case.

In the low-SNR regime, the capacity of a MIMO IM/DD Gaussian channel was characterized in \cite{ChaabanRezkiAlouini_MIMO_IMDD_Low_SNR_TCOM,LiMoserWangWigger} as follows.

\begin{theorem}[\cite{ChaabanRezkiAlouini_MIMO_IMDD_Low_SNR_TCOM,LiMoserWangWigger}]
\label{Thm:LowSNRMIMOAvgPeak}
The capacity of a MIMO IM/DD Gaussian channel with a peak constraint $\mathcal{A}$ and average constraint $\mathcal{E}=\alpha\mathcal{A}$  satisfies
\begin{align}\label{MIMO_LowSNR}
\lim_{\mathcal{A}\to0} \frac{ c^{\rm mimo}_{\G}(\mathcal{A},\alpha\mathcal{A})}{\frac{\eta\mathcal{A}^2}{2}}=1,
\end{align}
where
\begin{align*}
\eta=\max_{a_i: \sum_{i=1}^{n_{\rm t}}a_i\leq \alpha} \sum_{i=1}^{\nt}\sum_{j=1}^{\nt} \boldsymbol{g}_i^T\boldsymbol{g}_j\min\{a_i,a_j\}(1-\max\{a_i,a_j\}),
\end{align*}
and $\boldsymbol{g}_i$ is the $i$-th column of $\G$.
\end{theorem}

This statement was proved using the result in \cite{PrelovVanDerMeulen} for the channel capacity with weak inputs, and using the result in \cite{GuoShamaiVerdu} for the relation between mutual information and MMSE. Similar to the MISO case discussed in Theorem \ref{Thm:LowSNRMISOAvgPeak}, the asymptotic capacity $\frac{\gamma\mathcal{A}^2}{2}$ in Theorem \ref{Thm:LowSNRMIMOAvgPeak} is achievable using a maximally-correlated $\nt$-dimensional binary input distribution as shown in Table \ref{Tab:ProbX}.

\section{Multi-User IM/DD Gaussian Channels}
\label{Sec:MultiUser}
The statements in Sec. \ref{Sec:SISO_Cap}-\ref{Sec:CapacityBounds} on the capacity of the IM/DD Gaussian channel were used to develop results for a multi-user OWC system modelled as an IM/DD Gaussian broadcast channel (BC) or an IM/DD Gaussian multiple-access channel (MAC). In a $k$-user setting, the goal is to characterize the capacity region defined as the set of achievable rate tuples $(r_1,\ldots,r_k)$, where $r_i$ is the achievable rate of user $i$. Here, we review some recent results on this front, while restricting our attention to the 2-user case for simplicity.

\subsection{IM/DD Gaussian BC}
In a 2-user BC, we have a single transmitter and two receivers. The transmit signal is $X$ which satisfies $X\in[0,\mathcal{A}]$ and $\mathbb{E}[X]\leq\mathcal{E}$. The received signal of user $i$ is $Y_i=g_iX+Z_i$ where $Z_i\sim\mathcal{N}(0,1)$. We assume without loss of generality that $g_1\geq g_2$. 

We define the capacity region as follows. Let the set of message of user $i$ be denoted $\mathcal{W}_i=\{1,\ldots,2^{m_i}\}$. The transmitter wants to send a pair of messages $(w_1,w_2)$, which is uniformly distributed on $\mathcal{W}_1\times\mathcal{W}_2$. It uses an encoder to encode the message pair into a codeword $\x(w_1,w_2)$ of length $n$ satisfying the peak and average constraints, and sends it. Receiver $i$ uses a decoder to decode $\hat{w}_i$. This incurs a probability of error ${\sf p}_{{\sf e},n}=\mathbb{P}\{(w_1,w_2)\neq(\hat{w}_1,\hat{w}_2)\}$. The rate of user $i$ is $r_i=\frac{m_i}{n}$ (bits/transmission), and we call a rate pair $(r_1,r_2)$ achievable if there exists a sequence of codes (message sets, encoder, decoders) that satisfy ${\sf p}_{{\sf e},n}\to0$ as $n\to\infty$. The capacity region is the closure of the set of all achievable rate pairs $(r_1,r_2)$, and we denote it $\mathcal{C}^{\rm bc}_{g_1,g_2}(\mathcal{A},\mathcal{E})$. We express the capacity henceforth in nats/transmission, which can be converted to bits/transmission by dividing by $\log(2)$.

This BC is a stochastically-degraded BC \cite{ElgamalKim} since $X\to Y_1\to Y_2$ forms a Markov chain. {Note that the degradedness of the current BC holds since the channel to user 1 is better than that to user 2 ($g_1> g_2$). Generally, the single-aperture Gaussian BC is always degraded since we either have $g_1 \geq g_2$, i.e., $X\to Y_1 \to Y_2$ form a Markov chain, or $g_2<g_1$, i.e., $X\to Y_2 \to Y_1$ form a Markov chain.} The capacity of a degraded discrete-memoryless BC is known to be given by the convex-hull of the closure of the set of rate pairs $(R_1,R_2)\in\mathbb{R}_+^2$ satisfying \cite{Cover}
\begin{align}
\label{BCCap}
R_2 &\leq I (U; Y_2),\\
R_1 &\leq I (X; Y_1|U),\nonumber
\end{align}
for some distribution $\mathbb{P}_U(u)\mathbb{P}_{X|U}(x|u)\mathbb{P}_{Y_1Y_2|X}(y_1,y_2|x)$ over the set $\mathcal{U}\times\mathcal{X}\times\mathcal{Y}_1\times\mathcal{Y}_2$, where the cardinality of the auxiliary random variable $\mathcal{U}$ is bounded by $|\mathcal{U}| \leq \min\{|\mathcal{X}|, |\mathcal{Y}_1|, |\mathcal{Y}_2|\}$. 

This statement can be generalized to the IM/DD Gaussian BC with continuous alphabets using the discretization procedure explained in \cite[Sec. 3.4]{ElgamalKim}. In this case, we replace $\mathcal{X}$ by $\mathbb{R}_+$, $\mathcal{Y}_i$ by $\mathbb{R}$, and we define $\mathbb{P}_X(X)$ as a probability density function that satisfies $X\in[0,\mathcal{A}]$ and $\mathbb{E}[X]\leq\mathcal{E}$. Describing this capacity region in a simpler form is generally a difficult problem, since one needs to specify a good choice of $(U,X)$. However, capacity bounds were derived in \cite{ChaabanRezkiAlouini_OBC_GCW,ChaabanRezkiAlouini_TWC}. These bounds are presented next. 

Using a method devised by Bergmans in \cite{Bergmans}, the following capacity outer bound can be derived.

\begin{theorem}[\cite{ChaabanRezkiAlouini_TWC}]
The capacity region of the 2-user IM/DD Gaussian BC satisfies $$\mathcal{C}^{\rm bc}_{g_1,g_2}(\mathcal{A},\mathcal{E})\subseteq \overline{\mathcal{R}}^{\rm bc,\ell}_{g_1,g_2}(\mathcal{A},\mathcal{E})\triangleq \bigcup_{\rho\in[0,1]}\tilde{\mathcal{R}}_{g_1,g_2}^{\rm bc,\ell}(\mathcal{A},\mathcal{E},\rho),$$ where $\tilde{\mathcal{R}}_{g_1,g_2}^{\rm bc,\ell}(\mathcal{A},\mathcal{E},\rho)$ is the set of rate pairs $(r_1,r_2)$ satisfying
\begin{align}
0&\leq r_1 \leq \frac{1}{2}\log\left(1+\frac{g_1^2}{g_2^2}\left(e^{2\overline{r}_{g_2}^{\ell}(\rho\mathcal{A},\rho\mathcal{E})}-1\right)\right),\\
0&\leq r_2 \leq \overline{r}_{g_2}^{\ell}(\mathcal{A},\mathcal{E})-\overline{r}_{g_2}^{\ell}(\rho\mathcal{A},\rho\mathcal{E}),
\end{align}
with $\ell\in\{\rm lmw,0\}$.
\end{theorem}

A capacity inner bound was also derived in \cite{ChaabanRezkiAlouini_TWC}, by using superposition coding and truncated-Gaussian distributions. Namely, the transmitter sends $X=X_1+X_2$ where $X_i$ follows a truncated-Gaussian distribution. The peak constraint is split between $X_1$ and $X_2$ as $\mathcal{A}_1=\rho\mathcal{A}$ and $\mathcal{A}_2=(1-\rho)\mathcal{A}$, respectively. Then, we choose $X_i$ to be distributed according to the truncated-Gaussian distribution $\tilde{\mathbb{P}}_{\mu_i,\nu_i}^{\rm G}(x)$ as defined in \eqref{TGDist} with peak $\mathcal{A}_i$. The resulting mean $\tilde{\mu}_{i,\mathcal{A}_i}$ and variance $\tilde{\nu}^2_{i,\mathcal{A}_i}$ are as defined in \eqref{MuTilde} and \eqref{NuTilde}, respectively (where we indicate the dependence on $\mathcal{A}_i$ explicitly for clarity). Receiver 2 (the weaker receiver) decodes $X_2$, while receiver 1 (the stronger receiver) decodes both $X_2$ and $X_1$. 

Define $\boldsymbol{q}=(\mu_1,\nu_1,\mu_2,\nu_2)$, and $\mathcal{Q}$ as the set of $\boldsymbol{q}$ so that $\tilde{\mu}_{1,\mathcal{A}_1}+\tilde{\mu}_{2,\mathcal{A}_2}\leq\mathcal{E}$. Then, we can write the inner bound as follows. 

\begin{theorem}[\cite{ChaabanRezkiAlouini_TWC}]
\label{Thm:BCInBnd}
The capacity region of the 2-user IM/DD Gaussian BC satisfies $\mathcal{C}^{\rm bc}_{g_1,g_2}(\mathcal{A},\mathcal{E})\supseteq \mathcal{R}^{\rm bc}_{g_1,g_2}(\mathcal{A},\mathcal{E})$ where
$$\mathcal{R}^{\rm bc}_{g_1,g_2}(\mathcal{A},\mathcal{E})\triangleq {\rm co}\left(\bigcup_{\rho\in[0,1]}\bigcup_{\boldsymbol{q}\in\mathcal{Q}}\hat{\mathcal{R}}_{g_1,g_2}^{\rm bc}(\mathcal{A},\mathcal{E},\rho,\boldsymbol{q})\right),$$ where ${\rm co}(\cdot)$ denotes the convex hull, and $\hat{\mathcal{R}}_{g_1,g_2}^{\rm bc}(\mathcal{A},\mathcal{E},\rho,\boldsymbol{q})$ is the set of rate pairs $(r_1,r_2)$ satisfying
\begin{align*}
0&\leq r_1 \leq \frac{1}{2}\log\left(\frac{\nu_1^2}{\tilde{\nu}_{1,\mathcal{A}_1}^2}+g_1^2\nu_1^2\right)-\phi(\mathcal{A}_1,\mu_1,\nu_1),\\
0&\leq r_2 \leq \frac{1}{2}\log\left(\frac{\nu_2^2}{\tilde{\nu}_{2,\mathcal{A}_2}^2}+\frac{g_2^2\nu_2^2}{g_2^2\tilde{\nu}_{1,\mathcal{A}_1}^2+1}\right)-\phi(\mathcal{A}_2,\mu_2,\nu_2),
\end{align*}
where $\mathcal{A}_1=\rho\mathcal{A}$, $\mathcal{A}_2=(1-\rho)\mathcal{A}$, $\phi(\mathcal{A}_i,\mu_i,\nu_i)=\log(\eta_{i,\mathcal{A}_i})+\frac{1}{2}\left((\mathcal{A}_i-\mu_i)\tilde{\mathbb{P}}^{\G}_{\mu_i,\nu_i}(\mathcal{A}_i)+\mu_i\tilde{\mathbb{P}}^{\rm G}_{\mu_i,\nu_i}(0)\right)$, and $\eta_{i,\mathcal{A}_i}$ is as defined in \eqref{EtaTG} with $\mu$, $\nu$, and $\mathcal{A}$ replaced with $\mu_i$, $\nu_i$, and $\mathcal{A}_i$, respectively. 
\end{theorem}

Using these theorems, \cite{ChaabanRezkiAlouini_TWC} shows that outer bound $\overline{\mathcal{R}}^{\rm bc,lmw}_{g_1,g_2}(\mathcal{A},\mathcal{E})$ is nearly tight at high SNR, where it nearly meets the inner bound $\mathcal{R}^{\rm bc}_{g_1,g_2}(\mathcal{A},\mathcal{E})$. This asymptotic capacity region at high SNR can be approximated as given in the following theorem.

\begin{theorem}[\cite{ChaabanRezkiAlouini_TWC}]
\label{Thm:CapBCHighSNR}
The capacity region of the 2-user IM/DD Gaussian BC with a peak constraint $\mathcal{A}$ and an average constraint $\mathcal{E}=\alpha\mathcal{A}$ with $\alpha\leq\frac{1}{2}$ is within a gap $\delta$ of the region $\bigcup_{\rho\in[0,1]} {\mathcal{R}}^{\rm bc,h}_{g_1,g_2}(\mathcal{A},\mathcal{E},\rho)$ asymptotically at high SNR, where ${\mathcal{R}}^{\rm bc,h}_{g_1,g_2}(\mathcal{A},\mathcal{E},\rho)$ is the set of rate pairs $(r_1,r_2)$ satisfying
\begin{align}
0&\leq r_1 \leq \frac{1}{2}\log\left(1+c\rho^2g_1^2\mathcal{A}^2\right),\\
0&\leq r_2 \leq \frac{1}{2}\log\left(1+\frac{c(1-\rho)^2g_2^2\mathcal{A}^2}{c\rho^2g_2^2\mathcal{A}^2+1}\right),
\end{align}
where $c=\min\left\{\frac{1}{2\pi e},\frac{e\alpha^2}{2\pi}\right\}$, and $\delta=\log\left(\frac{3\sqrt{c}}{\alpha}\right)\leq 0.68$ nats/transmission.
\end{theorem}

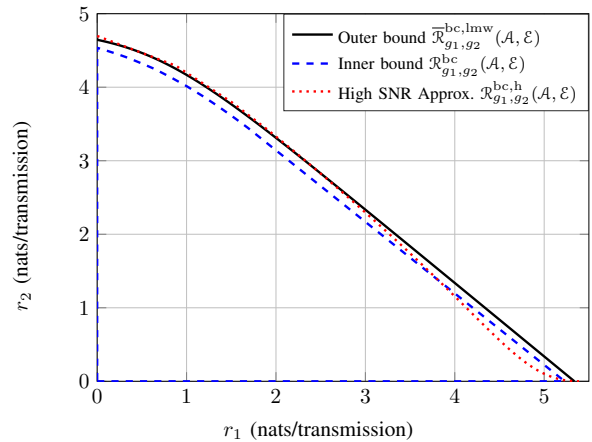
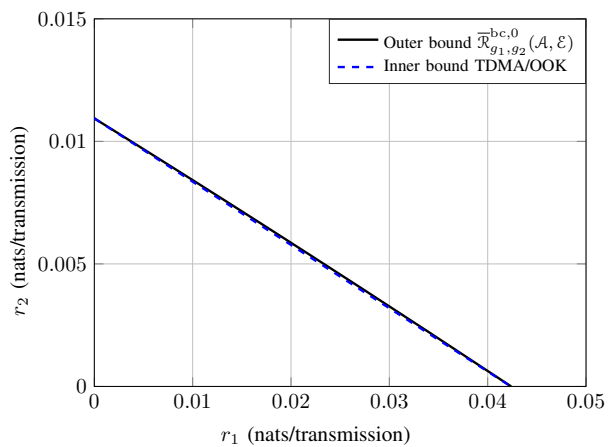
\begin{figure}[t]
%\centering
\begin{subfigure}[]{.46\textwidth}
\centering
\tikzset{every picture/.style={scale=.9}, every node/.style={scale=.9}}
%\input{matlab/Rate_reg_BC_g1_1_g2_p5}
% Minimal pgfplots version: 1.3
%  http://www.mathworks.com/matlabcentral/fileexchange/22022-matlab2tikz
\definecolor{mycolor1}{rgb}{0.92900,0.69400,0.12500}%
\begin{tikzpicture}

\begin{axis}[%
width=3.8in,
height=3in,
xmin=0,
xmax=5.5,
xlabel={$r_1$ (nats/transmission)},
xmajorgrids,
ymin=0,
ymax=5,
ylabel={$r_2$ (nats/transmission)},
ymajorgrids,
legend style={at = {(axis cs: 5.5,5)}, anchor = north east, legend cell align=left,align=left,draw=white!15!black}
]
\addplot [color=black,solid,line width=1.0pt]
  table[row sep=crcr]{%
0	4.64489391734489\\
8.01269021476076e-07	4.64489371702751\\
8.98765116571678e-07	4.64489369265346\\
1.00784840327334e-06	4.6448936653826\\
1.13041587226708e-06	4.64489363474068\\
1.26798418724292e-06	4.64489360034854\\
1.42323013542709e-06	4.64489356153697\\
1.59552685533525e-06	4.6448935184627\\
1.79031280577289e-06	4.64489346976608\\
2.00804413105784e-06	4.6448934153331\\
2.25220493938693e-06	4.6448933542927\\
2.52641762187065e-06	4.64489328573928\\
2.83482390983291e-06	4.6448932086374\\
3.18004367781988e-06	4.64489312233207\\
3.56771899518119e-06	4.64489302541275\\
4.00240660469293e-06	4.64489291674023\\
4.49061222339516e-06	4.64489279468805\\
5.03814035987758e-06	4.64489265780504\\
5.6522041298706e-06	4.64489250428786\\
6.3416976411073e-06	4.64489233191294\\
7.11501972470546e-06	4.64489213858046\\
7.98272570434674e-06	4.64489192165151\\
8.95631285428403e-06	4.64489167825163\\
1.0048716165021e-05	4.64489140514691\\
1.12744864646366e-05	4.64489109869944\\
1.26497269867261e-05	4.64489075488314\\
1.41928391269026e-05	4.64489036909734\\
1.59243248348863e-05	4.64488993621613\\
1.78670390184904e-05	4.64488945052528\\
2.00468335254787e-05	4.64488890556115\\
2.24926781955697e-05	4.64488829408048\\
2.5236945784197e-05	4.64488760798902\\
2.83161739984408e-05	4.64488683815105\\
3.1771207076096e-05	4.64488597435385\\
3.56478205749017e-05	4.64488500515147\\
3.99976371489595e-05	4.64488391763563\\
4.48783465127423e-05	4.64488269738062\\
5.03547570255117e-05	4.6448813281802\\
5.64996141811738e-05	4.64487979184279\\
6.33945504655522e-05	4.64487806795372\\
7.11311518981885e-05	4.64487613360821\\
7.9812217722903e-05	4.64487396309605\\
8.95531018532624e-05	4.64487152756567\\
0.000100483266346265	4.64486879463507\\
0.000112748010424669	4.64486572795867\\
0.000126510401636591	4.64486228674344\\
0.000141953497467872	4.64485842519207\\
0.000159282718103721	4.64485409190805\\
0.000178728676991799	4.64484922918579\\
0.000200550238175444	4.64484377224351\\
0.000225038005585952	4.64483764834738\\
0.000252505464778229	4.64483077902287\\
0.000283312155261255	4.64482307425482\\
0.00031788055599495	4.64481442825737\\
0.000356670358683517	4.6448047258998\\
0.000400197548673115	4.64479383792407\\
0.00044904127767377	4.64478161921267\\
0.000503851664260635	4.64476790682086\\
0.000565358514725461	4.64475251777424\\
0.000634381135974289	4.64473524658755\\
0.000711839373844828	4.64471586246977\\
0.00079876603171501	4.64469410617501\\
0.000896320833486784	4.64466968645569\\
0.00100580614627915	4.6446422760611\\
0.00112860901871142	4.64461152617069\\
0.00126630890485908	4.64457703932836\\
0.00142083545255207	4.6445383297825\\
0.00159424817107965	4.64449487849381\\
0.00178885948762008	4.64444610211186\\
0.00200726601719742	4.64439134487583\\
0.0022523837504126	4.6443298694622\\
0.00252748765896139	4.64426084663499\\
0.00283625630072534	4.64418334352864\\
0.00318227272580958	4.64409644836389\\
0.00357042671765424	4.64399891758197\\
0.00400602093662827	4.64388939905647\\
0.00449486352175726	4.64376640757969\\
0.00504347608424725	4.64362827093879\\
0.00565918188346725	4.64347310561577\\
0.00634956109771906	4.64329895165999\\
0.00712310273507225	4.6431036056069\\
0.00799106166378086	4.64288414686848\\
0.00896496334755123	4.64263756204545\\
0.0100577395278121	4.64236045180728\\
0.0112828646298323	4.64204924222655\\
0.0126543992025301	4.64170016600875\\
0.0141928005911978	4.6413077696668\\
0.0159182942923766	4.64086658117385\\
0.0178529624758985	4.6403705603573\\
0.0200156202648964	4.63981439521281\\
0.0224399395986893	4.63918881249556\\
0.0251572561970289	4.63848494506243\\
0.0281959993579489	4.63769445651312\\
0.0315940194926847	4.6368062905768\\
0.0353993272690245	4.63580636764252\\
0.0396503773178202	4.63468266926297\\
0.0443967246315351	4.63341972447893\\
0.0497052520035478	4.63199674190619\\
0.0556158865987395	4.63039931588027\\
0.0622133720078983	4.62859993687518\\
0.0695593806002355	4.62657602553234\\
0.0777322662578726	4.62429889369068\\
0.0868185292319437	4.62173563467286\\
0.0969026858221559	4.61885155079033\\
0.108077335641437	4.61560685016752\\
0.120459743478776	4.61195105779843\\
0.134110847229042	4.60784635051565\\
0.149176492105484	4.60322481250068\\
0.16575972909498	4.59802543986689\\
0.183944617152775	4.5921869342284\\
0.203833973161124	4.58563499016528\\
0.22553458242429	4.57828560415657\\
0.249127499737544	4.57005427261289\\
0.274680834828015	4.56085159434856\\
0.302210249793854	4.55059806064463\\
0.331733518016748	4.53920542221505\\
0.363276498663656	4.5265738750572\\
0.396726783353465	4.51265337171538\\
0.432032476487575	4.49736777130922\\
0.469025933933072	4.48069196818101\\
0.507645107211274	4.46255809680948\\
0.547547738568564	4.44303979317624\\
0.588727784152773	4.42206212754827\\
0.630946094690445	4.39967696542718\\
0.674013464471285	4.37592976028776\\
0.717919689010088	4.35078038489202\\
0.762424498123756	4.32433225648814\\
0.806166490041603	4.29741512744455\\
0.850035978934697	4.26951963416883\\
0.894573264898102	4.24029940963437\\
0.940040404790119	4.20955933649043\\
0.986526151595795	4.17721026147045\\
1.034139155724	4.14314662812352\\
1.08300125519786	4.10725011624361\\
1.13324112312176	4.0693939111861\\
1.18498978652876	4.02944633916046\\
1.23837872413292	3.98727301725312\\
1.29354031179613	3.94273734163751\\
1.35060989983526	3.89569959714951\\
1.40972891347604	3.84601500531272\\
1.47104878808581	3.79353077011135\\
1.53465289913126	3.73815477131698\\
1.60072659631391	3.67970951083366\\
1.66957010892794	3.6179141913576\\
1.7396897779753	3.55412633528359\\
1.80919772911706	3.49014203831143\\
1.88067834964867	3.42364721347181\\
1.95416610797954	3.35463589750704\\
2.02966952697716	3.28312898338859\\
2.10718365910048	3.20916248539872\\
2.18669582137124	3.13278158475394\\
2.26818523786138	3.05404032499112\\
2.35161294873958	2.973010819156\\
2.43694613637358	2.88975910505958\\
2.52414602926511	2.80435620559574\\
2.61316140243441	2.71688409597067\\
2.70393434996867	2.62742975807146\\
2.79640114375334	2.53608407277812\\
2.8904931751725	2.44294070896584\\
2.98613794026351	2.34809504393664\\
3.08326003260338	2.25164314342224\\
3.18178211307591	2.15368082244534\\
3.28162583115113	2.05430280174533\\
3.38271267798668	1.95360196852689\\
3.4849647571931	1.8516687451872\\
3.5883054642305	1.74859056550481\\
3.69266006994539	1.64445145452458\\
3.79795620761089	1.53933170597901\\
3.90412426596455	1.4333076494431\\
4.01109769315771	1.32645149840684\\
4.11881321828901	1.21883126994758\\
4.22721099836298	1.11051076658211\\
4.33623469917779	1.00154961107953\\
4.44583151889035	0.892003325437582\\
4.55595216291663	0.781923445803002\\
4.66655077847942	0.671357665794469\\
4.77758485658291	0.560350001425459\\
4.88901510853501	0.448940971588215\\
5.00080532340448	0.337167788824566\\
5.11291855636702	0.225068211121537\\
5.2253281546088	0.112669551258183\\
5.33800646070974	0\\
};
\addlegendentry{\footnotesize Outer bound $\overline{\mathcal{R}}_{g_1,g_2}^{\rm bc,lmw}(\mathcal{A},\mathcal{E})$};

\addplot [color=blue,dashed,line width=1.0pt]
  table[row sep=crcr]{%
0	0\\
5.22934128555411	0\\
1.91226105432256	3.22681862152976\\
1.68824406100644	3.44462163200366\\
1.54401433347992	3.57523837755178\\
1.36977234030916	3.72582332034134\\
1.18686138760892	3.87448456172064\\
1.00336060399891	4.01219013802402\\
0.823491995774965	4.13451623104742\\
0.641993815936819	4.24841529403014\\
0.488791172427624	4.32786862817209\\
0.342756898106002	4.40035266172644\\
0.227263276434908	4.4500259236086\\
0.0819678260533888	4.50670552702426\\
0.0460359126369007	4.51960395083787\\
0	4.5361230435337\\
0	4.53609714360841\\
0	4.53606453447569\\
0	4.53602347716219\\
0	4.53597178138365\\
0	4.53590668807686\\
0	4.53582489616809\\
0	4.53572150032954\\
0	4.53559150451844\\
0	4.53542846936262\\
0	4.53522152459003\\
0	0\\
};
\addlegendentry{\footnotesize Inner bound $\mathcal{R}_{g_1,g_2}^{\rm bc}(\mathcal{A},\mathcal{E})$};

\addplot[color = red, dotted, line width=1pt]
table[row sep=crcr]{
     0    4.6971\\
    0.8795    4.2925\\
    1.5035    3.7976\\
    1.8951    3.4324\\
    2.1778    3.1536\\
    2.3986    2.9295\\
    2.5797    2.7422\\
    2.7331    2.5811\\
    2.8661    2.4397\\
    2.9835    2.3134\\
    3.0887    2.1993\\
    3.1838    2.0949\\
    3.2707    1.9988\\
    3.3506    1.9096\\
    3.4246    1.8263\\
    3.4935    1.7481\\
    3.5580    1.6745\\
    3.6186    1.6048\\
    3.6757    1.5387\\
    3.7298    1.4757\\
    3.7810    1.4156\\
    3.8298    1.3582\\
    3.8763    1.3031\\
    3.9207    1.2503\\
    3.9633    1.1996\\
    4.0041    1.1507\\
    4.0433    1.1036\\
    4.0810    1.0582\\
    4.1174    1.0144\\
    4.1525    0.9721\\
    4.1863    0.9312\\
    4.2191    0.8917\\
    4.2509    0.8535\\
    4.2816    0.8165\\
    4.3115    0.7807\\
    4.3405    0.7461\\
    4.3686    0.7126\\
    4.3960    0.6802\\
    4.4227    0.6488\\
    4.4487    0.6185\\
    4.4740    0.5891\\
    4.4987    0.5608\\
    4.5228    0.5334\\
    4.5463    0.5070\\
    4.5693    0.4814\\
    4.5917    0.4568\\
    4.6137    0.4330\\
    4.6352    0.4101\\
    4.6563    0.3881\\
    4.6769    0.3669\\
    4.6971    0.3465\\
    4.7169    0.3269\\
    4.7363    0.3081\\
    4.7554    0.2900\\
    4.7741    0.2727\\
    4.7924    0.2562\\
    4.8104    0.2403\\
    4.8281    0.2252\\
    4.8455    0.2107\\
    4.8626    0.1970\\
    4.8794    0.1838\\
    4.8959    0.1713\\
    4.9122    0.1594\\
    4.9282    0.1481\\
    4.9439    0.1374\\
    4.9594    0.1273\\
    4.9747    0.1177\\
    4.9898    0.1086\\
    5.0046    0.1000\\
    5.0192    0.0919\\
    5.0336    0.0843\\
    5.0477    0.0771\\
    5.0617    0.0704\\
    5.0755    0.0641\\
    5.0891    0.0582\\
    5.1025    0.0527\\
    5.1158    0.0475\\
    5.1289    0.0427\\
    5.1418    0.0383\\
    5.1545    0.0341\\
    5.1671    0.0303\\
    5.1795    0.0268\\
    5.1918    0.0235\\
    5.2039    0.0205\\
    5.2159    0.0178\\
    5.2277    0.0153\\
    5.2394    0.0131\\
    5.2510    0.0110\\
    5.2624    0.0092\\
    5.2737    0.0076\\
    5.2849    0.0061\\
    5.2959    0.0049\\
    5.3068    0.0038\\
    5.3176    0.0028\\
    5.3283    0.0020\\
    5.3389    0.0014\\
    5.3494    0.0009\\
    5.3598    0.0005\\
    5.3700    0.0002\\
    5.3802    0.0001\\
    5.3902         0\\
};
\addlegendentry{\footnotesize High SNR Approx. $\mathcal{R}_{g_1,g_2}^{\rm bc,h}(\mathcal{A},\mathcal{E})$};

\end{axis}
\end{tikzpicture}%
\caption{$10\log_{10}(\mathcal{A})=30$, $\alpha=1/3$.}
\label{Fig:BCHigh}
\end{subfigure}
\begin{subfigure}[]{.46\textwidth}
\vspace{.4cm}
\centering
\tikzset{every picture/.style={scale=.9}, every node/.style={scale=.9}}
%\input{matlab/Rate_reg_BC_g1_1_g2_p5_A_m2dB}
%\vspace{-.4cm}
% Minimal pgfplots version: 1.3
%  http://www.mathworks.com/matlabcentral/fileexchange/22022-matlab2tikz
\pgfplotsset{scaled y ticks=false}
\pgfplotsset{scaled x ticks=false}
\begin{tikzpicture}

\begin{axis}[%
width=3.8in,
height=3in,
xmin=0,
xmax=.05,
xlabel={$r_1$ (nats/transmission)},
xmajorgrids,
xtick={0,.01,...,.05},
xticklabels={$0$, $0.01$, $0.02$, $0.03$, $0.04$, $0.05$},
ymin=0,
ymax=.015,
ylabel={$r_2$ (nats/transmission)},
ymajorgrids,
ytick={0,.005,...,.015},
yticklabels={$0$, $0.005$, $0.01$, $0.015$},
legend style={at = {(axis cs: .05,.015)}, anchor = north east, legend cell align=left,align=left,draw=white!15!black}
]
\addplot [color=black,solid,line width=1.0pt]
  table[row sep=crcr]{%
0	0.0109380151308016\\
0	0.0109380151308016\\
0	0.0109380151308016\\
0	0.0109380151308016\\
0	0.0109380151308016\\
0	0.0109380151308016\\
0	0.0109380151308016\\
0	0.0109380151308016\\
0	0.0109380151308016\\
0	0.0109380151308016\\
0	0.0109380151308016\\
0	0.0109380151308016\\
0	0.0109380151308016\\
0	0.0109380151308016\\
0	0.0109380151308016\\
0	0.0109380151308016\\
0	0.0109380151308016\\
0	0.0109380151308016\\
0	0.0109380151308016\\
0	0.0109380151308016\\
0	0.0109380151308016\\
0	0.0109380151308016\\
0	0.0109380151308016\\
0	0.0109380151308016\\
0	0.0109380151308016\\
0	0.0109380151308016\\
0	0.0109380151308016\\
0	0.0109380151308016\\
0	0.0109380151308016\\
0	0.0109380151308016\\
0	0.0109380151308016\\
0	0.0109380151308016\\
0	0.0109380151308016\\
0	0.0109380151308016\\
0	0.0109380151308016\\
0	0.0109380151308016\\
0	0.0109380151308016\\
0	0.0109380151308016\\
0	0.0109380151308016\\
4.44089209850062e-16	0.0109380151308015\\
4.44089209850062e-16	0.0109380151308015\\
4.44089209850062e-16	0.0109380151308015\\
4.44089209850062e-16	0.0109380151308015\\
8.88178419700124e-16	0.0109380151308014\\
8.88178419700124e-16	0.0109380151308014\\
1.33226762955019e-15	0.0109380151308013\\
1.33226762955019e-15	0.0109380151308013\\
1.77635683940025e-15	0.0109380151308011\\
2.22044604925031e-15	0.010938015130801\\
2.66453525910037e-15	0.0109380151308009\\
3.55271367880049e-15	0.0109380151308007\\
4.44089209850061e-15	0.0109380151308005\\
5.77315972805078e-15	0.0109380151308001\\
7.10542735760095e-15	0.0109380151307998\\
8.88178419700117e-15	0.0109380151307994\\
1.11022302462514e-14	0.0109380151307988\\
1.37667655053518e-14	0.0109380151307981\\
1.77635683940022e-14	0.0109380151307971\\
2.22044604925026e-14	0.010938015130796\\
2.79776202205532e-14	0.0109380151307946\\
3.50830475781537e-14	0.0109380151307928\\
4.44089209850043e-14	0.0109380151307905\\
5.55111512312547e-14	0.0109380151307877\\
7.0166095156305e-14	0.010938015130784\\
8.83737527601546e-14	0.0109380151307795\\
1.11022302462503e-13	0.0109380151307738\\
1.3988810110275e-13	0.0109380151307666\\
1.76303416310444e-13	0.0109380151307575\\
2.21600515715132e-13	0.0109380151307462\\
2.78888023785762e-13	0.0109380151307319\\
3.51274564991276e-13	0.0109380151307138\\
4.42312853010467e-13	0.010938015130691\\
5.56887869151668e-13	0.0109380151306624\\
7.01216862352757e-13	0.0109380151306263\\
8.82405259971296e-13	0.010938015130581\\
1.11111120304362e-12	0.0109380151305238\\
1.39888101102574e-12	0.0109380151304519\\
1.7608137170524e-12	0.0109380151303614\\
2.2168933355666e-12	0.0109380151302474\\
2.79110068389985e-12	0.0109380151301038\\
3.51363382832135e-12	0.0109380151299232\\
4.42357261929691e-12	0.0109380151296957\\
5.56887869148877e-12	0.0109380151294094\\
7.01083635585379e-12	0.0109380151290489\\
8.82582895648225e-12	0.0109380151285951\\
1.11111120303251e-11	0.0109380151280238\\
1.39879219316616e-11	0.0109380151273046\\
1.76099135270843e-11	0.0109380151263991\\
2.21693774444335e-11	0.0109380151252592\\
2.79096745706679e-11	0.0109380151238242\\
3.51363382821024e-11	0.0109380151220175\\
4.42339498343687e-11	0.0109380151197431\\
5.56874546444672e-11	0.0109380151168797\\
7.0106587197275e-11	0.0109380151132749\\
8.82587336470217e-11	0.0109380151087369\\
1.1111112029214e-10	0.0109380151030238\\
1.39880551566636e-10	0.0109380150958314\\
1.76099135242933e-10	0.0109380150867768\\
2.2169599484615e-10	0.0109380150753776\\
2.79098521993412e-10	0.010938015061027\\
3.51364270888332e-10	0.0109380150429605\\
4.42341274524427e-10	0.0109380150202163\\
5.56874546165574e-10	0.010938014991583\\
7.0106365108436e-10	0.0109380149555357\\
8.82586891679943e-10	0.0109380149101549\\
1.11111120181029e-09	0.0109380148530238\\
1.39880595799457e-09	0.0109380147811001\\
1.76099223781676e-09	0.0109380146905535\\
2.21695816768125e-09	0.010938014576562\\
2.79098476883428e-09	0.0109380144330554\\
3.51364180959379e-09	0.0109380142523911\\
4.42341317172355e-09	0.0109380140249483\\
5.56874721010272e-09	0.0109380137386148\\
7.01063691069869e-09	0.0109380133781424\\
8.82586929078227e-09	0.0109380129243343\\
1.11111110188134e-08	0.0109380123530238\\
1.39880598479357e-08	0.0109380116337866\\
1.7609923875426e-08	0.0109380107283206\\
2.2169580790382e-08	0.0109380095884063\\
2.79098483195467e-08	0.0109380081533394\\
3.51364174289161e-08	0.010938006346697\\
4.42341281798864e-08	0.0109380040722692\\
5.56874670895965e-08	0.0109380012089342\\
7.01063669040203e-08	0.0109379976042089\\
8.82586850090081e-08	0.0109379930661289\\
1.11111098632935e-07	0.0109379873530246\\
1.39880581757615e-07	0.0109379801606525\\
1.76099212620775e-07	0.0109379711059926\\
2.21695763669704e-07	0.0109379597068515\\
2.79098414421392e-07	0.0109379453561834\\
3.51364060957661e-07	0.0109379272897632\\
4.42341104811556e-07	0.0109379045454887\\
5.56874394018147e-07	0.0109378759121449\\
7.01063224478912e-07	0.0109378398649033\\
8.82586148583183e-07	0.0109377944841184\\
1.11110987655232e-06	0.010937737353101\\
1.39880405658732e-06	0.0109376654294206\\
1.76098933478594e-06	0.0109375748828864\\
2.21695321285433e-06	0.0109374608915768\\
2.79097713449363e-06	0.0109373173850574\\
3.51362949896685e-06	0.010937136721112\\
4.42339343963772e-06	0.010936909278773\\
5.568716029672e-06	0.0109366229459797\\
7.01058801244614e-06	0.0109362624745831\\
8.82579137947953e-06	0.0109358086683514\\
1.11109876562227e-05	0.0109352373607398\\
1.39878644689101e-05	0.0109345181279979\\
1.76096142583517e-05	0.0109336126690932\\
2.21690898016251e-05	0.0109324727662001\\
2.79090703081403e-05	0.0109310377171764\\
3.51351839362578e-05	0.0109292311033496\\
4.42321735186249e-05	0.0109269567205756\\
5.56843695388423e-05	0.0109240934570156\\
7.01014571649019e-05	0.0109204888450737\\
8.82509040667891e-05	0.0109159509444503\\
0.000111098767260623	0.0109102381246002\\
0.000139861038376352	0.0109030462033284\\
0.000176068239942659	0.0108939922579701\\
0.000221646678264263	0.0108825942491956\\
0.000279020625397107	0.0108682453757492\\
0.000351240785480379	0.0108501817998316\\
0.000442145750113624	0.0108274420329591\\
0.000556564824658078	0.0107988158330368\\
0.00070057268469392	0.0107627799127486\\
0.000881808883241131	0.0107174170695959\\
0.00110987836915663	0.0106603144849515\\
0.00139685299651628	0.0105884358616762\\
0.00175789860386683	0.0104979607264863\\
0.00221205770444104	0.0103840825535103\\
0.00278322419376755	0.0102407552971729\\
0.00350135370021719	0.0100603753720566\\
0.00440396106463883	0.0098333829905164\\
0.00553796444646046	0.00954776296444505\\
0.00696194277118751	0.00918842049552662\\
0.00874887800576339	0.00873640203448016\\
0.0109894533593876	0.00816792494299391\\
0.013795968723069	0.00745317248909002\\
0.0173069091920797	0.00655480285552827\\
0.021692152524861	0.00542611282513576\\
0.0271587153143315	0.00400878955932404\\
0.033956801860043	0.00223017904112653\\
0.0423857209933315	0\\
};
\addlegendentry{\footnotesize Outer bound $\overline{\mathcal{R}}_{g_1,g_2}^{\rm bc,0}(\mathcal{A},\mathcal{E})$};

\addplot [color=blue,dashed,line width=1.0pt]
  table[row sep=crcr]{%
0	0.0109374737227834\\
%0	0.00721259491727011\\
%0	0.00594760687824536\\
%9.52363898854003e-09	0.00594347705421372\\
%1.50939520882076e-08	0.00594154695267157\\
%2.39223012510337e-08	0.00593911753160059\\
%3.79142888196071e-08	0.00593605975763589\\
%6.00900964542461e-08	0.00593221133552113\\
%9.52363807815715e-08	0.00592736818233774\\
%1.50939482690404e-07	0.00592127374500784\\
%2.39222936793126e-07	0.00591360563210674\\
%3.7914274830797e-07	0.00590395891177797\\
%6.00900628366929e-07	0.00589182528705567\\
%9.52362974926402e-07	0.00587656719950713\\
%1.50939274656814e-06	0.00585738573565964\\
%2.39222416342777e-06	0.00583328102944281\\
%3.79141444684095e-06	0.00580300369019371\\
%6.01196595217957e-06	0.00576499568278832\\
%9.52733053516397e-06	0.0057173191202966\\
%1.50985086362354e-05	0.00565757173261039\\
%2.39277903801849e-05	0.00558278857114125\\
%3.79279155748868e-05	0.00548912691826509\\
%6.01146163248245e-05	0.00537256326059232\\
%9.52631775543722e-05	0.00522755352162685\\
%0.00015099001729757	0.00504755314132144\\
%0.000239276388073417	0.00482537831454444\\
%0.000379162286169743	0.00455286911953312\\
%0.000600824813102863	0.00422048661893282\\
%0.000951887087176084	0.00381991007752314\\
%0.00150778826180464	0.0033436546796568\\
%0.00238756780918981	0.00278855465517847\\
%0.00377878092101458	0.00216071816826036\\
%0.00597568879067079	0.00148367033501495\\
%0.00943778652305638	0.000811571143347845\\
%0.0148758075014443	0.000251754429478002\\
%0.0233735612080865	0\\
%0.0282403163385614	0\\
0.0423604649556197	0\\
%4.4221626360752e-10	0.0109386377853466\\
%2.90491408705407e-10	0.0109386377861647\\
%2.39225528275711e-10	0.0109386377864409\\
%0	0.0109374737227834\\
};
\addlegendentry{\footnotesize Inner bound TDMA/OOK};

\end{axis}
\end{tikzpicture}%
\caption{$10\log_{10}(\mathcal{A})=-2$, $\alpha=1/3$.}
\label{Fig:BCLow}
\end{subfigure}
\caption{capacity bounds for an IM/DD Gaussian BC with $g_1=1$ and $g_2=0.5$.}
\label{Fig:BC}
\end{figure}

Note that this relation shows that receiver 2 decodes while treating $X_1$ as noise, and $\frac{c(1-\rho)^2g_2^2\mathcal{A}^2}{c\rho^2g_2^2\mathcal{A}^2+1}$ can be thought of as a signal-to-interference-and-noise ratio (SINR). Fig. \ref{Fig:BCHigh} shows the outer and inner bounds $\overline{\mathcal{R}}_{g_1,g_2}^{\rm bc,lmw}$ and $\mathcal{R}_{g_1,g_2}^{\rm bc,lmw}$ along with the asymptotic region ${\mathcal{R}}_{g_1,g_2}^{\rm bc,h}$, for a channel with $g_1=1$, and $g_2=0.5$ at high SNR. This figure demonstrates Theorem \ref{Thm:CapBCHighSNR}. Note that the convex-hull of ${\mathcal{R}}_{g_1,g_2}^{\rm bc,h}$ provides a better approximation which nearly meets the outer bound $\overline{\mathcal{R}}_{g_1,g_2}^{\rm bc,lmw}$. However, the expression of ${\mathcal{R}}_{g_1,g_2}^{\rm bc,h}$ is easier to work with. 

For the low SNR regime, \cite{ChaabanRezkiAlouini_TWC} shows that the outer bound $\overline{\mathcal{R}}^{\rm bc,0}_{g_1,g_2}(\mathcal{A},\mathcal{E})$ is tight, where it meets the achievable rate of TDMA combined with coded OOK. Namely, the low-SNR asymptotic capacity is given as follows.

\begin{theorem}[\cite{ChaabanRezkiAlouini_TWC}]
The capacity region of the 2-user IM/DD Gaussian BC with a peak constraint $\mathcal{A}$ and an average constraint $\mathcal{E}=\alpha\mathcal{A}$ with $\alpha\leq\frac{1}{2}$ is given by the set of $(r_1,r_2)$ satisfying $r_i\geq0$ and 
\begin{align}
\frac{r_1}{g_1^2}+\frac{r_2}{g_2^2}\leq \frac{\alpha(1-\alpha)\mathcal{A}^2}{2}.
\end{align}
\end{theorem}

This result is shown in Fig. \ref{Fig:BCLow}. Note that \cite{ChaabanRezkiAlouini_TWC} also extends the  IM/DD Gaussian channel capacity lower bound in \cite{FaridHranilovic_SelectedAreas} given in Theorem \ref{Thm:FaridHranilovicLB} to the IM/DD Gaussian BC. The resulting inner bound is expressed as in \eqref{BCCap} with $U=X_2\in[0,\mathcal{A}_1]$ and $X=X_1+U$ with $X_1\in[0,\mathcal{A}_2]$, where $\mathcal{A}_1+\mathcal{A}_2=\mathcal{A}$, and $X_1$ and $X_2$ follow optimized discrete distributions with uniform spacing as in Theorem \ref{Thm:FaridHranilovicLB}.

\subsection{IM/DD Gaussian MAC}

Next, we consider an IM/DD Gaussian multiple access channel (MAC). In a 2-user MAC, we have two transmitters and one receiver. The transmit signals $X_i$, $i\in\{1,2\}$, satisfy $X_i\in[0,\mathcal{A}_i]$ and $\mathbb{E}[X_i]\leq\mathcal{E}_i$. The received signal is $Y=g_1X_1+g_2X_2+Z$ where $Z\sim\mathcal{N}(0,1)$. The capacity region is the set of achievable rate tuples $(R_1,R_2)$, defined similar to the BC. We denote the capacity region by $\mathcal{C}^{\rm mac}_{g_1,g_2}(\boldsymbol{\mathcal{A}},\boldsymbol{\mathcal{E}})$ where $\boldsymbol{\mathcal{A}}=(\mathcal{A}_1,\mathcal{A}_2)$ and $\boldsymbol{\mathcal{E}}=(\mathcal{E}_1,\mathcal{E}_2)$. 

The capacity region of a discrete-memoryless MAC is known to be given by the closure of the convex-hull of the set of $(R_1,R_2)\in\mathbb{R}_+^2$ satisfying~\cite{ElgamalKim}
\begin{subequations}
\label{MACCapReg}
\begin{align}
R_1&\leq I(X_1;Y|X_2)\\
R_2&\leq I(X_2;Y|X_1)\\
R_1+R_2&\leq I(X_1,X_2;Y),
\end{align}
\end{subequations}
for some input distributions $\mathbb{P}_{X_i}(x_i)$ on $\mathcal{X}_i$. This region is achievable by jointly decoding the two messages at the receiver, or using successive decoding combined with time-sharing. This statement can be also generalized to the IM/DD Gaussian MAC with continuous alphabets. Again, the question is how to choose $\mathbb{P}_{X_i}(x_i)$, and how to represent this region in a simpler form. One way to realize this is to derive capacity region outer and inner bounds, that allow us to draw further insights into asymptotic capacity and approximations. 

The following outer bounds was derived in \cite{ChaabanIbraheemyNaffouriAlouini_TWC} based on the IM/DD Gaussian channel capacity upper bounds. 

\begin{theorem}[\cite{ChaabanIbraheemyNaffouriAlouini_TWC}]
\label{Thm:MACOutBnd}
The capacity region $\mathcal{C}^{\rm mac}_{g_1,g_2}(\boldsymbol{\mathcal{A}},\boldsymbol{\mathcal{E}})$ of the 2-user IM/DD Gaussian MAC satisfies $\mathcal{C}^{\rm mac}_{g_1,g_2}(\boldsymbol{\mathcal{A}},\boldsymbol{\mathcal{E}})\subseteq\overline{\mathcal{R}}^{\rm mac,\ell}_{g_1,g_2}(\boldsymbol{\mathcal{A}},\boldsymbol{\mathcal{E}})$, where $\ell\in\{\rm lmw,0\}$, and $\overline{\mathcal{R}}^{\rm mac,\ell}_{g_1,g_2}(\boldsymbol{\mathcal{A}},\boldsymbol{\mathcal{E}})$ is defined as the set of rate pairs $(r_1,r_2)$ satisfying
\begin{align*}
0\leq r_i&\leq \overline{r}_{g_i}^{\ell}(\mathcal{A}_i,\mathcal{E}_i),\ i\in\{1,2\}\\
0\leq r_1+r_2&\leq \overline{r}_{1}^{\ell}(g_1\mathcal{A}_1+g_2\mathcal{A}_2,g_1\mathcal{E}_1+g_2\mathcal{E}_2).
\end{align*}
\end{theorem}

Here, the bound on $R_1+R_2$ is obtained by treating $g_1X_1+g_2X_2$ as a transmit signal with a peak constraint $g_1\mathcal{A}_1+g_2\mathcal{A}_2$ and average constraint $g_1\mathcal{E}_1+g_2\mathcal{E}_2$, and then using the single user bounds in Sec. \ref{Sec:CapacityBounds}. Asymptotic capacity expressions from Sec. \ref{Sec:AsymptoticCapacity} can be used to approximate the outer bound $\overline{\mathcal{R}}^{\rm max, \ell}_{g_1,g_2}(\boldsymbol{\mathcal{A}},\boldsymbol{\mathcal{E}}),$ in closed-form at high and low SNR. A closed-form outer bound for the average-constrained case was given in \cite{ZhouZhang_MAC} using an upper bounds for the single-user channel from \cite[(20)]{HranilovicKschischang}.

Capacity inner bounds were derived in \cite{ChaabanIbraheemyNaffouriAlouini_TWC} under both average and peak constraints, and in \cite{ZhouZhang_MAC} under either an average or a peak constraint only. An inner bound was derived in \cite{ChaabanIbraheemyNaffouriAlouini_TWC} using a truncated-Gaussian \eqref{TGDist}, where $X_i\sim\tilde{\mathbb{P}}^{\rm G}_{\mu_i,\nu_i}(x)$ with peak $\mathcal{A}_i$, mean $\tilde{\mu}_{i,\mathcal{A}_i}$ as defined in \eqref{MuTilde} (again indicating explicit dependence on $\mathcal{A}_i$ for clarity) satisfying $\tilde{\mu}_{i,\mathcal{A}_i}\leq\mathcal{E}_i$ and with variance $\tilde{\nu}_{i,\mathcal{A}_i}^2$ defined in \eqref{NuTilde}. Define $\boldsymbol{q}=(\mu_1,\nu_1,\mu_2,\nu_2)$ as the parameters of the truncated-Gaussian distribution of users 1 and 2, and $\mathcal{Q}$ as the set of $\boldsymbol{q}$ so that $\tilde{\mu}_{i,\mathcal{A}_i}$. Then, the following theorem presents a capacity region inner bound achievable under a truncated-Gaussian input distribution.

\begin{theorem}[\cite{ChaabanIbraheemyNaffouriAlouini_TWC}]
\label{Thm:MACInBnd}
The capacity region of the 2-user IM/DD Gaussian MAC satisfies $$\mathcal{C}^{\rm mac}_{g_1,g_2}(\boldsymbol{\mathcal{A,E}})\supseteq\mathcal{R}^{\rm mac}_{g_1,g_2}(\boldsymbol{\mathcal{A,E}})\triangleq \text{co}\left(\bigcup_{\boldsymbol{q}\in\mathcal{Q}}\hat{\mathcal{R}}^{\rm mac}_{g_1,g_2}(\boldsymbol{\mathcal{A,E}},\boldsymbol{q})\right),$$ where $\hat{\mathcal{R}}^{\rm mac}_{g_1,g_2}(\boldsymbol{\mathcal{A,E}},\boldsymbol{q})$ is the set of rate pairs $(r_1,r_2)$ satisfying
\begin{align*}
0\leq r_i&\leq\frac{1}{2}\log\left(\frac{\nu_i^2}{\tilde{\nu}_i^2}+\frac{\nu_i^2}{\sigma^2}\right)-\phi(\mathcal{A}_i,\mu_i,\nu_i),\ i\in\{1,2\},\\
0\leq r_1+r_2&\leq \frac{1}{2}\log\left(\frac{\nu_1^2\nu_2^2}{\tilde{\nu}_1^2\tilde{\nu}_2^2}+
\frac{\nu_1^2\nu_2^2}{\sigma^2\tilde{\nu}_2^2}+\frac{\nu_1^2\nu_2^2}{\tilde{\nu}_1^2\sigma^2}\right)\nonumber\\
&\qquad\qquad -\phi(\mathcal{A}_1,\mu_1,\nu_1)-\phi(\mathcal{A}_2,\mu_2,\nu_2),
\end{align*}
with $\phi_i(\cdot)$ is as defined in Theorem \ref{Thm:BCInBnd}.
\end{theorem}

In addition to this achievable lower bound, \cite{ChaabanIbraheemyNaffouriAlouini_TWC} provides an achievable inner bound based on uniformly-spaced discrete input distributions \eqref{InputDistKMassPoints}, and \cite{ZhouZhang_MAC} provide an inner bound achievable using a combination of discrete and continuous distributions. For the purpose of this tutorial, we present two asymptotic capacity statements next. At high SNR, the bounds in Theorems \ref{Thm:MACOutBnd} and \ref{Thm:MACInBnd} lead to the following statement at high SNR.

\begin{theorem}[\cite{ChaabanIbraheemyNaffouriAlouini_TWC}]
\label{Thm:CapMACHighSNR}
The capacity region of the 2-user IM/DD Gaussian MAC with $\mathcal{E}_i/\mathcal{A}_i=\alpha_i\leq\frac{1}{2}$ is within a gap $\delta\leq \frac{1}{2}\log\left(\frac{9e}{2\pi}\right)$ of the region $\mathcal{R}^{\rm mac,h}_{g_1,g_2}(\boldsymbol{\mathcal{A},\mathcal{E}})$ asymptotically at high SNR ($\mathcal{A}_1,\mathcal{A}_2\to\infty$), where $\mathcal{R}^{\rm mac,h}_{g_1,g_2}(\boldsymbol{\mathcal{A},\mathcal{E}})$ is the set of rate pairs $(r_1,r_2)$ satisfying
\begin{align}
0\leq r_1 &\leq \frac{1}{2}\log\left(1+c_ig_i^2\mathcal{A}_i^2\right), \ i\in\{1,2\}\\
0\leq r_1+r_2 &\leq \frac{1}{2}\log\left(1+c_{12}(g_1\mathcal{A}_1+g_2\mathcal{A}_2)^2\right),
\end{align}
where $c_i=\min\left\{\frac{1}{2\pi e},\frac{e\alpha_i^2}{2\pi}\right\}$ and 
$c_{12}=\min\left\{\frac{1}{2\pi e},\frac{e\alpha_{12}^2}{2\pi}\right\}$ with $\alpha_{12}=\frac{g_1\mathcal{E}_1+g_2\mathcal{E}_2}{g_1\mathcal{A}_1+g_2\mathcal{A}_2}$.
\end{theorem}

On the other hand, at low SNR, the outer bound $\overline{\mathcal{R}}^{\rm mac,0}_{g_1,g_2}(\boldsymbol{\mathcal{A,E}})$ is tight, where it matches the rate region achieved using coded OOK at both users (each user sends $X_i=\mathcal{A}_i$ with probability $\mathcal{E}_i/\mathcal{A}_i$ and $X_i=0$ with probability $1-\mathcal{E}_i/\mathcal{A}_i$), and using successive-cancellation decoding (SCD) at the receiver. This leads to the following theorem.

\begin{theorem}[\cite{ChaabanIbraheemyNaffouriAlouini_TWC}]
\label{Thm:CapMACLowSNR}
The capacity region of the 2-user IM/DD Gaussian MAC with $\mathcal{E}_i/\mathcal{A}_i=\alpha_i\leq\frac{1}{2}$ coincides asymptotically at low SNR with the set of $(r_1,r_2)$ satisfying $r_i\geq0$ and $r_i\leq \frac{\alpha_i(1-\alpha_i)g_i^2\mathcal{A}_i^2}{2}$.
\end{theorem}

\begin{figure}[t]
%\centering
\begin{subfigure}[]{.46\textwidth}
\centering
\tikzset{every picture/.style={scale=.9}, every node/.style={scale=.9}}
%\input{matlab/Rate_reg_MAC_g1_1_g2_p5_A30_alpha_1b3}
% Minimal pgfplots version: 1.3
%  http://www.mathworks.com/matlabcentral/fileexchange/22022-matlab2tikz
\begin{tikzpicture}

\begin{axis}[%
width=3.8in,
height=3in,
xmin=0,
xmax=6,
xlabel={$r_1$ (nats/transmission)},
xmajorgrids,
ymin=0,
ymax=5,
ylabel={$r_2$ (nats/transmission)},
ymajorgrids,
legend style={at = {(axis cs: 6,5)}, anchor = north east, legend cell align=left,align=left,draw=white!15!black}]
\addplot [color=black,solid,line width=1.0pt]
  table[row sep=crcr]{%
5.40285352045133	0\\
5.40285352045133	0.40147713073547\\
1.08323636234071	4.72109428884609\\
0	4.72109428884609\\
};
\addlegendentry{\footnotesize Outer bound $\overline{\mathcal{R}}^{\rm mac,lmw}_{g_1,g_2}(\boldsymbol{\mathcal{A,E}})$};

\addplot [color=blue,dashed]
  table[row sep=crcr]{%
5.31548645601963	1.72220535485934e-05\\
5.31548645601963	2.21206626402903e-05\\
2.3755352931007	2.93997328358157\\
0.964671424882908	4.32723952568932\\
0.852786264676785	4.41254360397798\\
0.790121428931212	4.45308893689618\\
0.728281680642569	4.48811675174353\\
0.68227168552795	4.50775675110807\\
0.63480273248971	4.52589123421689\\
0.557980792383012	4.5489348509929\\
0.371266326948265	4.58499523530225\\
0.166154233733483	4.60613884970628\\
0.000353813086634958	4.62233946146103\\
0.000275481595848781	4.62233946146103\\
0	4.62233946146103\\
0	0\\
5.31548645601963	0\\
5.31548645601963	1.72220535485934e-05\\
};
\addlegendentry{\footnotesize Inner bound $\mathcal{R}^{\rm mac}_{g_1,g_2}(\boldsymbol{\mathcal{A,E}})$};

\addplot [color=red,dotted,line width=1.0pt]
  table[row sep=crcr]{%
5.39021485854731	0\\
5.39021485854731	0.405459329558235\\
1.09857530710252	4.69709888100302\\
0	4.69709888100302\\
};
\addlegendentry{\footnotesize High SNR Approx. $\mathcal{R}^{\rm mac,h}_{g_1,g_2}(\boldsymbol{\mathcal{A,E}})$};

\end{axis}
\end{tikzpicture}%
\caption{$10\log_{10}(\mathcal{A})=30$.}
\label{Fig:MACHigh}
\end{subfigure}
\begin{subfigure}[]{.46\textwidth}
\vspace{.4cm}
\centering
\tikzset{every picture/.style={scale=.9}, every node/.style={scale=.9}}
%\input{matlab/Rate_reg_MAC_g1_1_g2_p5_A_m2dB_alpha_1b3}
%\vspace{-.4cm}
% Minimal pgfplots version: 1.3
%  http://www.mathworks.com/matlabcentral/fileexchange/22022-matlab2tikz
\definecolor{mycolor1}{rgb}{1.00000,0.00000,1.00000}%
\pgfplotsset{scaled y ticks=false}
\pgfplotsset{scaled x ticks=false}
\begin{tikzpicture}

\begin{axis}[%
width=3.8in,
height=3in,
xmin=0,
xmax=.05,
xlabel={$r_1$ (nats/transmission)},
xmajorgrids,
xtick={0,.01,...,.05},
xticklabels={$0$, $0.01$, $0.02$, $0.03$, $0.04$, $0.05$},
ymin=0,
ymax=.015,
ylabel={$r_2$ (nats/transmission)},
ymajorgrids,
ytick={0,.005,...,.015},
yticklabels={$0$, $0.005$, $0.01$, $0.015$},
legend style={at = {(axis cs: .05,.015)}, anchor = north east, legend cell align=left,align=left,draw=white!15!black}
]
\addplot [color=black,solid,line width=1.0pt]
  table[row sep=crcr]{%
0.0423857209933315	0\\
0.0423857209933315	0.0109380151308016\\
0.0423857209933315	0.0109380151308016\\
0	0.0109380151308016\\
};
\addlegendentry{Outer bound $\overline{\mathcal{R}}_{g_1,g_2}^{\rm mac,0}(\boldsymbol{\mathcal{A,E}})$};

\addplot [color=blue,only marks,mark=o,mark options={solid}]
  table[row sep=crcr]{%
0	0\\
0.0423606175945861	0\\
0.0423606175945861	0.00487673245126485\\
0.0423606175945861	0.0054652748017352\\
0.0423606175945861	0.00662828337380383\\
0.0423606175945861	0.0100534589095413\\
0.0414764947720105	0.0109375817321169\\
0.0228762763251087	0.0109375817321169\\
0	0.0109375817321169\\
0	0\\
};
\addlegendentry{Inner bound OOK/SCD};

\end{axis}
\end{tikzpicture}%
\caption{$10\log_{10}(\mathcal{A})=-2$.}
\label{Fig:MACLow}
\end{subfigure}
\caption{capacity bounds for an IM/DD Gaussian MAC channel with $g_1=1$,  $g_2=0.5$, $\mathcal{A}_1=\mathcal{A}_2=\mathcal{A}$, $\mathcal{E}_1=\mathcal{E}_2=\alpha\mathcal{A}$, and $\alpha=1/3$.}
\label{Fig:MAC}
\end{figure}
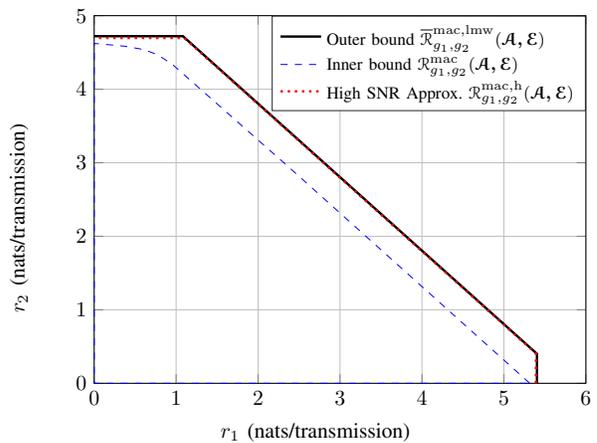
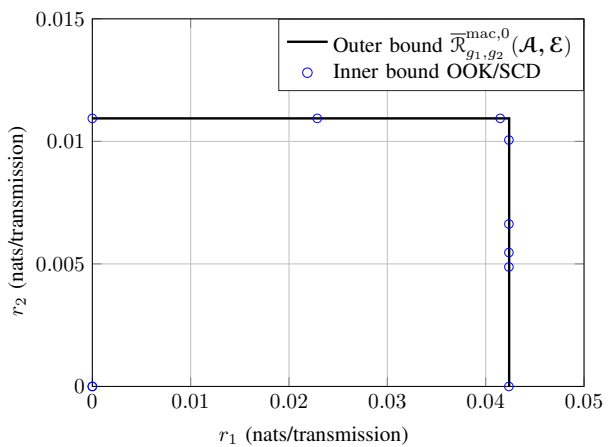

Figures \ref{Fig:MACHigh} and \ref{Fig:MACLow} show the capacity bounds for an exemplary channel with $g_1=1$, $g_2=0.5$, under $\mathcal{A}_1=\mathcal{A}_2=\mathcal{A}$, $\mathcal{E}_1=\mathcal{E}_2=\alpha\mathcal{A}$, and $\alpha=1/3$. The bounds in Fig. \ref{Fig:MACHigh} are within $<0.68$ nats which confirms Theorem \ref{Thm:MACInBnd}, and the bounds in Fig. \ref{Fig:MACLow} coincide which confirms the optimality of OOK/SCD at low SNR.

For moderate SNR, an inner bound based on discrete input distributions was given in \cite{ChaabanIbraheemyNaffouriAlouini_TWC}. Moreover, it was shown in \cite{SoltaniRezkiChaaban,MamandipoorMoshksarKhandani} that the IM/DD Gaussian MAC capacity-achieving distribution is discrete with a finite number of mass points.

\subsection{Other Multi-Terminal Channels}
Other multi-terminal IM/DD Gaussian channels were also studied in the literature. \ac{This includes the IM/DD Gaussian wiretap channel and the IM/DD Gaussian interference channel (IC). Next, we provide a brief overview of some works in this area.}

\subsubsection{The IM/DD Gaussian Wiretap Channel}
\ac{Due to the broadcast nature of OWC transmission (especially using LEDs), OWC remains susceptible to eavesdropping. This motivated several works that studied security in OWC \cite{YinHaas,ArafaPanayirciPoor,
QianChiZhaoObeedChaaban,
PanayirciYesilkayaCogalanPoorHaas,ChoChenCoon,
PhamPham_secureVLC}. The reader is referred to the following surveys for a more extensive overview of security aspects of OWC \cite{ZhangChaabanLampe,
ArfaouiSoltaniTavakkolnia,ObeedSalhabAlouiniZummo}.}

\ac{From an information theoretic perspective, \acc{there are several models that incorporate secure communications as an additional constraint to the classical reliability constraint, e.g., secure key agreement and the wiretap channel.} A wiretap channel consists of a transmitter, a receiver, and an eavesdropper which aims to intercept the communication and extract some of the transmitted information. Studying its capacity dates back to the 1970s \cite{Wyner_Wiretap}. Recently, several works studied the capacity of the IM/DD Gaussian wiretap channel as a model of OWC with an eavesdropper, and proposed transmission schemes and analysed their achievable rates.}

\ac{In an IM/DD Gaussian wiretap channel, the receiver and eavesdropper receive
\begin{align}
Y=gX+Z,\quad\text{and  } \quad Y_{{\rm e}}=g_{\rm e}X+Z_{{\rm e}},
\end{align}
respectively, where $X$ is the transmitted codeword symbol, $g$ and $g_{\rm e}$ are channel gains, and $Z$ and $Z_{{\rm e}}$ are Gaussian noises \acc{at the legitimate and the eavesdropper' receivers, respectively.} The transmitter wants to encode information into $X$ while ensuring reliability, i.e., ${\sf p}_{{\rm e},n}\to0$ as $n\to\infty$, and secrecy which requires the normalized mutual information between the message $W$ and $\Y_{\rm e}=(Y_{{\rm e},1},\ldots,Y_{{\rm e},n})$, i.e., $\frac{1}{n}I(W;\Y_{\rm e})$ to approach $0$ as $n\to\infty$. The secrecy capacity is defined as the largest rate $r$ under which these objectives can be achieved.}

\ac{The IM/DD Gaussian wiretap channel\acc{, being a scalar broadcast channel,} is stochastically degraded \cite{SoltaniRezki}. Hence, its secrecy capacity can be written as \cite{Wyner_Wiretap}
\begin{align}
\max_{\mathbb{P}_X} I(X;Y)-I(X;Y_{\rm e}),
\end{align}
where the maximization is with respect to all feasible input distributions. While this maximization is difficult to solve,} \cite{SoltaniRezki} proved that the optical input distribution is discrete. Moreover, to aid in the analysis of the secrecy capacity, capacity bounds for the IM/DD Gaussian wiretap channel were presented in \cite{ZhangChaabanLampe}. \ac{In general, imposing a security constraint reduces the capacity of the channel relative to the channel without a security constraint. However, as noted in \cite{SoltaniRezki}, there are cases where the security constraint does not impact capacity.} Transmission schemes and their achievable rates were also studied in \cite{thesis-Mostafa2017,WangLiuWangaWuLinCheng, ArfaouiZaidRezki,MostafaLampeTSP,MostafaLampeJSAC}.

\subsubsection{The IM/DD Gaussian Interference Channel}
In addition to the wirtetap channel, the capacity of the IM/DD Gaussian interference channel (IC) was also stydied in the literature. The IC consists of two transmitter-receiver pairs sharing the same transmission medium. \ac{The transmitters send $X_{1}$ and $X_{2}$, and the receivers receive
\begin{align}
Y_{j}=g_{jj}X_{j}+g_{kj}X_{k}+Z_{j}, \ \ j,k\in\{1,2\},\ j\neq k.
\end{align}
where $g_{jj}$ and $g_{kj}$ are channel gains, and $Z_j$ is Gaussian noise.}

Studying such a network is important for scenarios with multiple VLC cells \cite{AbdallaRahaimLittle}. \ac{However, the capacity of the IC remains an open problem to-date in general. Nonetheless, capacity inner and outer bounds can be derived to aid in studying the IC. To this end, the capacity of the IM/DD Gaussian IC was studied in \cite{ZhangChaaban}, which derived capacity bounds and studied the Han-Kobayashi transmission scheme applied in the IM/DD context \cite{HanKobayashi}.} Transmission schemes were also studied in \cite{MaLiHeYang}. \ac{Note that there is still significant room for improving capacity bounds for this channel, such as by using methods from \cite{ZhouZhang_MAC} \acc{which studies the IM/DD MAC, noting that MAC schemes are useful in an IC (cf. \cite{EtkinTseWang} for instance).}}

In addition to the aforementioned models, several other networks which combine RF and VLC links have been analysed in the literature \cite{ObeedSalhabZummoAlouini}. The reader is referred to \cite{ObeedSalhabAlouiniZummo} for a survey on related works.

\section{Summary}
\label{Sec:Summary}
We have discussed the capacity of IM/DD OWC systems modelled as Gaussian channels \ac{with real-valued, nonnegative, peak- and average-constrained inputs}. We started with a discussion on the channel model, which motivates the Gaussian channel assumption. Then, we discussed the capacity of the single user IM/DD  Gaussian channel in detail, by presenting the capacity and its numerical computation, capacity bounds, and asymptotic capacity expressions. Building on these results, we discussed the capacity of multi-aperture systems (SIMO, MISO, and MIMO) and also multi-user systems (broadcast and multiple access). It is important to note that this tutorial is by no means exhaustive. There has been a large amount of work on OWC in the past years, many of which are not covered here. However, we tried to cover some of the main advances that have been achieved in the information-theoretic direction of studying OWC systems.

This tutorial can be used for two purposes. First, it can be used as a reference that explains to the reader the main methods that are used in the literature for studying the capacity of the single-user IM/DD Gaussian channel. It also explains how the capacity of IM/DD Gaussian channel is different from the standard AWGN channel that is used to model RF systems. Second, it can be used as a guide for using these results to obtain capacity results for \ac{multi-terminal IM/DD Gaussian channels} by building on the results for single-user channels. 

The tutorial also shows that there is still room for additional contributions related to the capacity of IM/DD Gaussian channels. The capacity of the single-user channel is still unknown in closed-form {and sometimes not even computable}, and advances in this direction are important from an information-theoretic perspective. Moreover, existing results on multi-aperture and multi-user IM/DD Gaussian channels can be improved by considering tighter bounding techniques, and deriving simpler expressions that are amenable for further analysis of larger systems.

\end{document}